\documentclass{kapproc} 

\usepackage{amssymb,amsmath}
\usepackage{float}

\usepackage{procps} 

\usepackage[dvips]{graphicx}

\upperandlowercase

\kluwerbib 

\begin{document}


\booktitle[Soft X-ray Excess Emission from Clusters of Galaxies]
{Soft X-ray Excess Emission from Clusters of Galaxies\\ and Related Phenomena
}
\subtitle{}
\editor{Richard Lieu}
\edaffil{University of Alabama in Huntsville}

\editor{Jonathan Mittaz}
\edaffil{University of Alabama in Huntsville}


\halftitlepage


\titlepage


\tableofcontents




\begin{contributingauthors}
\begin{tabular}{lll}
W.I.Axford & U Ala Huntsville & ian@axford.org \\
M.Bonamente & NASA NSSTC Huntsville & bonamem@email.uah.edu \\
R.Cen & Princeton U. & cen@astro.princeton.edu \\
S.Chakrabarti & Boston U & supc@bu.edu \\
S.Cola-Francesco & INAF Roma & coa@coma.mporzio.astro.it \\
D.De Young & NOAO Tuscon & deyoung@noao.edu \\
K.Dolag & U Padova & kdolag@pd.astro.it \\
F.Durret & IAP (Paris France) & durret@iap.fr \\
K.Dyer & NRAO Socorro & kdyer@nrao.edu \\
T.Fang & Carnegie Mellon U & fangt@cmu.edu \\
A.Finoguenov & MPE Graching & alexis@xray.mpe.mpg.de \\
R.Fusco-Femiano & IASF-CNR & dario@rm.iasf.cnr.it \\
M.Henriksen & U Maryland & henrikse@umbc.edu \\
J.S.Kaastra & SRON Utrecht & j.s.kaastra@sron.nl \\
R.Lieu & U Ala Huntsville & lieur@cspar.uah.edu \\
F.J.Lockman & NRAO Green Bank & jlockman@nrao.edu \\
D.Lumb & ESA ESTEC & dlumb@rssd.esa.int \\
D.McCammon & W Wiscinsin Madison & mccammon@wisp.physics.wisc.edu \\
J.P.D.Mittaz & U Ala Huntsville & mittazj@email.uah.edu \\
D.Nagai & U Chicago & daisuke@addjob.uchicago.edu \\
J.Nevalainen & Harvard CfA/SAO & jukka@head-cfa.harvard.edu \\
F.Nicastro & Harvard CfA/SAO & fnicastro@cfa.harvard.edu \\
R.Petre & NASA/GSFC & robert.Petre-1@nassa.gov \\
L.A.Phillips & Caltech & phillips@tapir.caltech.edu \\
W.Sanders & U Wisconsin Madison & sanders@physics.wisc.edu \\
B.Savage & U Wisconsin Madison & savage@astro.wisc.edu \\
K.Sembach & STScI & sembach@stsci.edu \\
M.Shull & U Colorado & mshull@spitzer.colorado.edu \\
J.Slavin & Harvard CfA/SAO & jslavin@cfa.harvard.edu \\
S.Snowden & NASA/GSFC & snowden@riva.gsfc.nasa.gov \\
Y.Takahashi & U Ala Huntsville & yoshi@cosmic.uah.edu \\
T.Tripp & Princeton U Observatory & tripp@astro.princeton.edu\\
\end{tabular}
\end{contributingauthors}

%

\begin{preface}
Since the discovery in 1995 by the EUVE mission that some clusters of
galaxies  exhibit an EUV and soft X-ray emission component separate from the
radiation of the hot virialized intracluster gas, the phenomenon was
confirmed to be real by the ROSAT, BeppoSAX, and most recently the XMM-Newton
missions.  To date, approximately 30 \% of the total sample of
X-ray bright clusters located in directions of low Galactic HI absorption
are found to have this soft excess behavior.

The theoretical development has principally revolved around two scenarios.
First is the originally proposed thermal interpretation, which invokes large
quantities of a warm ($T \sim$ 10$^6$ K) gas - the amount is
significant compared with the total baryonic matter budget in clusters.
In current version of this model  the warm gas is believed to be located 
primarily at the outer parts of a cluster where the density of 
the hot virialized gas is low.  It may in fact correspond to 
that portion of the
the Warm Hot Intergalactic Medium (WHIM) found in the vicinity of 
galaxy clusters - the nodes of high over density 
where filaments of warm intergalactic matter
intersect.  

In the second proposed scenario the cluster soft excess is interpreted as
signature of a large population of cosmic rays, the electron component of which
undergoes inverse Compton scattering with the cosmic microwave background to
produce EUV and soft X-ray photons.  The motive underlying behind this approach
is to avoid enlisting too much extra baryonic matter, although it turns out
that the cosmic rays introduce a pressure which is directly proportional to the
measured EUV brightness of clusters, and which in many cases can be large
enough to reach equipartition levels between relativistic particles and the
thermal hot gas.  Our present understanding of the role played by cosmic rays
in the soft excess phenomenon is that at the centers of some clusters where the
excess is extremely bright the cause is likely to be non-thermal, because the
amount of thermal warm gas required to account for this excess is unacceptably
large, and in any case the necessary emission lines that bear witness to their
existence are absent.  It is entirely plausible that the pressure of this
central cosmic ray component, which primarily resides with the protons, is
responsible for the discontinuation of the cooling flow, while the secondary
electrons are the cause of the soft excess emission as they scatter off the
microwave background.

This is the first meeting specifically devoted to the topic of cluster
soft excess and related phenomena.  It calls together a group of some
40 scientists, mostly experts in the field, to present their findings
on the observational aspects of emission from clusters, signature of
the WHIM in the cluster, intergalactic, and local environment,  theoretical
modeling of the WHIM in cosmological hydrodynamic simulations, theory
and observations of cluster cosmic rays, magnetic fields, radio data
and the use of the Sunyaev-Zeldovich effect as a means of constraining the
important parameters.  We are particularly grateful to F.J. Lockman and
S.L. Snowden for presenting their latest picture of the cold gas distribution
in the interstellar medium, as this affects our ability to model 
extragalactic EUV and soft X-ray data via Galactic absorption by HI and HeI.
Papers given on future hardware concepts to further the spatial and
spectral diagnosis
of diffuse soft X-ray emission in general are also included.

Finally we express our thanks to Dana Ransom and Clay Durret at
the University Relations of UAH for their design (both hardcopy and
world-wide-web versions) of the conference logo and on-line
registration facilities, to the UAH Physics Department secretaries
Carolyn Schneider and Cindi Brasher for their help in printing the
conference program and arranging the mini reception, and to
the UAH computer support staff Joyce Looger and Wayne Tanev for
their sterling effort in providing computer services to the
delegates during the meeting.  Lastly, Richard Lieu is indebted to his
wife Anna for looking after the general well being of participants
throughout this event.
\prefaceauthor{Richard Lieu \& Jonathan Mittaz}
\end{preface}

\kluwerprintindex




\part[The Cluster soft excess]{The Cluster soft excess}






\articletitle[The extreme ultraviolet excess emission in five
clusters of galaxies revisited]{The extreme ultraviolet excess emission in five
clusters of galaxies revisited}

\chaptitlerunninghead{EUV excess emission in five clusters}

\author{Florence Durret\altaffilmark{1}, Eric Slezak\altaffilmark{2},
        Richard Lieu\altaffilmark{3}, Sergio Dos Santos\altaffilmark{4}, 
	Massimiliano Bonamente\altaffilmark{3}}

\altaffiltext{1}{Institut d'Astrophysique de Paris, CNRS, 98bis Bd Arago, 
75014 Paris, France}
\altaffiltext{2}{Observatoire de la C\^ote d'Azur, B.P. 4229, F-06304 Nice  
Cedex 4, France}
\altaffiltext{3}{Department of Physics, University of Alabama, Huntsville AL 
35899, USA}
\altaffiltext{4}{SAp CEA, L'Orme des Merisiers, B\^at 709, 
F-91191 Gif sur Yvette Cedex, France}

\begin{abstract}

Evidence for excess extreme ultraviolet (EUV) emission over a tail of
thermal X-ray bremsstrahlung emission has been building up
recently. In order to improve the signal to noise ratio in the EUV, we
have performed a wavelet based image analysis for five clusters of
galaxies observed both at EUV and X-ray energies with the EUVE and
ROSAT satellites respectively. The profiles of the EUV and X-ray
reconstructed images all differ at a very large confidence level and
an EUV excess over a thermal bremsstrahlung tail is detected in all
five clusters (Abell 1795, Abell 2199, Abell 4059, Coma and Virgo) up
to large radii. Our results tend to suggest that the EUV excess is
probably non thermal in origin.

\end{abstract}

\section{Introduction}

The Extreme Ultraviolet Explorer (EUVE) has detected emission from a
few clusters of galaxies in the $\sim$70-200 eV energy range. By order
of discovery, these were: Virgo (Lieu et al. 1996b, Bergh\"ofer et
al. 2000a), Coma (Lieu et al. 1996a), Abell 1795 (Mittaz et al. 1998),
Abell 2199 (Bowyer et al. 1998), Abell 4059 (Bergh\"ofer et al. 2000b)
and Fornax (Bowyer et al. 2001).  

Excess EUV emission relative to the extrapolation of the X-ray
emission to the EUV energy range was then detected in several
clusters, suggesting that thermal bremsstrahlung from the hot ($\sim
10^8$K) gas responsible for the X-ray emission could not account
entirely for the EUV emission.  This excess EUV emission can be
interpreted as due to two different mechanisms: thermal radiation from
a warm ($10^5-10^6$K) gas, as first suggested by Lieu et al. (1996a),
or inverse Compton emission of relativistic electrons or cosmic rays
either on the cosmic microwave background or on stellar light
originating in galaxies (Hwang 1997; Bowyer \& Bergh\"ofer 1998;
Ensslin \& Biermann 1998; Sarazin \& Lieu 1998; Ensslin et al. 1999),
or both. The main difficulty with the thermal model is that since gas
at typical cluster densities and in the temperature range $10^5-10^6$K
cools very rapidly, a source of heating is necessary.  Besides, recent
results obtained with FUSE and XMM-Newton (see various contributions
in these Proceedings) tend to indicate smaller quantities of warm gas
and much smaller cooling flow rates than predicted from simulations.
On the other hand, non thermal models (Bowyer \& Bergh\"ofer 1998,
Sarazin \& Lieu 1998, Ensslin et al. 1999, Lieu et al. 1999, Atoyan \&
V\"olk 2000, Brunetti et al. 2001) seem able to account for the excess
EUV emission, but require very high cosmic ray pressure which may
exceed the hot gas pressure (e.g. Lieu et al. 1999, Bonamente, Lieu \&
Mittaz 2001).

We present here a new analysis of EUV and X-ray images for five
clusters, based on their wavelet denoising. Our aims are: to confirm
the detection of EUV emission as far as possible from the cluster
center; to derive accurate EUV and X-ray profiles; to confirm the
reality and radial distributions of the soft excesses over thermal
bremsstrahlung in these clusters.

\section{The data and image analysis}

Five clusters were observed with the EUVE satellite: Abell 1795, Abell
2199, Abell 4059, Coma and Virgo, and their ROSAT PSPC images were
retrieved from the archive (see details in Table~\ref{d1-tbl:mainchar}).

\begin{table*}
\centering
\caption{Exposure times and main cluster characteristics.}
\begin{tabular}{lrrlrr}
\hline
Cluster  & EUVE total  & ROSAT  exp. & Redshift & kT$_{\rm X}$ & Scaling \\
         & exp. time (s)  &  time (s)&          & (keV)        & (kpc/superpx) \\
 \hline		    		     
 A 1795  & 158689    &   33921       & 0.063    & 5.9          & 32.7     \\
 A 2199  &  93721    &   34633       & 0.0299   & 4.1          & 15.8     \\
 A 4059  &  145389   &    5225       & 0.0478   & 4.0          & 25.0     \\
 Coma    &   60822   &   20112       & 0.023    & 8.7          & 12.2     \\
 Virgo   &  146204   &    9135       & 0.003    & 2.4          &  1.6     \\
\hline
\end{tabular}
\label{d1-tbl:mainchar}
\end{table*}

In order to have comparable spatial resolutions in the EUV and X-rays,
all the images were rebinned to a ``superpixel'' size of 0.3077
arcminutes (18.5 arcseconds).  The linear scale per superpixel at the
cluster distance, estimated with H$_0=50$~km~s$^{-1}$~Mpc$^{-1}$ and
q$_0$=0 is given in Col. 6 of Table~\ref{d1-tbl:mainchar}.

\begin{figure}
\centering
\mbox{\includegraphics[width=12cm,clip]{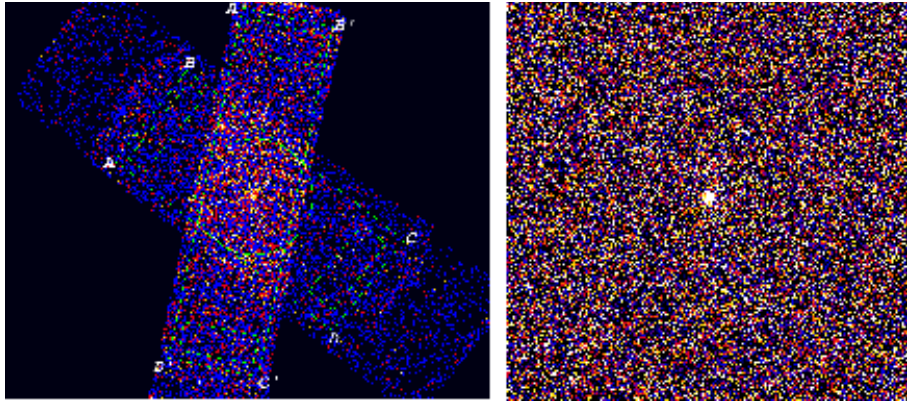}}
\caption[A1795 polygons]{Abell 1795: left: initial EUVE image, right:
``large'' EUVE image (see text). } 
\label{d1-fig:a1795}
\end{figure}

We first estimated the background level in several rectangles at the
extremities of the EUV images. For all the clusters, the rectangular
shape of the EUVE images, and in some cases the fact that we had two
rectangular images (see Fig.~1) was a problem for any spatial analysis
involving an isotropic examination of the data at scales larger than a
fraction of the shorter side.  In order to fill in the gaps and
investigate without noticeable edge effects a scale as large as the
shorter side of the EUVE rectangle, we created for each cluster a
``large'' square EUV image, keeping unchanged the central part of the
image (a circle of $r\sim 16$~arcmin radius) and replacing each pixel
outside this area by a random number drawn from a Poisson
distribution, the variance of which is obtained by adding in
quadrature the variances in each of the rectangles. The final image
thus obtained for Abell 1795 is displayed in Fig.~\ref{d1-fig:a1795}. Of
course, profiles were only derived for radii smaller than the half
width of the EUV detector, i.e.  for $r<16$~arcmin. Any features found
within regions smaller than the PSFs (1 arcmin and 2.5 arcmin radii in
the EUV and X-rays respectively) are of course unreliable.

The ROSAT data were reduced in the usual way using Snowden's software
(Snowden et al. 1994). Since the geometry of these images is circular
we had no problems here with the detector shape. The background was
estimated as a mean value far from the edges of the detector and
subtracted to the images before deriving the ``raw'' profiles (but not
before making the wavelet analysis, since background subtraction is
actually part of this process).

In order to increase the signal to noise ratio in the derived
profiles, we applied part of the wavelet vision model described in
detail in Ru\'e \& Bijaoui (1997), following these steps: first, a
discrete wavelet transform of the image was performed up to about 1/3
the size of the square image; second, the noise was removed from the
data; third a positive image where the background has been subtracted
was restored. Profiles were then derived for the EUV and X-ray
emissions, both on the raw (background subtracted) images and on the
wavelet reconstructed ones, to check that the wavelet analysis and
reconstruction did not modify the shape of the profile but only
improved the signal to noise ratio.  A point source (the star
$\gamma$Tau) was analyzed in a similar way to see how the instrument
point spread functions (PSF) could influence our results. The
resulting PSFs drawn in the EUV and X-rays shown in Fig.~2 imply that
no spatial information is reliable for radii smaller than
$\sim$2.5~arcmin.

\section{Results}

\begin{figure}
\centering
\mbox{\includegraphics[width=11cm,clip]{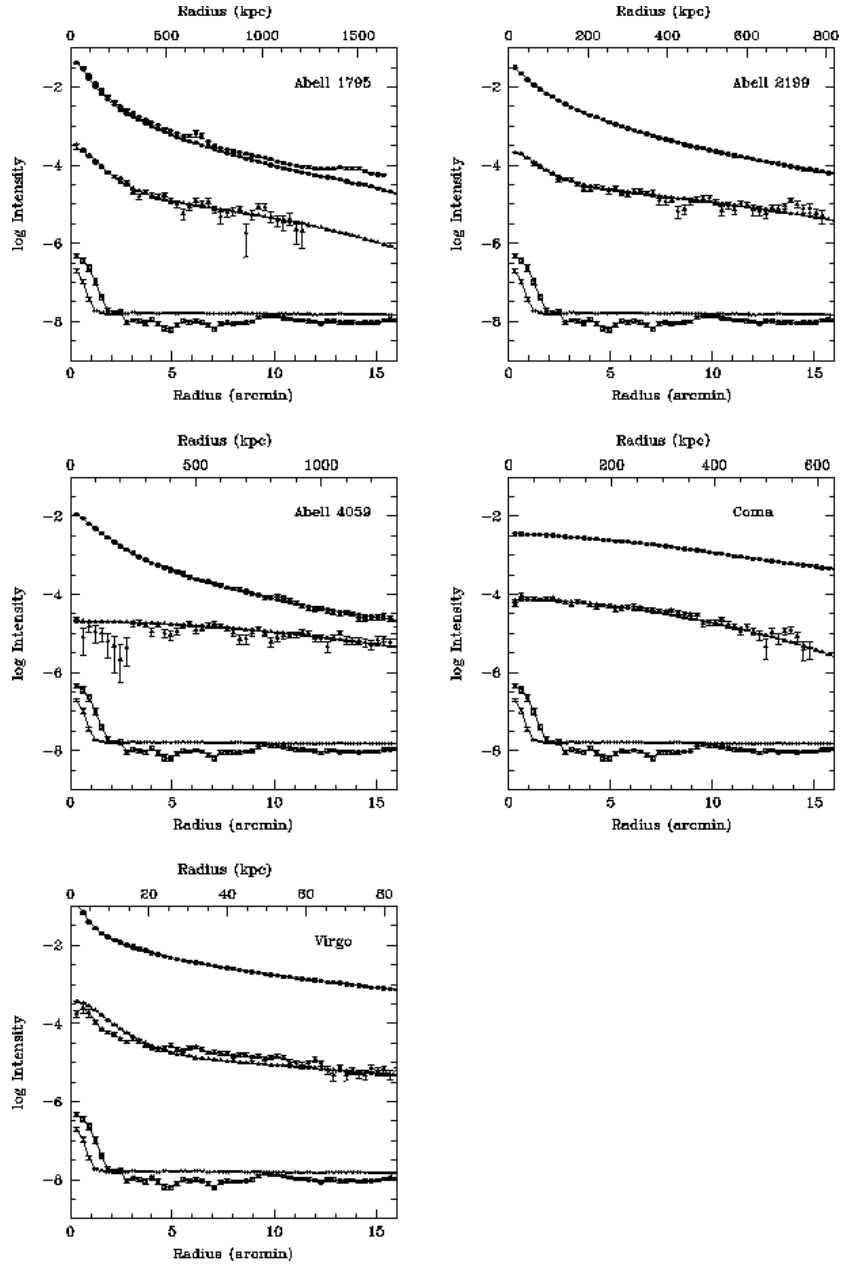}}
\caption{X-ray (top curves) and EUV (middle curves) profiles for all
clusters, obtained from the raw background subtracted data (data with
the largest rms error bars) and after a wavelet analysis followed by a
reconstruction. The error bars for the profiles derived from the
wavelet reconstructed images are drawn but they are too small to be
clearly visible. The bottom lines show the EUV (x's) and ROSAT PSPC
(empty squares) instrument PSFs. }
\label{d1-fig:profils}
\end{figure}

X-ray and EUV profiles are displayed for the five clusters in
Fig.~\ref{d1-fig:profils}. They show EUV emission as far as 16~arcmin
radius (i.e. the central circle, see Section~2), comparable or larger
than previously published values by other authors.

We performed a Kolmogorov-Smirnov test on the EUV and X-ray profiles
drawn from the wavelet reconstructed images, after normalizing the EUV
and X-ray profiles to the same innermost pixel value.  In all
clusters, the probability that the two profiles are issued from the
same parent population is smaller than 0.1\%, implying that
statistically the EUV and X-ray profiles differ.

\begin{figure}
\centering
\mbox{\includegraphics[width=11cm,clip]{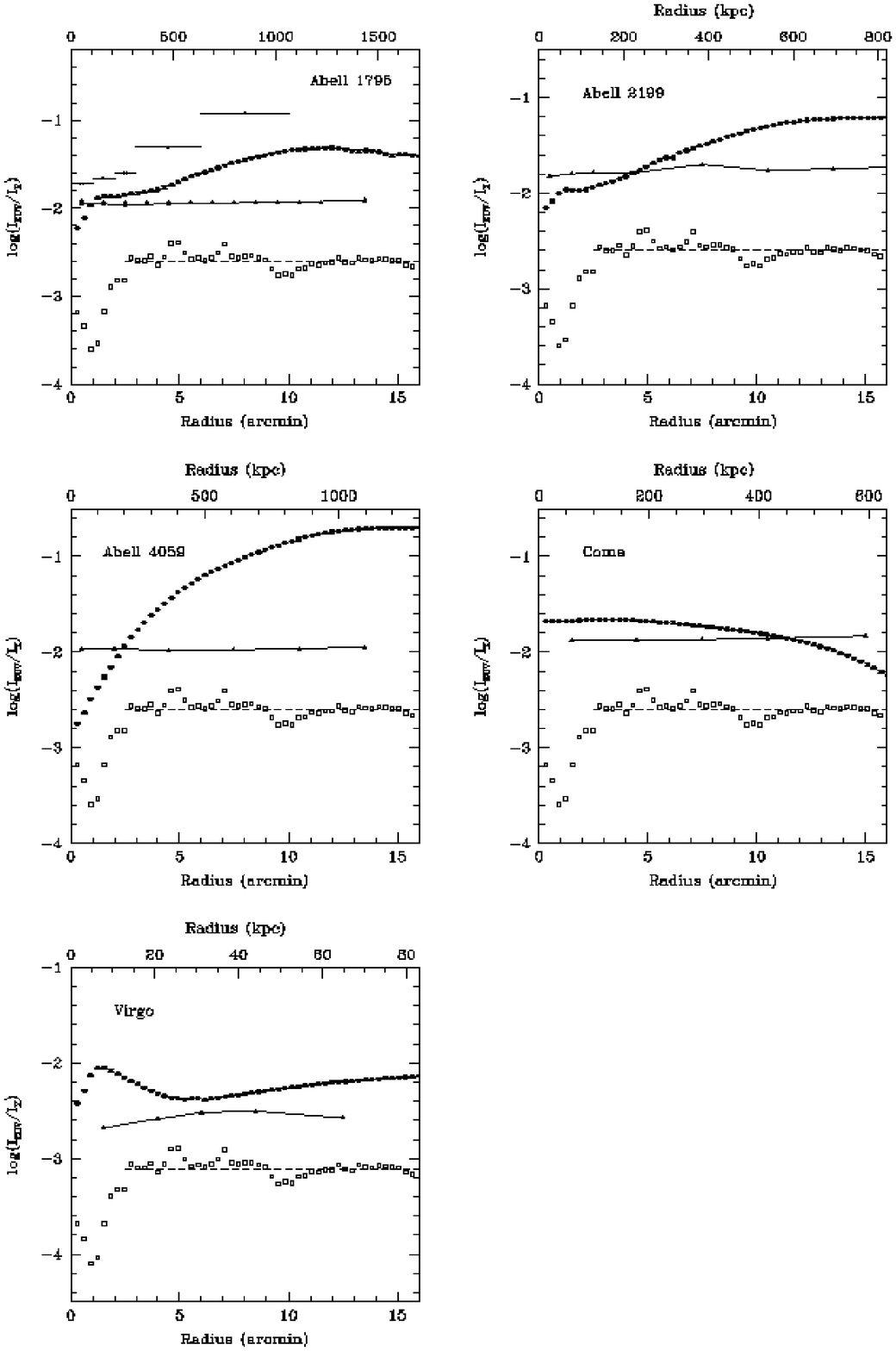}}
\caption{EUV to X-ray intensity ratio: observed (filled squares),
predicted by thermal bremsstrahlung emission from the X-ray emitting
gas (filled triangles), observed by Mittaz et al. (1998) in Abell 1795
(x's) and EUV to X-ray PSF ratio (empty squares) and mean value
(dashed line). }
\label{d1-fig:rapp}
\end{figure}

The observed EUV to X-ray intensity ratios (in the same concentric
ellipses) are displayed for the five clusters in Fig.~\ref{d1-fig:rapp}
together with the ratios predicted in the hot gas bremsstrahlung
hypothesis. Note that for the three clusters in which XMM-Newton has
shown the existence of a temperature gradient (Abell 1795, Coma and
Virgo), these ratios remain virtually unchanged when this gradient is
taken into account. All clusters show an EUV excess over hot thermal
bremsstrahlung. This excess strongly increases with radius in Abell
1795, Abell 2199 and Abell 4059, and somewhat in Virgo. On the other
hand, in Coma it remains roughly constant up to $r\sim 10$~arcmin, then
decreases radially.

\begin{figure}
\centering
\mbox{\includegraphics[width=7cm,clip]{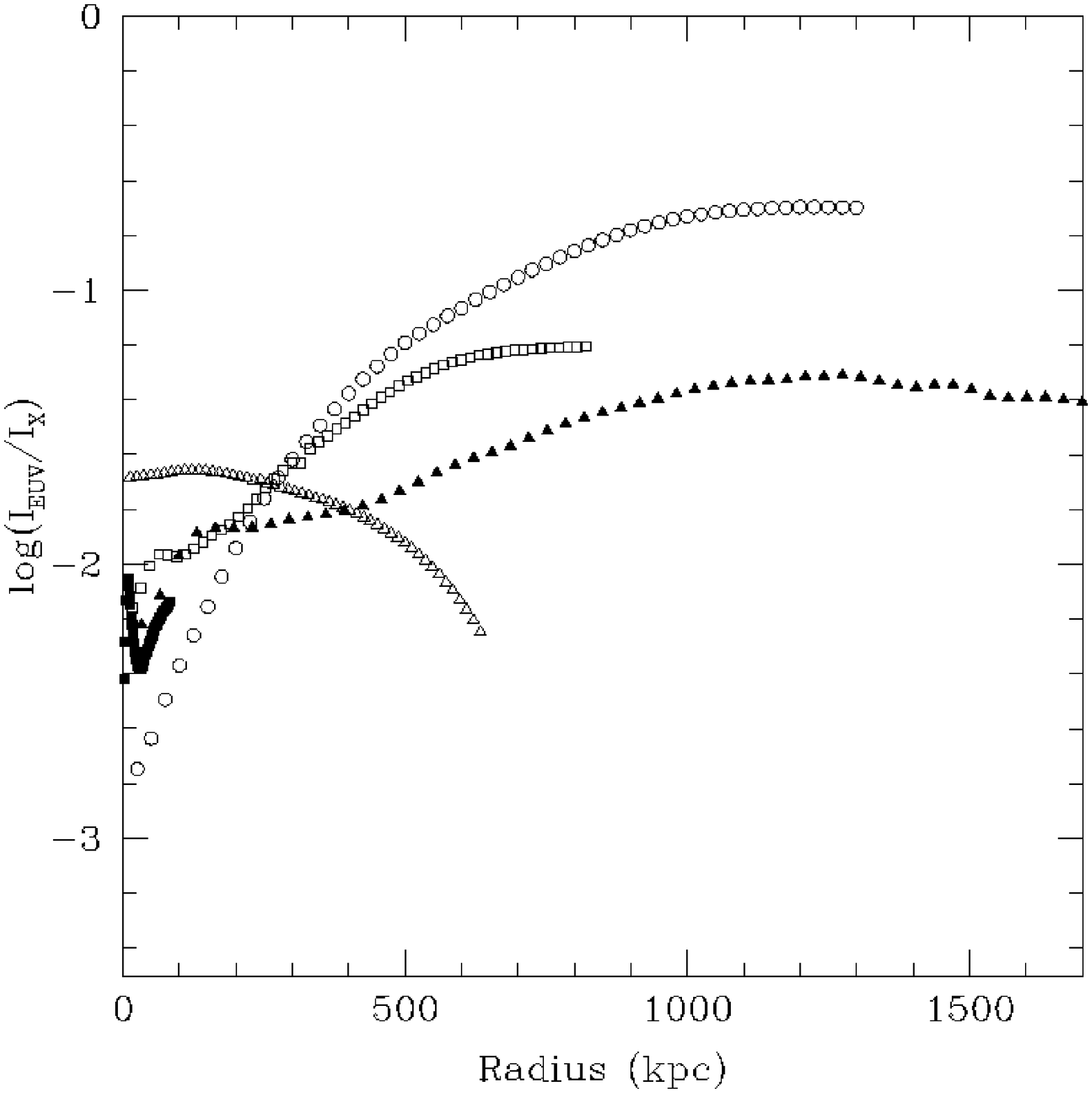}}
\caption{EUV to X-ray ratios for the five clusters in our sample,
with the radius expressed in physical units (kpc). The symbols are the
following: Abell 1795: filled triangles, Abell 2199: empty squares, Abell
4059: empty circles, Coma: empty triangles, Virgo: filled
squares. Error bars are omitted for clarity. }
\label{d1-fig:rappkpc}
\end{figure}

The ratios of the EUV to X-ray intensities for all clusters with radii
expressed in physical units (kpc) are shown in Fig.~\ref{d1-fig:rappkpc}.
Note that the radial extent of the EUV and X-ray emission in Virgo is
comparable to the other clusters when the radius is expressed in
arcminutes, but becomes extremely small in physical distance units,
due to the very small redshift of the cluster: slightly more than 80
kpc.

\section{Discussion}

We have shown unambigously the existence of a EUV excess in all five
clusters of our sample.  In the first three clusters (Abell 1795,
Abell 2199 and Abell 4059), the EUV to X-ray intensity ratios show a
possible deficiency of EUV emission over a bremsstrahlung tail in the
very central regions (but this may be an effect of the ratio of the
EUV to X-ray PSFs), and a clear EUV excess beyond a few arcmin. As
suggested by several authors (Bowyer et al. 1999; Lieu et al. 2000),
the first of these features, if real, can be interpreted as due to
excess absorption within the cluster core due to the fact that in the
cooler central regions some metals are not fully ionized and these
ions absorb part of the soft X-ray flux, in agreement with the
existence of cooling flows in these three clusters. However,
XMM-Newton has detected much weaker emission lines than expected from
{\sl bona fide} isobaric cooling flow models, implying that there is
significantly less ``cool'' gas than predicted (see e.g. Kaastra et
al. 2001 and these proceedings). Therefore it is no longer
straightforward to interpret excess absorption at the center of these
clusters as due to the presence of cooler gas.

The only cluster (Coma) for which no EUV dip is seen in the very
center is both the hottest one by far and the only one with no
``cooling flow'' whatsoever. Since the ratio of the EUV to X-ray PSFs
shows a dip for radii smaller than about 2.5 arcmin, the absence of
such a dip in Coma suggests that in fact there is a significant EUV
excess in the central regions of Coma.

Virgo shows a roughly constant EUV excess, somewhat stronger in the
very center, suggesting as for Coma that the EUV excess is in fact
strong in this zone. Images with higher spatial resolution are
obviously necessary to analyze the EUV excess at these small radii.

The mere presence of an EUV excess in all five clusters of our sample
indicates that a mechanism other than bremsstrahlung from the hot gas
responsible for the X-ray emission is needed to account for this
emission.  Our data does not allow us to discriminate between the
various mechanisms proposed in the literature to account for the EUV
excess. However, in view of the most recent results obtained with
XMM-Newton suggesting that there is much less warm gas in the central
regions of clusters than previously believed (see in particular
several presentations in these proceedings and references therein), it
seems likely that this EUV excess is probably of non thermal origin,
at least in the central regions.

A full description of the method and results presented here can be
found in Durret et al. (2002).
\\

\noindent{\bf Acknowledgments}

We are very grateful to M. Traina for her invaluable help in obtaining
the up-to-date version of the multiscale vision model package used
throughout.  We acknowledge discussions with D. Gerbal, G. Lima Neto,
G.~Mamon and J. Mittaz, and thank Jelle Kaastra for several interesting
suggestions.











\articletitle[Soft (~1 keV) X-ray Emission in Galaxy Clusters]{Soft (~1 keV) X-ray Emission in Galaxy Clusters}



\author{Mark Henriksen}

\affil{University of Maryland, Baltimore County}

\begin{abstract}

In this paper, we present results from our program of analysis of
the broad band spectra of galaxy clusters. For two clusters, Abell 754
and IC 1262, we find evidence of a soft excess that can be modeled either
as thermal emission or as non-thermal emission. For IC 1262, the thermal
interpretation is non-physical and the non-thermal model has further support
from a radio halo that we propose exists in the cluster center. For Abell 754,
we modeled the ~1 keV thermal component as emission from the IGM of several galaxy groups that may lie
within 300 kpc of the central region of the cluster. We further argue that this interpretation
is more likely than either elliptical galaxies or intercluster gas as the origin of the emission.
However, it is also possible that it is non-thermal in origin. Under the non-thermal
interpretation, it is likely to originate from primary cosmic-rays accelerated at the
shock front formed during the ongoing merger. Finally, we present the first observational
L$_{nt}$ versus T${x}$ plot. The plot shows a weak (~90\%) correlation that indicates
at least some of the non-thermal components detected in clusters must result from the
formation of the cluster in order to provide this energy relationship.

\end{abstract}

\section{Introduction: History of Non-thermal emission Detection}

Over the first several decades of X-ray studies of galaxy clusters using broad band spectra,
increasingly sensitive limits have been placed on the amount of non-thermal X-ray emission
from these objects.  Most of the early studies were restricted to clusters known to have diffuse radio halos 
since they have a population of non-thermal electrons. A wide range of data were used in this work (Rephaeli \& Gruber
1988 using {\it HEAO1-A4}; Henriksen 1998 using {\it ASCA} and {\it HEAO 1-A2};
Rephaeli, Ulmer, \& Gruber 1994 using OSSE on {\it CGRO}, and
EGRET, Shreekumar et al. 1996). 
The upper limit on
non-thermal X-ray emission together with the radio
spectral parameters were used to place a lower limit on the average
magnetic field, $<$B$>$. Lower limits to
$<$B$>$ in the range of $\ge$0.1--0.2$\mu$G were found.
The EGRET observation provided the
highest lower limit on the average cluster field (0.4$\mu$G) from these early
observations with a flux upper limit of 4$\times$10$^{-8}$ ph cm$^{-2}$
s$^{-1}$ at $>$100 MeV.
\smallskip
The next generation of hard X-ray observatories have given new and
more sensitive limits on and several detections of non-thermal emission. For example, hard X-ray
emission is detected above $~$25 keV with the {\it SAX} PDS  
identified with the flat spectrum, $\alpha_r$ = 0.8 radio
source in A2256 (Fusco-Femiano et al. 2000). 

\smallskip
Observations are needed at energies $>$ 25 keV to
detect a flat, non-thermal X-ray source in hot, $\sim$7 keV clusters. Consequently,
the {\it SAX} PDS has been the primary detector used to find hard X-ray 
components in A2256, A2199 (Kaastra et al. 1999), and Coma (Fusco-Femiano et al. 2000). In Abell 2199, the
non-thermal component was also detected in the 8 - 10 keV range in the MECS (Kaastra et al. 1999).
Non-thermal emission from cooler, less luminous clusters such as Abell 1750, Abell 1367, or IC 1262
can be detected with the RXTE PCA and the BeppoSaxMECS that provide high signal-to-noise up to 15 keV.
For Abell 1367, the PCA
is quite sensitive to non-thermal emission from the steep spectrum radio relic, $\alpha_r$ = 1.9, and gave
a stringent upperlimit on non-thermal emission, 3.3$\times$10$^{-3}$ ph cm$^{-2}$ keV$^{-1}$
s$^{-1}$ at 1 keV, provided that the hot gas component is 2 phase. The lowerlimit on the average
magnetic field from the relic is very high, 0.84$\mu$G (Henriksen \& Mushotzky 2001) and challenges
the early notion that the average cluster magnetic field was in the range of 0.1 - 0.2; low enough to be
created by galaxy wakes in the intracluster medium and points toward a more energetic
process such as merger for magnetic field aplification. A low cluster temperature makes it possible to
use the MECS to isolate and analyze small regions of the cluster and thereby minimize contamination from point sources
while localizing the source of non-thermal emission. For a relatively steep spectrum source, as is typical
of non-thermal radio halos, the non-thermal emission may also be detected as a soft excess over the cluster's
thermal component. While the source flux of the powerlaw component is much higher in the low energy data and
the sensitivity of the detectors is much superior to that at high
energy, this strategy is complicated. In this case, the soft emission may represent the cooler phase of a two-phase intracluster
medium. These competing hypotheses are explored in detail in the rest of this paper.

\section{IC 1262: A Soft Component: Non-thermal Emission}

We found preliminary indications of diffuse, non-thermal emission in IC 1262, a cool cluster (see Figure 1).
using the BeppoSax MECS and ROSAT PSPC detectors (Hudson, Henriksen, \& Colafrancesco 2003).  By fitting a
6 arcmin ($\sim$360 ${\it h_{50}^{-1}}$ kpc) region with a single mekal model including photoelectric absorption,
we find a temperature of 2.1 - 2.3 keV, and abundance of 0.45 - 0.77 (both 90\% confidence). 

\begin{figure}[ht]

\includegraphics[width=12cm, height=12cm]{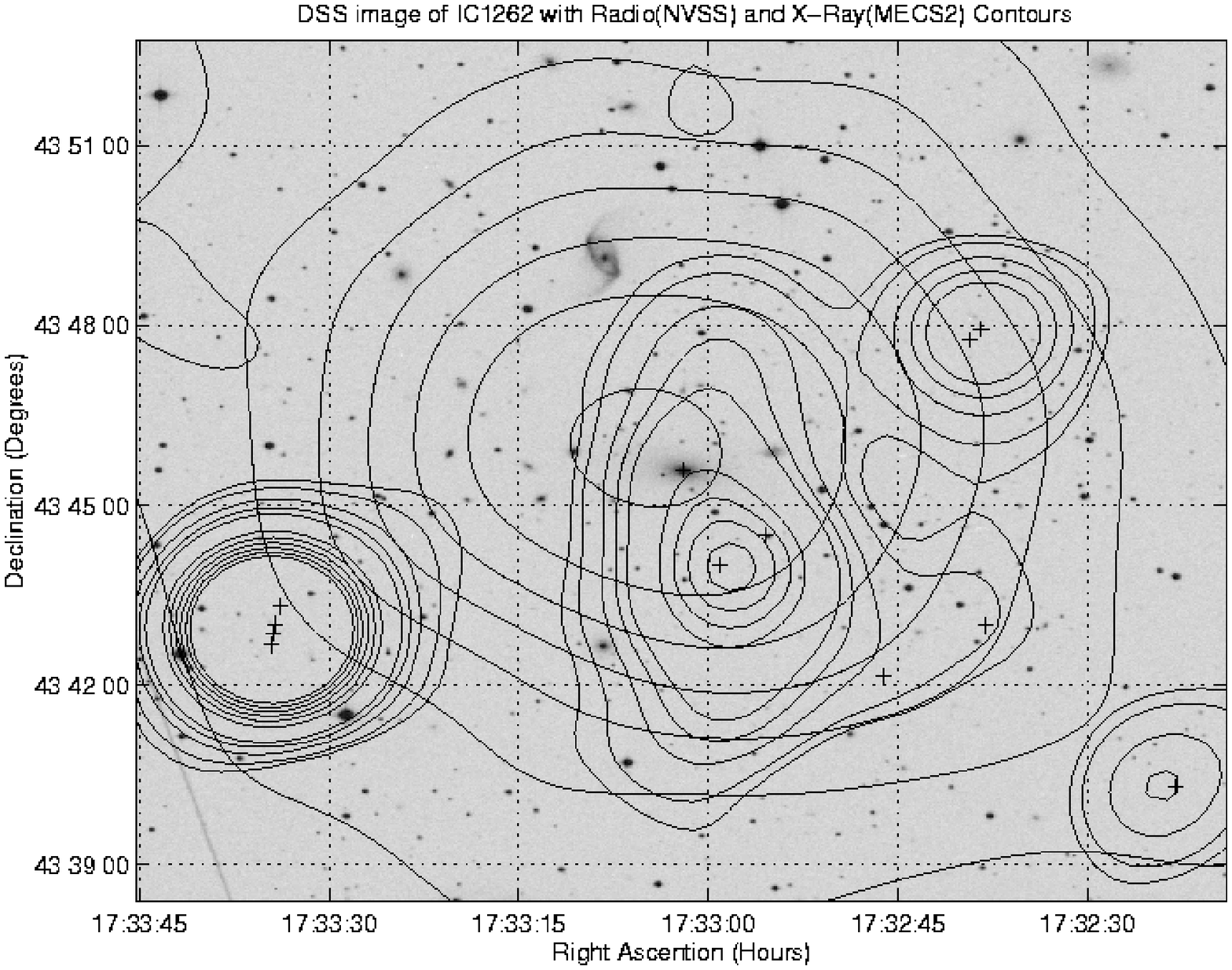}
\caption{The North-South running contours near the center are radio intensity contours that
delineate a radio halo.The other contours are X-ray and are the emission from the MECS.}

\end{figure}

The
addition of a power-law component provides a statistically significant improvement (ftest = 90\%)to the fit. 
The addition of a second thermal component also improves the fit though it is physically
implausible based on the following argument. The two temperature components are such that the
hottest component has a luminosity that is significantly lower than that predicted using the luminosity
temperature relationship (Horner et al. 2000) to predict its temperature. Thus it is unphysical.
 Additional evidence of diffuse, non-thermal emission comes from
the NRAO VLA Sky Survey (NVSS) and Westerbork Northern Sky Survey (WENSS) radio measurements, where excess
diffuse, radio flux is observed after point source subtraction.  The radio excess can be fit with a simple
power-law, the spectral index of $\sim$1.8 which is consistent with the non-thermal X-ray emission spectral
index.  The steep spectrum is typical of diffuse emission and the size of the radio source implies that it
larger than the cD galaxy and not due to a discreet source. 
The power-law component has a photon index 
($\Gamma_{x}$) of 0.5 - 2.8 and a non-thermal flux of
(2.2  - 7349.5) $\times$ 10$^{-6}$ photons cm$^{-2}$ s$^{-1}$ over the 1.5 - 10.5 keV range in the Medium
Energy Concentrator spectrometer (MECS) detector. 

\section{A Soft X-ray Component in Abell 754: Thermal Interpretation}

A large region of the BeppoSax MECS and ROSAT PSPC observations were selected
for analysis along with the PDS. Together they offer
a broad energy band of 0.5 - 200 keV with reasonable signal to noise. The MECS has the
best hard X-ray coverage of the recent imaging instruments. For comparison, while the ASCA GIS
has a comparable bandwidth, the MECS has a factor of 2 higher effective area at 8 keV. For a 
hot cluster such as Abell 754, the thermal components will dominate these spectra and the
PDS is crucial for detecting the non-thermal component at high energy. However, the PDS
is most sensitive to flat spectrum sources which are more likely to be AGN 
than diffuse X-ray emission based on radio characteristics. The PSPC may be sensitive 
to the steeper powerlaw that characterizes diffuse non-thermal emission; a component
that may dominate the spectrum at low energy. 

Our use of the PDS is to constrain an active galaxy that contaminates the emission.
The residuals show that a single mekal model fit
gives significant residual emission appears around 100 keV and below 1 keV. The PDS
field of view contains the BL LAC object, 26W20, that is also known to be a variable
non-thermal X-ray source (Silverman, Harris, \& Junor, 1998).
After modeling 26W20 in the PDS, the residuals around 100 keV are eliminated.
A better fit to the lower energy data is provided by adding either a non-thermal component or
a second thermal component in addition to the high temperature thermal component. 
The second component is either a thermal component 
with temperature 0.75 - 1.03 keV and
7.0$\times$10$^{43}$) ergs s$^{-1}$ luminosity which is 7.2\% of the the hot thermal
component or non-thermal with 
$\alpha \simeq$2.3 and bolometric (0.1 - 100 keV) X-ray luminosity of 2.4$\times$10$^{42}$ or 0.25\% of 
the thermal. There are a number of possible sources for cool thermal gas including: galaxies and
intercluster gas. 

Proposed sources of low temperature 
thermal emission of galactic origin include the integrated
emission from gas associated with several elliptical galaxies or groups within the observed region of the cluster 
(Henriksen \& Silk 1994). A number of small scale structures exist in the Abell 1367 cluster that are significantly
cooler than their surroundings suggesting that the early-type galaxy coronae can survive in the intracluster
medium (Sun \& Murray  2002). We evaluated this hypothesis using the following simulations.
The {\it Chandra} temperature map maintains approximately
6000 counts per region to allow spectral fitting of each region. The smallest regions
are then 1 x 1 arcmin and the largest are 2 x 2 arcmin. At the redshift of Abell 754, 1 arcmin = 73 kpc (H$_{0}$ = 65). 
Recent XMM observations of NGC 5044, an elliptical galaxy in a group, will help to
evaluate the importance of a single, large galaxy or group in providing the
soft emission. These observations show that the 20 kpc region centered on the galaxy has 
temperature components of 0.7 and 1.1 keV, very similar to the cool thermal component
in Abell 754. The luminosity from the entire group (r $<$ 250 kpc)
is 1.8$\times$10$^{43}$ ergs s$^{-1}$ (David et al (1994). 
Simulations (Henriksen \& Hudson 2003) were performed to obtain an emission weighted 
temperature map for comparison with the Chandra map. The simulations consist 
of Abell 754 cluster emission with an embedded group whose gas is described by the NGC 5044 
parameters (Mulchaey et al. 1996) 
The group is embedded at radii ranging from 100 to 400 kpc, which places it within the
region of the BeppoSax data analyzed. The
parameters for the group are kT = 0.98 keV, core radius = 28 kpc, ion density = 9.6e-3 cm$^{-3}$. The Abell 754 parameters are
1.85e-3 cm$^{-3}$ (Abramopoulos \& Ku 1981) and kT = 10 keV. The resulting simulated emission weighted temperature is 
(4.8, 5.1, 3.5, 2.5 keV) for a 
large pixel matched to the Chandra map and with a luminosity
fraction of (24.6,22.8,37.5,50.4\%) for the relative contributions of group versus cluster for a distance of (100,200,300,400 kpc).
The emission weighted temperature at 400 kpc , 2.5 keV, is significantly
less than 3.9 keV, the lowest temperature measured in the Chandra temperature. The group component begins
to dominate the cluster component (luminosity fraction $>$ 50\%) at this radius. This implies that
the group(s) must be located within 300 kpc, otherwise the measured emission weighted temperature will contradict the
Chandra temperature map constraint. 

Because the luminosity of a single group is too low compared with
the cool component we measured, we simulated the temperature map with 4 groups within 8' or a 580 kpc analyzed region. This would match 
the observed soft component luminosity. The groups must be further constrained to be within 300 kpc to avoid violating the constraint from
the temperature map (e.g., the lowest measured temperature regions).
The simulated temperature map gives one pixel slightly below 3.9 keV which does not violate the temperature
map lower temperature limit. Thus we conclude that groups, like NGC 5044 may be the source of the cool component.

Elliptical galaxies have a hot gaseous halo of ~1 keV. A King model using Abell 754 parameters (Abramopoulos \& Ku 1983),
with central surface density of galaxies of 92 Mpc$^{2}$ and the 
X-ray core radius for a King model (0.71 Mpc), gives 52 galaxies within 
the region. A typical elliptical fraction is 80\%.
This would give 42 elliptical galaxies. To give the cool component would require and average L$_x$ of 1.7e-42 ergs s$^-1$ per
elliptical galaxy. The distribution of normal early-type galaxies (Eskridge, Fabbiano, \& Kim 1995) shows that 10$^{41}$ ergs s$^{-1}$
is typical and that only a small fraction have X-ray luminosities a factor of 10 higher. Thus it is implausible that
this large cool component is from elliptical galaxies only. 

Another possibility is a diffuse
baryonic halo surrounding the cluster. 
Evidence for a similar, diffuse component in the 0.5 - 1 keV  range have been reported with several different data sets and authors.
Henriksen \& White (1966) using HEAO-1 and Einstein data with a similar broad band coverage but with much lower effective area reported 0.5 - 1 keV
gas beyond the central cooling flow region in several clusters and showed that the amount of the cool component
did not correlate with the amount of cool central gas. There are several recent papers reporting cool emission
within the cluster atmosphere.



Kaastra et al. (2003) report a soft X-ray component that is visible as an excess in the 0.4 - 0.5 keV range that
is attributed to the warm hot IGM.
Bonamente, Joy, and Lieu (2003) report evidence for
a diffuse baryonic halo around the Coma cluster that extends to 2.8 Mpc. Their spectral analysis of ROSAT PSPC observations
show that the radial temperature profile of this component is consistently 0.25 keV with 10\% uncertainty. This suggests that
the intercluster component is substantially cooler than the Abell 754 component since our temperature
component is consistently higher. Nevalainen et al. (2003) report
soft excesses in XMM observations of the Coma, Abell 2199, Abell 1795, and Abell 3112 in the cluster's inner 0.5 Mpc, a similar 
region to that analyzed in our BeppoSax and PSPC observations. The characteristic
temperature is 0.6 - 1.3 keV, which is also consistent with the Abell 754 cool component. They point out that their
 density, 10$^{-4}$ - 10$^{-3}$
cm$^{-3}$ is well above those expected from the intercluster medium (Dave 2001) in the core region of the cluster.
The density of the Abell 754 cool component is similarly high compared to the WHIGM and comparable
to the Coma, Abell 2199, and Abell 3122 clusters. Based on our analysis, we suggest that the soft 0.5 - 1 keV component in 
the central region of clusters, that is in excess of the projected intercluster component, is not
due to the integrated emission from elliptical galaxy halos or a central cD galaxy but rather
it is due to the integrated emission of the intergalactic medium of several embedded groups.Since most groups
have a higher abundance than clusters, a test of this hypothesis would be to look for a high abundance
in the soft component.

\subsection{Soft X-ray Emission in Abell 754: Non-thermal Interpretation}

The Abell 754 cluster shows evidence for a merger in its radio structure, X-ray surface brightness, and temperature
map.
The X-ray peak is surrounded by a peanut shaped, high surface brightness region. This region
is cooler than the surrounding atmosphere by 30\%, though it does not show a temperature discontinuity.
Comparison of the morphology of X-ray center in the merger simulations (Roetigger, Stone, \& Mushotzky 1998) to the that seen in the Chandra
image shows a similarity in the steep density gradient to the EAST. The gradient is formed by ram-pressure
as the subcluster hits the dense primary cluster core. However, the observed surface brightness
also shows a steep gradient to the NW. This peanut shaped X-ray region isn't produced by the
thermal gas motions as are shown in the merger simulations suggesting that another process contributes to
giving that morphology. For example, the radio halo may be expanding and compressing the X-ray gas
on the West side

The results of our analysis of the various X-ray data sets brings together new details of the dramatic cluster merger
in Abell 754. Simulations of cluster merger (Takizawa \& Naito 2000) trace the spatial distribution and time
evolution of synchrotron radiation
due to electrons accelerated at the shock fronts that form during a cluster merger. The morphology of the radio halo
and its observed location relative to the shocked cluster gas depends on the viewing angle relative to
the merger axis and the age of the merger. When the viewing perpendicular to the merger axis, the radio and
thermal X-ray have different spatial relationships at different times in the merger. The shocked gas in Abell 754
West of the cluster center and the radio halo are co-spatial in the observations. That morphology seems to be consistent with the simulation
morphology
after maximum contraction. The simulations show two outward traveling shocks when the radio and shocked gas are co-spatial. 
While viewing nearly along the line of sight could
also account for the shocked gas and radio being co-spatial, that viewing would also make both regions appear
to be in the center of the cluster, which is not the case since they are West of center. Thus, we are likely
viewing the merger perpendicular to the merger axis.

The Eastern radio source is not coincident with shocked gas and requires altering this merger scenario. Also, the 
X-ray peak region is not hot. However, this is consistent with the simulations of an off-center merger (Roettiger, Stone, \&
Mushotzky, 1998) that was hypothesized based on the ASCA temperature map (Henriksen \& Markevitch 1996). A significant
feature that is predicted and that we see inthe observations is a band of cool
gas is produced running through the center. This gas is a mixture of pre-shocked primary cluster gas and cooler subcluster
gas in the simulations.
The Eastern radio source may be from an earlier merger or accretion event. 
Radio halos loose their energy due to inverse-Compton cooling and have a relatively short lifetime compared
to the non-thermal hard X-ray component as shown in the simulations of Takizawa \& Naito (2000). These simulations
show that the radio halo fades quickly after the most contracting epoch, when the magnetic field is strongest, during
the cluster merger. On the other hand, the 10 - 100 keV hard X-ray component reaches a peak after the radio when the
cosmic-ray density peaks. Both occur only when there are signs of a merger present. 
Thus, it would be unlikely for the radio source to still be around after the
thermal signs of the merger are gone. This would make the "previous merger" hypothesis unlikely. Ohno, Takizawa, \& Shibata (2002)
however find that turbulent Alfven waves may reaccelerate cosmic-rays that originated at the shock front thereby extending
the lifetime of the radio halo. The turbulent gas mixing would also diminish the regions of hot gas from the shock front
gas Secondary electrons produced by proton-proton collisions followed by pion
decay may produce a radio halo with a longer lifetime (Blasi \& Colafrancesco 1999). However, the spectral index
is predicted to be flatter than we have observed from the X-ray. Thus, the present merger in Abell 754 and the radio
structure, together with a hypothesized earlier merger to create the Eastern relic,
 make a non-thermal interpretation of the soft X-ray component physically plausible.
\begin{figure}[ht]

\includegraphics[width=10cm, height=10cm]{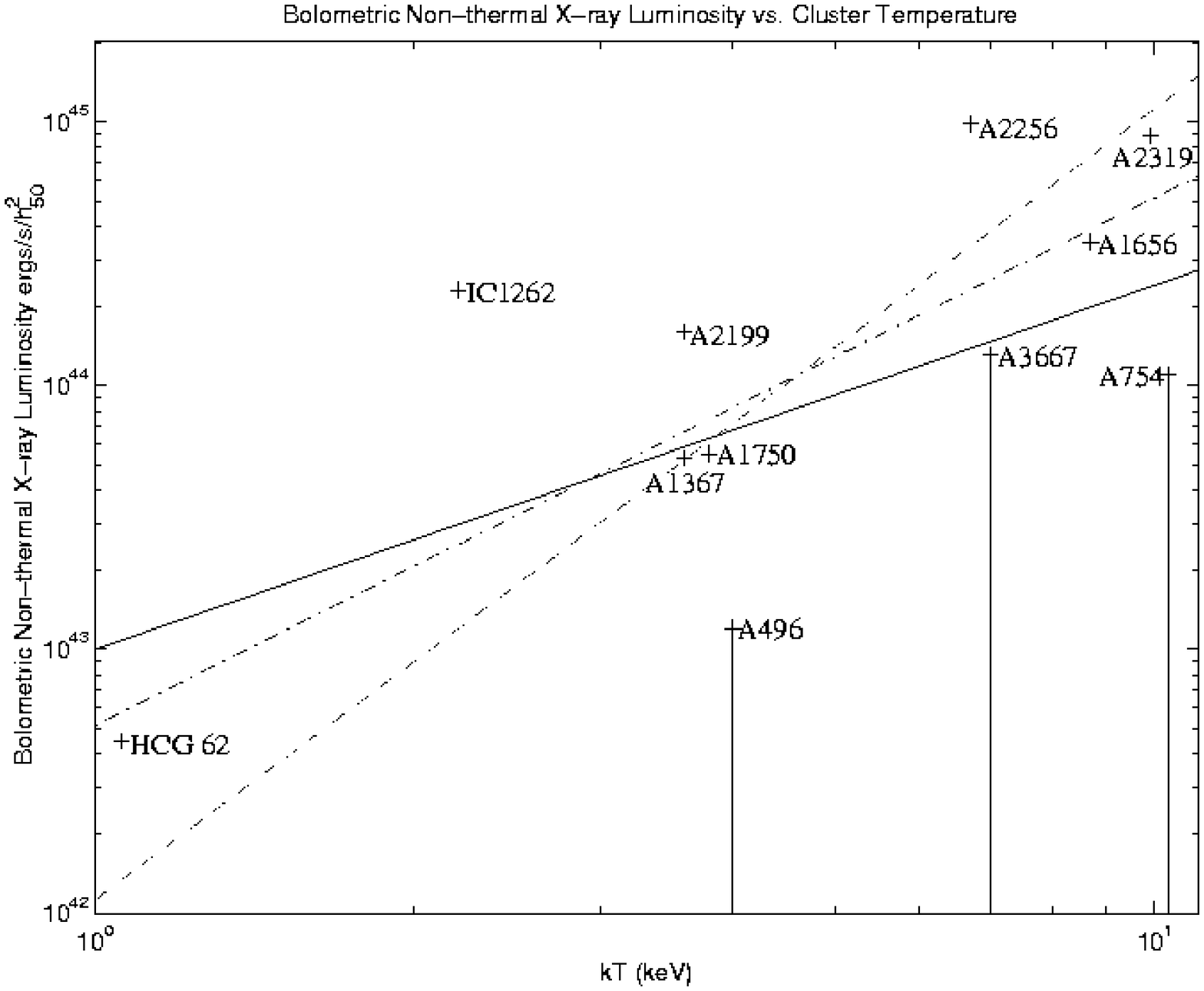}
\caption{Clusters for which non-thermal X-ray measurements are available versus their emission weighted temperature.}

\end{figure}

In Figure 2, we show a plot of the non-thermal luminosity versus cluster temperature. The non-thermal X-ray
luminosities represent all of the detections in the literatur. The figure shows that there is a correlation, as predicted by simulations 
of non-thermal emission by cosmic rays accelerated during cluster mergers (Miniati et al. 2001);
but that there is a great deal of scatter. As this is a first generation plot of these two quantities, some of the
scatter may be due to differences in analysis between authors. However, it does indicate that at least some of the
detections must be real and not due to AGN, which would not show the correlation.\\




\noindent{\bf Acknowledgments}

I thank Danny Hudson and Dr. Eric Tittley for
contributions to this work.

\newcommand{\bapj}{Astrophys. Journal }
\newcommand{\bapjl}{Astrophys. Journal Letters }
\newcommand{\bapjs}{Astrophys. Journal Supplements }
\newcommand{\baap}{Astron. and Astrophys. }
\newcommand{\baraa}{Annual Revue of Astron. and Astrophys. }
\newcommand{\baj}{Astronomical Journal }
\newcommand{\bmnras}{Monthly Not. of the Royal Astron. Society }
\articletitle{A massive halo of warm baryons in the Coma cluster}

\author{M. Bonamente\altaffilmark{1}, M.K. Joy\altaffilmark{2} and 
R. Lieu\altaffilmark{1}}
\altaffiltext{1}{Department of Physics, University of Alabama in Huntsville}
\altaffiltext{2}{NASA/ MSFC Huntsville, Al.}

\begin{abstract}
Several  deep PSPC observations of the Coma cluster reveal a  very large-scale halo 
of soft X-ray emission, 
substantially in excess of the well known radiation from the
hot intra-cluster medium.
The excess emission, previously reported in the central region of the cluster using 
lower-sensitivity EUVE and ROSAT data, is
now evident out to a radius of 2.6  Mpc, 
demonstrating that the soft excess radiation
from clusters is a phenomenon  of cosmological significance.
The X-ray spectrum at these large radii cannot be modeled non-thermally, but is consistent
with the original scenario of thermal emission from warm gas at $\sim$ $10^6$ K. The mass of
the warm gas is on par with that of the hot X-ray emitting plasma,  
and significantly more massive if the warm gas resides in low-density
filamentary structures. 
\end{abstract}

\section{Introduction}
In 1996, Lieu et al. (1996) reported the discovery of excess EUV and soft X-ray emission
above the contribution from the hot ICM in the Coma cluster; their
conclusions were based upon {\it Extreme UltraViolet Explorer (EUVE)} Deep Survey data (65-200 eV)
and {\it ROSAT} PSPC data (0.15-0.3 keV).
In this  paper we present the analysis of a mosaic of {\it ROSAT} PSPC observations around the
Coma cluster, revealing a very diffuse soft X-ray halo extending to
considerably larger distances than reported in the previous studies.
The spectral analysis of PSPC data reported in this paper indicates that the emission
is very likely thermal in nature.

The redshift to the Coma cluster is $z$=0.023 (Struble and Rood 1999). Throughout this paper 
we assume a Hubble constant of $H_0=72$  km s$^{-1}$ Mpc$^{-1}$ (Freedman et al. 2001), 
and all quoted uncertainties are at the 68\% confidence level.

\section{The ROSAT PSPC data}

Coma is the nearest rich galaxy cluster, and the X-ray emission from its
hot ICM reaches an angular radius
of at least 1 degree (e.g., White et al. 1993, Briel et al. 2001). 
Here we show how  several off-center PSPC observations provide a reliable measurement of the background
and reveal a very extended halo 
of soft X-ray excess radiation, covering a
region  several megaparsecs in extent in the Coma cluster.

The RASS maps of the diffuse X-ray background (Snowden et al. 1997) are
suitable to measure the extent of the soft X-ray emission (R2 band, $\sim$ 0.15-0.3 keV) 
around the Coma cluster and to compare it with that 
of the higher-energy X-ray emission (R7 band, $\sim$ 1-2 keV). In Fig. 1 we show a radial
profile of the surface brightness of PSPC bands R2 and R7 centered on Coma: 
the higher-energy emission has reached a constant background value at a radius of $\sim$ 1.5 degrees, while the
soft X-ray emission persists out to a radius of $\sim$ 3 degrees. 
Thus, the RASS data indicate that the soft emission is more extended than the 1-2 keV emission, which 
originates primarily from the hot phase of the ICM.

\begin{figure}[h]
       \includegraphics[angle=90,width=4in]{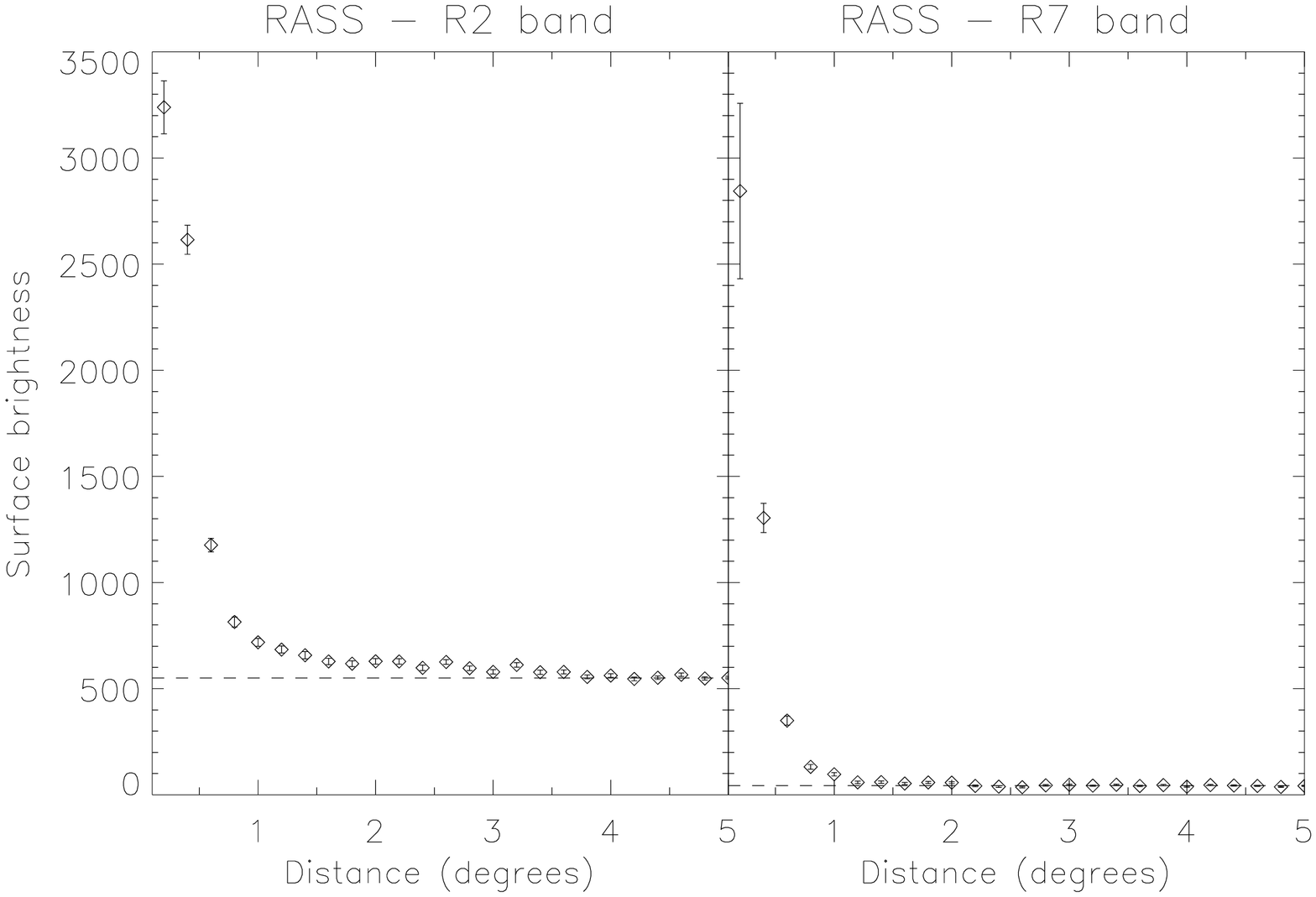}
       \caption{Radial profiles
of soft X-ray emission (R2 band, 0.15-0.3 keV) and higher energy X-ray emission
(R7 band, 1-2 keV) in the Coma cluster. Surface brightness units are
$10^{-6}$ counts s$^{-1}$ arcmin$^{-2}$ pixel$^{-1}$ (Snowden et al. 1998)}
       \end{figure}

The all-sky survey data is based on very short exposures ($\sim$ 700 sec); therefore, for detailed studies
of the soft X-ray emission in the Coma cluster, we use four pointed PSPC observations.
Four additional deep PSPC observations are used to determine the local
background.
We show in section 3 below that the distribution of $N_H$ is essentially constant within 5 degrees
of the cluster center. Therefore, the off-source PSPC fields located 2.5-4 degrees away from the cluster center
provide an accurate measurement of the soft X-ray background, and are also far enough away from the cluster
center to avoid being contaminated by cluster X-ray emission (Fig. 1).

\section{Galactic HI absorption in the direction of the Coma cluster}

Knowledge of the Galactic absorption is essential to determine the intrinsic X-ray emission from extragalactic
objects, particularly at energies $\leq$ 1 keV.
Throughout this paper, we use the absorption cross sections of Morrison and McCammon (1983) in our models.

We use two methods to determine the distribution of  neutral hydrogen in the Coma cluster.
First, we use the radio measurements of Dickey and Lockman (1990) and of Hartmann and Burton (1997).
The  radio measurements are in excellent agreement, and show the measured HI column density 
varying smoothly from 9 $\times 10^{19}$ cm$^{-2}$ to 11  $\times 10^{19}$ cm$^{-2}$
within a radius of 5 degrees from the center of the Coma cluster.
Second, we employ the far-infrared IRAS data, and use  the correlation between 100 $\mu$m IRAS flux and HI column density
(Boulanger and Perault 1988). The slope of the correlation is $1.2 \times 10^{20}$ cm$^{-2}$ / (MJy sr$^{-1}$),
and the offset is determined by fixing the central $N_H$ value to 9 $\times 10^{19}$ cm$^{-2}$, which is
well established from independent radio measurements of the center of the Coma cluster (Dickey and Lockman 1990,
Lieu et al. 1996, Hartmann and Burton 1997).
The radial variation of $N_H$ inferred from the IRAS data is in extremely good agreement with the
radio measurements. The data indicate that\\
 (a) the HI column density within the central 1.5 degree of the
cluster is constant (9$\pm$1 $\times 10^{19}$ cm$^{-2}$), and \\
(b) in the region where the off-source background fields are located (2.5-4 degrees from the
cluster center), the HI column density is between
9 $\times 10^{19}$ cm$^{-2}$ and 11 $\times 10^{19}$ cm$^{-2}$. An $N_H$ variation of this magnitude has a negligible effect
on the soft X-ray flux in the PSPC R2 band (cf. Fig. 2 in Snowden et al. 1998).

\section{Spectral analysis}

The 4 PSPC Coma observations were  divided into concentric annuli centered at
R.A.=12h59'48", Dec.=25$^o$57'0" (J2000), 
and the spectra were coadded to reduce the statistical errors.
The pointed PSPC
data were reduced according to the prescriptions of Snowden et al. (1994). 
For each of the 4 off-source fields in Fig. 1, a spectrum was extracted after removal of 
point sources. The spectra were statistically consistent 
with one another within at most $\sim$ 10 \% 
point-to-point fluctuations. The off-source spectra were  therefore coadded,
and a 10\% systematic uncertainty in the background was included 
in the error analysis. Further details on the data analysis can be found in Bonamente et al. (2002).

\subsection{Single temperature fits}
Initially, we fit the spectrum of each annulus in XSPEC, using a single-temperature MEKAL plasma model 
and the WABS Galactic absorption model.
The results of the single temperature fit are given in Table 1 of Bonamente, Joy and Lieu (2003). If the 
neutral hydrogen column density is fixed at the Galactic value (9 $\times 10^{19}$ cm$^{-2}$, see section 3),
the fits are statistically unacceptable (reduced $\chi^2$ ranging from 3.1 to 9.7).
Allowing the neutral hydrogen column density to vary results in an
unrealistically low $N_H$ for all of the annuli, and also produces statistically 
unacceptable fits (reduced $\chi^2$ ranging from 2.5 to 4.2).
We conclude that a single temperature plasma model does not adequately describe the spectral data,
particularly at energies below 1 keV.
Therefore, in the analysis that follows, we fit only the high energy portion of the spectrum (1-2 keV)
with a single temperature plasma model, and introduce an additional model component to account for the 
low energy emission.

\subsection{Modelling the hot ICM}
To fit the high energy portion of the spectrum, we apply a MEKAL model to the data between 1 and 2 keV, and a 
photoelectric absorption model with $N_H=9 \times 10^{19}$ cm$^{-2}$. The metal abundance is fixed at 0.25 solar
for the central 20 arcmin region (Arnaud et al. 2001), and at 0.2 solar in the outer regions. 
The spectra are also subdivided into quadrants, in order to obtain 
a more accurate temperature for each region of the cluster.
The results of the `hot ICM' fit are given in Table 2 of Bonamente, Joy and Lieu (2003), and are consistent with the results
previously derived from the PSPC data by Briel and Henry (1997),
and with  recent XMM measurements (Arnaud et al. 2001).
In addition, the temperature found at large radii is in agreement with the composite
cluster temperature profile of De Grandi and Molendi (2002).

\subsection{Soft excess emission}
The measured  fluxes in the soft  X-ray band can now be compared with the
hot ICM model predictions in the 0.2-1 keV band. The results are shown in
Fig. 2 and in Table  2 of Bonamente, Joy and Lieu (2003). 
The error bars reflect the uncertainty in the hot ICM temperature
and the uncertainty in the Galactic HI column density 
($N_H=9\pm1 \times 10^{19}$ cm$^{-2}$).
The soft excess component  is detected with high statistical significance 
throughout the 90' radius of the pointed PSPC data, which corresponds to
a radial distance of 2.6  Mpc. The soft excess emission
(Fig. 2, left panel) is much more extended than that of the hot ICM (Fig. 2, right panel),
in agreement with the conclusions drawn from the all-sky survey data (Fig. 1).

\begin{figure}[h]
       \includegraphics[angle=90,width=4in]{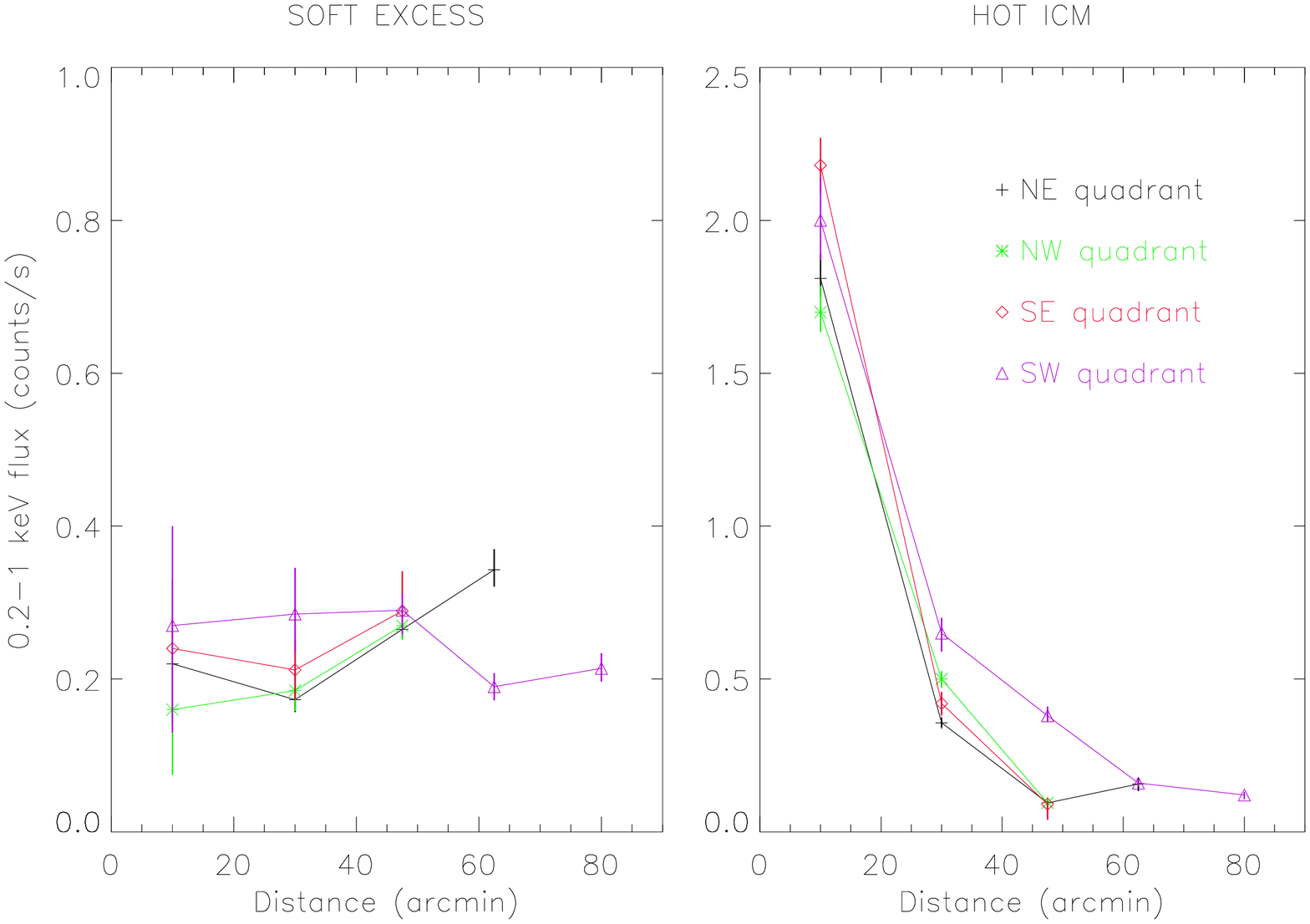}
       \caption{The radial distribution of the soft excess emission and of emission from the hot ICM}
\end{figure}

\subsection{Low energy non-thermal component}
Having established the hot ICM temperature for each quadrant, we now consider additional components
in the spectral analysis. First, we add a power law non-thermal component, which predominantly 
contributes to the low energy region of the spectrum (Sarazin and Lieu 1998). The neutral
hydrogen column density was fixed at the Galactic value ($N_H=9 \times 10^{19}$ cm$^{-2}$).
The results of fitting the hot ICM plus power law models to the annular regions are shown in Table 3 
of Bonamente, Joy and Lieu (2003).
The reduced $\chi^2$ values are poor: the average $\chi^2_{red}$ is 1.48, and the worst case value is 1.89;
we conclude that the combination of a low energy power law component and the hot ICM thermal model does not adequately
describe the PSPC spectral data.

\subsection{Low energy thermal component}
Finally, we consider a model consisting of a `hot ICM' thermal component (section 4.2)
and an additional low-temperature thermal component. As before, the neutral hydrogen column density 
was fixed at the Galactic value (see section 3). 
The results of fitting the hot ICM plus warm thermal models are shown again in Table 3
of Bonamente, Joy and Lieu (2003). The reduced $\chi^2$ values are
significantly improved relative to the previous case:
the average reduced $\chi^2$ is 1.24 and the worst case value is 1.45. In every region the 
fit obtained with a warm thermal component was superior to the fit using a non-thermal component, as
indicated by inspection of the $\chi^2_{red}$ values and by an F-test (Bevington 1969) on the
two $\chi^2$ distributions (Table 3 of Bonamente, Joy and Lieu 2003).

\section{Interpretation}

The spectral analysis of section 4 indicates that the excess emission can be explained
as thermal radiation from diffuse warm gas. The non-thermal model
appears viable only in a few quadrants, and will not be further considered in this paper.

\subsection{A warm phase of the ICM}
If the soft excess emission originates from a warm phase of the intra-cluster medium,
 the ratio of the emission integral of the hot ICM and  the 
emission integral of the warm gas (Table 3 of Bonamente, Joy and Lieu 2003) 
can be used to measure the relative mass of the two phases.
The emission integral is defined  as
\begin{equation}
I=\int n^2 dV \;,
\end{equation}
where $n$ is the gas density and $dV$ is the volume of emitting region (Sarazin 1988).
The emission integral is readily measured by fitting the X-ray spectrum. 

The emission integral of each quadrant determines the average density of the  gas
in that region, once
the volume of the emitting region is specified (Eq. 1). Assuming that each 
quadrant corresponds to a sector of a spherical shell, the density
in each sector can be calculated. 
The density of the warm gas ranges from $\sim$ 9$\times 10^{-4}$ cm$^{-3}$ 
to $\sim 8 \times 10^{-5}$ cm$^{-3}$, and the  density of the hot gas
varies from $1.5 \times 10^{-3}$ cm$^{-3}$ 
to $6 \times 10^{-5}$ cm$^{-3}$.

We assume that both the warm gas and the hot gas are distributed in spherical shells of constant density.
 Since the emission integral is proportional to $n^2 dV$ and the mass is proportional
to $n dV$, the ratio of the warm-to-hot gas mass is
\begin{equation}
\frac{M_{warm}}{M_{hot}}=\frac{\int n_{warm} dV}{\int n_{hot} dV} =  \frac{\int dI_{warm}/n_{warm}}{\int dI_{hot}/n_{hot}} 
\end{equation} 
We evaluate Eq. 2 by summing  the values of $I_{hot}/n_{hot}$ and
$I_{warm}/n_{warm}$ for all regions (Tables 2 and 3 of Bonamente, Joy and Lieu 2003),
and conclude that  $M_{warm}/M_{hot}$=0.75 within a radius of 2.6 Mpc.

\subsection{Warm filaments around the Coma cluster}
It is also possible that the warm gas is distributed in extended low-density
filaments rather then being concentrated near the cluster center like the hot ICM.
Recent large-scale hydrodynamic  simulations 
(e.g., Cen et al. 2001, Dav\'{e} et al. 2001, Cen and Ostriker 1999) indicate that
this is the case, and that 30-40 \% of the present epoch's baryons reside in these
filamentary structures. Typical filaments feature a temperature of T$\sim10^5-10^7$ K,
consistent with our results, and density of $\sim 10^{-5}-10^{-4}$ cm$^{-3}$ 
(overdensity of $\delta \sim 30-300$, Cen et al. 2001).

The ratio of mass in warm filaments to mass in the hot ICM is
\begin{equation}
\frac{M_{fil}}{M_{hot}}= \frac{\int n_{fil} dV_{fil}}{\int n_{hot} dV}= \frac{\int dI_{warm}/n_{fil}}{\int dI_{hot}/n_{hot}}.
\end{equation}

Assuming a filament density of $n_{fil}=10^{-4}$ cm$^{-3}$, Eq. (3) yields the conclusion that $M_{fil}/M_{hot}$= 3 within
a radius of 2.6 Mpc; the ratio will be even larger if the filaments are less dense.
The warm gas is therefore  more massive than the hot ICM if it is distributed in low-density
filaments. More detailed mass estimates require precise knowledge of the filaments spatial distribution.

\section{Conclusions}

The analysis of deep PSPC data of the Coma cluster reveals a large-scale
halo of soft excess radiation, considerably more extended than previously thought.
The PSPC data indicate that the excess emission is due to warm gas
at T$\sim 10^6$ K, which may exist either as
a second phase of the intra-cluster medium,
or in diffuse filaments outside the cluster. Evidence in favor of the latter scenario
is provided by the fact that the spatial extent of the
soft excess emission is significantly greater than  that of the hot ICM.

The total mass of the Coma cluster within 14 Mpc is 1.6$\pm0.4 \times 10^{15} M_{\bigodot}$
(Geller, Diaferio and Kurtz).
The mass of the  hot ICM is
$\sim 4.3 \times 10^{14} M_{\bigodot}$ within 2.6 Mpc (Mohr, Mathiesen and Evrard 1999).
The present detection of soft excess emission out to a distance of 2.6 Mpc from the cluster's
center implies that the warm gas has a mass of at least $ 3 \times 10^{14} M_{\bigodot}$,
or considerably  larger if the gas is in very low density filaments.
The PSPC data presented in this paper 
therefore lends observational support to the current theories of large-scale formation and evolution (e.g,
Cen and Ostriker 1999),
which predict that a large fraction of the current epoch's baryons are in
a diffuse warm phase of the intergalactic medium.

\articletitle{Soft X-ray Excess Emission In Three Clusters Of Galaxies Observed With XMM-Newton}
\author{J.Nevalainen,R.Lieu,M.Bonamente,D.Lumb}

\begin{figure}[h]
\includegraphics[width=1.2\textwidth,viewport=130 155 540 564,clip]{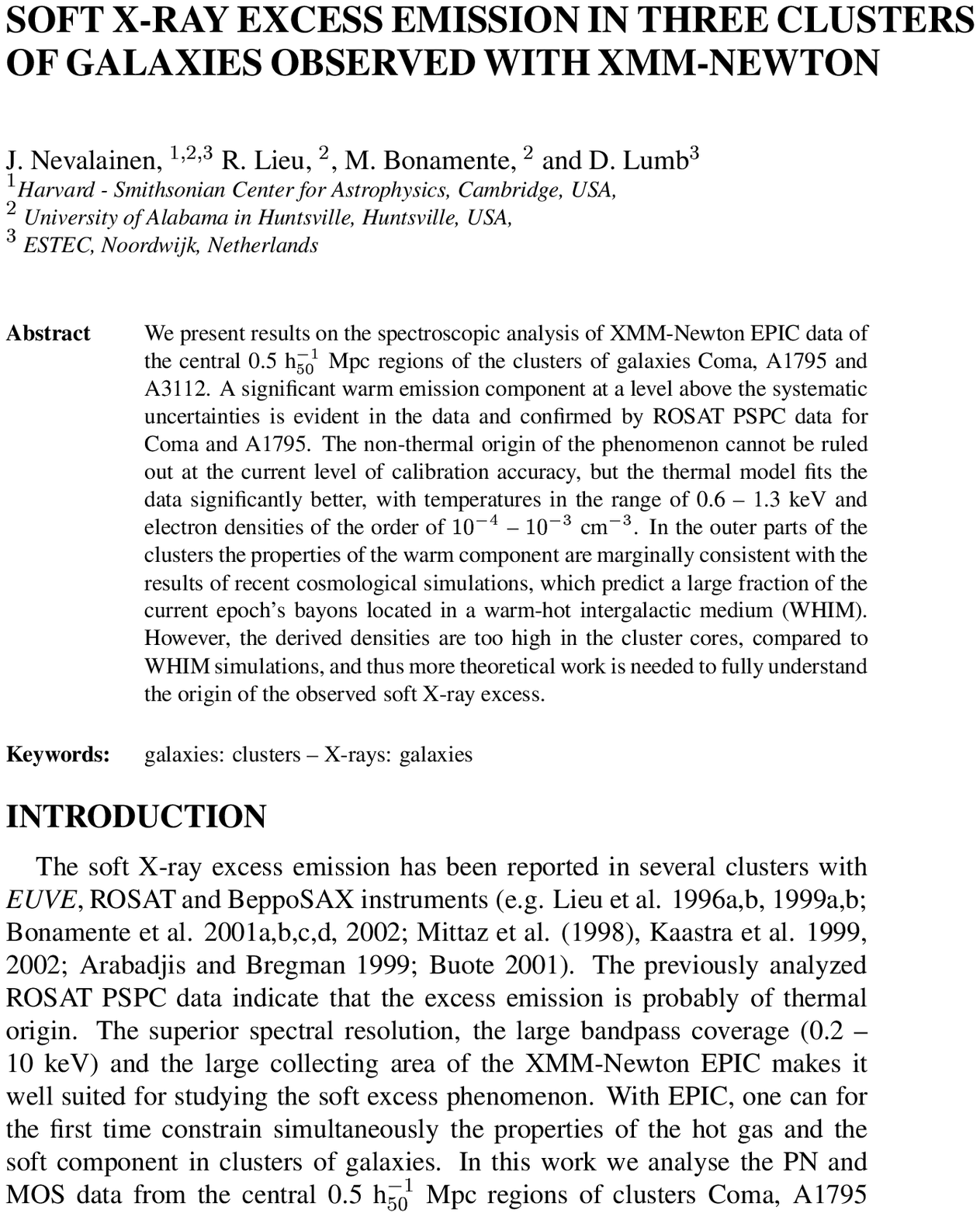}
\end{figure}
\begin{figure}[h]
\includegraphics[width=1.2\textwidth,viewport=130 120 540 690,clip]{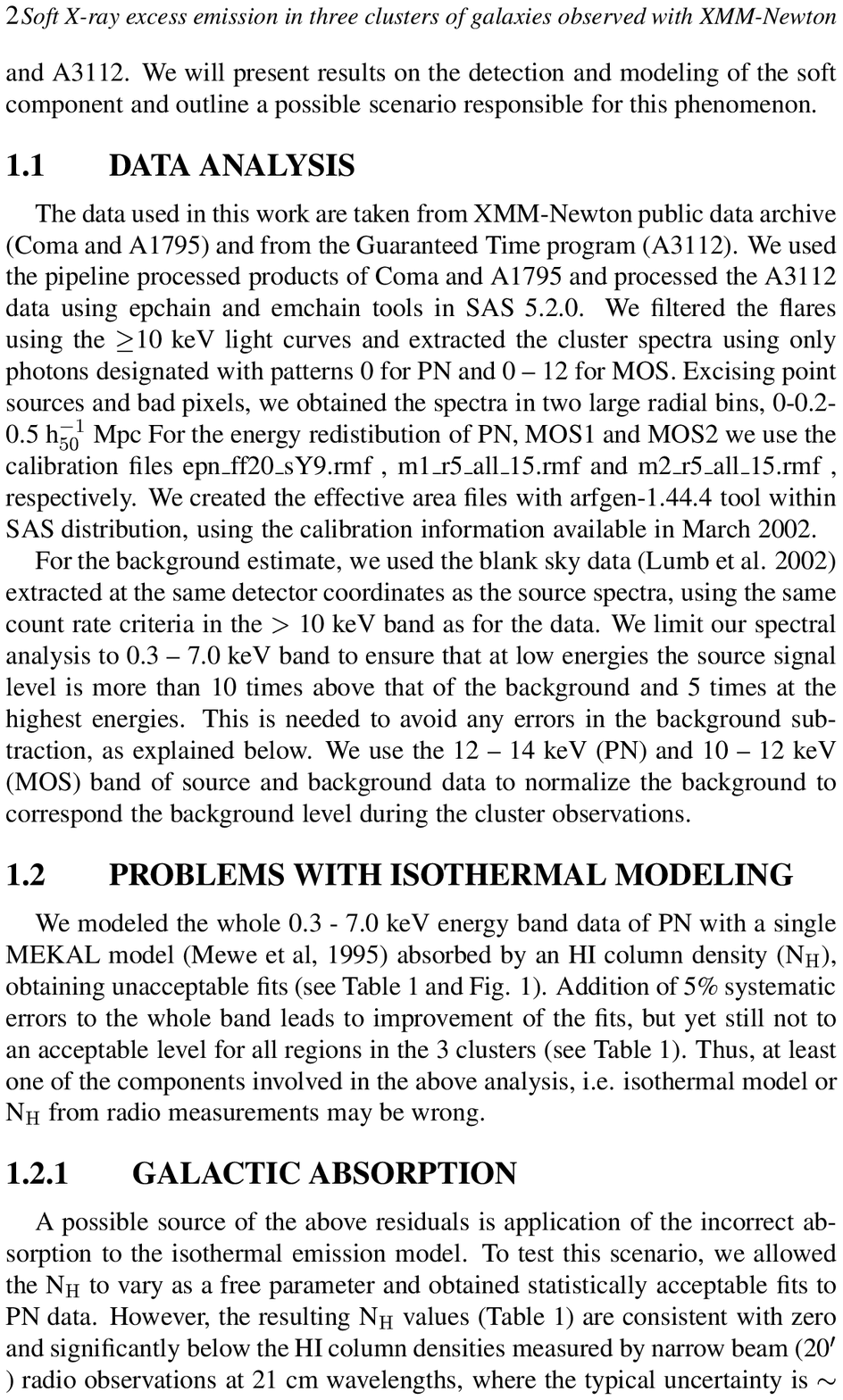}
\end{figure}
\begin{figure}[h]
\includegraphics[width=1.2\textwidth,viewport=130 120 540 690,clip]{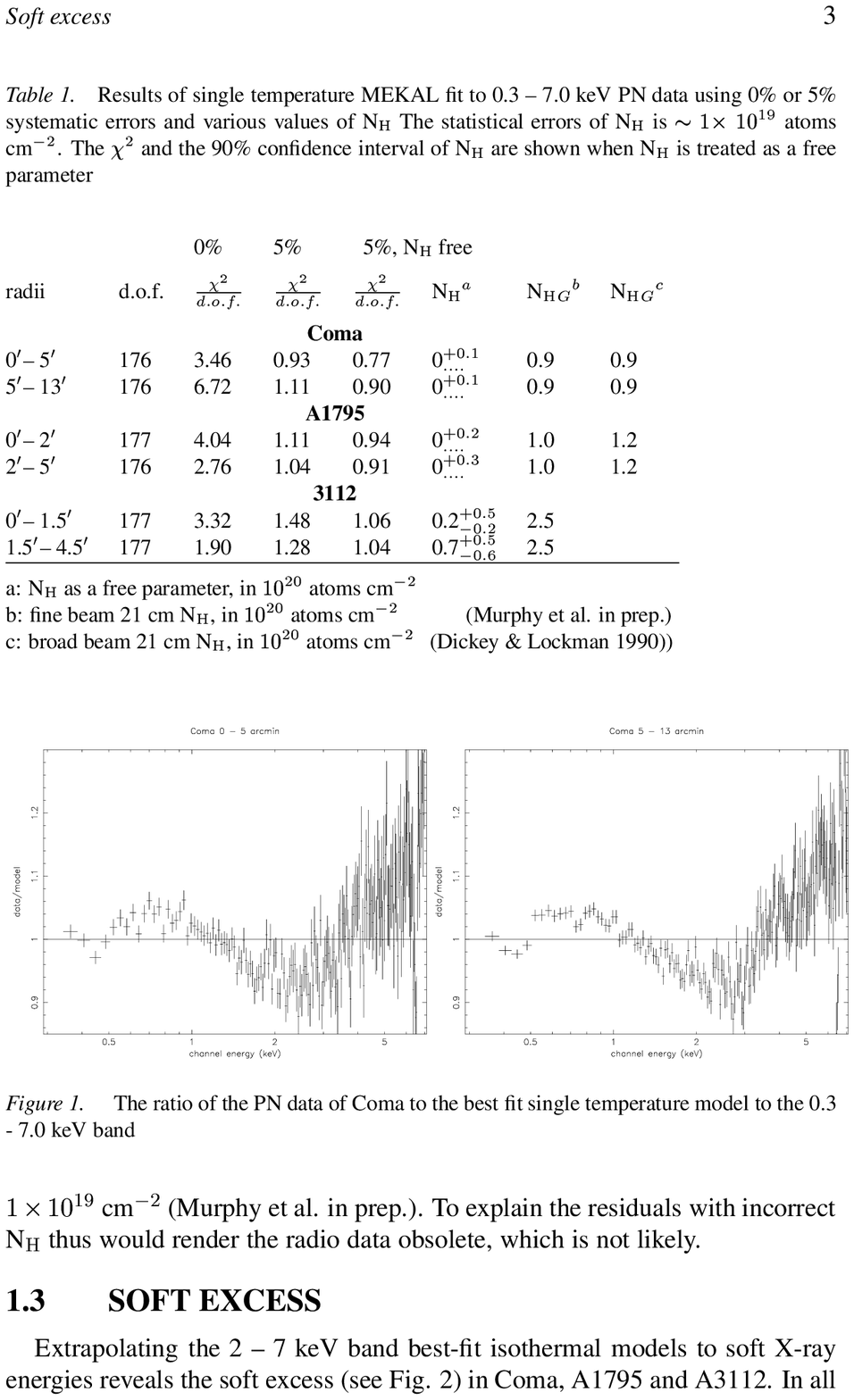}
\end{figure}
\begin{figure}[h]
\includegraphics[width=1.2\textwidth,viewport=130 120 540 690,clip]{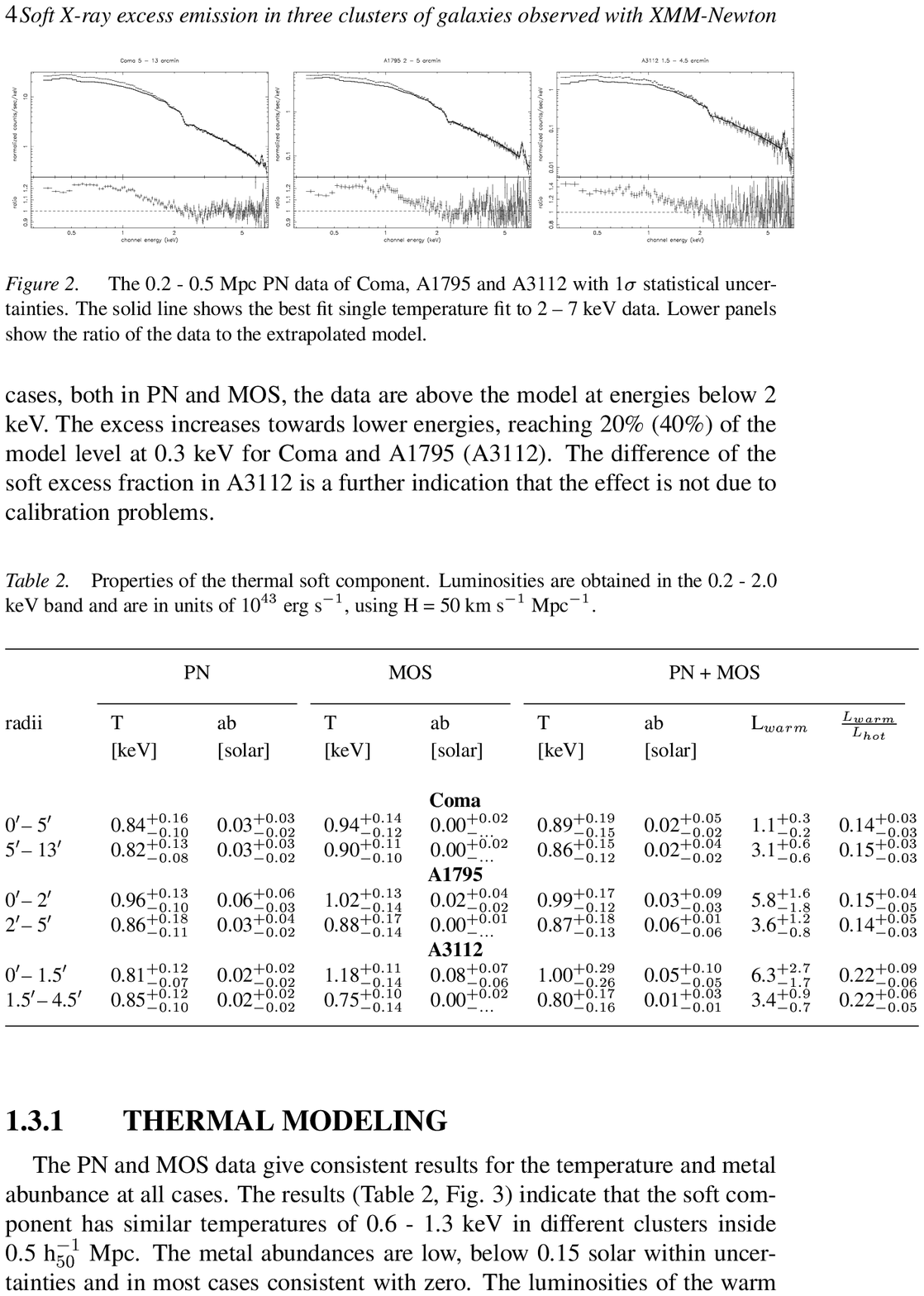}
\end{figure}
\begin{figure}[h]
\includegraphics[width=1.2\textwidth,viewport=130 120 540 690,clip]{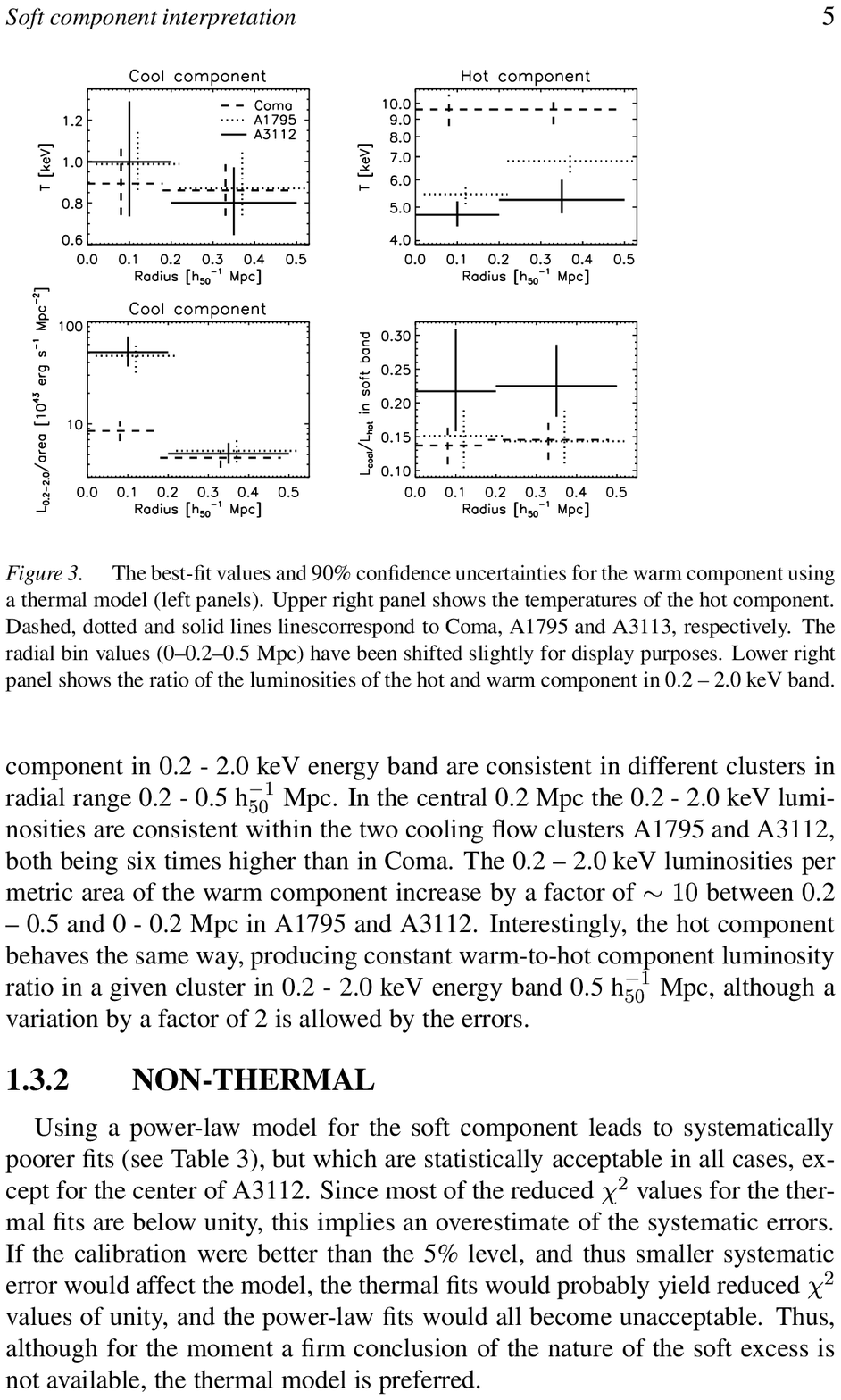}
\end{figure}
\begin{figure}[h]
\includegraphics[width=1.2\textwidth,viewport=130 120 540 690,clip]{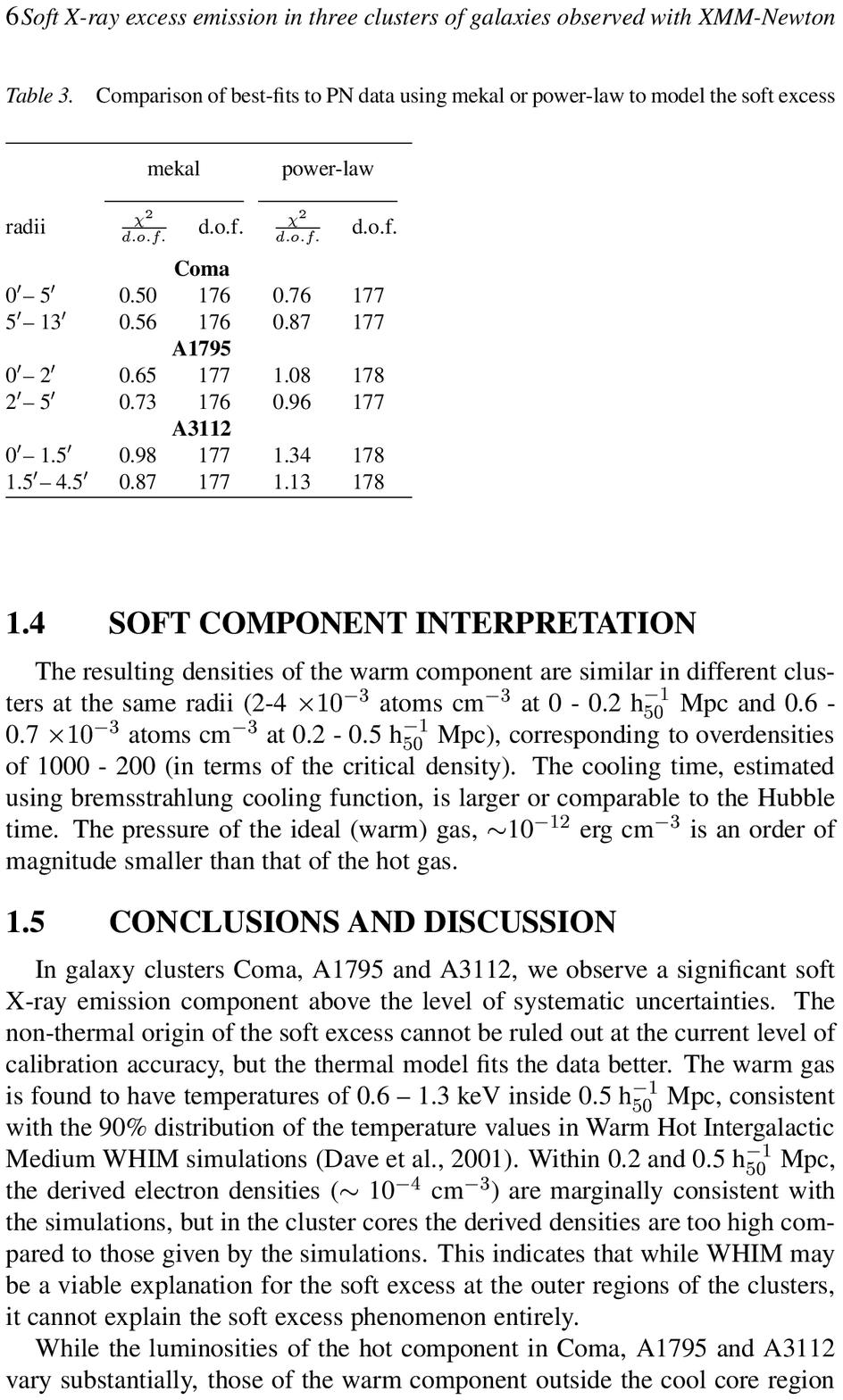}
\end{figure}
\begin{figure}[h]
\includegraphics[width=1.2\textwidth,viewport=130 120 540 690,clip]{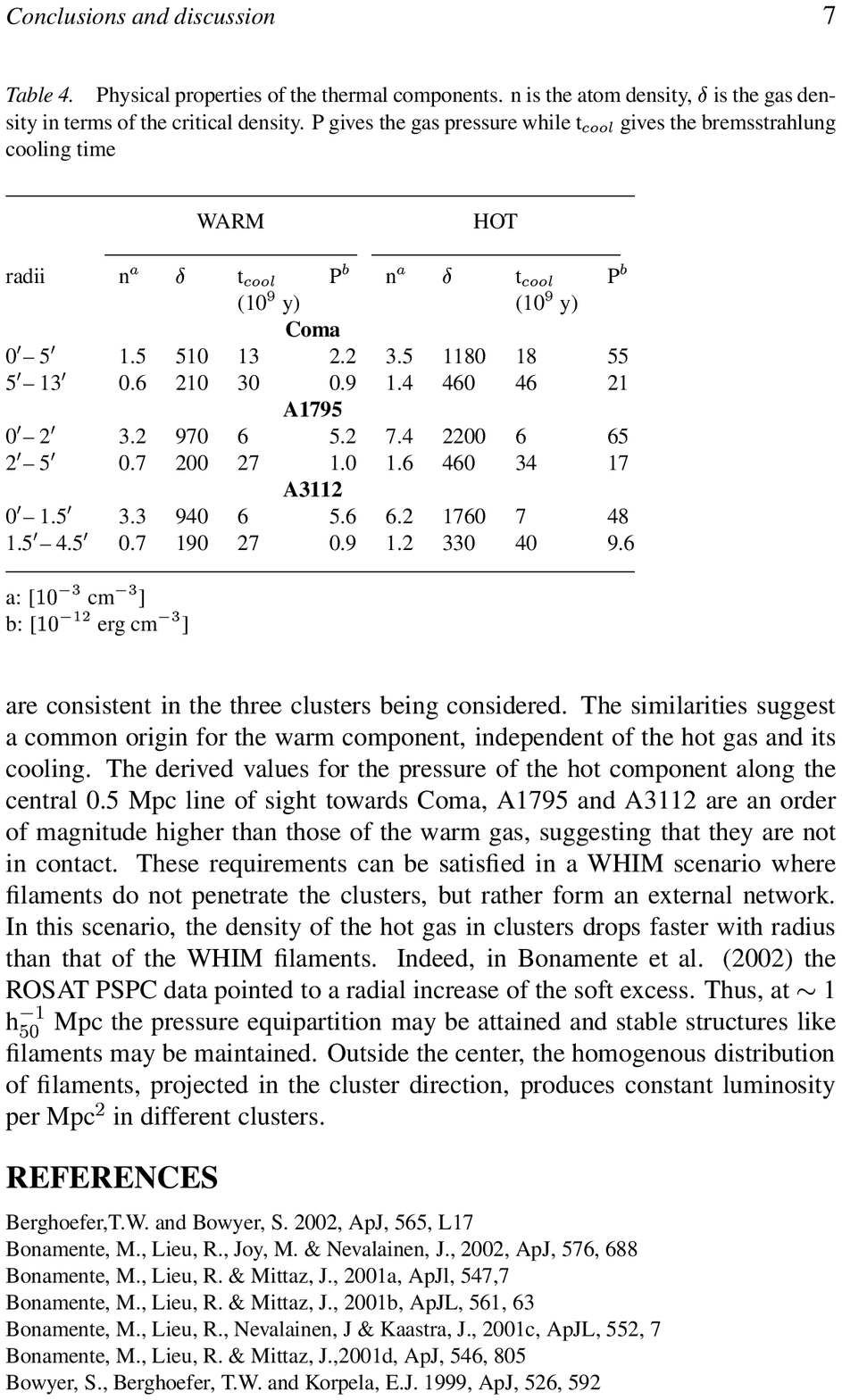}
\end{figure}
\begin{figure}[h]
\includegraphics[width=1.2\textwidth,viewport=130 120 540 690,clip]{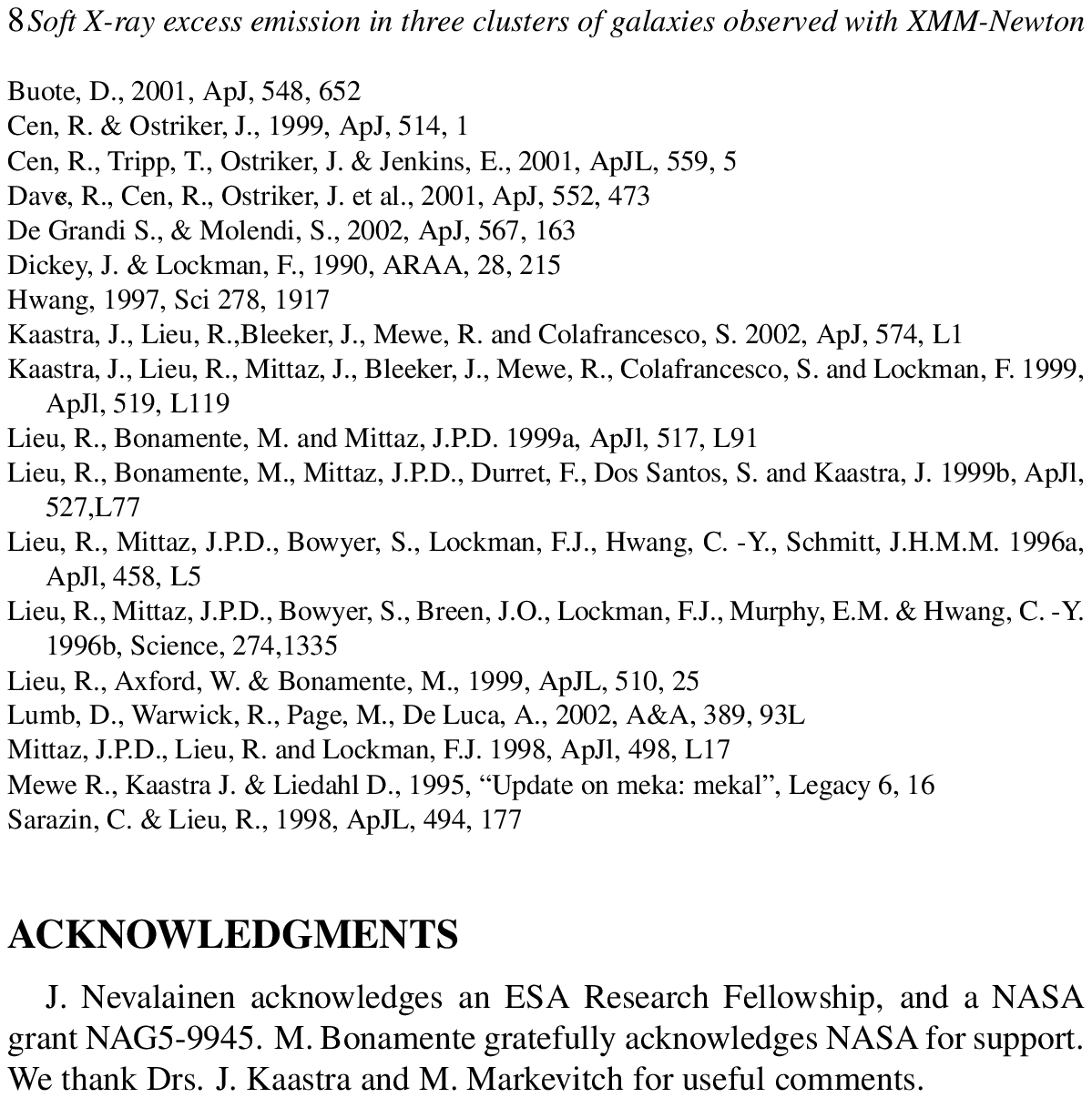}
\end{figure}







\articletitle[XMM-Newton discovery of O VII emission from warm gas in
clusters of galaxies]{XMM-Newton discovery of O VII emission from warm gas in
clusters of galaxies}

\chaptitlerunninghead{XMM-Newton discovery of O VII}

\author{Jelle S. Kaastra\altaffilmark{1}, R. Lieu\altaffilmark{2}, 
T. Tamura\altaffilmark{1}, F.B.S. Paerels\altaffilmark{3} 
and J.W.A. den Herder\altaffilmark{1}}

\altaffiltext{1}{SRON National Institute for Space Research, Utrecht, The Netherlands}
\altaffiltext{2}{University of Alabama at Huntsville, USA}
\altaffiltext{3}{Columbia University, New York, USA}

\begin{abstract}
XMM-Newton recently discovered O~VII line emission from $\sim$ 2 million K gas
near the outer parts of several clusters of galaxies.  This emission is
attributed to the Warm-Hot Intergalactic Medium.  The original sample of
clusters studied for this purpose has been extended and two more clusters with a
soft X-ray excess have been found.  We discuss the physical properties of the
warm gas, in particular the density, spatial extent, abundances and temperature.
\end{abstract}


\section{Introduction}

Soft excess X-ray emission in clusters of galaxies was first discovered using
EUVE DS and Rosat PSPC data in the Coma and Virgo cluster (Lieu et al.
1996a,b).  It shows up at low energies as excess emission above what is expected
to be emitted by the hot intracluster gas, and it is often most prominent in the
outer parts of the cluster.  However, a serious drawback with the old data
(either EUVE or Rosat) concerns their spectral resolution, which does not exist
for the EUVE DS detector, and is very limited for the Rosat PSPC at low energies
where the width of the instrumental broadening causes a significant
contamination of the count rate by harder photons.

The forementioned reasons render it very difficult for firm conclusions about
the nature of the soft excess emission to be deduced from the original data
alone.  For example, already in the first papers it was suggested that the
emission may have a thermal origin, but is also consistent with it having a
power law spectrum caused by Inverse Compton scattering of the cosmic microwave
background on cosmic ray electrons (Sarazin \& Lieu 1998).

With the launch of XMM-Newton it is now possible to study the soft excess
emission with high sensitivity and with much better spectral resolution using
the EPIC camera's of this satellite.  The high resolution Reflection Grating
Spectrometer (RGS) of XMM-Newton has proven to be extremely useful in studies of
the central cooling flow region, but due to the very extended nature of the soft
excess emission, the RGS is not well suited to study this phenomemon.

\section{XMM-Newton observations}

XMM-Newton has by now observed a large number of clusters.  We investigated the
presence of soft excess emission in a sample of 14 clusters of galaxies.  This
work has been published by Kaastra et al.  (2003a).  In that paper the details
of the data analysis are given.  Briefly, much effort was devoted to subtracting
properly the time-variable soft proton background, as well as the diffuse X-ray
background.  We made a carefull assesment of the systematic uncertainties in the
remaining background, since a proper background subtraction has been one of the
contentious issues in the discussion around the discovery of the soft excess in
EUVE and Rosat data.  In a similar way, the systematic uncertainties in the
effective area and instrumental response of the EPIC camera's were carefully
assessed and quantified.  In the spectral fitting procedures, both the
systematic uncertainties in the backgrounds and the instrument calibration were
taken into account.  Spectra were accumulated in 9 concentric annuli between 0
and 15 arcmin from the center of the cluster.

The original sample used by Kaastra et al (2003a) has been extended to 21
clusters using archival data (see also Kaastra et al.  2003c).  These additional
clusters were analyzed in exactly the same way as the original 14 clusters.  The
spectra were initially analyzed using a two temperature model for the hot gas,
with the second temperature of the component fixed at half that of the first
component.  From experience with cooling flow analysis (Peterson et al.  2003;
Kaastra et al 2003b) we learnt that such a temperature parameterization is
sufficient to characterise fully the cooling gas in the cores of clusters, while
in the outer regions it is an effective method to take the effects of eventual
non-azimuthal variations in the annular spectra into account.  We note that the
temperature of the coolest "hot gas" component in all cases where we detect a
soft excess is much higher than the (effective) temperature of the soft excess.
This is due to the now well-known fact that the emission measure distribution of
the cooling flow drops off very rapidly.  In fact, our models for the cooling
flow predict no significant emission from O VII ions in the cooling plasma (at
least below the detection limit of XMM-Newton).

The presence of a soft excess in this sample of clusters was tested by formally
letting the Galactic absorption column density be a free parameter in the
spectral fitting.  Of the 21 clusters, 5 have apparent excess absorption.  All
these 5 clusters are located in regions where dust etc.  is important, or they
have a very compact core radius such that the temperature gradients in the core
are not fully resolved by XMM-Newton and therefore the spectra are highly
contaminated.

While the excess absorption in 5 clusters can be fully explained, the absorption
deficit in 7 of these clusters cannot be explained by uncertainties in the
calibration, background emission or foreground absorption, but only by the
presence of an additional emission component.  In fact, in several of the
clusters the best-fit column density is zero!

\begin{figure}
{\includegraphics[height=\hsize,width=0.30\vsize,angle=-90]{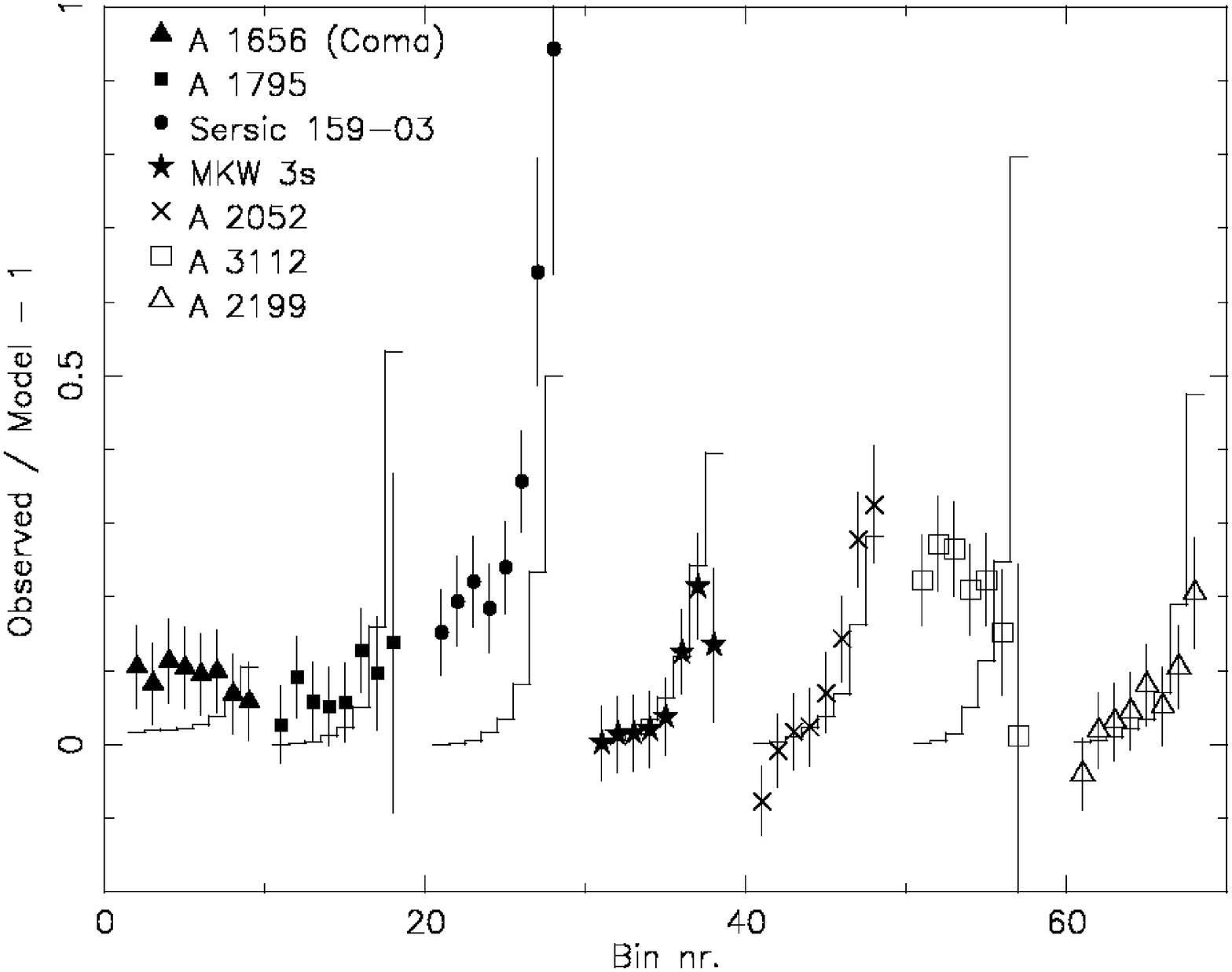}}
\caption{Soft excess in the 0.2--0.3~keV band as compared with a two temperature
model (data points with error bars).  The solid histogram is the predicted soft
excess based upon scaling the sky-averaged XMM-Newton background near the
cluster with the relative enhancement of the soft X-ray background derived from
the 1/4~keV PSPC images.  For the $k$th cluster the values for annulus $j$ are
plotted at bin number $10(k-1)+j$.}\label{k-fig:soft}
\end{figure}

In Fig.~\ref{k-fig:soft} we show the soft excess in the 0.2--0.3~keV band for the
seven clusters with a significant soft component.  These clusters are Coma,
A~1795, S\'ersic~159$-$03, MKW~3s, A~2052 (see also Kaastra et al.  2003a);
A~3112 (see also Nevalainen et al.  2003 and Kaastra et al.  2003c) and A~2199
(paper in preparation).

The field of view of XMM-Newton is relatively small ($\sim$15 arcmin radius) and
therefore all these relatively nearby clusters fill the full field of view.  For
this reason, only a sky-averaged soft X-ray background (obtained from deep
fields) as well as the time variable soft proton background (which is relatively
small at low energies) were subtracted from the XMM-Newton data.  Using the
1/4~keV Rosat PSPC sky survey data (Snowden et al.  1995), maps at 40 arcmin
resolution were produced to estimate the average soft X-ray background in an
annulus between 1--2 degrees from the cluster.  Most of these seven clusters
show an enhanced 1/4~keV count rate in this annulus (as compared to the typical
sky-averaged 1/4~keV count rate).  This, combined with the decreasing density of
the hot gas in the outer parts of the cluster causes the apparent soft excess
with increasing relative brightness at larger radii in Fig.~\ref{k-fig:soft}.  The
figure shows complete consistency of the XMM-Newton data with the PSPC 1/4~keV
data in this respect for A~1795, S\'ersic~159$-$03, MKW~3s, A~2052 and A~2199.
We show below that in these clusters this large-scale soft X-ray emission is due
to thermal emission from the (super)cluster environment.

However, there is an additional soft component in Coma, A~3112, S\'ersic
159$-$03 and perhaps A~1795.  The fact that this component is above the
prediction from the large scale PSPC structures implies that its spatial extent
is at most 10--60 arcmin.  We shall return to this component in Sect.~5.

\section{Emission from the Warm-Hot Intergalactic Medium}

\begin{figure}
\resizebox{\hsize}{!}{\includegraphics[angle=0]{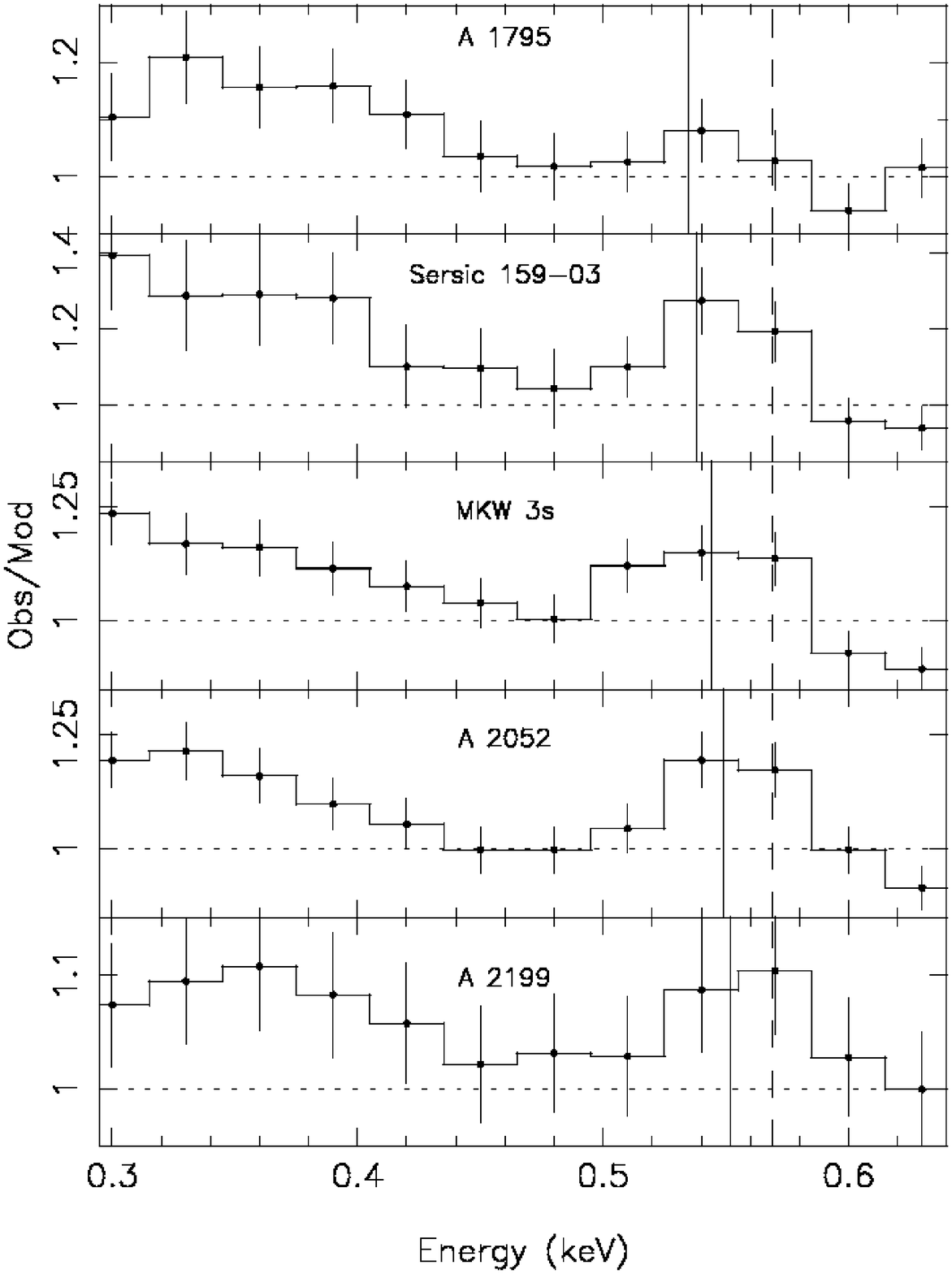}}
\caption{Fit residuals with respect to a two temperature model for the outer
4--12 arcmin part of six clusters.  The position of the O~VII triplet in the
cluster restframe is indicated by a solid line and in our Galaxy's rest frame by
a dashed line at 0.569~keV (21.80~\AA).  The fit residuals for all instruments
(MOS, pn) are combined.  The instrumental resolution at 0.5~keV is $\sim$60~eV
(FWHM).}
\label{k-fig:o7}
\end{figure}

In the previous section we found that five clusters show a soft excess in the
0.2--0.3~keV band at a spatial scale of at least 1--2 degrees, combining our
XMM-Newton spectra with PSPC 1/4~keV imaging.  It is not obvious a priori
whether this excess emission is due to emission from the cluster region or
whether it has a different origin, for example galactic foreground emission.
Here the spectral resolution of the EPIC camera's is crucial in deciding which
scenario is favoured.  In Fig.~\ref{k-fig:o7} we show the fit residuals of the fit
with two hot components and Galactic absorption only, in the outer 4--12 arcmin
region combining all three EPIC camera's.  The fit residuals show two distict
features:  a soft excess below 0.4--0.5~keV, and an emission line at
$\sim$0.56~keV.  This emission line is identified as the O~VII triplet, and
detailed spectral fitting shows that both phenomena (soft excess and O~VII line)
can be explained completely by emission from a warm plasma with a temperature of
0.2~keV (see Kaastra et al.  2003a for more details).  Thus, the soft excess has
a thermal origin.  Moreover, the centroid of the O~VII triplet (which is
unresolved) agrees better with an origin at the redshift of the cluster than
with redshift zero.  This clearly shows that the thermal emission has an origin
in or near the cluster, although a partial contribution from Galactic foreground
emission cannot be excluded in all cases.  We also note that in A~1795 and
A~2199 the O~VII line is relatively weak and needs more confirmation.

Taking this additional soft thermal component into account in the spectral
fitting yields fully acceptable fits.  In fact, in the energy band below 1~keV,
the soft component contributes 20--40~\% of the X-ray flux of the outer (4--12
arcmin) part of the cluster!

We identify this component as emission from the Warm-Hot Intergalactic Medium
(WHIM).  Numerical models (for example Cen \& Ostriker 1999, Fang et al.  2002)
show that bright clusters of galaxies are connected by filaments that contain a
significant fraction of all baryonic matter.  Gas falls in towards the clusters
along these filaments, and is shocked and heated during its accretion onto the
cluster.  Near the outer parts of the clusters the gas reaches its highest
temperature and density, and it is here that we expect to see most of the X-ray
emission of the warm gas.

\section{Properties of the warm gas}

The temperature of the warm gas that we find for all our clusters is 0.2~keV.
The surface brightness of the warm gas within the field of view of the
XMM-Newton telescopes is approximately constant, with only a slight enhancement
towards the center for some clusters (a stronger increase towards the center for
S\'ersic 159$-03$ is discussed in the next section).  In Table~1 we list the
central surface brightness $S_0$ as estimated from our XMM-Newton data,
expressed as the emission measure per solid angle.  We use
$H_0=70$~km\,s$^{-1}$\,Mpc$^{-1}$ throughout this paper.  Using the known
angular distance to the cluster we estimate the same quantity in units of
m$^{-5}$ (see also Table~1).  We then use a simplified model for the geometry of
the emitting warm gas, namely a homogeneous sphere with uniform density.  The
radius $R$ of this sphere is estimated from the radial profile of the Rosat PSPC
1/4~keV profile around the cluster, and is also listed in Table~1.  We find
typical radii of 2--6~Mpc, i.e.  the emission occurs on the spatial scale of a
supercluster.  From this radius and the emission measure, the central hydrogen
density $n_{\mathrm H}(0)$ is estimated.  We find typical densities of the order
of 50-150~m$^{-3}$.  Assuming a different density profile (for example
$n(r)=n(0)[1+(r/a)^2]^{-1}$ for $r<R$ and $n=0$ for $r>R$) yields central
densities that are only 20--50~\% larger.  These densities are 200--600 times
the average baryon density of the universe.  We also estimate the total hydrogen
column density, which is typically $1.6-2.8\times 10^{25}$~m$^{-2}$.  Using then
the measured oxygen abundances from our XMM-Newton data (essentially determined
by the ratio of the O~VII triplet to the soft X-ray excess), which are typically
0.1 times solar, we then derive total O~VII column densities of the order of
$0.4-0.9\times 10^{21}$~m$^{-2}$.  These column densities and the typical sizes
of the emitting regions are similar to those as calculated for the brightest
regions in the simulations of Fang et al.  (2002).  We have taken here a solar
oxygen abundance of $8.5\times 10^{-4}$ and an O~VII fraction of 32~\%,
corresponding to a plasma with a temperature of 0.2~keV.

\begin{table}
\caption{Properties of the warm gas}
\begin{tabular*}{\textwidth}{@{\extracolsep{\fill}}lccccc}
\hline
Parameter & A~1795 & S\'ersic 159$-03$ & MKW~3s & A~2052 & A~2199 \\
\hline
Redshift & 0.064 & 0.057 & 0.046 & 0.036 & 0.030 \\
Scale (kpc/arcmin) & 71 & 64 & 52 & 41 & 35 \\
$S_0$\,$^a$ & 49$\pm$37  & 71$\pm$41 & 84$\pm$18 & 73$\pm$15 & 28$\pm$13 \\
$S_0$\,$^b$ & 10   & 18  & 33  & 46  & 24 \\
$R$ (arcmin) & 80  & 60  & 60 & 80 & 60 \\
$R$ (Mpc)    & 5.7 & 3.8 & 3.1 & 3.3 & 2.1 \\
$n_{\mathrm H}(0)$ (m$^{-3}$) & 45 & 80 & 120 & 140 & 120 \\
$N_{\mathrm H}(0)$ ($10^{24}$~m$^{-2}$) & 16 & 19 & 23 & 28 & 16 \\
Abundance O & 0.08$\pm$0.05 & 0.08$\pm$0.03 & 0.09$\pm$0.03 & 0.12$\pm$0.03 & 
         0.16$\pm$0.05 \\
$N_{\mathrm O VII}(0)$ ($10^{20}$~m$^{-2}$) & 4 & 4 & 6 & 9 & 7 \\
$M_w$ ($10^{15}$~M$_{\odot}$) & 1.2 & 0.6 & 0.5 & 0.7 & 0.2 \\
$M_{\mathrm A}$ ($10^{15}$~M$_{\odot}$) & 1.1 & 0.5 & 0.5 & 0.4 & 0.6 \\
\hline
\end{tabular*}
\begin{tablenotes}
$^a$Surface brightness, expressed as emission measure per
solid angle ($10^{68}$~m$^{-3}$arcmin$^{-2}$).

$^b$Surface brightness in $10^{26}$~m$^{-5}$.
\end{tablenotes}
\end{table}

Finally, we determined the total mass of the warm gas ($M_w$, Table~1).  This
mass is for most clusters comparable to the total cluster mass $M_A$ within the
Abell radius (2.1~Mpc for our choice of $H_0$) as derived by Reiprich \&
B\"ohringer (2002) for the same clusters.

We make here a remark on MKW~3s and A~2052.  These clusters are separated by
only 1.4 degree and both belong to the southernmost extension of the Hercules
supercluster (Einasto et al.  2001).  A~2199, at 35 degrees to the North, is at
the northernmost end of the same supercluster.  The redshift distribution of the
individual galaxies in the region surrounding MKW~3s ($z=0.046$) and A~2052
($z=0.036$) shows two broad peaks centered around the redshifts of these
clusters, but galaxies with both redshifts are found near both clusters.
Therefore this region has a significant depth (43~Mpc) as compared to the
projected angular separation (4.3~Mpc).  The relative brightness of the warm gas
near these clusters (as seen for example from the value of $S_0$) is then
explained naturally if there is a filament connecting both clusters.  In that
case we would see the filament almost along its major axis.

\section{Non-thermal emission?}

Apart from the large scale, extended emission from the warm gas some clusters
also exhibit a centrally condensed soft excess component (Fig.~3).  In MKW~3s,
A~2052 and A~2199 this central enhancement is relatively weak.  It could be a
natural effect of the enhanced filament density close to the cluster centers.
In A~1795 and S\'ersic 159$-03$ the enhancement is much larger.  It is unlikely
that this emission component for the latter two clusters also originates from
projected filaments in the line of sight - the high surface brightness would
necessitate filaments of length far larger than a cluster's dimension.  Another
possibility is warm gas within the cluster itself.  In order to avoid the rapid
cooling which results from this gas assuming a density sufficient to secure
pressure equilibrium with the hot virialized intracluster medium, it should be
be magnetically isolated from the hot gas.  Yet another viable model for the
central soft component is non-thermal emission.  We note that the soft excess in
the center of Coma and A~3112 also possibly has a non-thermal origin, as there
is no clear evidence for oxygen line emission in their spectra.  Clearly, deeper
spectra and in particular a higher spectral resolution is needed to discriminate
models.  At this conference, several new mission concepts have been presented
that may resolve these issues in the near future.

\begin{figure}
{\includegraphics[height=\hsize,width=0.55\vsize,angle=-90]{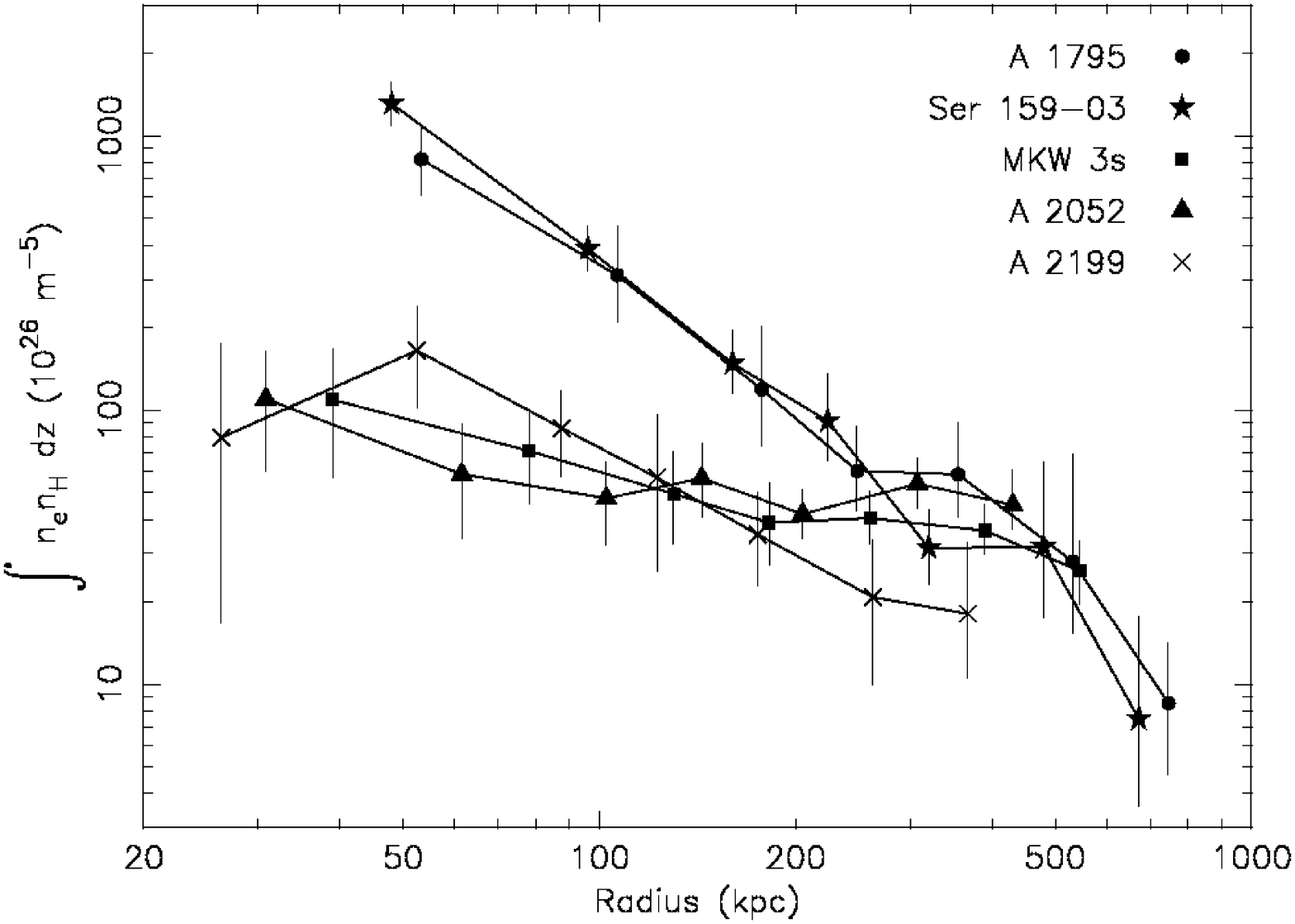}}
\caption{Emission measure integrated along the line of sight for five
clusters of galaxies.}
\label{k-fig:sb}
\end{figure}

\vspace{0.1cm}
\noindent{\bf Acknowledgments}
This work is based on observations obtained with XMM-Newton, an ESA science
mission with instruments and contributions directly funded by ESA Member States
and the USA (NASA).  SRON is supported financially by NWO, the Netherlands
Organization for Scientific Research.

\articletitle{XMM-Newton discovery of an X-ray filament in Coma.}


\author{A. Finoguenov\altaffilmark{1}, U.G.Briel\altaffilmark{1} and J.P.Henry\altaffilmark{2}}
\altaffiltext{1}{Max-Planck-Institut fuer extraterrestrische Physik,\\
Giessenbachstra\ss e, 85748 Garching, Germany}


\altaffiltext{2}{Institute for Astronomy, University of Hawaii,\\
2680 Woodlawn Drive, Honolulu, Hawaii 96822, USA}

\begin{abstract}
XMM-Newton observations of the outskirts of the Coma cluster of
galaxies confirm the existence of warm X-ray gas claimed previously and
provide a robust estimate of its temperature ($\sim0.2$ keV) and oxygen
abundance ($\sim0.1$ solar). Associating this emission with a 20 Mpc infall
region in front of Coma, seen in the skewness of its galaxy velocity
distribution, yields an estimate of the density of the warm gas of $\sim 50
\rho_{critical}$. Our measurements of gas mass associated with the warm
emission strongly support its nonvirialized nature, suggesting that we are
observing the warm-hot intergalactic medium (WHIM). Our measurements provide
a direct estimate of the O, Ne and Fe abundance of the WHIM.  Differences
with the reported Ne/O ratio for some OVI absorbers hints to different
origin of the OVI absorbers and the Coma filament. 
\end{abstract}

\begin{keywords}
clusters: individual: Coma ---
cosmology: observations --- intergalactic medium --- large-scale structure
\end{keywords}

\section*{Introduction}
The total amount of baryons found in the local Universe accounts only for
10\% of the values implied from the primordial nucleosynthesis, observed at
high-redshift (Fukigita, Hogan, Peebles 1998 and references therein) and
within the local closed box systems, such as clusters of galaxies (White,
Briel, Henry 1993). It has been predicted by Cen \& Ostriker (1999) that
most of the missing baryons reside in the warm-hot intergalactic medium
(WHIM), owing to the shocks driven by the large-scale structure formation
that results in an order of magnitude drop in the star-formation rate
density (Nagamine et al. 2001).

Since then, several surveys have been undertaken to find the WHIM.  Tripp,
Savage, Jenkins (2000) have claimed a substantial fraction of the missing
baryons in the form of OVI absorbers. However, their observations lack
information on the ionization equilibrium as well as O abundance required to
link the abundance of OVI ions to underlying O mass and then to the amount
of baryons. First Chandra grating results (Nicastro et al. 2002, Mathur,
Weinberg, Chen 2002) on detection of OVII absorption lines allowed, for the
first time, an estimate of the ionization state of the gas, suggesting that
OVI absorbers account for over 80\% of the local baryons. Still, such claims
rely on the assumed O abundance.

Searches for the warm X-ray emission with ROSAT, have found some association
of soft X-ray emission with galactic filaments (Scharf et al. 2000;
Zappacosta et al. 2002). Based on the prediction that hotter gas will be in
denser regions, search for the X-ray emission from the regions connecting
clusters of galaxies has been a separate, although related, topic. Briel \&
Henry (1995) set upper limits with stacked analysis of ROSAT all-sky survey
data on the emitting gas density to be less than $7.4\times10^{-5}h^{1/2}$
cm$^{-3}$, assuming a temperature of 0.5 keV and an iron abundance of 0.3
solar. 

In this {\it Contribution} we carry out a spectroscopic study of the warm
X-ray gas in the outskirts of the Coma cluster, using the mosaic of
observations carried out by the XMM-Newton. We adopt the Coma cluster
redshift of 0.023, $H_{\rm o}=70$~km~s$^{-1}$~Mpc$^{-1}$, and quote
errorbars at the 68\% confidence level. One degree corresponds to 1.67 Mpc.

\section{Observations}\label{f-s:res}

The initial results of the XMM-Newton performance
verification observations of the Coma cluster are reported in Briel et al.
(2001 and references therein). In addition to the
observations reported there, three additional observations have been carried
out, completing the planned survey of the Coma cluster. 

The Coma observations reported here are performed with EPIC pn detector,
using its medium filter, which leads to a different detector countrate from
the cosmic and galaxy emission in the soft band, compared to the usual thin
filter.  We used a 100 ksec observation of the point source APM08279+5255,
performed with the same (medium) filter, to measure these two diffuse
emission components. To ensure the compatibility of the X-ray background, we
excise from the spectral extraction region the APM08279+5255 quasar and the
X-ray sources in Coma identified with the Coma galaxies (which otherwise
increase the apparent CXB flux by 10\%). Since there is a mismatch in the
$N_H$ toward the Coma and the APM fields, we use an additional background
observation, obtained with the filter wheel closed, to extract the
foreground (the Local Bubble and the Milky Way halo, e.g. Lumb et al. 2002)
and CXB spectra in APM field. We then use the same spectral shape and
normalization, but change the $N_H$ from $3.9\pm0.3\times10^{20}$ cm$^{-2}$
(APM) to $9\times10^{19}$ cm$^{-2}$ in the direction of the Coma
cluster. Since some foreground components, like the Local Bubble, are not
subject to the absorption, we introduce a separate unabsorbed component of
0.07 keV temperature (Lumb et al. 2002) in both the Coma and the APM
fields. We have also checked that the derived results on the soft X-ray
emission from Coma do not depend on the possible large-scale variation in
the intensity of the 0.07 keV component.

\begin{figure}[ht]
\includegraphics[width=12cm]{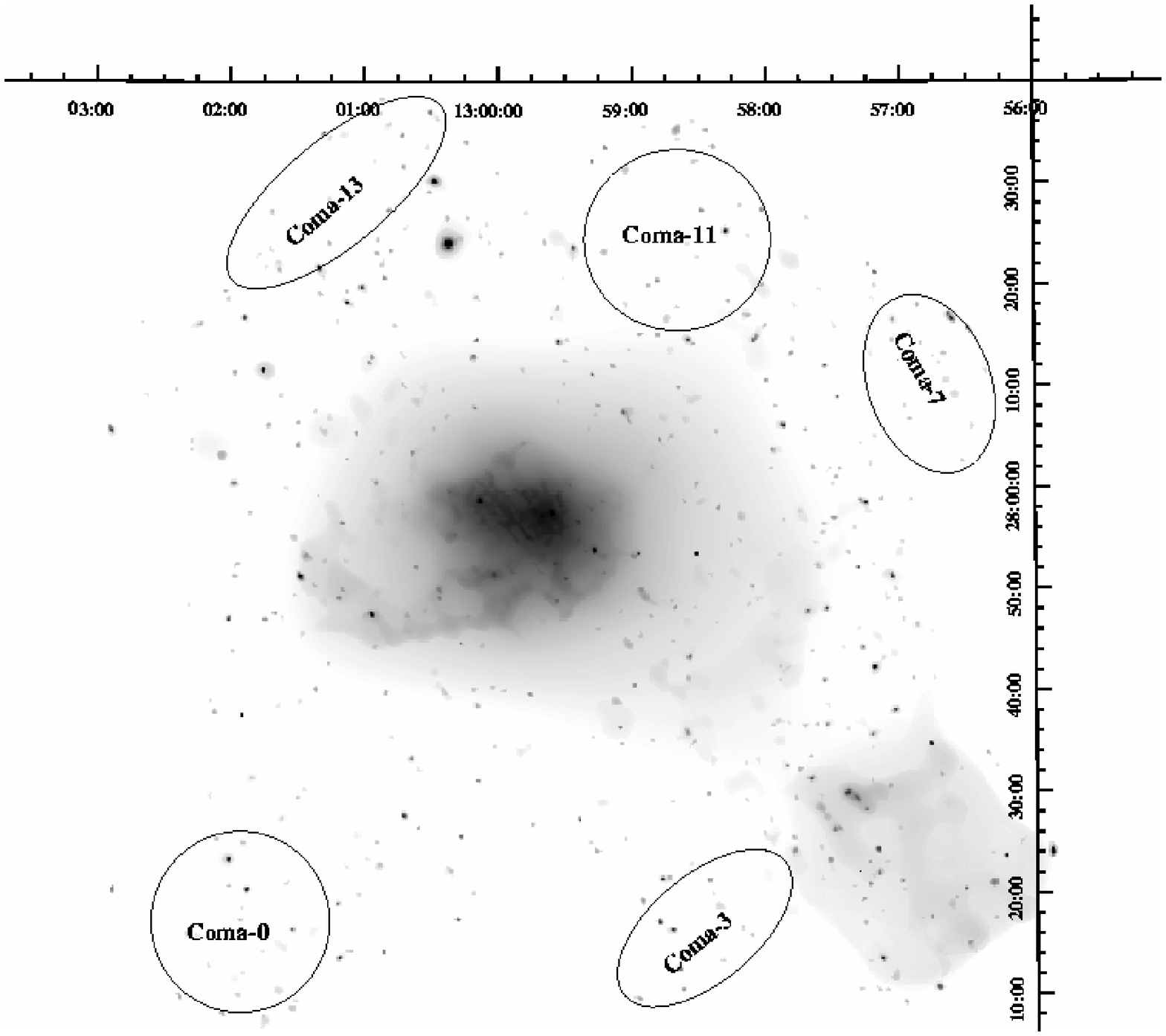}
\caption{An image of the Coma cluster in the 0.5--2
  keV band, showing the positions of the spectral extraction regions (solid
  ellipses) with names according to the XMM-Newton survey notation. The
  coordinate grid marks R.A., Dec. (J2000.0).}
\label{f-f:imh}
\end{figure}

To improve on the detection statistics of the warm emission and to reduce
the systematic effects of subtraction of the cluster emission, we
concentrate on the {\it outskirts} of the Coma cluster.  In Fig.\ref{f-f:imh}
we show the location of the spectral extracting areas.  We select regions
$\sim40'$ from the Coma center to the North-West, North, North-East, South
and South-East. The South-West direction is complicated by the presence of
the infalling subcluster (NGC4839), while some other pointings were affected
by background flares, which after screening lead to insufficient exposures.

\begin{table}[!b]
{
\footnotesize
{\renewcommand{\arraystretch}{0.9}\renewcommand{\tabcolsep}{0.09cm}
\caption{\footnotesize
Characteristics of the warm emission around Coma
\label{f-t:warm}}

\begin{tabular}{cccccc}
 \hline
Location & $kT_{Coma}$ & $kT_{warm}$ & $Z/Z_{\odot}^{\dag}$ &
$\rho^{\flat}/\rho_{crit}$ & $M_{gas}^{\flat}$\\
(Field) & keV & keV &  &  &  $10^{12}M_{\odot}$\\
\hline
Coma--0 & $15\pm6$ & $0.19\pm0.01$ &$0.06\pm0.02$  &$65\pm8$  &$4.6\pm0.5$\\
Coma--3 & $5.8\pm0.9$& $0.19\pm0.02$  &$0.04\pm0.02$ &  $73\pm10$ &$4.5\pm0.6$ \\
Coma--7 & $10\pm3$ & $0.22\pm0.02$  & $0.07\pm0.03$  &  $40\pm11$ & $2.4\pm0.6$\\
Coma--11 & $16\pm2$ & $0.24\pm0.01$  & $0.09\pm0.01$ & $85\pm9$ &$6.0\pm0.1$ \\
Coma--13 & $3.1\pm0.5$ & $0.17\pm$0.01 & $0.12\pm0.04$ & $54\pm12$ &$2.4\pm0.6$ \\ 
\hline 
\end{tabular}
\begin{enumerate}
\item[$^{\flat}$]{\footnotesize ~ Estimated by assuming a 90 Mpc distance,
projected length of 20 Mpc, area of the extraction regions in
Fig.\ref{f-f:imh} and $n_e=5.5\times 10^{-7}$ cm$^{-3}$ in correspondence to
$\rho_{crit}$ for h=0.7}

\item[$^{\dag}$]{\footnotesize ~ Assuming a solar abundance ratio.}

\end{enumerate}
}}
\vspace*{-1.cm}
\end{table}

\section{Results}

\begin{figure}[ht]
\includegraphics[width=8.cm]{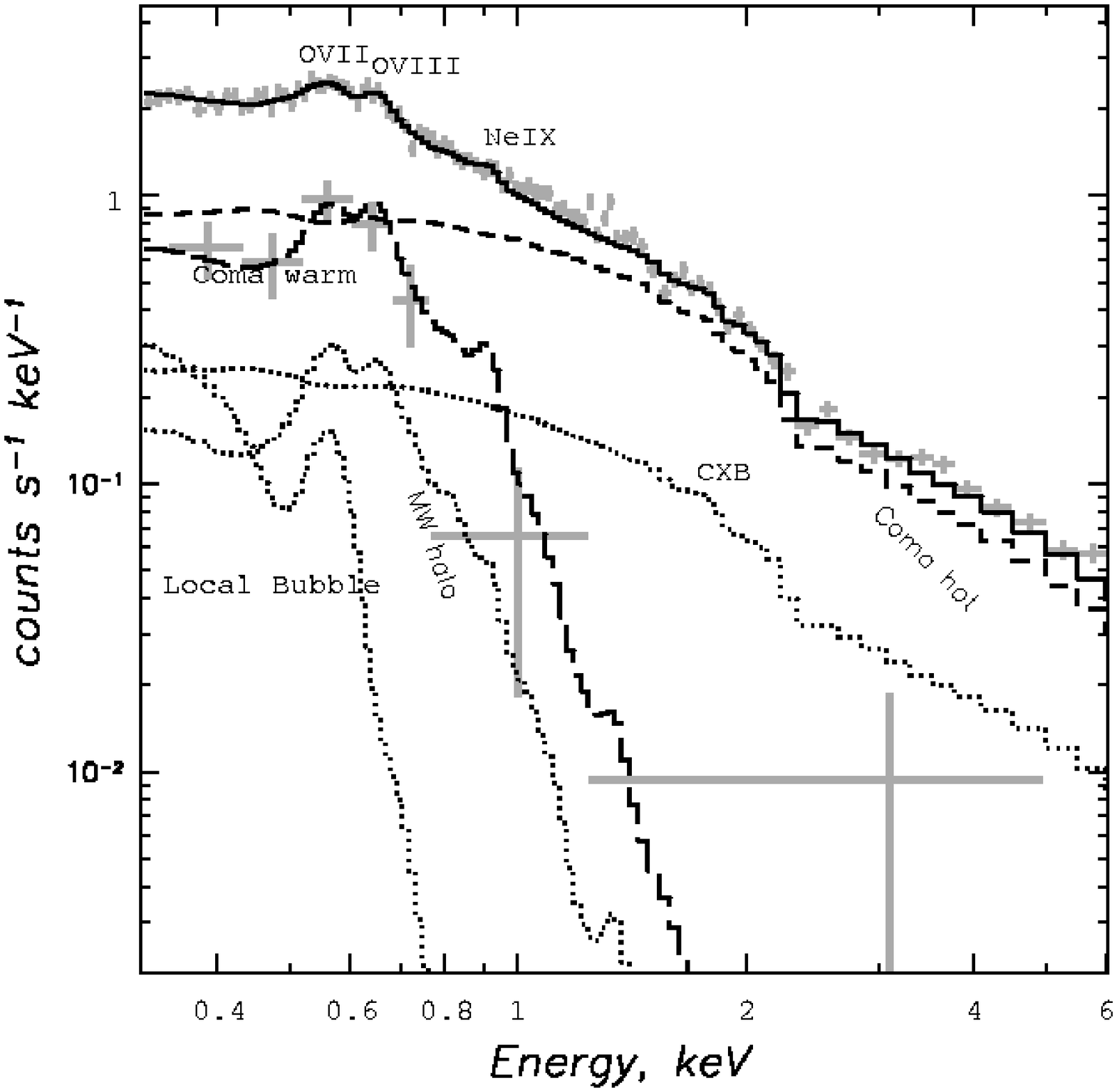}
\caption{Detailed decomposition of the pn
  spectrum (small grey crosses) in the Coma-11 field into foreground and
  background components, obtained from the analysis of the APM field, plus
  hot and warm emission from the Coma cluster. Large crosses is a result of
  the in-field subtraction of Coma emission and detector background,
  possible due to differences in the spatial distributions between the Coma
  warm, Coma hot and detector background components.}
\label{f-f:spepn}
\end{figure}

Our major results are shown in Fig.\ref{f-f:spepn} and listed in Table
\ref{f-t:warm}. We find that the soft excess in Coma on spatial scales of
$30'-50'$ is characterized by thermal plasma emission, as indicated by our
discovery of O lines, with a characteristic temperature of $\sim0.2$ keV,
varying with position on the sky. The highest temperature of the warm
component corresponds to the hot spot in Coma, found in ASCA observations
(Donnelly et al. 1999), and identified as an accreting zone of the Coma
cluster. 

At the moment, it is difficult from the O line redshift alone to decide
whether this emission is galactic or extragalactic. However, some apparent
differences with the galactic emission are already noticeable. The soft
component is centered on Coma and has an amplitude exceeding the variation
of the underlying galactic emission (Bonamente, Joy, Lieu 2003). An O
abundance of 0.1 times solar is much lower than the galactic value of one
solar (Markevitch et al. 2002; Freyberg \& Breitschwerdt 2002). The
temperature variations of the Coma warm component are not seen in the quiet
zones of galactic emission at high latitudes.  The observed warm component
exceeds the level of the galactic emission by a factor of up to 5, so to
significantly change the derived O abundance of the Coma warm emission a
possible variation in the metallicity for the galactic emission should be by
a factor of 3, which is not observed.

\section{Interpretation}

\begin{figure}[ht]
\includegraphics[width=8.cm]{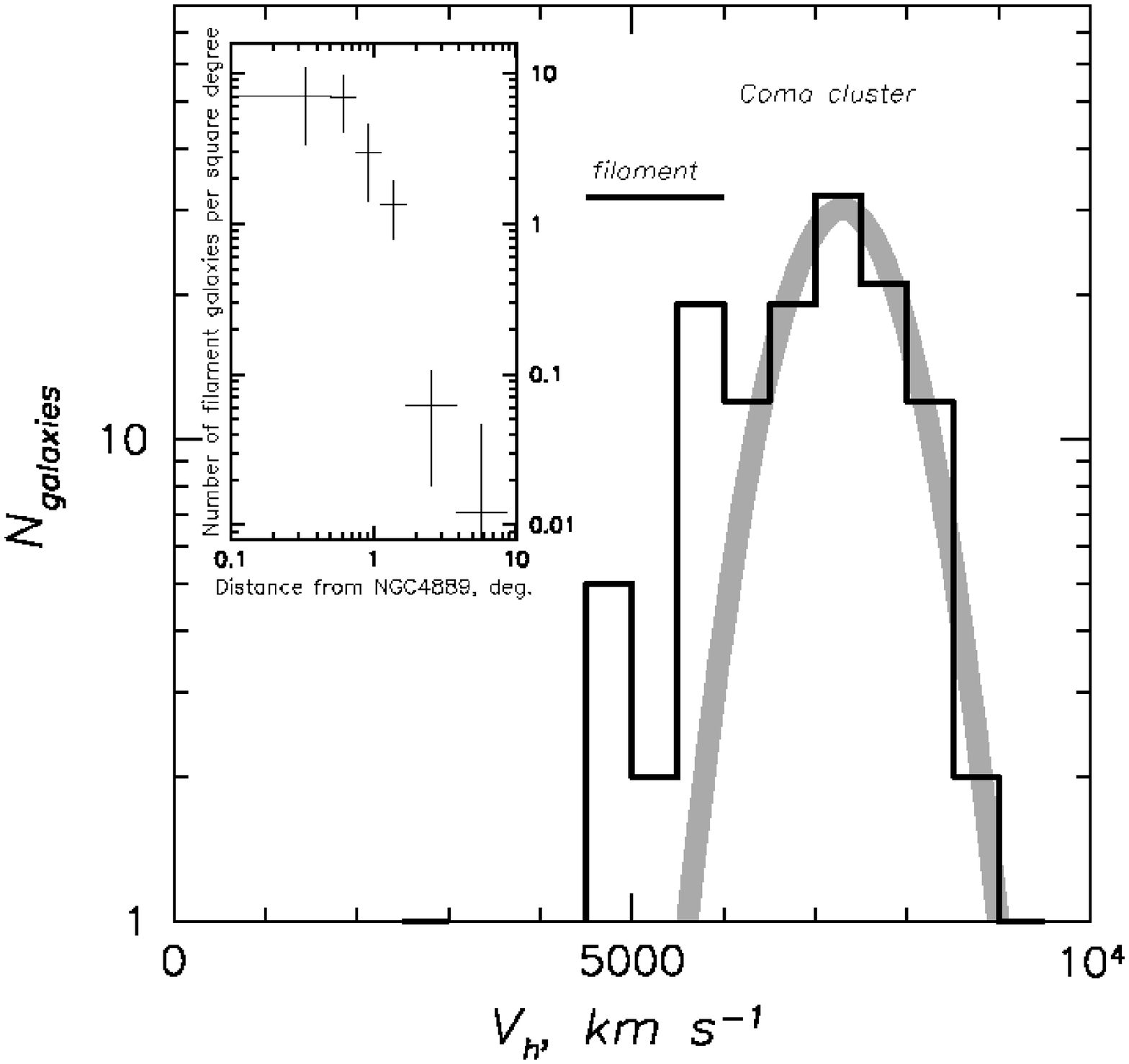}
\caption{Galaxy distribution in the direction to the Coma cluster from
CfA2 survey. The central 2 degrees in radius field centered on NGC4889 is
shown, excluding the 60 degrees cone in the South-West direction from the
center, to avoid the effect of the NGC4839 subcluster. An excess of galaxies
over the Gaussian approximation of the velocity dispersion of the Coma
cluster is seen in the $4500-6000$ km s$^{-1}$ velocity range. The insert
shows the spatial distribution of the excess galaxies in the $4500-6000$ km
s$^{-1}$ velocity bin over the galaxies in the $8500-10000$ km s$^{-1}$
velocity bin, this time out to 10 degrees.}
\label{f-f:ghst}
\end{figure}

In the rest of this {\it Contribution} we present an interpretation assuming
an extragalactic origin of the warm component. With the CCD-type
spectrometry we can place limits on the possible redshifts of the warm
component, which for the Coma-11 field is determined as
$0.007\pm0.004\pm0.015$ (best-fit, statistical and systematic error). Any
extragalactic interpretation should therefore concentrate on the large-scale
structure in front of the Coma cluster.  We have carried out an analysis of
the CfA2 galaxy catalog (Huchra, Geller, Corwin 1995) towards the Coma
cluster, excising the South-West quadrant, where strong influence of the
infalling subcluster NGC4839 on the velocity distribution has been suggested
(Colles \& Dunn 1996). Fig.\ref{f-f:ghst} illustrates our result, indicating a
significant galaxy concentration in front of the Coma cluster, with
velocities lying in the $4500-6000$ km s$^{-1}$ range. The sky density of
the infall zone is shown in the insert of Fig.\ref{f-f:ghst}. It was
constructed from the number of galaxies in the $4500-6000$ km s$^{-1}$
velocity range minus the number of galaxies in the $8500-10000$ km s$^{-1}$
bin, corresponding to a similar difference in the velocity dispersion from
the cluster mean. By excising the galaxies at velocities lower than 6000 km
s$^{-1}$, we recover a velocity dispersion of $\sim700$ km s$^{-1}$. Using
the $M-\sigma$ relation of Finoguenov, Reiprich, Boehringer (2001), we find
that such velocity dispersion is more in accordance with mass of the Coma
cluster. The excess of galaxies exhibits a flat behavior within 0.8 degrees
(1.3 Mpc), followed by a decline by a factor of 5 within 1.7 degrees (2.8
Mpc) and a subsequent drop by two orders of magnitude within 10 degrees
(16.7 Mpc) The soft emission from the Coma is detected to a 2.6 Mpc distance
from the center (Bonamente et al. 2003) in a remarkable correspondence to
the galaxy filament.

In the absence of the finger of god effect, validated below by our
conclusion on the filamentary origin of this galaxy concentration, the
infall zone is characterized by a 20 Mpc projected length. This length
assumption affects the density estimates, while the O abundance is only
based on the assumption of collisional equilibrium. We have verified the
later assumption by studying the ionization curves for OVII and OVIII
presented in Mathur et al. (2002). We have concluded that the assumption of
pure collisional ionization is valid for our data, since $n_e>10^{-5}$
cm$^{-3}$ and $T>2\times10^{6}$ K (0.17 keV). An advantage of our
measurement is that we also determine the temperature by the
continuum. Lower temperatures, which at densities near $10^{-5}$ cm$^{-3}$
result in similar line ratios for OVII and OVIII, fail to produce the
observed (H-e and He-e) bremsstrahlung flux at 0.7--1 keV.

An important question we want to answer is whether the detected emission
originates from a group or from a filament. We cannot decide from the
overdensity of the structure ($\sim50\rho_{crit}$), as it suits
both. However, the implied mass of the structure exceeds the mass of the
Coma cluster (see also Bonamente et al. 2003). On the other hand the
temperature of the emission is $\sim0.2$ keV, almost two orders of magnitude
lower than that of the Coma cluster. It takes a few hundred groups with
virial temperature of 0.2 keV to make up a mass of the structure, which
leads to an overlapping virial radii if we are to fit them into the given
volume of the structure. Thus we conclude that the observed structure is a
filament.

For the Coma-11 field, where the statistics are the highest, we investigated
the effect of relaxing the assumption of solar abundance ratios. We note
that an assumption for C abundance is important for overall fitting and a
solar C value would affect our results. When left free, however, the C
abundance tends to go to zero. Also, there is no systematic dependence of
our results on the assumed C abundance as long as the C/O ratio is solar or
less, which is the correct assumption from the point of view of chemical
enrichment schemes and observations of metal-poor stars in our
Galaxy. Significant abundance measurements are: O/O$_{\odot}=0.14\pm0.02$,
Ne/Ne$_{\odot}=0.14\pm0.06$, Fe/Fe$_{\odot}=0.04^{+0.03}_{-0.01}$ (assuming
Fe$_{\odot}/H=4.26\times10^{-5}$ by number), indicating that Fe is
underabundant by a factor of three in respect to the solar Fe/O ratio,
implying a dominant contribution of SN II to Fe enrichment.  Element
abundances for the hot emission obtained from our XMM data on the center of
Coma are Mg/Mg$_{\odot}=0.36\pm0.12$, Si/Si$_{\odot}=0.45\pm0.08$,
S/S$_{\odot}=0.01\pm0.01$, Fe/Fe$_{\odot}=0.20\pm0.03$, also suggesting
prevalence of SN II in Fe enrichment, although Fe abundance is a factor of 5
higher compared to the filament. The filament Ne/O ratio reveals a subtle
difference with the OVI absorbers. Nicastro et al. (2002) reports Ne/O $=2$
times solar, while we observe a lower (solar) ratio at 95\%
confidence. Lower ratios seem to be a characteristic of enrichment at $z>2$
(Finoguenov, Burkert, Boehringer 2003), while WHIM originates at $z<1$
(Cen \& Ostriker 1999).  While we would rather have more observational
evidence on the dispersion of O to alpha element ratios, we believe that
this is a major source of systematics in interpreting the element abundance
of X-ray filaments as a universal value.

To illustrate the point, we calculate the amount of baryons traced by the
OVI absorbers:

\begin{equation}
\Omega_b^{OVI} = 0.020h_{70}^{-1} \left( {0.14 \over 10^{[O/H]}}\right) \left({\langle
[f(OVI)]^{-1}\rangle \over 32}\right)
\end{equation}

where the original value of $\Omega_b(OVI) = 0.0043h_{70}^{-1}$ of Tripp et
al. (2000) has been scaled for a ionization equilibrium implied by
measurements of Mathur et al (2002) and using the measurements of the O
abundance reported here. The formal errorbar is 0.006, mostly from the
uncertainty in the estimate for ionization. If, on the other hand, our
measurements of Ne abundance are used, and the Ne/O ratio for OVI absorbers
is taken from Nicastro et al. (2002),

\begin{equation}
\Omega_b^{OVI} = 0.040h_{70}^{-1} \left( {0.14 \over 10^{[Ne/H]}}
 { 10^{[Ne/O]}  \over 2}\right)
\left({\langle [f(OVI)]^{-1}\rangle \over 32}\right)
\end{equation}

So, with $\Omega_b^{\rm total}=0.039$, there is no room left for other major
components of local baryons, $Ly_\alpha$ absorbers ($0.012\pm0.002$) and
stars and clusters of galaxies ($\sim0.006$; e.g., Finoguenov et al. 2003
and references therein). If O is depleted onto dust grains in the OVI
absorbers, as suggested by Nicastro et al. (2002) as an explanation of the
high Ne/O ratio, the solar Ne/O ratio in our observations simply
indicates dust sputtering, then the baryon budget of OVI absorbers follows the
calculation in Eq.2 and yields, consequently, unacceptably high values.

\begin{chapthebibliography}{}
\par Bonamente, M., Joy, M., Lieu, R. 2003, ApJ, in press
(astro-ph/0211439) 
\par Briel, U.G., Henry, J.P. 1995, A\&A, 302, 9
\par Briel, U.G., Henry, J.P., Lumb, D.H., Arnaud, M., Neumann, D.,
et al. 2001, A\&A, 365, L60 
\par Cen, R., Ostriker, J.P. 1999, ApJ, 514, 1
\par Colles, M., Dunn, A.M. 1996, ApJ, 458, 435
\par Donnelly, R.H., Markevitch, M., Forman, W., Jones, C.,
Churazov, E., Gilfanov, M. 1999, ApJ, 513, 690 
\par Finoguenov, A., Reiprich T., Boehringer, H. 2001, A\&A, 368, 749
\par Finoguenov, A., Burkert, A., Boehringer, H. 2003b, ApJ, submitted
\par Freyberg, M. J., Breitschwerdt, D. 2002, Proc. of JENAM 2002,
in press
\par Fukugita, M., Hogan, C.J., Peebles, P.J.E. 1998, ApJ, 503, 518
\par Huchra, J.P., Geller, M.J., Corwin, H.G.Jr. 1995, ApJS, 99,391
\par Lumb, D.H., Warwick, R.S., Page, M., De Luca, A.  2002, A\&A,
389, 93
\par Markevitch, M., Bautz, M.W., Biller, B., et al. 2002, ApJ, in
press, (astro-ph/0209441)
\par Mathur, S., Weinberg, D.H., Chen, X. 2002, ApJ, in press,
 astro-ph/0206121
\par Nagamine, K., Fukugita, M., Cen, R., Ostriker, J.P. 2001, ApJ,
558, 497
\par Nicastro, F., Zezas, A., Drake, J., Elvis, M., Fiore, F.,
Fruscione, A., Marengo, M., Mathur, S., Bianchi, S. 2002, ApJ, 573, 157
\par Scharf, C., Donahue, M., Voit, G.M., Rosati, P., Postman,
M. 2000, ApJ, 528, L73
\par Tripp, T.M., Savage, B.D., Jenkins, E.B. 2000, ApJ, 534, L1
\par White, S.D.M., Briel, U.G., Henry, J.P. 1993, MNRAS, 261, L8
\par Zappacosta, L., Mannucci, F., Maiolino, R., Gilli, R., Ferrara,
A., Finoguenov, A., Nagar, N. M., Axon, D. J. 2002, A\&A, 394, 7
\end{chapthebibliography}                                                          

\kluwerprintindex










\articletitle{The cluster Abell 85 and its X-ray filament revisited by
Chandra and XMM-Newton}

\chaptitlerunninghead{X-ray filament in Abell 85}

\author{Florence Durret\altaffilmark{1}, Gast\~ao B. Lima Neto\altaffilmark{2},
        William R. Forman\altaffilmark{3}, Eugene Churazov\altaffilmark{4}} 

\altaffiltext{1}{Institut d'Astrophysique de Paris, 98bis Bd Arago, 75014 
Paris, France}
\altaffiltext{2}{Instituto Astron\^omico e Geof\'{\i}sico/USP, Av. Miguel 
Stefano 4200, S\~ao Paulo/SP, Brazil}
\altaffiltext{3}{Harvard Smithsonian Center for Astrophysics, 60 Garden St, 
Cambridge MA 02138, USA}
\altaffiltext{4}{MPI f\"ur Astrophysik, Karl Schwarzschild Strasse 1, 85740 Garching, 
Germany}

\begin{abstract}
We have observed with XMM-Newton the extended 4~Mpc filament detected
by the ROSAT PSPC south east of the cluster of galaxies Abell~85.  We
confirm the presence of an extended feature and find that the X-ray
emission from the filament is best described by thermal emission with
a temperature of $\sim2$ keV. This is significantly lower than the
ambient cluster medium, but is significantly higher than anticipated
for a gas in a weakly bound extended filament. It is not clear whether
this is a filament of diffuse emission or a chain of several groups of
galaxies.
\end{abstract}

\section{X-ray and galaxy filaments}

Numerical simulations of structure formation in cold dark matter (CDM)
cosmological models (e.g. Frenk et al. 1983, Jenkins et al. 1998)
predict that galaxies form preferentially along filaments, and that
clusters are located at the intersections of these filaments. The
detection of filaments visible in X-rays can therefore be a test to
scenarios of large scale structure formation.

Various types of observations indeed seem to indicate that clusters
remember how they formed. Some clusters show preferential orientations
at various scales; in Abell 85, for example, the central cD galaxy,
the brightest galaxies, the overall galaxy distribution and the X-ray
gas distribution all show elongations along the same direction (Durret
et al. 1998b). At even larger scales, the probability for no alignment
between two clusters separated by a few Mpc has been found to be very
small (Chambers et al. 2002).

Bridges of material (galaxies and/or groups of galaxies) have been
detected between several clusters (West et al. 1995, West \&
Blakeslee 2000), though they are not very easy to spot since they
require the measurements of numerous galaxy redshifts.

When ROSAT all sky survey background fluctuations were correlated with
the Abell cluster catalogue, no filament was detected, implying an
upper limit to the electron density consistent with that predicted by
hydrodynamical simulations (Briel \& Henry 1995). A few X-ray
filaments or at least highly elongated X-ray emitting regions were
detected however in a few clusters such as Coma (Vikhlinin et
al. 1997), Abell~2125 (Wang et al. 1997), Abell~85 (Durret et
al. 1998b), the core of the Shapley supercluster (Kull \& B\"ohringer
1999), Abell ~1795 (Fabian et al. 2001, Markevitch et al. 2001), and
between Abell~3391 and Abell~3395 (Tittley \& Henriksen 2001).

\section{The X-ray filament in Abell~85}

The complex of clusters Abell~85/87/89 is a well studied structure,
dominated by Abell~85 at a redshift of $z=0.056$ (Durret et al. 1998b)
(at $z=0.056$, 1 arcmin $= 90 h_{50}^{-1}\,$kpc). Abell~87 is not
detected in X-rays, neither by the ROSAT PSPC nor in our XMM-Newton
image, so it is probably not a real cluster, but perhaps a
concentration of several groups. This would agree with Katgert et
al. (1996) who find two different redshifts in that direction, one
coinciding with that of Abell 85 and the other at z=0.077. Abell~89
comprises two background structures and will not be discussed further
(see Durret et al. 1998b for details).

The ROSAT PSPC image of Abell~85 has revealed an elongated X-ray
structure to the southeast of the main cluster Abell~85. This filament
is at least 4~Mpc long (projected extent on the sky), and it is not
yet clear whether this is a filament of diffuse emission or a chain
made by several groups of galaxies. \cite{d2-Durret98X} have shown that
after subtracting the modelled contribution of Abell~85 the remaining
emission has the appearance of being made of several groups, each
having an X-ray luminosity of about $10^{42}$ erg~s$^{-1}$.  Kempner
et al. (2002) have recently observed Abell~85 with Chandra; they
present data concerning the X-ray concentration at the northernmost
end of the filamentary structure, but the filament itself is outside
their field.

\begin{figure}[!htb]
\centering
\mbox{\includegraphics[width=8.5cm,clip]{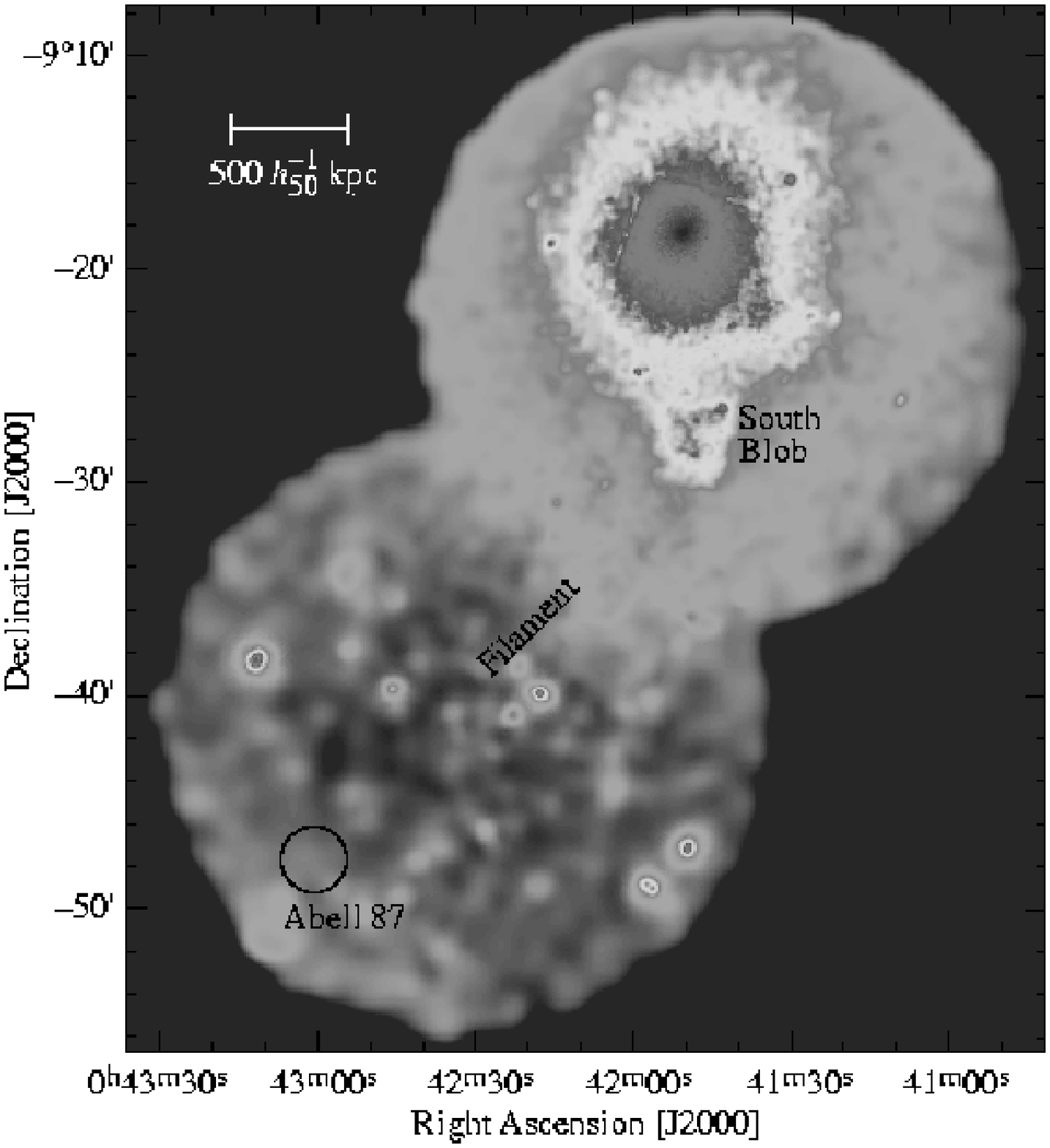}}
\caption{Cleaned, background subtracted, exposure map corrected
image obtained from the two MOS1 and MOS2 XMM-Newton exposures. The
image has been smoothed with a Gaussian of $\sigma=5\,$arcsec.  }
\label{d2-fig:rawimage}
\end{figure}

\begin{figure}[!htb]
\centering \mbox{\includegraphics[width=8.5cm]{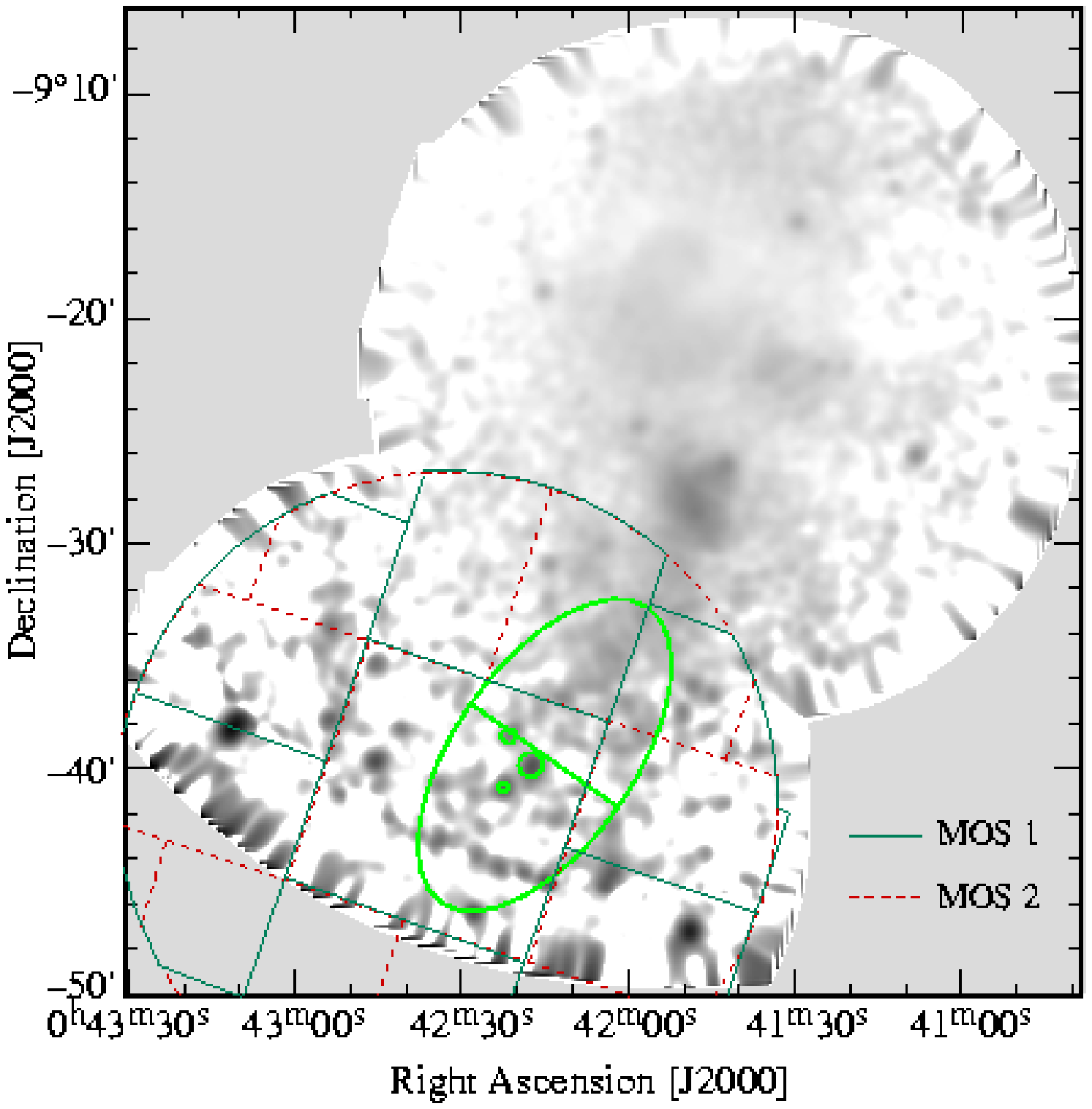}}
\caption{Merged XMM-Newton image obtained after subtracting an
azimuthal average of the X-ray emission of the overall cluster.  The
ellipse, excluding the three circles, shows the region where we
extracted the events for the spectral analysis. The straight line
dividing the ellipse in two defines the ``north'' and ``south'' halves
of the filament. The borders of MOS1 and MOS2 are also shown (thin
lines).}
\label{d2-fig:imminusmodel}
\end{figure}

We have reobserved Abell~85 with XMM-Newton. The cleaned, background
subtracted and exposure map corrected merged MOS1+2 image is shown in
Fig.~\ref{d2-fig:rawimage}. The image obtained after subtracting an
azimuthal average of the X-ray emission of the overall cluster is
shown in Fig.~\ref{d2-fig:imminusmodel}. Besides the main cluster and the
south blob the filament is clearly visible towards the south east.

Events were extracted inside an elongated region of elliptical shape
centered at R.A. $=0^{\rm h}42^{\rm m}15.0^{\rm s}$ and Decl. $=
-09^{\circ}39^{\prime}24^{\prime\prime}$ (J2000), with major and minor
axis of 8 and 4~arcmin, respectively. Three circular regions
corresponding to point sources (see Fig.~\ref{d2-fig:imminusmodel}) were
excluded.

The background was taken into account by extracting spectra (for MOS1
and MOS2) from the EPIC blank sky templates described by
\cite{d2-Lumb02}. We have applied the same filtering procedure to the
background event files and extracted the spectra in the same
elliptical region of the filament in detector coordinates. Finally,
the spectra have been rebinned so that there are at least 30 counts
per energy bin.

\begin{figure}[!htb]
\centering
\mbox{\includegraphics[width=8.4cm,clip]{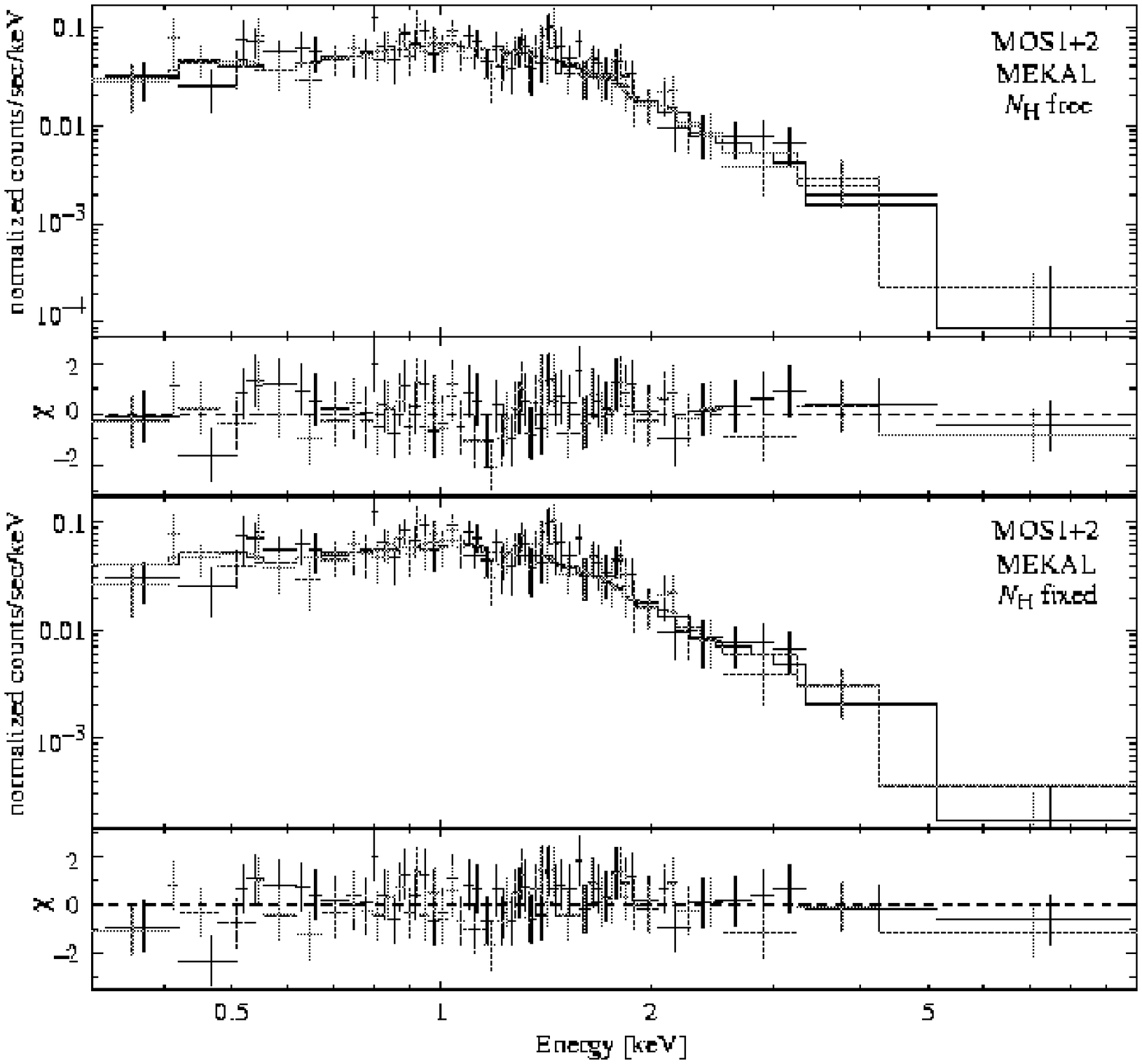}}
\caption{XMM-Newton MOS1 and MOS2 rebinned spectra of the filament
region with the best MEKAL fits superimposed. Top: fit with the
hydrogen column density free. Bottom: fit with $N_{\rm H}$ fixed at
the galactic value at the position of the filament.}
\label{d2-fig:spectre}
\end{figure}

The count rates after background subtraction are
$0.1096\pm0.0055$~cts/s and 0.0899$\pm$0.0057 cts/s for the MOS1 and
MOS2 respectively, corresponding to total source counts of 1354 and
1115.  The spectral fits were done with XSPEC 11.2, with data in the
range [0.3--10~keV], simultaneously with MOS1 and MOS2. A MEKAL model
(Kaastra \& Mewe 1993, Liedahl et al. 1995) with photoelectric
absorption given by \cite{d2-Balucinska} was used to fit the spectral
data. Since the spectra were rebinned, we have used standard
$\chi^{2}$ minimization.  Fig.~\ref{d2-fig:spectre} shows the MOS1 and
MOS2 spectra, together with the best MEKAL fits and residuals.

\begin{table}[!htb]
\centering
\caption{Results of the spectral fits for the filament with a MEKAL
model. Error bars are 90\% confidence limits.}
\begin{tabular}{c c c c}
\hline
$kT$ [keV] & $Z$ [solar] & $N_{\rm H}$ [$10^{20}\,$cm$^{-2}$] & $\chi^{2}/$dof\\
\hline
$1.9_{-0.4}^{+0.5}$ & $0.05_{-0.05}^{+0.13}$ & $8.2_{-3.6}^{+4.1}$ & 194.0/192\\
$2.4_{-0.3}^{+0.5}$ & $0.17_{-0.12}^{+0.17}$     &   3.16 (fixed)  & 199.5/193\\
\hline
\end{tabular}
\label{d2-tbl:spectralfits}
\end{table}

Table~\ref{d2-tbl:spectralfits} summarizes results of the spectral fits
for the filament. If $N_{\rm H}$ is left free to vary in the fit, it is
not well constrained, since its 90\% confidence interval is $(4.6 \leq
N_{\rm H} \leq 12.3) \times10^{20}\,$cm$^{-2}$, and the corresponding
X-ray temperature is in the range $1.5 \leq kT \leq 2.4\,$keV. For the
metallicity, we only have an upper limit $Z \leq 0.18\ Z_{\odot}$,
which is almost the same as the mean metallicity obtained in the fit
with fixed $N_{\rm H}$.

The galactic neutral hydrogen column density in the direction of the
filament is $N_{\rm H}=3.16\times 10^{20}\,$cm$^{-2}$ (Dickey \&
Lockman 1990). When $N_{\rm H}$ is fixed to this value, a MEKAL fit to
the filament spectrum gives a gas temperature of $2.0 \leq kT \leq
2.8\,$keV and a metallicity of $0.04\leq Z \leq 0.33 Z_\odot$ (90\%
confidence range).

We also attempted to fit a power-law to the spectra.  The resulting
photon index is in the interval 2.5--3.0 and 1.8--2.0 when $N_{\rm H}$
is left free to vary or is fixed, respectively.  The MEKAL and power
law fits are of comparable statistical quality when $N_{\rm H}$ is
left free to vary; however, when $N_{\rm H}$ is fixed, the MEKAL fit
is significantly better. Since the fitted value of $N_{\rm H}$ for the
power law is unreasonably large and the value of $\chi^2$ unacceptable
when $N_{\rm H}$ is fixed at the Galactic value, we reject the power
law as a possible spectral model and favour thermal emission.

Visual inspection of the filament (Fig.~\ref{d2-fig:imminusmodel})
reveals a North $\rightarrow$ South gradient. We defined two halves of
the filament region and analysed them in the same way as the whole
ellipse.  Within error
bars, we find no gradient of temperature or metal abundance.

\section{A more general X-ray view of Abell~85}

\begin{figure}[!htp]
\centering
\mbox{\includegraphics[width=8.5cm,clip]{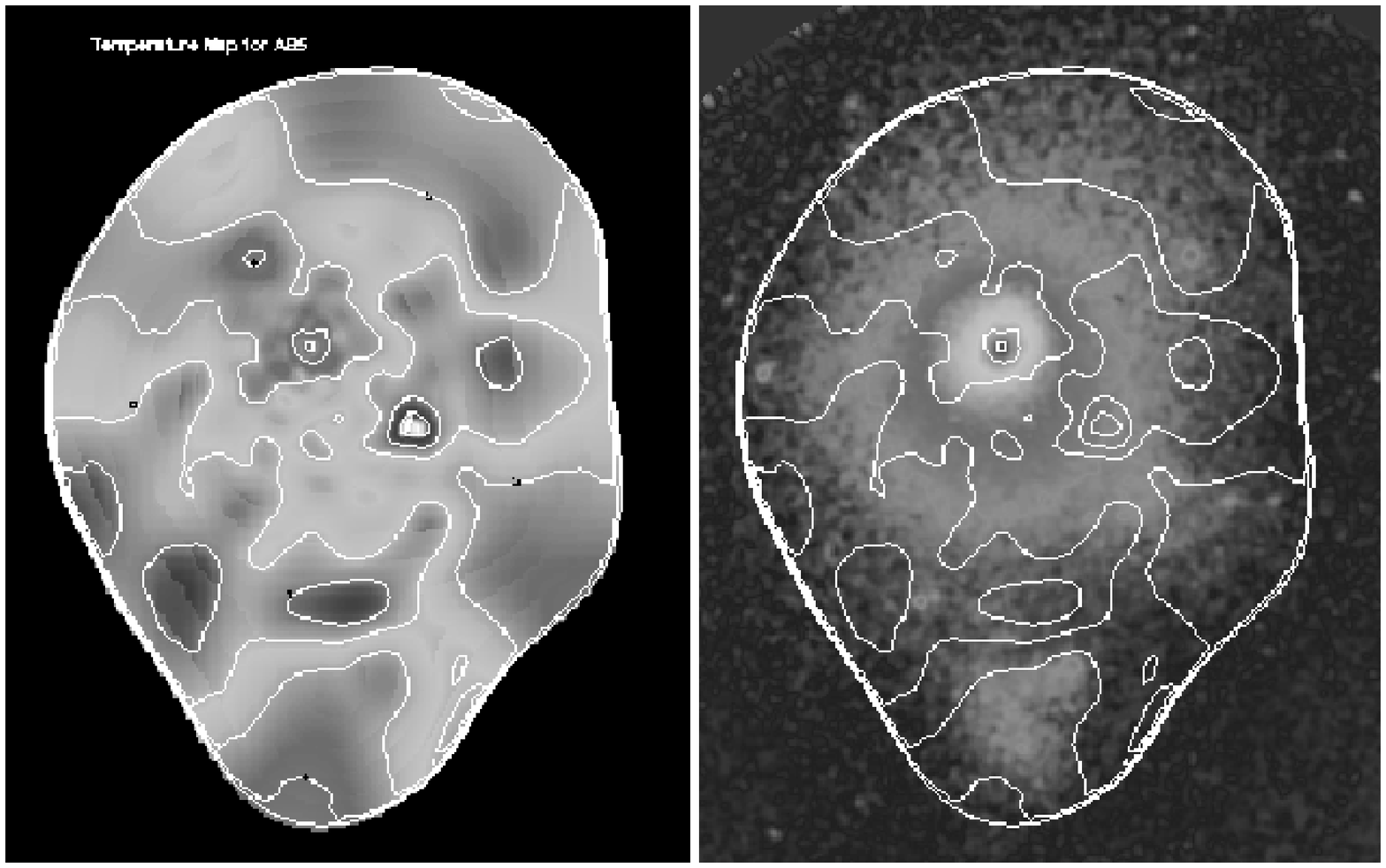}}
\caption{Left: temperature map with iso-temperature contours
superimposed. Lower temperatures are 4--5~keV (lighter) and highest
temperatures are $\sim 7$ keV (darker). The map is adaptively
smoothed. Right: central part of Fig.~1 with the
iso-temperature contours from the temperature map superimposed.}
\label{d2-fig:tmodel}
\end{figure}

The XMM temperature map of Abell~85 is displayed in Fig.~\ref{d2-fig:tmodel}.
The three expected cool
regions (the central cD where a cooling flow is observed, the south
clump and the southwest clump) are clearly seen. The region just above
the southern clump is hot, in agreement with Kempner, Sarazin \&
Ricker (2002), and with the ASCA temperature map of Donnelly et
al. (2003, in preparation). This agrees with the idea that in the
impact region where material from the filament is falling onto
Abell~85 a shock may occur and the gas is compressed and hotter
(Durret et al. 1998b).

\begin{figure}[!htp]
\centering
\mbox{\includegraphics[width=8.5cm]{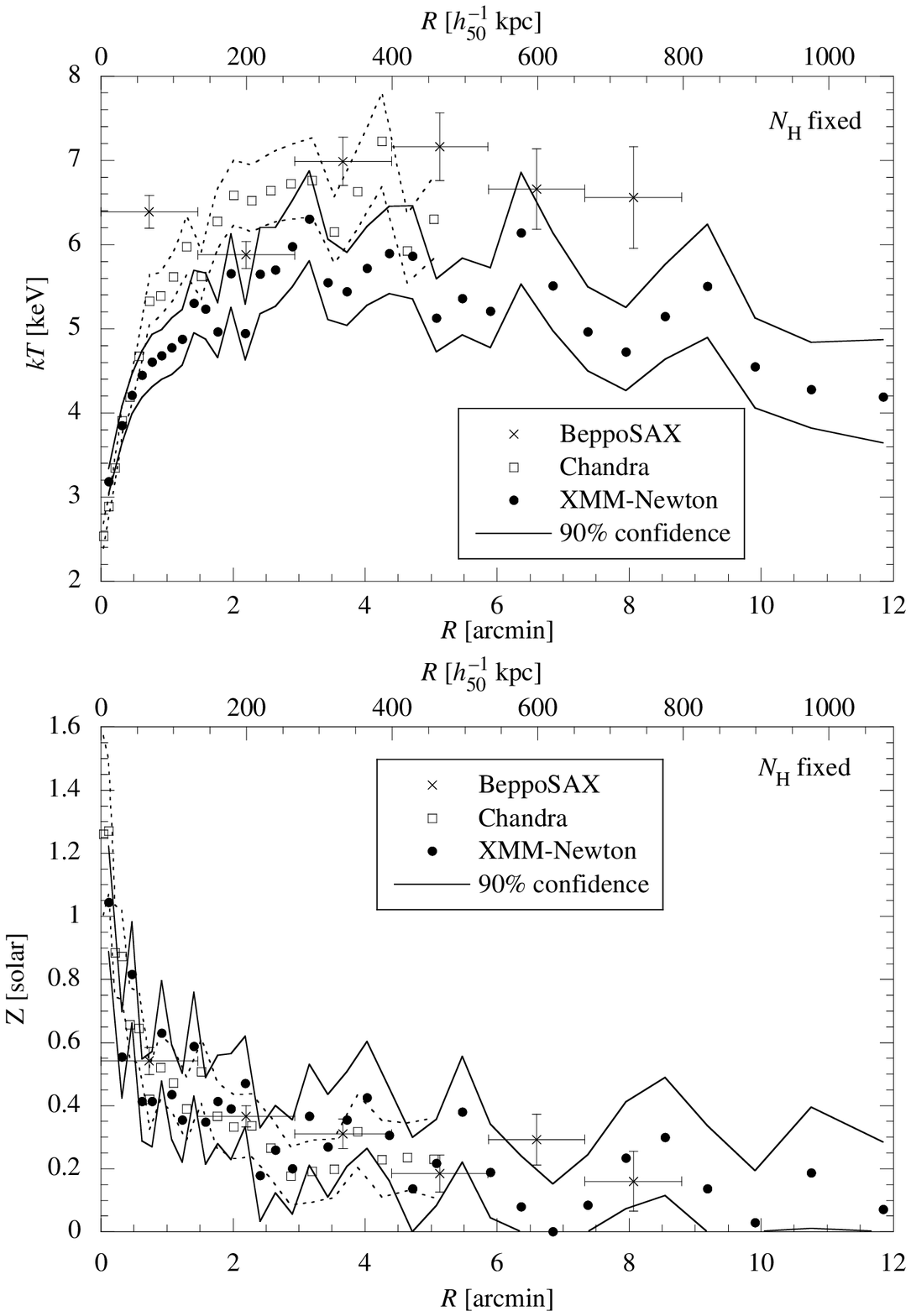}}
\caption{XMM-Newton (black circles and full lines), Chandra (empty
squares and dotted lines) and Beppo-Sax (x's with error bars)
temperature (top) and metallicity (bottom) profiles. All the spectral
fits were done with the hydrogen column density fixed.}
\label{d2-fig:profils}
\end{figure}

Fig.~\ref{d2-fig:profils} shows the radial profiles of temperature and
metal abundance for the overall cluster, computed in circular rings
around the cluster center both from Chandra and Beppo-Sax
data. Results appear quite different when the hydrogen column density
is fixed (to the Lockman \& Dickey 1990 value) or when it is left free
to vary. A clear temperature drop and metallicity rise are observed
towards the center.

\section{Discussion and conclusions}

Our XMM-Newton observations confirm the existence of a highly
elongated filamentary like structure extending from the South blob to
the south east of Abell~85 along the direction defined by all the
structures pointed out by Durret et al. (1998b).  The fact that the
spatial structure of the X-ray filament detected by XMM-Newton cannot
be exactly superimposed to that obtained from ROSAT data shows that it
is still difficult to determine exactly its structure.

However, we have shown that the X-ray spectrum from this structure is
most likely thermal and its temperature is about 2.0 keV, consistent
with that of groups. This value is notably cooler than that of the
main cluster: the temperature map by Markevitch et al. (1998) shows
the presence of gas at about 5 keV in the region at a distance from
the cluster center at least as far as the northern part of the
ellipse. So, we appear to be seeing cool gas as it enters the cluster
core. A full description of the X-ray properties of the filament can
be found in Durret et al. (2003).

In an attempt to characterize substructure in the filament region from
the galaxy distribution, we have analysed the galaxy distribution in
this zone.  There are 18 galaxies in our redshift catalogue with
redshifts in the cluster velocity range (Durret et al. 1998a) and
coinciding with the position of the X-ray filament. The velocity
histogram of these 18 galaxies shows a pair of galaxies with
velocities around 14060 km~s$^{-1}$ and two peaks, both including 8
galaxies: the first one is centered on 15770 km~s$^{-1}$ with a
dispersion of 235 km~s$^{-1}$, and the second one is centered on 17270
km~s$^{-1}$ with a dispersion of 380 km~s$^{-1}$. However, neither of
these two ``groups'' of galaxies is concentrated on the sky. Instead,
their galaxy positions cover the entire filament, so we cannot say we
are detecting any structure in the galaxy velocity distribution.

\vspace{0.1cm}
\noindent{\bf Acknowledgments}

We acknowledge help with the XMM-Newton SAS software from S\'ebastien
Majerowicz and Sergio Dos Santos. F.D. and G.B.L.N. acknowledge
financial support from the \textsc{usp/cofecub}. G.B.L.N. acknowledges
support from FAPESP and CNPq. W. Forman thanks the
Max-Planck-Institute f\"ur Astrophysik for its hospitality during the
summer of 2002 and acknowledges support from NASA Grant
NAG5-10044. This work is based on observations obtained with
XMM-Newton, an ESA science mission with instruments and contributions
directly funded by ESA Member States and the USA (NASA).





\part[Absorption line studies of out Galaxy and the WHIM]{Absorption line studies of out Galaxy and the WHIM}







\articletitle{Chandra Detection of X-ray Absorption from Local Warm/Hot Gas}

\author{T. Fang\altaffilmark{1,2}, C. Canizares\altaffilmark{2},
K. Sembach\altaffilmark{3}, H. Marshall\altaffilmark{2},
J. Lee\altaffilmark{2}, D. Davis\altaffilmark{4}} 
\altaffiltext{1}{Dept. of Physics,Carnegie Mellon Univ., 5000 Forbes Ave.,
  Pittsburgh, PA 15213}
\altaffiltext{2}{MIT, Center for Space Research, 70 Vassar St., Cambridge, MA
  02139}
\altaffiltext{3}{STScI, 3700 San Martin Drive, Baltimore, MD 21218}
\altaffiltext{4}{GSFC, Code 661., GLAST SSC, Greenbelt, MD 20771}


\begin{abstract}
Recently, with the {\sl Chandra} X-ray Telescope we have detected several local X-ray absorption lines along lines-of-sight towards distant quasars. These absorption lines are produced by warm/hot gas located in local intergalactic space and/or in our Galaxy. I will present our observations and discuss the origin of the X-ray absorption and its implications in probing the warm/hot component of local baryons.
\end{abstract}

\section{Introduction}

The cosmic baryon budget at low and high redshift indicates that a
large fraction of baryons in the local universe have so far escaped
detection (e.g., Fukugita, Hogan, \& Peebles~1998). While there is
clear evidence that a significant fraction of these ``missing
baryons'' (between 20-40\% of total baryons) lie in photoionized,
low-redshift Ly$\alpha$ clouds (Penton, Shull, \& Stocke~2000), the
remainder could be located in intergalactic space with temperatures of
$10^{5}-10^{7}$ K (warm-hot intergalactic medium, or WHIM). Resonant
absorption from highly-ionized ions located in the WHIM gas has been
predicted based on both analytic studies of structure formation and
evolution (Shapiro \& Bahcall~1981;Perna \& Loeb~1998;Fang \& Canizares~2000) and cosmic
hydrodynamic simulations (Hellsten, Gnedin, \& Miralda-Escud{\'
e}~1998;Cen \& Ostriker~1999a;Dav{\' e} et al.~2001;Fang, Bryan, \&
Canizares~2002). Recent discovery of O VI absorption lines by
the Hubble Space Telescope ({\sl HST}) and the Far Ultraviolet
Spectroscopic Explorer ({\sl FUSE}) (see, e.g., Tripp \& Savage~2000) indicates that there may be
a significant reservoir of baryons in O VI absorbers. While
Li-like O VI probes about $\sim 30-40\%$ of the WHIM gas (Cen et
al.~2001;Fang \& Bryan~2001), the remaining $\sim 60-70\%$ is hotter
and can only be probed by ions with higher ionization potentials, such
as H- and He-like Oxygen, through X-ray observation.

Recently, with {\sl Chandra} Low Energy Transmission Grating Spectrometer (LETGS) we detected resonance absorption lines from H- and He-like Oxygen in the X-ray spectra of background quasars, namely PKS~2155-304 and 3C~273. The detected lines can be categorized into (1) those at $z \approx 0$ and (2) one redshifted intervening system. In this paper, we will discuss these detections and their implications for the physical properties of the hot gases that give rise to these absorption features.

\vbox{
\begin{center}
\begin{tabular}{cccc}
\multicolumn{4}{c}{Table 1: Fitting parameters of the X-ray absorption Lines~~~} \\
\hline
\hline
& \multicolumn{2}{c}{PKS~2155-304} & \multicolumn{1}{c}{3C~273} \\
\cline{2-3}\cline{4-4}
& O~{\sc viii} Ly$\alpha$ &  O~{\sc vii} He$\alpha$ & O~{\sc vii} He$\alpha$ \\
\hline
$\lambda_{obs}$       & $20.02_{-0.015}^{+0.015}$            & $21.61_{-0.01}^{+0.01}$       & $21.60_{-0.01}^{+0.01}$ \\ 
$cz({\rm km\ s^{-1}})$ & $16624\pm237$      & $112_{-138}^{+140}$  & $-26_{-140}^{+140}$ \\ 
Line Width$^{a}$      &  $< 0.039$                & $<0.027$             & $<0.020$\\ 
$\rm Line\ Flux^{b}$   & $4.8_{-1.9}^{+2.5}$       & $5.5_{-1.7}^{+3.0}$  & $4.2_{-0.9}^{+1.8}$ \\
EW (m${\rm \AA}$)      & $14.0_{-5.6}^{+7.3}$      & $15.6_{-4.9}^{+8.6}$ & $28.4_{-6.2}^{+12.5}$ \\ 
SNR              & 4.5                       & 4.6                  & 6.4 \\
\hline
\end{tabular}

\parbox{5in}{
\vspace{0.1in}
\small\baselineskip 9pt
\footnotesize
a. 90\% upper limit of the line width $\sigma$, in units of $\rm\AA$.\\
b. Absorbed line flux in units of $\rm 10^{-5}~photons~cm^{-2}s^{-1}$.\\
}
\end{center}
\normalsize
\centerline{}
}

\section{Data Reduction}

PKS~2155-304 and 3C~273 are bright extragalactic X-ray sources used as {\sl Chandra} calibration targets. They were observed
with the {\sl Chandra} LETG-ACIS (the observations ids for PKS~2155 are 1703, 2335, 3168; and the ids for 3C~273 are 1198, 2464, 2471). For detailed data analysis, we refer to Fang et al.~2002. We
found all continua are well described by a single power law absorbed by
Galactic neutral hydrogen.

After a blind search for any statistically significant absorption features, several absorption features with S/N~$>$~4 were detected in the spectra of both quasars in the 2--42~\AA\, region of the LETGS spectral bandpass (Figure~1). These
features were subsequently fit in ISIS (Houck \& Denicola~2000).

\section{Discussion}

\subsection{PKS~2155-304}

The absorption feature at $\sim 21.6$~\AA\, was reported by Nicastro
et al.~(2002) in the LETGS-HRC archival data. We concentrate on the
absorption feature which appears at $20.02~{\rm \AA}$ (619~eV).
Considering cosmic abundances and oscillator strengths  for different
ions, O VIII Ly$\alpha$ is the only strong candidate line between
18 and 20~\AA\,,  the measured wavelength de-redshifted to the
source. It is plausible that the 20~\AA\ absorption is due to
O VIII Ly$\alpha$ in a known intervening system at $cz\approx
16,734\rm\ km~s^{-1}$. With {\sl HST}, Shull et al.~(1998) discovered
a cluster of low metallicity H I Ly$\alpha$ clouds along the
line-of-sight (LOS) towards PKS~2155-304, most of which  have
redshifts between $cz = 16,100\ {\rm km\ s^{-1}}$ and $18,500\ {\rm
km\ s^{-1}}$. Using 21 cm images from the {\sl Very Large Array} ({\sl
VLA}), they detected a small group of four H I galaxies offset
by $\sim 400-800\ {\rm h_{70}^{-1}}$ kpc from the LOS, and suggested
that the H I Ly$\alpha$ clouds could arise from gas associated
with the group (We use $\rm H_0 = 70h_{70}~km~s^{-1}Mpc^{-1}$
throughout the paper).

\begin{figure}[h]
\includegraphics[height=1.5in,width=5.5in]{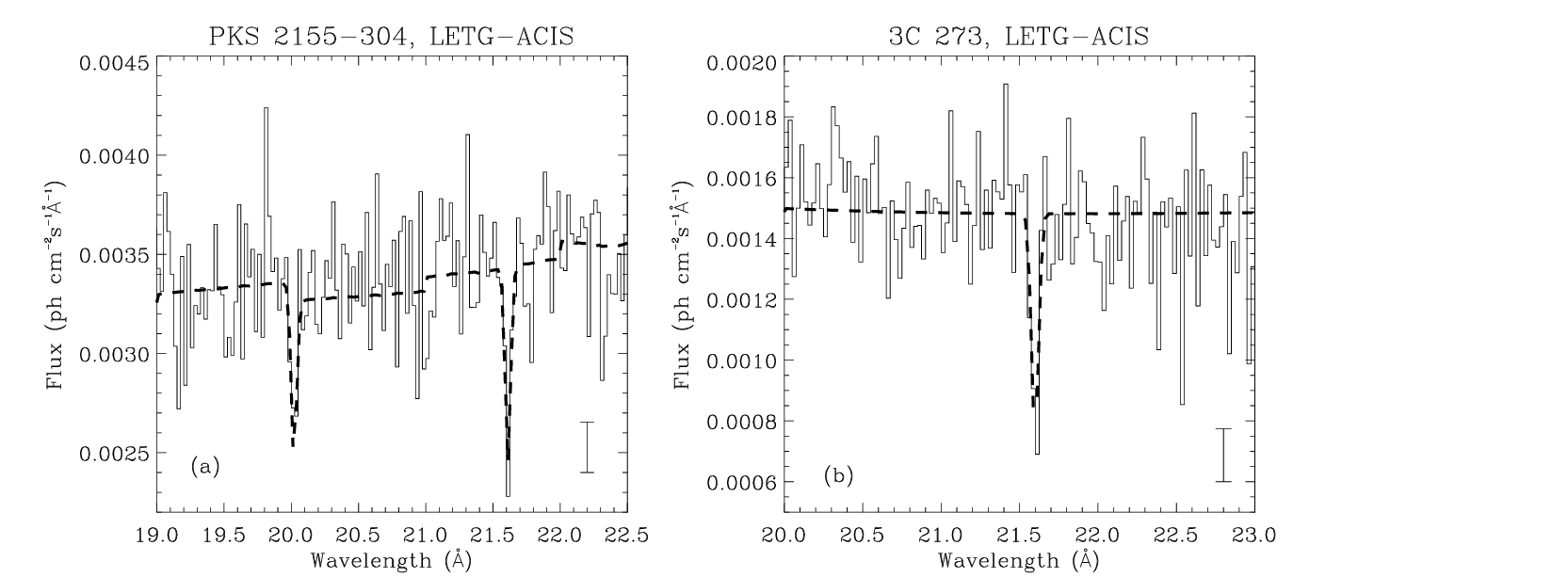}
\caption{The {\sl Chandra} LETG-ACIS spectra of (a) PKS~2155 and (b)
3C273. The dashed lines are the fitted spectra. The average
$1\sigma$ error bar plotted on the right-bottom of each panel is based
on statistics only.}
\end{figure}

Taking the absorption line to be O VIII Ly$\alpha$, we estimate
the column density is $\rm N(O~VIII) \sim 9.5 \times 10^{15}~cm^{-2}$
if the line is unsaturated. We can constrain the density of the
absorbing gas, assuming it is associated with the intervening galaxy
group. Since the line is unresolved, a lower limit of $n_b>
(1.0\times10^{-5}\ {\rm cm}^{-3})\ Z_{0.1}^{-1}f_{0.5}^{-1}l_{8}^{-1}$
can be obtained. Here $Z_{0.1}$ is the metallicity in units of 0.1
solar abundance, $f_{0.5}$ is the ionization fraction in units of 0.5,
and $l_{8}$ is the path length in units of ${\rm 8h_{70}^{-1}Mpc}$. A
more reasonable estimate of the  path length comes from the mean
projected separation of $\sim$1 Mpc for the  galaxies in the group,
which gives $n_{b} \approx 7.5 \times 10^{-5}\ {\rm
cm}^{-3}~Z_{0.1}^{-1}f_{0.5}^{-1}$. This implies a range of baryon
overdensity ($\delta_{b} \sim$ 50 - 350) over the cosmic mean $\left<n_{b}\right> = 2.14\times
10^{-7}~\rm cm^{-3}$. Interestingly,
Shull et al.(1998) estimate an overdensity for the galaxy group of
$\delta_{gal}\sim$100.

In the case of pure collisional ionization, temperature is the only
parameter  of importance over a wide range of density so long as the
gas is optically thin.  The O VIII ionization fraction peaks at
0.5, and exceeds 0.1 for temperatures $T \sim 2-5 \times10^{6}$
K. Using CLOUDY (Ferland et al.~1998) we find that photoionzation by
the cosmic UV/X-ray background is not important for $n_{b} > 10^{-5}\
{\rm cm^{-3}}$. CLOUDY calculations of the column density ratios
between other ions and O VIII also show that $T \gtrsim 10^{6.4}$ K.

Assuming  conservative upper limits of $\rm Z \lesssim 0.5Z_{\odot}$ and an 
O VIII ionization fraction of $f \lesssim 0.5$ , and a path
length of $\Delta z \lesssim 0.116$, we estimate $\rm \Omega_{b}(O~VIII)
\gtrsim 0.005h_{70}^{-1}$. This is about 10\% of the total baryon
fraction, or about 30-40\% of the WHIM gas, if the WHIM gas contains
about 30-40\% of total baryonic matter. This baryon fraction is
consistent with the prediction from Perna \& Loeb~(1998) based on a
simple analytic model.

\subsection{3C~273}

Based on the detected line equivalent width ($W_{\lambda}$) and
non-detection of local O VII He$\beta$ line at $18.6288\AA$, we
estimate the column density $\rm N(O~{VII}) = 1.8_{-0.7}^{+0.2}
\times 10^{16}~cm^{-2}$ if the line is unsaturated\footnote{Rasmussen et
  al.(2002) also reported the detection of a similar feature with {\sl
    XMM}-Newton in this conference.}. The O VII
indicates the existence of the gas with high temperatures $\sim
10^{6}$ K. The extremely high temperatures imply that O VII is
unlikely to to be produced in the nearby interstellar medium (ISM),
although we cannot rule out the possibility of an origin from
supernova remnants. It is more plausible that the He-like Oxygen is
produced in distant, hot halo gas, or even in the Local Group (LG).

{\bf 3.2.1 Local Group Origin?}\\
We can constrain the density of the absorbing gas, assuming it is
associated with the Local Group. The path length can be set equal to
the distance to the boundary of the Local Group, where the gas begins
to participate in the Hubble flow. Assuming a simple
geometry for the Local Group, the path length is $\sim 1$ Mpc (see the
following text). Adopting the 90\% lower limit of the O VII
column density, this gives $n_{b} > (2.2\times10^{-5}\ {\rm cm}^{-3})\
Z_{0.2}^{-1}f_{1}^{-1}l_{1}^{-1}$, where $Z_{0.2}$ is the metallicity
in units of 0.2 solar abundance, $f_{1}$ is the ionization fraction, and $l_{1}$ is the path length in units of 1 Mpc.

Assuming spherical symmetry and isothermality, a model
characterizing the distribution of the Local Group gas is given by the
standard $\beta$-model. We adopt a simplified geometry model of the
LG, where the LG barycenter is located along the line connecting M31
and the Milky Way, at about 450 kpc away from our Galaxy (Rasmussen
\& Pedersen~2001). At about $r \sim 1200$ kpc the gravitational
contraction of the Local Group starts to dominate the Hubble flow,
and this was defined as the boundary of the Local Group (Courteau \&
van den Bergh~1999).  Based on this simple model we estimate the
column density of O VII by integrating the O VII number
density $n_{O~{VII}}$ along this path length. We find a tight upper
limit of Local Group temperature $\rm T \leq 1.2\times10^6$ K; at temperatures
higher than $1.2\times10^6$ K, the O VII ionization fraction drops quickly
and O VIII starts to dominate. We also find that the temperature
of the Local Group should be higher than $2.3\times10^5$ K. To satisfy the
observed O VII column density we find that the gas distribution
should have a rather flat core with $r_{c} \geq 100$ kpc.

{\bf 3.2.2 Hot Halo Gas?}\\
Strong O VI absorption ($\rm \log N(O~VI) = 14.73\pm0.04$) along
the sight line towards 3C~273 was detected  with {\sl FUSE} between
-100 and +100 $\rm km\ s^{-1}$ (Sembach et al.~2001). This absorption
probably  occurs in the interstellar medium of the Milky Way disk and
halo. Absorption  features of lower ionization species are also
present at these velocities. O VI absorption is also detected
between +100 and +240 $\rm km\ s^{-1}$  in the form of a broad,
shallow absorption wing extending redward of the  primary  Galactic
absorption feature with $\rm \log N(O~VI) = 13.71$. The O VI
absorption wing  has been attributed to hot gas flowing out of the
Galactic disk as part of a "Galactic chimney" or "fountain" in the
Loop IV and North Polar Spur regions  of the sky. Alternatively, the
wing might be remnant tidal debris from  interactions of the Milky Way
and smaller Local Group galaxies (Sembach et al.~2001). It is possible
to associate the {\sl Chandra}-detected O VII absorption  with
these highly ionized metals detected by {\sl FUSE}. Here we discuss
several scenarios:

{\it (1). The O VII is related to the primary O VI
feature}: In this case, $\rm \log N(O~{VI}) = 14.73$, and  $\rm
\log [N(O~{VII})/N(O~{VI})]\sim1.5$, assuming $\rm N(O~{VII}) = 1.8\times10^{16}\ cm^{-2}$.  This is  within about a factor of
2 of the O~{\sc vii}/O~{\sc vi} ratio observed for the  PKS 2155-304
absorber and is consistent with the  idea that the gas is radiatively
cooling from a high temperature (Heckman et al.~2002).  This possibility is appealing since the  centroids of
the O~{\sc vi} and O~{\sc vii} absorption features are similar  ($\sim
6\pm10\ \rm km\ s^{-1}$  versus $-26\pm140\ \rm km\ s^{-1}$), and the
width of the resolved O~{\sc vi} line (FWHM $\sim$ 100 $\rm km\
s^{-1}$) is consistent with a broad O~{\sc vii} feature.  However,
this possibility also has drawbacks that the predicted O~{\sc viii}
column density is too high and the amount of C~{\sc iv} predicted is
too low.

{\it (2). The O VII is related to the O VI wing}: In this
case, the O~{\sc vii} is associated only with the  O~{\sc vi}
absorption "wing".  This seems like a reasonable possibility.   Then
$\rm \log [N(O~{VII})/N(O~{VI})]\sim2.5$. In collisional
ionization equilibrium, this would imply a temperature of $>10^6$ K
(Sutherland \& Dopita~1993). The non-detection of O VIII
Ly$\alpha$ absorption requires the temperatures lower than
$\sim 10^{6.3}$ K.

{\it (3). The O VII is related to none of the O VI}: In
this case, the temperature should be high enough to
prevent the production of O VI ions. We can reach the similar conclusions to those in situation (2).

We thank members of the MIT/CXC team for their support. This work is supported in part by contracts NAS 8-38249 and SAO SV1-61010. KRS acknowledges financial support through NASA contract NAS5-32985 and Long Term Space Astrophysics grant NAG5-3485.








\def\ion#1#2{#1\,#2}
\def\HI{\ion{H}{I}}
\def\Halpha{H$\alpha$}
\def\CII{\ion{C}{II}}
\def\CIV{\ion{C}{IV}}
\def\NV{\ion{N}{V}}
\def\OI{\ion{O}{I}}
\def\OV{\ion{O}{V}}
\def\OVI{\ion{O}{VI}}
\def\SiIV{\ion{Si}{IV}}
\def\kms{km\,s$^{-1}$}
\def\cmm#1{cm$^{-#1}$}
\def\deg{$^\circ$}
\def\arcmin{$^\prime$}
\def\dex#1{10$^{#1}$}
\def\tdex#1{$\times$10$^{#1}$}
\def\babs{$\vert$$b$$\vert$}
\def\zabs{$\vert$$z$$\vert$}
\def\zabskpc{$\vert$$z$(kpc)$\vert$}


\articletitle{THE FUSE SURVEY OF \OVI\ IN THE GALACTIC HALO}

\author{B.D. Savage\altaffilmark{1}, K.R. Sembach \altaffilmark{2},
        B.P. Wakker\altaffilmark{1}, P. Richter\altaffilmark{3},
        M. Meade\altaffilmark{1}, E.B Jenkins\altaffilmark{4},
        J.M. Shull\altaffilmark{5}, H.W. Moos\altaffilmark{6}, and
        G. Sonneborn\altaffilmark{7}}

\altaffiltext{1}{Department of Astronomy, University of Wisconsin, Madison, WI}
\altaffiltext{2}{Space Telescope Science Institute, Baltimore, MD}
\altaffiltext{3}{Osservatoria Astrofisico di Arcetri, Firenzi, Italy}
\altaffiltext{4}{Princeton University Observatory, Peyton Hall, Princeton, NJ}
\altaffiltext{5}{Center for Astrophysics and Space Astronomy, University of
                 Colorado, Boulder, CO}
\altaffiltext{6}{Department of Physics and Astronomy, Johns Hopkins University,
                 Baltimore, MD}
\altaffiltext{7}{Laboratory for Astronomy and Solar Physics, NASA, Greenbelt, MD}


\begin{abstract}
We summarize the results of the Far-Ultraviolet Spectroscopic Explorer (FUSE)
program to study \OVI\ in the Milky Way halo. Spectra of 100 extragalactic
objects and two distant halo stars are analyzed to obtain measures of \OVI\ 
absorption along paths through the Milky Way thick disk/halo. Strong \OVI\
absorption over the velocity range from $-$100 to 100 \kms\ reveals a widespread
but highly irregular distribution of \OVI, implying the existence of substantial
amounts of hot gas with $T$$\sim$3\tdex5 K in the Milky Way thick disk/halo. The
overall distribution of \OVI\ can be described by a plane-parallel patchy
absorbing layer with an average \OVI\ mid-plane density of $n_o$(\OVI) =
1.7\tdex{-8} \cmm3, an exponential scale height of $\sim$2.3 kpc, and a
$\sim$0.25 dex excess of \OVI\ in the northern Galactic polar region. The
distribution of \OVI\ over the sky is poorly correlated with other tracers of
gas in the halo, including low and intermediate velocity \HI, \Halpha\ emission
from the warm ionized gas at $\sim$\dex4 K, and hot X-ray emitting gas at
$\sim$\dex6 K. The \OVI\ has an average velocity dispersion, $b$=60 \kms\ and
small standard deviation of 15 \kms. Thermal broadening alone cannot explain the
large observed profile widths. A combination of models involving the radiative
cooling of hot fountain gas, the cooling of supernova bubbles in the halo, and
the turbulent mixing of warm and hot halo gases is required to explain the
presence of \OVI\ and other highly ionized atoms found in the halo.
\end{abstract}


\section{Introduction}
Absorption line observations of the highly ionized lithium-like atoms of \OVI,
\NV, and \CIV\ provide valuable information about gas in interstellar space with
temperatures of $\sim$3\tdex5 to \dex5 K. Among these three species, \OVI\ is
especially important because of the large cosmic abundance of oxygen and the
large energy (113.9 eV) required to convert \OV\ into \OVI. Studies of
interstellar \OVI\ have been hampered by the difficulties involved in making
observations in the far-UV wavelength range of its resonance doublet at 1031.93
and 1037.62 \AA. Instruments operating efficiently at wavelengths shortward of
$\sim$1150 \AA\ require reflecting optics with special coatings (such as LiF or
SiC) and windowless detectors. Although the Copernicus satellite successfully
observed interstellar \OVI\ toward many stars in the Galactic disk (Jenkins
1978a,b), the studies were limited to stars with visual magnitudes m$_V$$<$7 and
did not yield information about the extension of \OVI\ away from the Galactic
plane. 

Except for brief observing programs with the Hopkins Ultraviolet Telescope (HUT;
Davidsen 1993) and the spectrographs in the Orbiting and Retrievable Far and
Extreme Ultraviolet Spectrometers (ORFEUS; Hurwitz \& Bowyer 1996; Hurwitz et
al. 1998; Widmann et al. 1998; Sembach, Savage, \& Hurwitz 1999), the study of
\OVI\ in the Milky Way halo has required the high throughput capabilities of the
Far-Ultraviolet Spectroscopic Explorer (FUSE) satellite launched in 1999 (Moos
et al. 2000; Sahnow et al. 2000).


\section{Observations and Reductions}
The FUSE \OVI\ catalog paper (Wakker et al.\ 2003) contains the full details of
the FUSE observations and reductions of Galactic \OVI\ absorption toward 100
extragalactic sources and two halo stars. (The \OVI\ $\lambda$1037.62 line is
often confused by blending with \CII* $\lambda$1037.02 and the H$_2$ (5-0) R(1)
and P(1) lines at 1037.15 and 1038.16 \AA. The \OVI\ $\lambda$1031.93 line is
usually relatively free of blending, since the H$_2$ (6-0) P(3) and R(4) lines
are often relatively weak and are displaced in velocity by $-$214 and 125 \kms\
from the rest \OVI\ velocity.

\begin{figure}[p]
\centerline{\includegraphics[width=\textwidth]{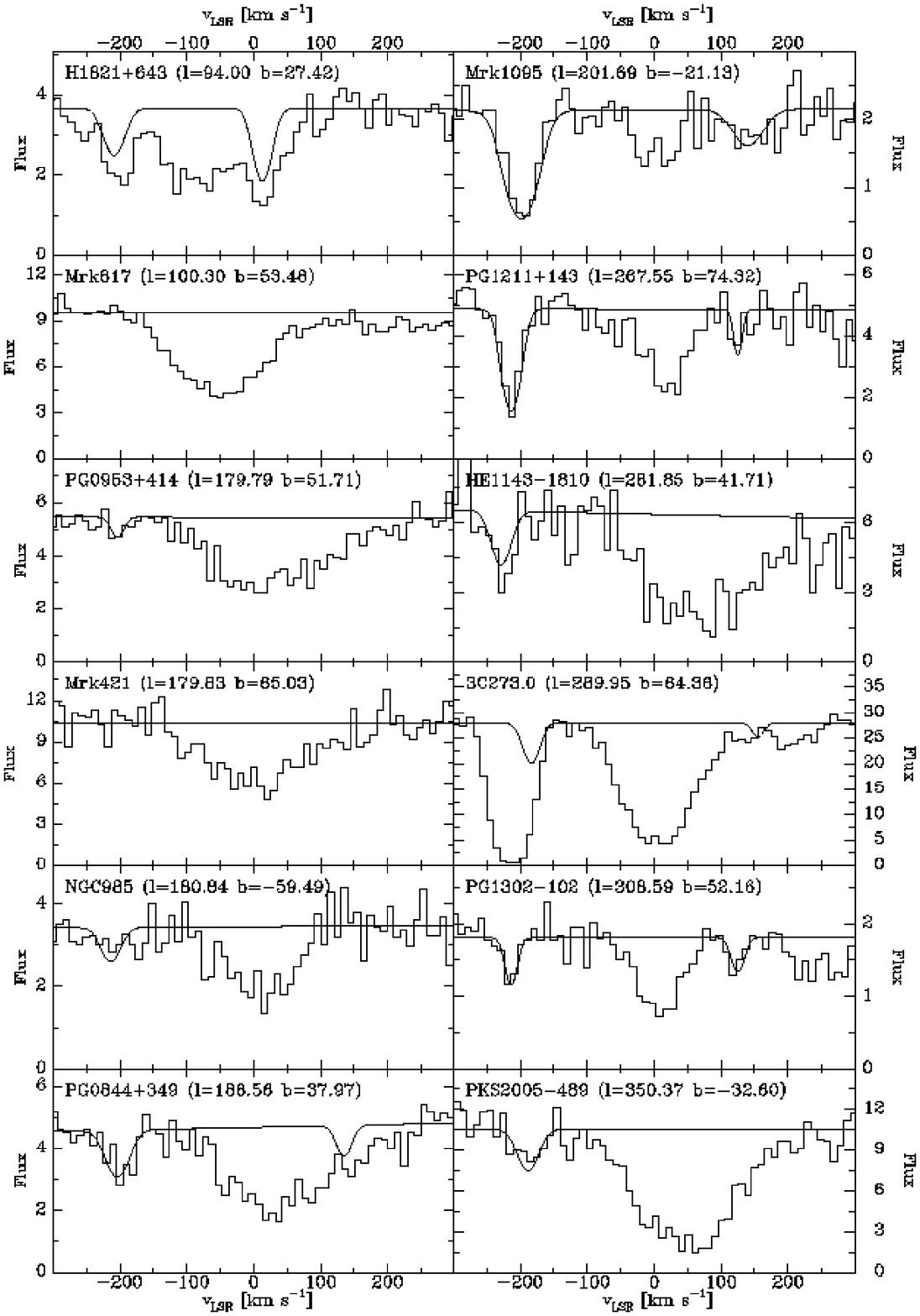}}\caption{
A sample of the \OVI\ $\lambda$1031.93 absorption line profiles from Wakker et
al. (2003) is displayed. Flux (\dex{-14} erg \cmm2 s$^{-1}$ \AA$^{-1}$) is 
plotted against LSR velocity with the solid line showing the continuum placement
including an estimate of the contaminating H$_2$ (6-0) R(3) and R(4) absorption
when necessary (from Savage et al.\ 2003).
}\end{figure}

Information about \OVI\ in the Milky Way halo (see Savage et al. 2003) has
mostly been derived from the relatively blend-free \OVI\ $\lambda$1031.93 line.
Figure 1 shows \OVI\ $\lambda$1031.93 line profiles versus LSR velocity for a
sample of 12 objects. The continuum placement is generally quite reliable since
most of the objects are AGNs, which usually have well-defined power-law
continua. Along sight lines where the H$_2$ (6-0) P(3) and R(4) lines at 1031.19
\AA\ and 1032.35 \AA\ blend with \OVI, the expected strength of the
contamination is based on an analysis of other H$_2$ J=3 and 4 absorption lines
in the spectrum. The estimated H$_2$ absorption line profiles have been included
in Figure 1. For \OVI\ at 3\tdex5 K (the temperature at which \OVI\ peaks in
abundance) the thermal Doppler contribution to the \OVI\ line width corresponds
to $b$=17.7 \kms\ (FWHM=29.4 \kms), where $b$ is the standard Doppler spread
parameter. The expected thermal width of the \OVI\ line is comparable
to the FUSE resolution of 20--25 \kms. Since the \OVI\ lines are usually
resolved, reliable column densities can be obtained using the apparent optical
depth method (Savage \& Sembach 1991). The \OVI\ column densities, velocities
and line widths can be found in Savage et al.\ (2003).


\section{Separating Low Velocity Thick \OVI\ Disk Absorption from High Velocity
Absorption}
The lines of sight to the objects observed in the FUSE halo gas \OVI\ program
pass through Milky Way disk gas, thick disk/halo gas, intermediate-velocity
clouds (IVCs), high-velocity clouds (HVCs), and may even sample intergalactic
\OVI\ in the Local Group of galaxies. At high Galactic latitude the dividing
lines in velocity between high, intermediate, and low velocity are generally at
$\vert$v$_{\rm LSR}$$\vert$ = 90 and 30 \kms, although the effects of Galactic
rotation must also be considered. The \OVI\ absorption lines trace a complex set
of processes and phenomena involving hot gas in the Milky Way disk, thick
disk/halo, and beyond. We refer to \OVI\ in the Milky Way disk-halo interface
extending several kpc away from the Galactic plane as the ``thick disk \OVI''.

The \OVI\ profiles exhibit a diversity of strengths and kinematical behavior.
Disk, and thick disk \OVI\ are clearly detected within the general velocity
range $-$90 to 90 \kms\ toward 91 of the 102 survey objects. In addition, the
absorption profiles reveal \OVI\ at high velocities with $\vert$v$_{\rm
LSR}$$\vert$ ranging from $\sim$100 to 400 \kms\ in $\sim$60\% of the sight
lines. The \OVI\ absorption at high-velocity is considered by Sembach et al.\
(2003).


\section{Column Density Distribution for the Thick Disk Absorption}
The relatively strong \OVI\ absorption associated with the thick disk of
Galactic \OVI\ has log N(\OVI) between 13.85 and 14.78. The equivalent column
density perpendicular to the Galactic plane, log [N(\OVI)sin\babs], ranges from
13.45 to 14.68. Figures 2a and 2b show histograms of the distribution of log
N(\OVI) and log [N(\OVI)sin\babs]. The distribution of \OVI\ on the sky is quite
irregular. In Figure 3 the \OVI\ measurements are superposed on a grey scale
representation of the 0.25 keV X-ray background as recorded by ROSAT (Snowden et
al. 1997).

\begin{figure}[t]
\centerline{\includegraphics[width=\textwidth]{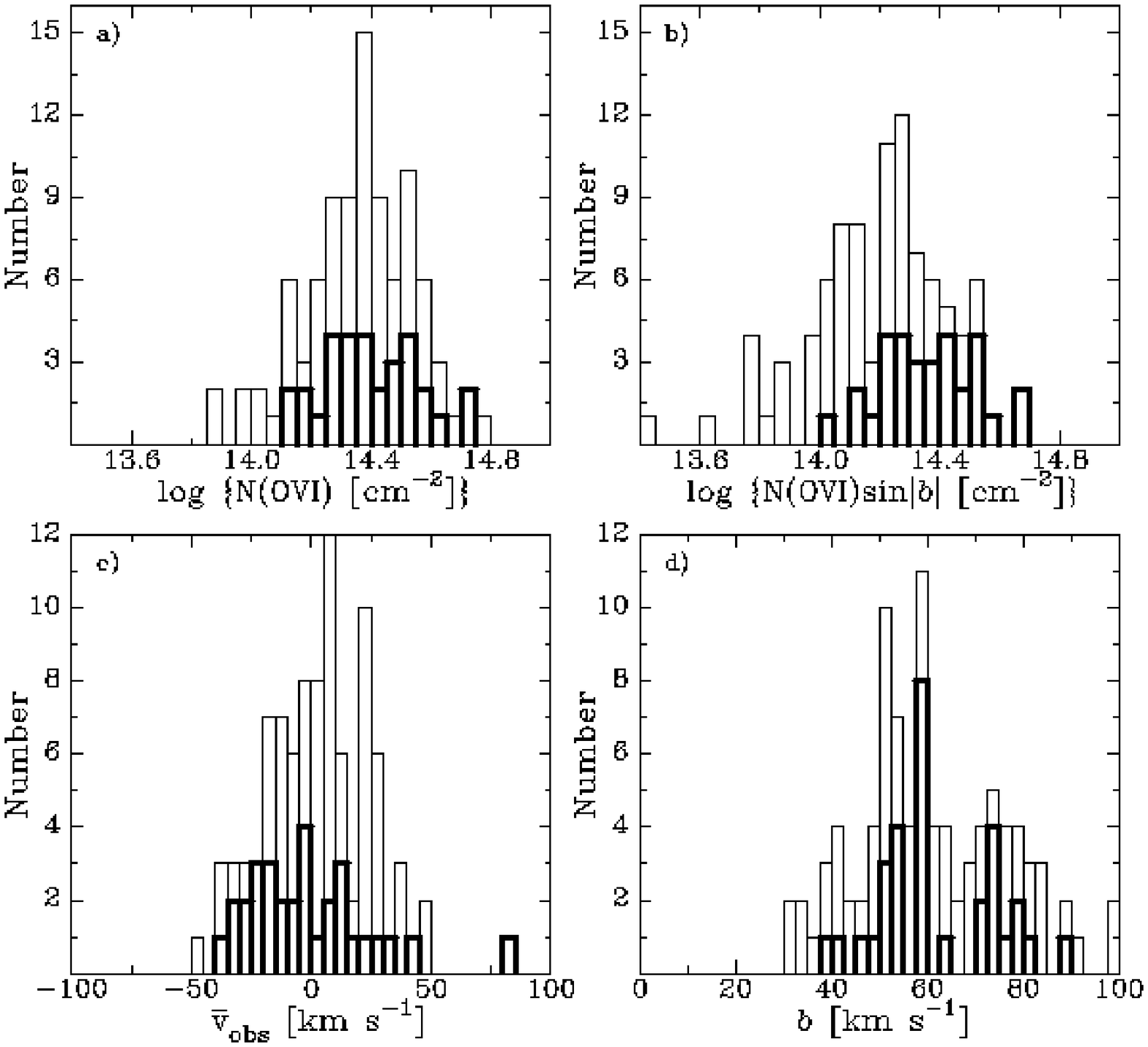}}\caption{
Number distribution of log N(\OVI), log [N(\OVI)sin\babs], $v_{\rm obs}$ (\kms),
and $b$ (\kms) for \OVI\ absorption associated with the thick disk of the Milky
Way are shown in (a), (b), (c), and (d), respectively. In the histograms for log
N(\OVI) and log [N(\OVI)sin\babs] upper limits are indicated as detections. In
the various panels the upper histogram is for the full object sample while the
heavy line histogram is for survey objects with \babs$>$45\deg\ (from Savage et
al.\ 2003).
}\end{figure}

\begin{figure}[t]
\centerline{\includegraphics[angle=270,width=\textwidth]{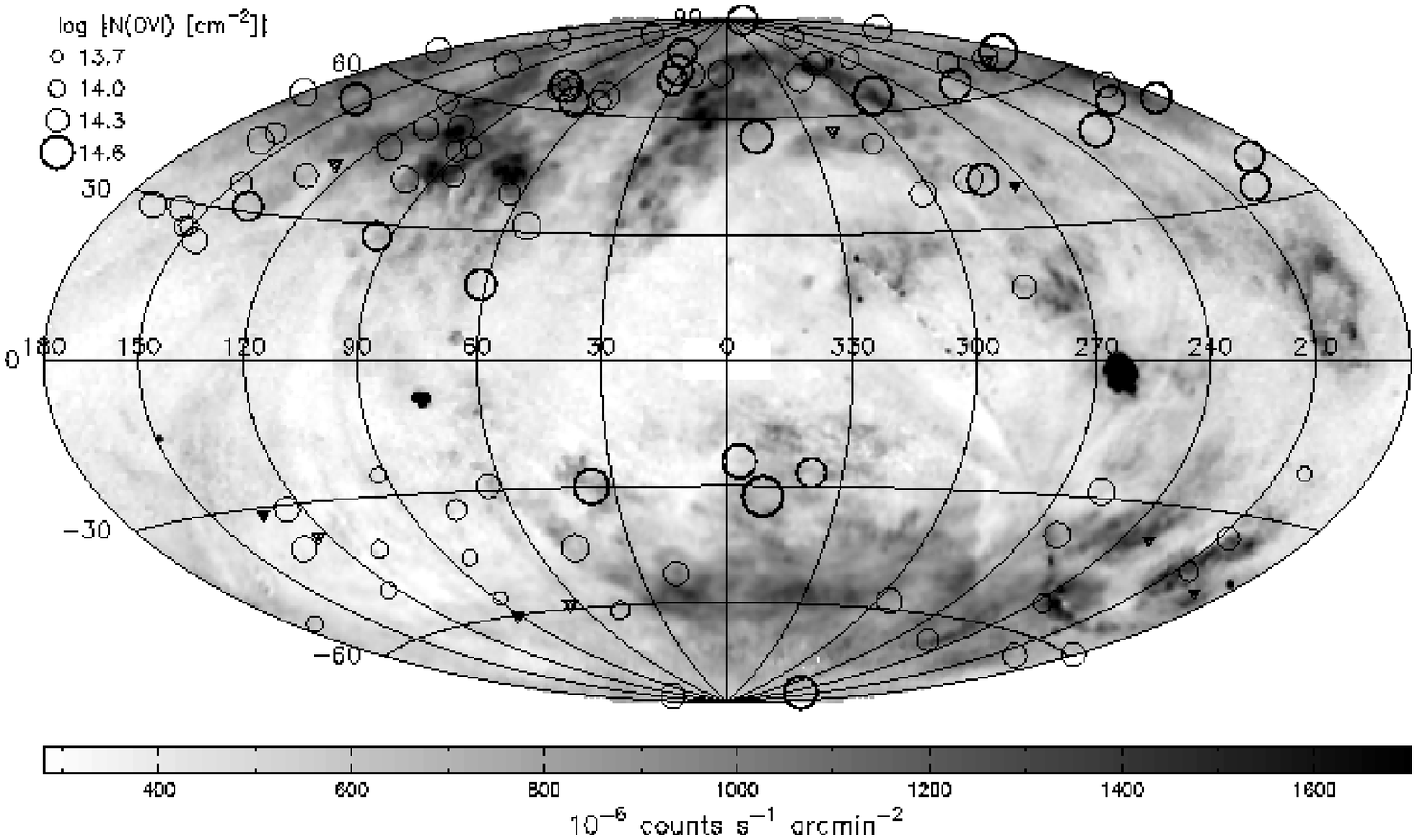}}\caption{
Values of log N(\OVI) for the thick disk of the Milky Way are represented as
circles displayed in these aitoff projections of the sky with the Galactic
center at the center of the figures and Galactic longitude increasing to the
left. The circle size is proportional to log N(\OVI) according to the code
shown. Upper limits to log N(\OVI) are denoted with triangles with a size
proportional to the limit. The grey scale shows the 0.25 keV X-ray sky diffuse
background count rate as measured by the ROSAT Satellite (Snowden et al.\ 1997)
(from Savage et al.\ 2003).
}\end{figure}


\section{The Extension of \OVI\ Away from the Galactic Plane}
The large \OVI\ column densities measured along extragalactic sight lines imply
that a substantial amount of \OVI\ is situated in the Galactic thick disk at
large distances away from the plane of the Galaxy. Figure 4 shows log
[N(\OVI)sin\babs] versus log \zabskpc\ for various data samples, as detailed in
the figure caption. There is a considerable spread in the values of log [N(O
VI)]sin\babs] at all values of \zabs. The data points for the LMC and SMC
directions at \zabs=27 and 50 kpc, respectively, show the spread in values of
log[N(\OVI)sin\babs] over the relatively small angular extents of these two
galaxies. In contrast the spread for the Copernicus observations and the
extragalactic observations illustrate the variation in the values of log
[N(\OVI)sin\babs] extending over most of the sky.

\begin{figure}[t]
\centerline{\includegraphics[width=\textwidth]{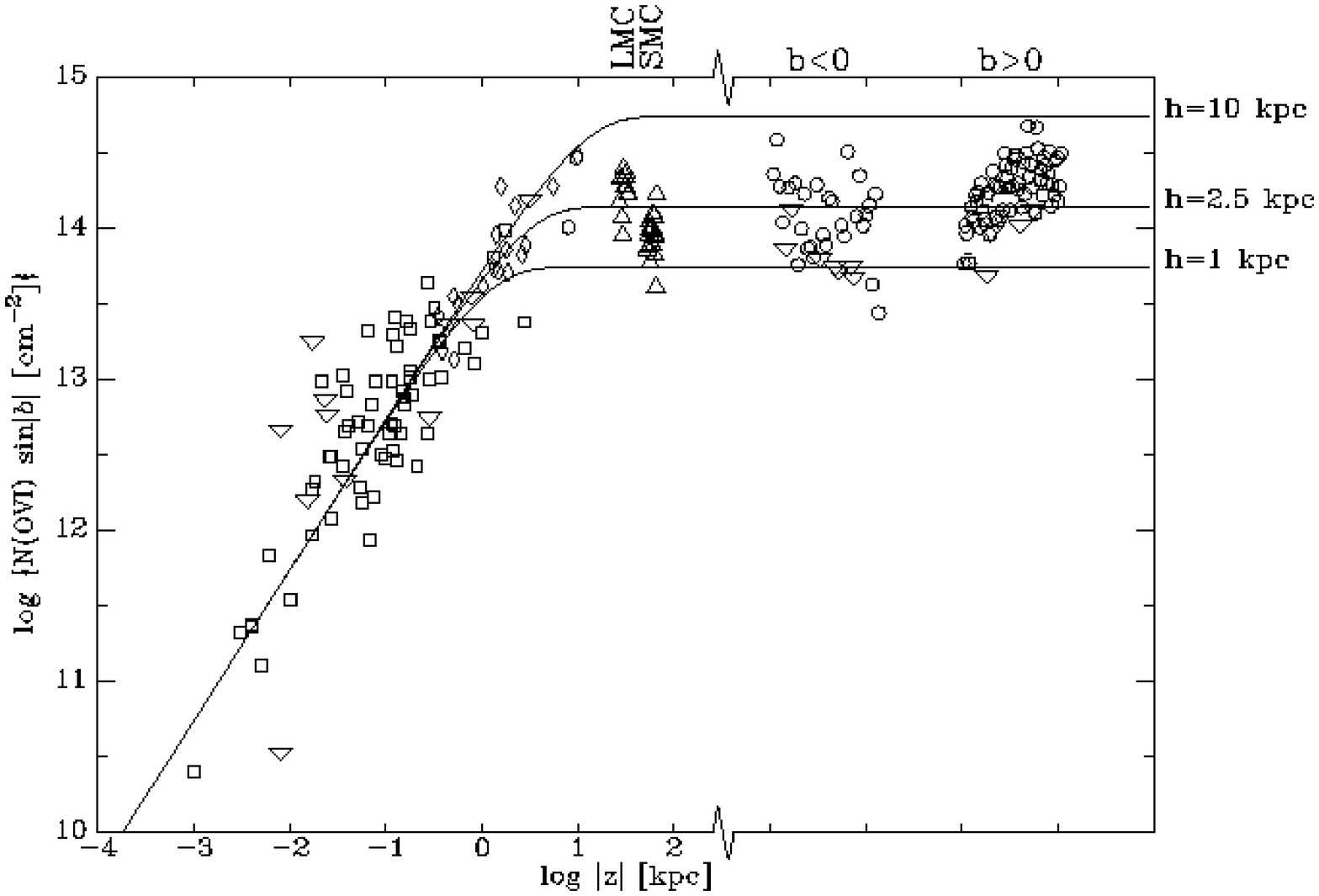}}\caption{
log [N(\OVI)sin\babs] for gas in the Milky Way disk and thick disk is plotted
against log \zabskpc\ for (1) stars from the Copernicus sample of stars in the
Galactic disk (Jenkins 1978a, open squares); (2) the FUSE 22 halo star survey
(Zsargo et al. 2003, open diamonds); (3) Milky Way absorption toward stars in
the LMC and SMC (Howk et al.\ 2002 and Hoopes et al.\ 2002, open upward pointing
triangles); and (4) the 100 extragalactic and two Galactic stellar lines of
sight analyzed by us (Savage et al.\ 2003, open circles, with the 100
extragalactic objects plotted on the right hand side of the figure in the region
beyond the break in the log \zabskpc\ axis and the two stars plotted near log
\zabs$\sim$1.0; 3$\sigma$ upper limits are plotted as the downward pointing
triangles). The three solid lines show the expected behavior of the log
N(\OVI)sin\babs\ versus log \zabskpc\ distribution for a smoothly distributed
exponentially stratified plane parallel Galactic atmosphere with an \OVI\
mid-plane density $n_o$(\OVI)=1.7\tdex{-8} \cmm3 and exponential scale heights
1.0, 2.5 and 10 kpc (adapted from Savage et al.\ 2003).
}\end{figure}

The three solid lines in Figure 4 show the expected behavior of the log
[N(\OVI)sin\babs] versus log \zabskpc\ distribution for a smoothly distributed,
exponentially stratified plane-parallel Galactic atmosphere with an \OVI\
mid-plane density $n_o$(\OVI)= 1.7\tdex{-8} \cmm3\ and exponential scale heights
of 1.0, 2.5, and 10 kpc. The use of a mid-plane density of 1.7\tdex{-8} \cmm3\
appears well justified since the value represents an average extending over
$\sim$220 stars in the combined Copernicus and FUSE disk star sample (Jenkins et
al. 2001). The large irregularity in the distribution of \OVI\ and the
enhancement in the amount of \OVI\ over the northern Galactic hemisphere
introduces a complication when trying to estimate a Galactic scale height for
\OVI.

The data points in Figure 4 are consistent with an \OVI\ scale height in the
range between 1.0 and 10 kpc. The observations are not well fitted by a
symmetric plane parallel model for the distribution of \OVI\ extending away from
the Galactic plane. One can imagine a wide range of possible models to fit the
observed values of log [N(\OVI)sin\babs] that are more complicated. The simplest
model that improves the fit is a superposition of a plane-parallel patchy
absorbing layer with an exponential scale height of $\sim$2.3 kpc, a mid-plane
density of 1.7\tdex{-8} \cmm3, and a $\sim$0.25 dex excess of \OVI\ at the
higher northern Galactic latitudes.


\section{Kinematics of the Thick Disk \OVI}
The kinematic properties of the \OVI\ absorption provide useful information
about the various physical processes controlling the distribution of gas at
$T$$\sim$3\tdex5 K in the Galactic halo. Two simple measures of the kinematics
of the \OVI\ absorption studied by Savage et al. (2003) are the average LSR
velocity, and the velocity dispersion of the absorption as measured by the
Doppler-spread parameter $b$. The distributions of the values of $v_{\rm obs}$
and $b$ are shown in Figures 2c and 2d. In these figures the upper (light line)
histogram is for the complete sample, while the lower (heavy line) histogram is
for high latitude objects (\babs$>$45\deg). At high latitudes, where the effects
of Galactic rotation are small, the values of $v_{\rm obs}$ exhibit a large
spread in velocity with an average of 0 \kms\ and a standard deviation of 22
\kms. It is interesting that the \OVI\ at high latitudes in each Galactic
hemisphere is moving with positive and negative velocity with nearly equal
frequency. For the full sample, $b$(min)=30 \kms, $b$(max)=99 \kms,
$b$(median)=59 \kms, $b$(average)=61 \kms, and a the standard deviation on $b$
is 15 \kms. The \OVI\ absorption lines are much wider than the absorption
expected from thermal Doppler broadening alone since gas at $T$$\sim$3\tdex5 K,
the temperature at which \OVI\ is expected to peak in abundance, has
$b$(\OVI)=17.7 \kms. Inflow, outflow, Galactic rotation, and turbulence
therefore also affect the profiles.


\section{\OVI\ Versus Other ISM Tracers}
The relationship between the column density of \OVI\ and other ISM tracers was
studied for \HI, \Halpha, soft X-ray emission, and non-thermal radio emission.
In all cases the amount of \OVI\ is poorly related to the amount of these other
ISM tracers. One example is shown in Figure 3 where log N(\OVI) is compared to
the 0.25 keV soft X-ray background count rate from Snowden et al. (1997) at a
resolution of 36\arcmin. The correspondence between the X-ray sky brightness and
N(\OVI) is poor. This is not too surprising given the 0.25 Kev diffuse X-ray
background is a complex superposition of a non-uniform local bubble component
and halo and extragalactic background components (Snowden et al. 1998, 2000)
experiencing attenuation due to photoelectric absorption occurring in cooler
foreground gas. 


\section{Origin of \OVI\ and Other Highly Ionized Species in the Galactic Halo}
The FUSE \OVI\ survey observations reported here provide important new insights
into the distribution and kinematics of highly ionized gas in the Milky Way
halo. The observations confirm the basic validity of Spitzer's (1956) prediction
that the ISM of the Galaxy contains a hot gas phase that extends well away from
the Galactic plane. Theories for the origin of highly ionized gas in the Milky
Way halo must explain the distribution, ionization, kinematics, and support of
the gas. For reviews of the models see Spitzer (1990), McKee (1993), and Savage
(1995). The ionization of \SiIV\ and \CIV\ is likely either from electron
collisions in a hot gas, photoionization, or some combination of both processes.
With their high ionization threshold, \OVI\ and \NV\ are more likely to be
produced by collisional ionization in hot gas. However, non-equilibrium
ionization effects will probably be important because collisionally ionized gas
cools very rapidly in the temperature range (1--5)\tdex5 K. Such transition
temperature gas might occur within the cooling gas of a ``Galactic fountain''
(Shapiro \& Field 1976; Edgar \& Chevalier 1986; Shapiro \& Benjamin 1991), in
the conductively heated interface region between the hot and cool interstellar
gas (Ballet, Arnaud, \& Rothenflug 1986), in radiatively cooling SN bubbles
(Slavin \& Cox 1992, 1993) or in turbulent mixing layers (TMLs) where hot gas
and warm gas are mixed by turbulence to produce gas with non-equilibrium
ionization characteristics (Begelman \& Fabian 1990; Slavin, Shull, \& Begelman
1993). The heating and ionization of the gas could also occur through magnetic
reconnection processes (Raymond 1992; Zimmer, Lesch, \& Birk 1997).  With a
number of processes probably contributing to the \OVI\ found in the halo, it
will be difficult to clearly identify the most important processes.


\section{Summary}
FUSE far-UV spectra of 100 extragalactic objects and two halo stars are used to
obtain measures of \OVI\ far-UV absorption along paths through the Milky Way
halo. High velocity \OVI\ absorption with $\vert$$v$$\vert$$>$100 \kms\ is seen
along $\sim$60\% of the sight lines (see Sembach et al. 2003). In this paper we 
review the results found in the study the Milky Way thick-disk \OVI\ absorption
that covers the approximate velocity range from $-$100 to 100 \kms\ (see Savage
et al. 2003). The results are:

1. There exists a widespread but highly irregular distribution of \OVI\ in the
Galactic thick disk, implying the existence of substantial amounts of hot gas
with $T$$\sim$3\tdex5 K. The integrated \OVI\ logarithmic column density through
the halo, log N(\OVI), ranges from 13.85 to 14.78 with an average of 14.38.

2. Averages of log [N(\OVI)sin\babs] over Galactic latitude reveal a $\sim$0.25
dex excess of \OVI\ to the north Galactic polar region compared to the rest of
the Galaxy.

3. The observations are not well described by a simple symmetrical
plane-parallel patchy distribution of \OVI\ absorbing structures. The simplest
departure from such a model that provides a better fit to the observations is a
plane-parallel patchy absorbing layer with an average \OVI\ mid-plane density of
$n_o$(\OVI)=1.7\tdex{-8} atoms \cmm3\ and a scale height of $\sim$2.3 kpc
combined with a $\sim$0.25 dex excess of \OVI\ absorbing gas in the northern
Galactic polar region.

4. The \OVI\ is poorly correlated with other ISM tracers of gas in the halo,
including low and intermediate velocity \HI, \Halpha\ emission from the warm
ionized medium, and hot 0.25 keV X-ray emitting gas.

5. The \OVI\ profiles range in velocity dispersion, $b$, from 30 to 99 \kms,
with an average value of 61 \kms\ and standard deviation of 15 \kms. The thermal
Doppler component of the broadening cannot explain the large observed profile
widths since gas at $T$$\sim$3\tdex5 K, the temperature at which \OVI\ is
expected to peak in abundance in collisional ionization equilibrium, has
$b$(\OVI)=17.7 \kms. Inflow, outflow, Galactic rotation, and turbulence
therefore also affect the profiles.

6. The \OVI\ average absorption velocities for thick disk gas toward high
latitude objects (\babs$>$45\deg) range from $-$37 to 82 \kms, with a high
latitude sample average of 0 \kms\ and a standard deviation of 21 \kms. Thick
disk \OVI\ is observed to be moving both toward and away from the plane with
roughly equal frequency.

7. The broad high positive velocity \OVI\ absorption wings extending from
$\sim$100 to $\sim$250 \kms\ seen in the spectra of 21 objects may be tracing
the outflow of gas into the halo, although we can not rule out a more distant
origin.

8. A combination of models involving the radiative cooling of hot gas in a
Galactic fountain flow, the cooling of hot gas in halo supernova bubbles, and
the turbulent mixing of warm and hot halo gases appears to be required to
explain the highly ionized atoms found in the halo. If the origin of the \OVI\
is dominated by a fountain flow, a mass flow rate of approximately 
1.4\,M$_\odot$\,yr$^{-1}$ to each side of the Galactic disk for cooling hot gas
with an average density of \dex{-3} \cmm3\ is required to explain the average
value of log [N(\OVI)sin\babs\ found in the southern Galactic hemisphere.

This work is based on data obtained for the Guaranteed Time Team by
NASA-CNES-CSA FUSE mission operated by Johns Hopkins University. Financial
support to U.S.\ participants has been provided by NASA contract NAS5-32985.
K.R.S.\ acknowledges additional financial support through NASA Long Term Space
Astrophysics grant NAG5-3485. B.P.W.\ acknowledges additional support from NASA
grants NAG5-9179, NAG5-9024, and NGG5-8967.









\newcommand{\degr}{^\circ}
\def\fdg{\hbox{$.\!\!^\circ$}}



\articletitle{The FUSE Survey of  High Velocity O VI
 in the Vicinity of the Milky Way}

\author{K.R. Sembach\altaffilmark{1}, B.P. Wakker\altaffilmark{2}, B.D. Savage\altaffilmark{2}, P. Richter\altaffilmark{3}, M. Meade\altaffilmark{2},
J.M. Shull\altaffilmark{4}, E.B. Jenkins\altaffilmark{5}, H.W. Moos\altaffilmark{6}, and G. Sonneborn\altaffilmark{7}}
\altaffiltext{1}{Space Telescope Science Institute, 3700 San Martin Dr., Baltimore, MD 21218}
\altaffiltext{2}{Astronomy Dept., University of Wisconsin, 475 N.Charter St., Madison, WI 53706}
\altaffiltext{3}{Osservatorio Astrofisico di Arcetri, Largo 
        E. Fermi 5, Florence, Italy}
\altaffiltext{4}{Center for Astrophysics and Space Astronomy, University of Colorado, Boulder, CO  80309}
\altaffiltext{5}{Princeton University Observatory, Peyton Hall, 
        Princeton, NJ  08544}
\altaffiltext{6}{Department of Physics \& Astronomy, Johns Hopkins 
        University, Baltimore, MD  21218}
\altaffiltext{7}{Laboratory for Astronomy and Solar Physics, 
        NASA/GSFC, Greenbelt, MD 20771}

\chaptitlerunninghead{High Velocity O\,{\sc vi}}

\begin{abstract}
We describe an extensive FUSE survey of  high velocity O\,{\sc vi} 
absorption along $\sim100$ complete sight lines through the Galactic halo.  
The high velocity O\,{\sc vi} traces tidal interactions with the 
Magellanic Clouds, accretion of gas, outflowing material from the Galactic 
disk, warm/hot gas interactions in a highly extended Galactic corona, and 
intergalactic gas in the Local Group. 
 Approximately 60\%  of the sky (and perhaps as much as 85\%)  is 
covered by high velocity hot H$^+$ associated with the high velocity 
O\,{\sc vi}.
Some of the O\,{\sc vi} may be produced at the boundaries between warm 
clouds and a hot, highly-extended Galactic corona or Local Group medium. 
A hot Galactic corona or Local Group 
medium and the prevalence of high velocity O\,{\sc vi}
are expected in various galaxy formation scenarios.  
Additional spectroscopic data in the coming years will help to determine
the ionization properties of the high velocity clouds and 
discriminate
between the multiple types of high velocity O\,{\sc vi} features
found in this study.  
\end{abstract}

\section{Introduction}
Observational information about the highly ionized gas in the vicinity
of galaxies is required for complete descriptions of galaxy formation
and evolution. In this article, we outline a program we have 
conducted with the Far 
Ultraviolet Spectroscopic Explorer (FUSE) to study the hot
gas in the vicinity of the Milky Way.  The study is described in detail 
in a series of three articles devoted to probing the highly ionized 
oxygen (O\,{\sc vi}) absorption along complete paths through the Galactic 
halo and Local Group.  The articles include a catalog of the spectra and basic 
observational information (Wakker et al. 2003), a study of the hot
gas in the Milky Way halo (Savage et al. 2003), and an investigation of the
highly ionized high velocity gas in the vicinity of the Galaxy (Sembach
et al. 2003).  Here, we summarize the high velocity gas results.  A 
companion paper by Savage et al. (this volume) summarizes the Milky
Way thick disk/halo results.

The O\,{\sc vi} $\lambda\lambda1031.926, 1037.617$ doublet lines
 are the best UV resonance lines to use for 
kinematical investigations of hot ($T \sim 10^5-10^6$\,K) gas in the 
low-redshift universe. 
X-ray spectroscopy of the interstellar or intergalactic gas in 
higher ionization lines (e.g., O\,{\sc vii}, O\,{\sc viii}) is possible with
XMM-Newton and the Chandra X-ray Observatory for a small number 
of sight lines toward AGNs and QSOs, but the spectral resolution 
(R~$\equiv \lambda/\Delta\lambda < 400$)
is modest compared to that afforded by FUSE (R~$\sim 15,000$).  While 
the X-ray lines provide extremely useful information about the amount of 
gas at temperatures greater than $10^6$\,K, the interpretation of where that
gas is located, or how it is related to the $10^5-10^6$\,K gas traced by
O\,{\sc vi}, is hampered at low redshift by the complexity of the hot ISM and 
IGM along the sight lines observed.

\section{Properties of the High Velocity O\,{\sc vi}}

We have conducted a study of the highly ionized high velocity gas in the 
vicinity of the Milky Way using an extensive set of FUSE data.  
We summarize the results for  the sight lines toward 100 AGNs/QSOs and two 
distant halo stars in this section (see Sembach et al. 2003; 
Wakker et al. 2003).  For the purposes of this study, gas with $|v_{LSR}|
\stackrel{_>}{_\sim} 100$ km~s$^{-1}$ is typically identified as ``high 
velocity'', while lower velocity gas is 
attributed to the Milky Way disk and halo.  A sample 
spectrum from the survey is shown in Figure~1.

\begin{figure}[h!]
\includegraphics{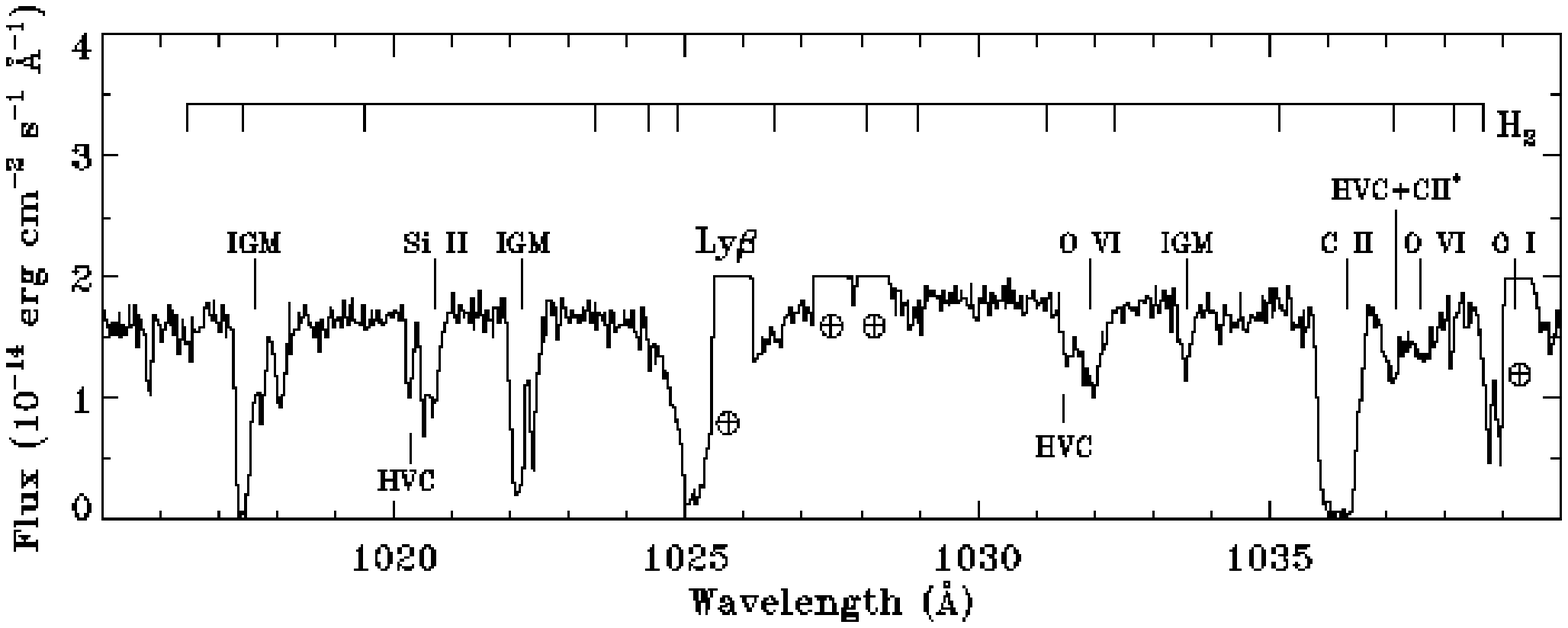}
\vspace{2.2in}
\caption{\small
A portion of the FUSE spectrum of PG\,1259+593 in the 1015--1040\,\AA\
spectral region. Prominent interstellar and intergalactic lines are 
indicated.  Although weak along this sight line, the wavelengths of 
common H$_2$ lines are indicated at the top of the figure.
 The Si\,{\sc ii} HVC
and O\,{\sc vi} HVC absorption features identified below the spectrum
trace gas in Complex~C.  H\,{\sc i} and O\,{\sc i}
airglow emission lines ($\oplus$) have been truncated for clarity.}
\end{figure}

\subsection{Detections of High Velocity O\,{\sc vi}}
We have identified approximately 85 individual 
high velocity O\,{\sc vi} features along the 102 sight lines in our sample.
A critical part of this identification process involved detailed consideration
of the absorption produced by O\,{\sc vi} and other species 
(primarily H$_2$) in the thick disk and halo of the Galaxy, as well as the 
absorption produced by low-redshift intergalactic absorption lines of 
H\,{\sc i} and ionized metal species. 
Our careful process
of identifying the high velocity features, and the possible complications
involved in these identifications,
are described by Wakker et al. (2003).
We searched for absorption in a velocity 
range of $\pm1200$ km~s$^{-1}$ centered on the O\,{\sc vi} 
$\lambda1031.926$ line.  With few exceptions, the high velocity O\,{\sc vi}
absorption is confined to $|v_{LSR}| \le 400$ km~s$^{-1}$,  
indicating that the O\,{\sc vi} features observed are either associated with 
the Milky Way or nearby clouds within the Local Group.  

We detect high velocity O\,{\sc vi} $\lambda1031.926$ absorption 
with total equivalent widths $W_\lambda > 30$ m\AA\ at 
$\ge 3\sigma$ 
confidence along 59 of the 102 sight lines surveyed.  For the highest 
quality sub-sample of the dataset, the high velocity detection frequency 
increases to 22 of 26 sight lines.  Forty of the 59 
sight lines have high velocity O\,{\sc vi} $\lambda1031.926$ absorption
with  $W_\lambda > 100$ m\AA, and 27 have $W_\lambda > 150$~m\AA.  
Converting these O\,{\sc vi} equivalent width detection frequencies
 into estimates of $N$(H$^+$) in the 
hot gas indicates
that $\sim60$\% of the sky (and perhaps as much as $\sim85$\%) 
is covered by hot ionized hydrogen at a level of 
$N({\rm H}^+)  \stackrel{_>}{_\sim} 8\times10^{17}$ cm$^{-2}$,
assuming an ionization fraction f$_{\rm O\,VI} < 0.2$ and a gas
 metallicity similar to that of the Magellanic Stream ($Z\sim0.2-0.3$).
This detection frequency of 
hot H$^+$ associated with the  high velocity O\,{\sc vi} is
 larger than the value of $\sim37$\% found for high velocity warm neutral 
gas with 
$N$(H\,{\sc i})$ \sim 10^{18}$ cm$^{-2}$ traced through 21\,cm emission
(Lockman et al. 2002).

\subsection{Velocities}

The high velocity O\,{\sc vi} features have velocity centroids ranging 
from $-372 < v_{LSR} < -90$ km~s$^{-1}$ to 
$+93 < v_{LSR}
< +385$ km~s$^{-1}$. There are an additional 6  
confirmed or very likely ($>90$\% confidence) detections and 2 tentative 
detections of O\,{\sc vi} 
between $v_{LSR} = +500$ and +1200 km~s$^{-1}$; these very high velocity
features probably trace intergalactic gas beyond the Local Group. 
Most of the high velocity O\,{\sc vi} features have velocities incompatible 
with those of Galactic rotation (by definition).  The dispersion about the 
mean of the high velocity O\,{\sc vi}
centroids  decreases when the velocities are converted from the
Local Standard of Rest (LSR) into the Galactic Standard of Rest (GSR) and 
the Local Group Standard of Rest (LGSR) reference frames.   While this 
reduction is expected if the 
O\,{\sc vi} is associated with gas in a highly extended Galactic corona or 
in the Local Group,  it {\it does not} provide sufficient proof by itself of an
extragalactic location for the high velocity gas.  Additional information,
such as the gas metallicity or ionization state, is needed to constrain the 
cloud locations.

\begin{figure}[h!]
\includegraphics{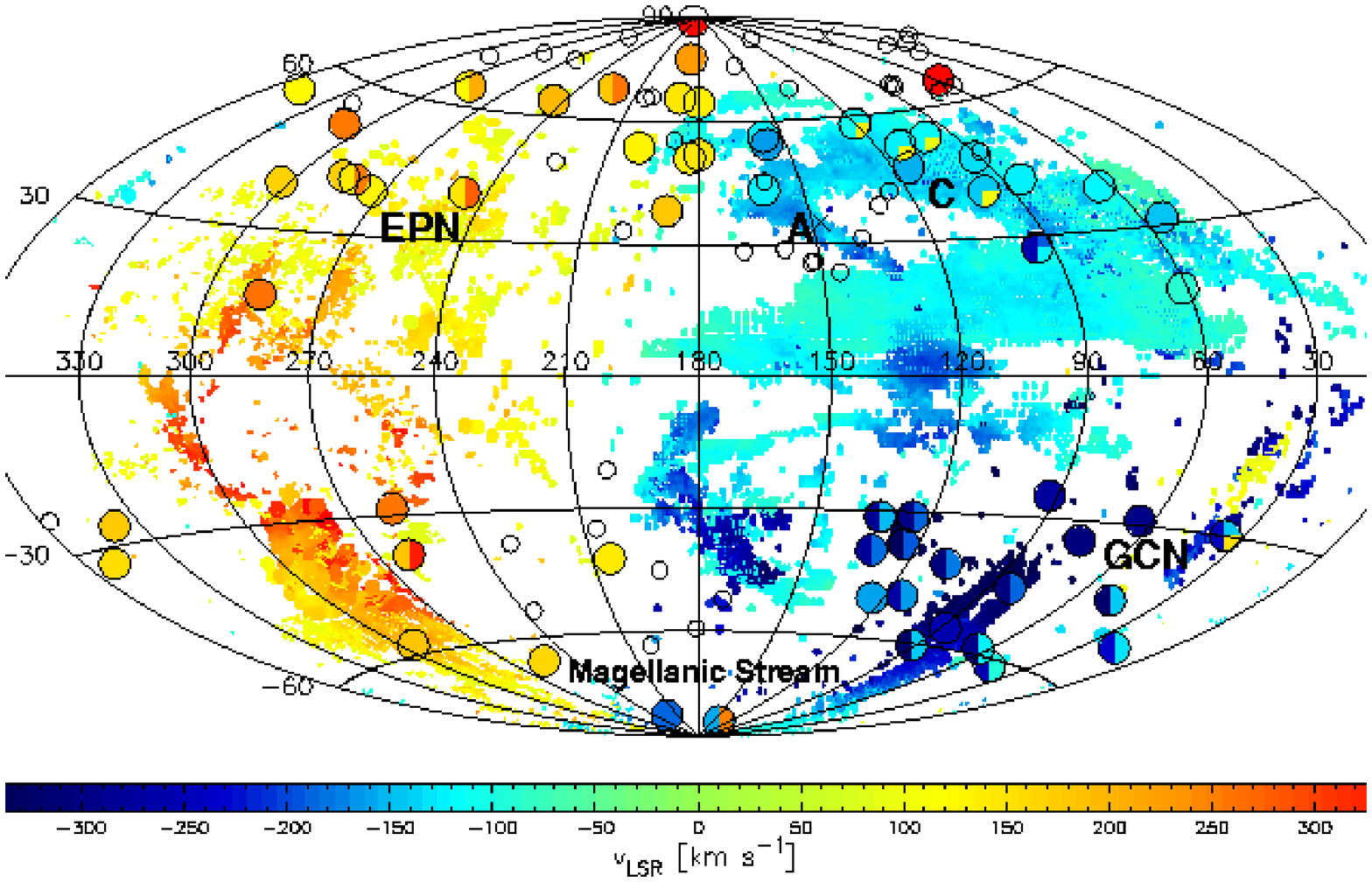}
\vspace{2.7in}
\caption{\small All-sky Hammer-Aitoff 
projection of the high velocity O\,{\sc vi} 
features (circles) superimposed on the high velocity H\,{\sc i}
 sky observed in 21\,cm emission.  
The velocity color coding is the same for both species.
Small circles indicate directions where no
high velocity O\,{\sc vi} was detected. (From Sembach et al. 2003 and
Wakker et al. 2003.)}
\end{figure}

We plot the locations of the high velocity O\,{\sc vi} features in 
Figure~2 with filled colored circles indicating the velocities of the 
O\,{\sc vi}  (blue/green = negative 
velocities, orange/red = positive velocities).  Small open circles 
(or ``X'' marks for the two stellar sight lines) denote 
directions where no high velocity O\,{\sc vi} is detected).  In the same 
figure, the distribution of high velocity H\,{\sc i} 21\,cm emission
is also shown at the finer spatial resolution afforded by the H\,{\sc i}
data (see Wakker et al. 2003).  
Several large structures or groups of 
H\,{\sc i} high velocity clouds are visible
in Figure~2, including:  1) the Magellanic Stream, which passes 
through the south Galactic pole and extends up to $b \sim -30\degr$,
with positive velocities for 
$l \gtrsim 180\degr$ and 
negative velocities for $l \lesssim 180\degr$; 
2) high velocity cloud Complex~C, which covers a large portion of the 
northern Galactic sky between $l=30\degr$ and $l = 150\degr$ and has 
velocities of roughly --100 to --170 \kms; 3) the extreme positive 
velocity clouds in the northern sky (EPN), which are located in the general
region $180\degr \lesssim l \lesssim 330\degr, b \approx 30\degr$; and 4)
the Galactic center negative velocity (GCN) clouds located near $l \sim45\degr,
-60\degr \lesssim b \lesssim -30\degr$.

Several key points about the relationship of the high velocity 
O\,{\sc vi} and H\,{\sc i} in Figure~2 are worth highlighting:
\smallskip
\\
\noindent
1) There is excellent velocity correspondence between the H\,{\sc i}
and O\,{\sc vi} in Complex~C, with sight lines passing near to Complex~C,
but not through the high velocity H\,{\sc i},
showing no O\,{\sc vi} absorption. 
\smallskip
\\
\noindent
2) Toward Complex~A, which is near Complex~C,
there is a close pair of sight lines exhibiting an
O\,{\sc vi} detection (Mrk\,106: $l=161\fdg1, b=42\fdg9\degr$) 
and a non-detection 
(Mrk\,116: $l=160\fdg5, b=44\fdg8$).  At the 4--10 kpc
distance of Complex~A (Wakker 2001), 
the sight lines are separated by $\sim140-350$ pc.
\smallskip
\\
\noindent
3) The high velocity component toward NGC\,3310 ($l = 156\fdg6, 
b = +54\fdg1$) has a velocity similar to that of Complexes A and C, even 
though there is no H\,{\sc i} 21\,cm emission detected along the sight line 
at these velocities.
\smallskip
\\
\noindent
4) The H\,1821+643 sight line ($l=94\fdg0, b=27\fdg4$) contains O\,{\sc vi} absorption at the 
velocities of the Outer Arm as well as at more negative velocities.  The progression of velocities between the Outer Arm
and Complex~C is relatively smooth.
\smallskip
\\
\noindent
5) In the $l < 180\degr, b < 0\degr$ quadrant of the sky there are often two 
negative velocity components, with the highest negative velocity components
occurring in the Magellanic Stream 
or the extension of the Magellanic Stream
at velocities typical of the Stream.  Components with $\bar{v}
\sim -120$ \kms\ concentrate to longitudes less than those of the Stream,
whereas components at $l \sim120\degr-140\degr$ have velocities 
typical of the Stream.
\smallskip
\\
\noindent
6) There is good velocity correspondence between the H\,{\sc i}
and O\,{\sc vi} velocities in the positive velocity portion of the 
Magellanic Stream at $l > 180\degr$.  The O\,{\sc vi}
features off the main axis of the 
Stream at these longitudes 
have  velocities similar to those of the H\,{\sc i} near the Stream.
\smallskip
\\
\noindent
7) In the $l > 180\degr, b > 0\degr$ quadrant of the sky there are positive 
velocity O\,{\sc vi} and H\,{\sc i} features, sometimes at similar velocities.
In some cases, the O\,{\sc vi} features have substantially higher velocities
than the H\,{\sc i}.
The +277 \kms\ feature toward ESO\,265-G23 ($l = 285\fdg9, b=+16\fdg6$)
has a velocity and location close to that of H\,{\sc i} in the leading arm of 
the Magellanic Stream identified by Putman et al. (1998);  H\,{\sc i} at
similar velocities is seen $\sim1\degr$ away.  
\smallskip
\\
\noindent
8) At $l \sim 180\degr, b > 0\degr$ there is high velocity O\,{\sc vi} with 
$\bar{v} \sim +150$ \kms.  Some of these features are broad 
absorption wings extending from the lower velocity absorption produced by
the Galactic thick disk/halo.
\smallskip
\\
\noindent
9) There may be high velocity H\,{\sc i} near the +143 \kms\ O\,{\sc vi}
feature toward PKS\,0405-12 ($l = 204\fdg9, b = -41\fdg8$).
\smallskip
\\
\noindent
10) High velocity O\,{\sc vi} features toward  Mrk\,478 ($l=59\fdg2,
b=+65\fdg0$, $\bar{v} \approx +385$ \kms), 
NGC\,4670 ($l=212\fdg7, b = +88\fdg6$, $\bar{v} \approx +363$ \kms), 
and Ton\,S180 
($l=139\fdg0, b=-85\fdg1$, $\bar{v} \approx +251$ \kms) stand out as 
having particularly unusual
velocities compared to those of other O\,{\sc vi} features in similar regions 
of the sky. These features  may be 
located outside the Local Group (i.e., in the IGM).
\smallskip
\\
\noindent
11) Sight lines that contain both negative and positive high velocity 
features include Mrk\,509 ($l=36\fdg0, b=-29\fdg9$), Ton\,S180 ($l=139\fdg0, b=-85\fdg1$), 
and several Complex~C sight lines
(PG\,1259+593, PG\,1351+640, Mrk\,817, Mrk 876: $l\sim85\degr-120\degr,
b\sim40\degr-60\degr$).

\subsection{Column Densities}
The high velocity O\,{\sc vi} features have logarithmic column densities
(cm$^{-2}$)
of 13.06 to 14.59, with an average of $\langle \log N \rangle = 
13.95\pm0.34$ and a median of 13.97 (see Figure 3, left panel).   
The average high velocity O\,{\sc vi} column density is a factor of 
2.7 times lower 
than the typical low velocity O\,{\sc vi} column density found for the same
sight lines 
through the thick disk/halo of the Galaxy (see Savage et al. 2003).

\begin{figure}[ht!]
\includegraphics{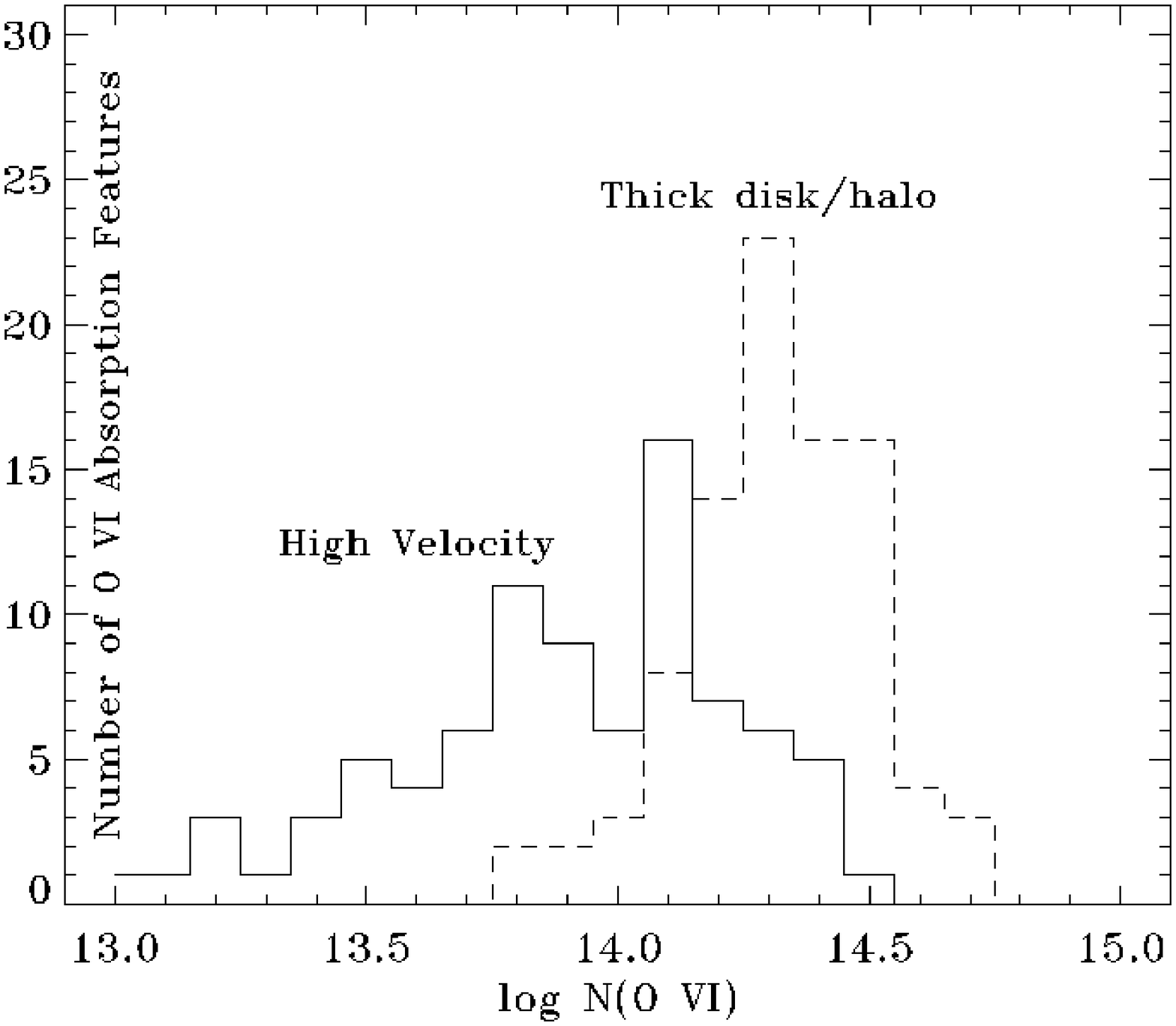}
\includegraphics{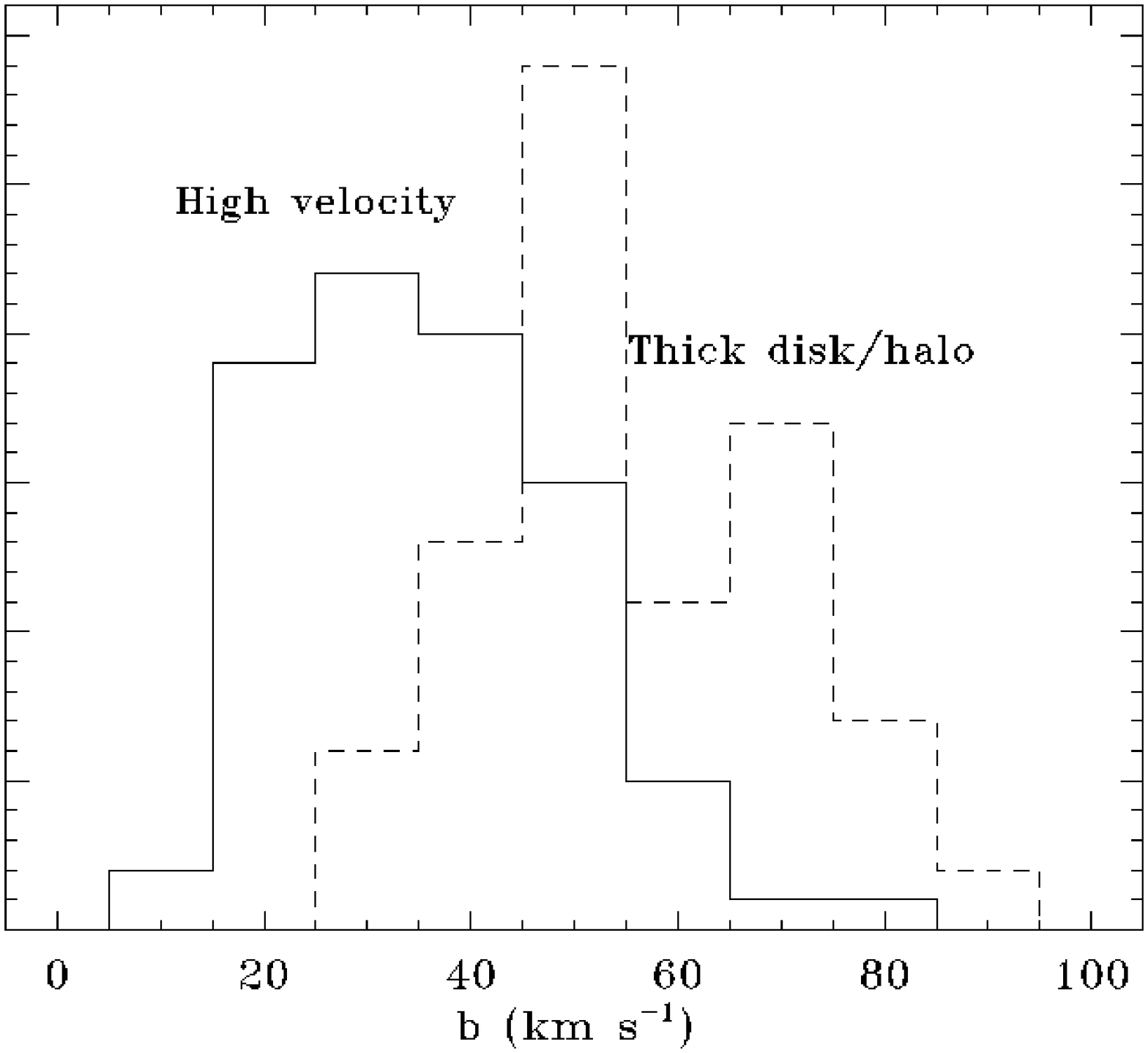}
\vspace{2.5in}
\caption{\small
Histograms of the high velocity O\,{\sc vi} column densities and line
widths (solid lines). The bin sizes are 0.10 dex and 10 \kms, respectively.
For comparison, the distributions for the O\,{\sc vi} absorption
arising in the thick disk and halo of the Galaxy are also shown 
(dashed lines) (from Sembach et al. 2003).}
\end{figure}

\subsection{Line Widths}
The line widths of the high velocity O\,{\sc vi} features range from
$\sim$16 km~s$^{-1}$ to $\sim$81 km~s$^{-1}$, with an average of 
$\langle {\rm b} \rangle = 40\pm14$ km~s$^{-1}$ (see Figure 3, right panel).
The lowest values of b are close to the thermal width of 17.1 km~s$^{-1}$
expected for O\,{\sc vi}
at its peak ionization fraction temperature of $T = 2.8\times10^5$\,K in
collisional
ionization equilibrium (Sutherland \& Dopita 1993).
The higher values of b 
require additional non-thermal broadening mechanisms or gas
temperatures significantly larger than $2.8\times10^5$\,K.

\section{Origin of the High Velocity O\,{\sc vi}}

One possible explanation for some of the high velocity O\,{\sc vi} is that
transition temperature gas arises at the boundaries between cool/warm 
clouds of gas and a very hot ($T > 10^6$\,K) Galactic corona or Local Group
medium.   Sources of the high velocity material might include infalling or 
tidally disturbed galaxies.  
 A hot, highly extended ($R > 70$ kpc)
corona or Local Group medium might be left over from the formation of the 
Milky Way or Local Group, or may be the result of continuous accretion of 
smaller galaxies over time.  
N-body simulations
of the tidal evolution and structure 
of the Magellanic Stream favor a low-density medium ($n < 10^{-4}$ cm$^{-3}$)
for imparting
weak drag forces to deflect some of the Stream gas and providing a possible
explanation for the absence of stars in the Stream (Gardiner 1999). 
Moore \& Davis (1994) also postulated a hot,
low-density corona to provide ram pressure stripping of some of the Magellanic
Cloud gas.
Hydrodynamical simulations of clouds moving through a hot, low-density 
medium show that weak bow shocks develop on the leading edges of the 
clouds as the gas is compressed and heated (Quilis \& Moore 2001).  
Even if the clouds are 
not moving at 
supersonic speeds relative to the ambient medium, some viscous or turbulent
stripping of the cooler gas likely occurs.

An alternative explanation for the O\,{\sc vi} observed at high velocities 
may be that the clouds and any associated H\,{\sc i} fragments are simply 
condensations within
large gas structures falling onto the Galaxy.
Cosmological structure formation models predict large numbers of cooling 
fragments embedded in dark matter, and some of these structures should be 
observable in O\,{\sc vi} absorption as the gas passes through the 
$T=10^5-10^6$\,K
temperature regime.  The simulations suggest that $\sim30$\% 
of the hot gas should be  detectable 
in O\,{\sc vi} absorption, while the remaining $\sim70\%$ may be visible in
O\,{\sc vii} and higher ionization stages  (Dav\'e et al. 2001).

The  tenuous hot Galactic corona or 
Local Group gas may be revealed through X-ray absorption-line 
observations of O\,{\sc vii}. The column density of O\,{\sc vii} in the hot gas
is given by $N$(O\,{\sc vii}) = (O/H)$_\odot$ $Z$ f$_{\rm O\,VII}$ $nL$,
where $Z$ is the 
metallicity of the gas, f is the ionization fraction, and $L$ is the 
path length, and (O/H)$_\odot = 5.45\times10^{-4}$ (Holweger 2001).  
At $T \sim 10^6$\,K,  f$_{\rm O\,VII} \approx 1$ 
(Sutherland \& Dopita 1993). For $n = 10^{-4}$ cm$^{-3}$,  
$N$(O\,{\sc vii}) $\sim 2\times10^{16}~Z$~($L / 100$ kpc) (cm$^{-2}$).
Preliminary results 
(Nicastro et al. 2002; Fang et al. 2003; Rasmussen et al. 2003) 
demonstrate that O\,{\sc vii} absorption is detectable near zero velocity at a 
level consistent with the presence of a large, nearby reservoir of hot gas.

\section{Are the O\,{\sc vi} HVCs Extragalactic Clouds?}

Some of the high velocity O\,{\sc vi} clouds may be extragalactic 
clouds, based on what we currently know about their ionization 
properties.  However, claims that
essentially  {\it all} of the O\,{\sc vi} HVCs are extragalactic 
entities associated with an extended Local Group filament based on
kinematical arguments alone appear to be untenable.
 Such arguments
fail to consider the selection biases inherent in the O\,{\sc vi} 
sample, the presence of neutral (H\,{\sc i}) and lower ionization 
(Si\,{\sc iv}, C\,{\sc iv})
gas associated with some of the O\,{\sc vi} HVCs, and the known ``nearby''
locations for at least two of the primary high velocity complexes 
in the sample ---  the Magellanic Stream is circumgalactic tidal debris, and 
Complex~C is probably interacting with the Galactic corona.  Furthermore, the  
O\,{\sc vii} X-ray absorption measures used to support an extragalactic 
location have not yet been convincingly tied to either the O\,{\sc vi} HVCs
or to a Local Group location.  The O\,{\sc vii} absorption may well have a 
significant Galactic component in some directions
(see Fang et al. 2003). The Local Group filament 
interpretation (Nicastro et al. 2003) may be suitable for some of the 
observed high velocity 
O\,{\sc vi} features, but it clearly fails in 
other particular cases (e.g., the Magellanic Stream)
or for the whole ensemble of high velocity O\,{\sc vi} 
features
in our sample.  For example, the ``Local Supercluster Filament'' model
(Kravtsov et al.  2002) predicts average O\,{\sc vi}
velocity centroids higher than those 
observed ($\langle \bar{v} \rangle \sim 1000$ 
\kms\ vs. $\langle \bar{v} \rangle < 400$ \kms) and average line widths 
higher than
those observed (FWHM $\sim 100-400$ \kms\ vs. FWHM $\sim 30-120$ \kms).
Additional absorption and emission-line
observations of other ions at ultraviolet wavelengths would provide
valuable information about the physical conditions, ionization, and 
locations of the O\,{\sc vi} clouds.







\newcommand\aj{AJ}%
\newcommand\araa{ARA\&A}%
\newcommand\apj{ApJ}%
\newcommand\apjl{ApJ}%
\newcommand\apjs{ApJS}%
\newcommand\aap{A\&A}%
\newcommand\aaps{A\&AS}%
\newcommand\mnras{MNRAS}%


\articletitle{Ionization of High Velocity Clouds in the Galactic
Halo}

\author{Jonathan D. Slavin}
\affil{Harvard-Smithsonian Center for Astrophysics}
\email{jslavin@cfa.harvard.edu}

\begin{abstract}
Recent observations of H$\alpha$ emission from high velocity clouds (HVCs)
have raised the question of the ionization source for these clouds, which are
believed to be fairly distant from ionizing radiation sources in the galactic
disk. Obs
ervations with FUSE of \OVI\ absorption in HVCs also present us with
questions regarding their source of ionization and the relation to the hot gas
in the halo. We discuss sources for the ionization of the H$\alpha$ emitting gas
and the \OVI\ containing gas. In particular we examine if interactions between
warm ionized gas in the HVCs and the hot gas of the surrounding galactic halo
could explain both the highly ionized gas and the ionization of cooler gas in
the clouds.  \end{abstract}

\section{Introduction}

New \emph{FUSE} and \emph{WHAM} data show that gas with a wide range of
ionization exists in high velocity clouds. In some cases the \OVI\ absorption
and H$\alpha$ emission apparently come from the same cloud. These
observations raise a number of interrelated questions.
\begin{itemize}
\item Assuming the H$\alpha $ comes from warm, photoionized gas, what is
the source of ionizing flux?
\item Is the \OVI\ gas photoionized or collisionally ionized?
\item 
Is the \OVI\ gas related to the H$\alpha$ emitting gas and H\,I gas?
\item What is the relationship of these HVCs to the hot Galactic halo gas?
\end{itemize}
The answers to these questions have the potential to tell us much
about the Galactic halo and the intergalactic medium.

\section{Warm Ionized Gas in HVCs}

One surprise revealed by very sensitive H$\alpha $ observations carried
out with the Wisconsin H$\alpha $ Mapper is the brightness of many
HVCs. H$\alpha $ emission has been detected from clouds in
 complexes
M, A and C (\cite{TRH98}) and toward 4 of 5 compact HVCs observed
(\cite{Tea02}).

In nearly every direction looked at where there is significant high-velocity
H\,I 21 cm emission, associated H$\alpha $ emission has been found.
H$\alpha $ intensities for the HVCs are $\sim 0.1$ R ($=10^{5}/4\pi $ photons
cm$^{-2}$ s$^{-1}$ sr$^{-1}$) -- if the emission is from optically thick
clouds at $T\approx 10^{4}$ K that are photoionized, this implies a Lyman
continuum flux of $\Phi_{\mathrm{LyC}}=2.1\times
 10^{5}$ photons cm$^{-2}$
s$^{-1}$. These H$\alpha $ intensities are too high to be explained by
photoionization from the metagalactic ionizing flux.

The warm ionized HVCs require a substantial Lyman continuum flux implying
they are near a galactic source of ionizing radiation or have a diffuse
source local to the halo.

\section{Highly Ionized Gas in HVCs}

\emph{FUSE} has revealed that the HVCs contain highly ionized gas in addition
to H\,I and the ionized hydrogen (\cite{Sea02}). High velocity \OVI\ ha
s been
seen in absorption in the majority of lines of sight observed towards
extragalactic objects.

85 \OVI\ features were observed including many associated with H\,I HVCs.
Most of the \OVI\ cannot be produced by photoionization because the required
ionization parameter is too low (i.e. the density is too low). The size of
regions implied by the required densities and column densities is too large
for the regions to be coherent in velocity.  The \OVI\ is also very unlikely
to be in quiescent hot gas since
 such gas either has little \OVI\ (if
$T\gtrsim 10^{6}$ K) or has a very high cooling rate (if $T\sim 10^{5.5}$ K).
The existence of \OVI\ HVCs associated with H$\alpha$ and H\,I high-velocity
gas points to a multi-phase structure in HVCs.

\section{Sources for the Ionization of HVCs}

Both the H$^{+}$ and the O$^{+5}$ require substantial energy sources
for their creation and maintenance in the Galactic halo. Being far
from the Galactic disk requires either that the energy is transported
from the disk or th
at some sort of \emph{in situ} source be present.
Possible sources for photoionization include Lyman continuum flux from early
type stars escaping the disk, diffuse emission from cooling hot gas in
the disk and halo, the metagalactic ionizing flux, shocks between HVCs and halo
gas, and EUV/soft X-ray emission at the boundaries between warm/cold clouds
and hot halo gas. The latter is a potentially important \emph{in situ}
source that we discuss in detail below.

\section{Sources for Hot Gas in the Halo}

The \OVI\ associated with high velocity H\,I almost certainly has as its
energy source hot gas in the halo. Possible sources for the hot gas are
superbubbles and supernovae that break out of the H\,I disk, accretion shocks
during galaxy formation (part of the ``warm hot intergalactic medium''),
infall kinetic energy (again leading to shocks but in the current epoch), and
a Galactic wind (most probably originating at the center of the Galaxy).  The
hot halo loses energy via radiative cooling and adiabatic expan
sion.

\section{Escape of Stellar Radiation}

Several authors have estimated the fraction of O star radiation that
leaks from the Galactic disk (e.g., \cite{MC93}, \cite{DSF00}, \cite{BHM99}).
These calculations depend on many factors including the scale height and
morphology of the H\,I disk, the magnetic and cosmic ray pressure in the halo,
superbubble evolution, and star formation history in OB associations.
Consequently estimates for the escape fraction cover a broad range, $\sim
1-10$\%.  The added unc
ertainty of the distance to HVC gas adds to the
uncertainty in whether escaping stellar radiation can account for its
ionization.

These uncertainties and the association of H$\alpha$ HVCs with \OVI\ HVCs
prompts us to look at the possibility that radiation generated in the
interaction of hot halo gas and cooler gas could be a significant source of
ionizing flux.

\section{Ionizing Flux from Evaporative Cloud Boundaries}

Evaporation of cool clouds via thermal conduction from surrounding hot gas has
long be
en proposed to occur in the Galactic disk and halo (e.g., \cite{CM77}).
As the gas in such interfaces is heated and accelerated, it also emits
prodigiously in the EUV. If this occurs around HVCs, it does not need to have
a high luminosity in ionizing photons to provide significant ionization since
it is generated in a thin layer that surrounds the warm gas.  Because of this
geometry a large fraction ($\sim$half) of the radiation is captured by the
cloud.

For steady flow evaporation of a spherical cloud, th
e energy equation is
\begin{equation} \frac{1}{4\pi r^{2}} \frac{d}{dr}\left[\frac{5}{2} \dot{M}
  c^{2} g + 4\pi r^{2}q\right] = -L(T)n_{e}n_{H},
\end{equation}
where $\dot{M}$ is the mass loss rate, $c$ is the sound speed, $g = 1 + 
\mathcal{M}^2/5$ ($\mathcal{M}$ is the mach number), $q$ is the (radial) heat
flux and $L(T)$ is the radiative cooling coefficient.  Inward heat flux is
balanced by outward enthalpy flux reduced by cooling losses.

An important parameter in studies of evaporating clouds is the
 saturation
parameter (see, e.g., \cite{CM77}),
\begin{equation}
  \sigma _{0}\equiv \frac{2 \kappa _f T_f}{25\phi \rho_f c_f^3 R_{cl}}
  \propto \frac{T_f^3}{P R_{cl}},
\end{equation}
where $\kappa_f$ is the classical (Spitzer) thermal conductivity
at the asymptotic temperature, $T_f$, $R_{cl}$ is the cloud radius,
$\phi$ is the saturation fudge factor, $c_{f}$ is the sound speed
at $T_{f}$ and $P$ is the (thermal) pressure.

For highly saturated conduction (as occurs for high $T_f$ and low $P$ as
expected
 in the halo) we find, using detailed numerical calculations of
cloud evaporation, that the ionizing flux goes as
\begin{equation}
  \Phi _{\mathrm{LyC}}\propto P^{1.1},
\end{equation}
with essentially no dependence on temperature of the hot gas (as long as
$\sigma_0$ is high).

\section{\OVI\ from Evaporative Cloud Boundaries}

\begin{figure}[ht]
  \includegraphics[width=\textwidth]{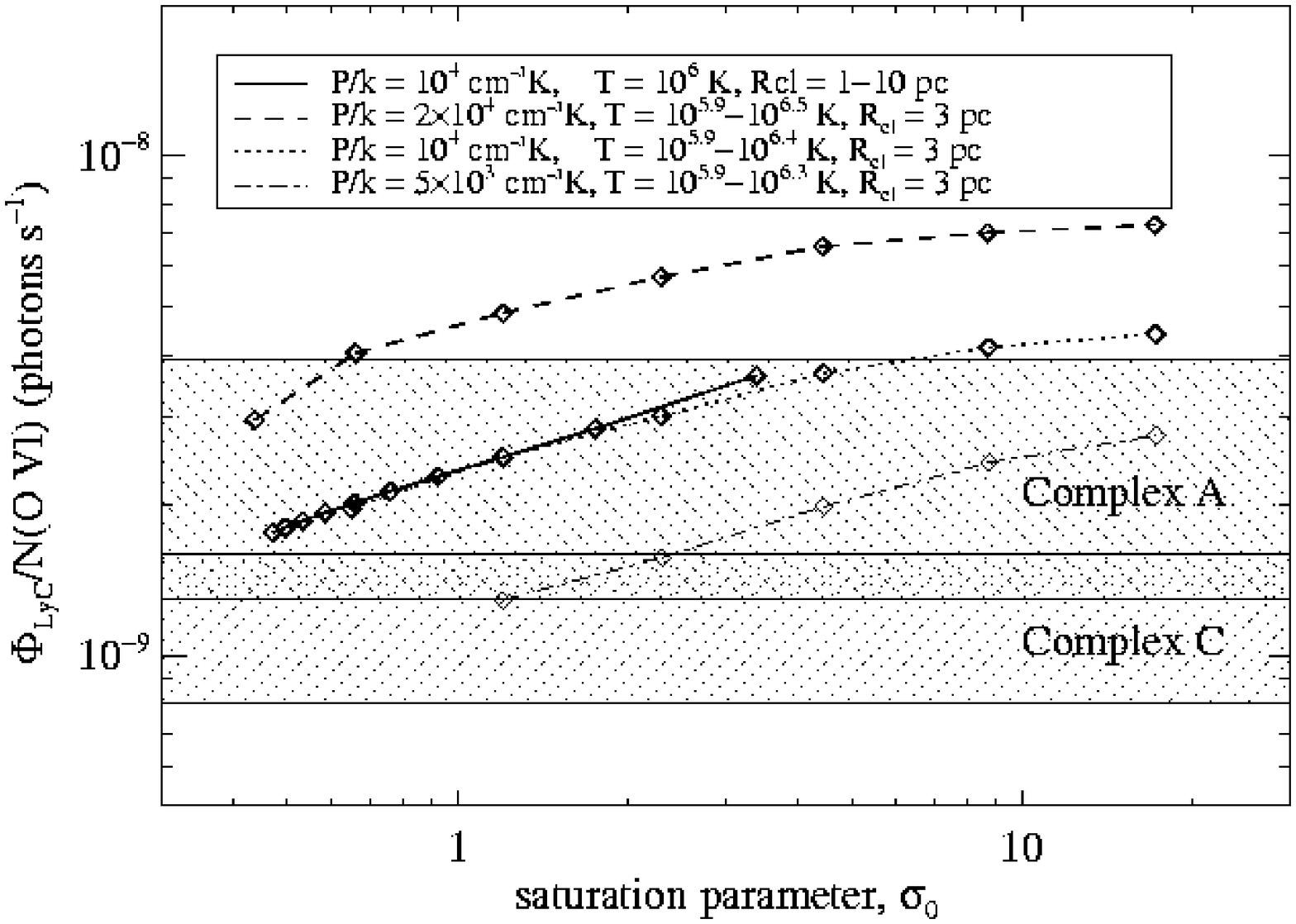}
  \caption{Dependence of the ratio, $\Phi_\mathrm{LyC}/N($O\,VI$)$, on the
  saturation parameter, $\sigma_0$, for evapo
rating clouds. Shown are curves
  for a range of assumptions for the pressure, temperature and cloud radius.
  Note that the curves can be characterized by just two parameters, $P$ and
  $\sigma_0$. The regions labeled Complex A and Complex C show the 1-$\sigma$
  uncertainty in the ratio as derived from H$\alpha$ and \OVI\ observations.
  \label{sl-fig:cldevap}} 
\end{figure}

In the boundary of an evaporating cloud \OVI\ is generated as
O is ionized from a low ionization state to beyond O$^{+5}$ (assuming
$T_{f}\gtrsim 10^{5.9}$ K). We expect $N($\OVI$)\
propto \dot{M}$
but $\dot{M}$ depends on $T_{f}$, $R_{cl}$, and $\sigma _{0}$. For highly
saturated evaporation, we find a pressure dependence
\begin{equation}
  N(\mathrm{O}\, \mathrm{VI})\propto P^{4/9}
\end{equation}
Figure \ref{sl-fig:cldevap} shows the dependence of
$\Phi_{\mathrm{LyC}}/N($\OVI$)$ on $\sigma_0$ for a variety of assumptions for
pressure, temperature and cloud radius. It can be seen that varying $R_{cl}$
or varying $T_h$ gives the same results dictated only by the value of
$\sigma_0$ and $P
$. For the specific case of a $3$ pc radius cloud with
$P=5000$ cm$^{-3}$ K, $T_{f}=10^{6}$ K, and $Z=1.0$, we find
\begin{eqnarray}
  N(\mathrm{O\,VI}) & = & 1.3\times 10^{13}\; \mathrm{cm}^{-2} \\
  \Phi_{\mathrm{LyC}} & = & 2.0\times 10^{4} \mathrm{photons}\;
  \mathrm{cm}^{-2}\; \mathrm{s}^{-1}.  
\end{eqnarray}

Because the predicted values for $N($\OVI$)$ and $I_{\alpha}$ are well below
the observed values, each HVC would need to consist of ma
ny ($\sim 10$)
evaporating clouds along a given line of sight which, given the angular size
of the HVCs, is plausible.  Calculations for lower pressures are difficult
numerically because the degree of saturation becomes very high.  The scaling
of $N($\OVI$)$ and $I_{\alpha}$ with $P$ for runs of high saturation is very
linear, however, giving us confidence in extrapolating to lower $P$ values.

We have carried out limited testing of the metallicity dependence and have
found that $\Phi_{\mathrm{LyC}}$ has a 
weak dependence on $Z$, $\sim
Z^{0.06}$, while $N($\OVI$)\propto Z$. 

So $\Phi_{\mathrm{LyC}}/N($
\OVI$)$ goes roughly as $\sim Z^{0.94}$.  The weak
dependence of $\Phi_\mathrm{LyC}$ on $Z$ is in part due to the fact that a
large fraction of the ionizing flux comes from the He\,II 304\AA\ line
and in part due to the complex interplay of non-equilibrium ionization effects
and cooling in cloud boundaries. In general the gas in evaporative outflows
is significantly under-ionized relative to collisional ionizati
on equilibrium
(CIE).  Under-ionized gas tends to have a substantially higher emissivity than
CIE gas. Also, the evaporation rate is decreased by radiative cooling in the
flow.  Thus there is something of a self-regulation process wherein for lower
$Z$ clouds, the radiative cooling would be expected to be lower, yet the
reduced cooling rate allows for a higher mass loss rate which in turn leads to
a flow that is farther from CIE and thus has enhanced radiative cooling. The
net result is a weak dependence on
 $Z$ of the cloud boundary temperature
profile, mass loss rate and total ionizing flux.

\section{An Alternative Interface Model: Turbulent Mixing Layers}

Another, and possibly more likely form for the interaction of hot
halo gas and cool HVCs is in turbulent mixing layers (TML). The TML
model (\cite{BF90,SSB93}) proposes that shear flows between hot gas and cool
clouds results in Kelvin-Helmholtz instabilities.  Turbulent mixing rapidly
occurs in the shear layer leading to gas at an intermediate temperatu
re,
$\bar{T}$, between the hot gas temperature, $T_{h}$, and the cloud
temperature, $T_{c}$.  After mixing the gas rapidly cools and radiates.
\OVI\ is generated in the cooling gas as is EUV radiation.  Given the high
relative speed of HVCs and gas coupled to Galactic rotation, substantial shear
is likely.

We have further explored models of the type proposed by \cite{SSB93}
with an eye to their application to the ionization of HVCs.  In these TML
models the layer is assumed to have reached a steady state i
n which
cooling in the layer balances enthalpy flux into the layer.  As a result,
$\Phi_{\mathrm{LyC}}\propto P$. The flux depends on $\bar{T}$, but does not
depend sensitively on $T_{h}$.  \OVI\ is independent of $P$, (due to the
steady state assumption) but depends strongly on both $\bar{T}$ and $T_{h}$.
Because the metal lines, particularly \OVI, are important coolants,
$N($\OVI$)$ has a weak dependence on $Z$. TMLs produce relatively small
amounts of \OVI\ and $\Phi_{\mathrm{LyC}}$ per layer, but we exp
ect the
clouds to contain many such layers -- each HVC would be a cluster of cloudlets
that are being turbulently disrupted.

\begin{figure}[ht]
  \includegraphics[width=\textwidth]{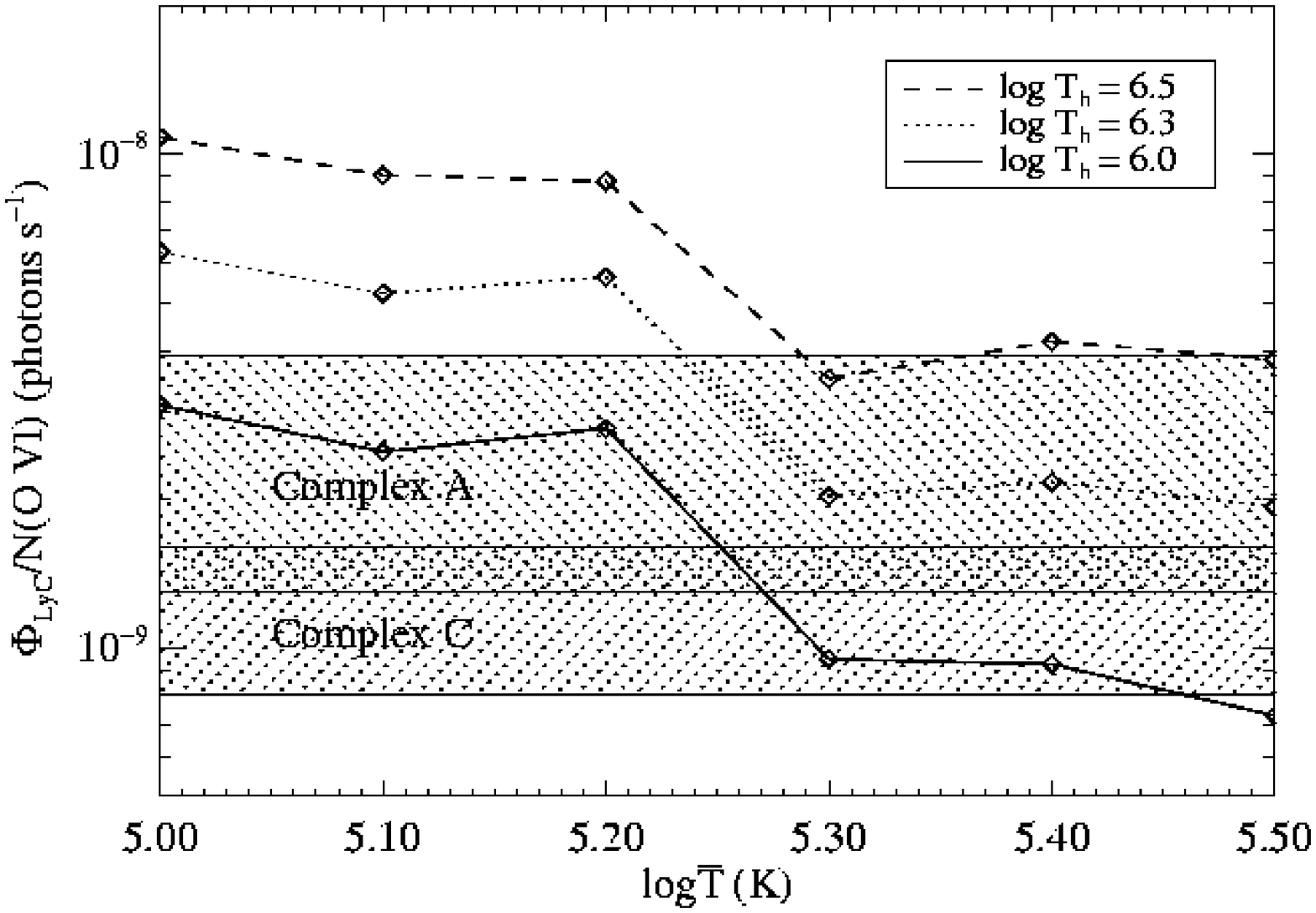}
  \caption{Dependence of the ratio, $\Phi_\mathrm{LyC}/N($O\,VI$)$, on the
  mixing temperature, $\bar T$, for turbulent mixing layers. The various
  curves are for different values of the hot gas temperature.  For all cases
  the $T_c = 10^4$ K and $P/k = 1000$ cm$^{-3}$ K.\label{sl-fig:mxlyr}}
\end{figure}

\section{Comparison with the Data}

To date we have analyzed two cases in which there is H$\alpha $ emission
observed and \OVI\ absorption that are clearly associated in an HVC,
though there are potentially many more. For Complex C we have 
$I_{\alpha}=0.13\pm 0.03$ R which implies $\Phi _{LyC}=2.7\times 10^{5}$
cm$^{-2}$ s$^{-1}$ and $N($\OVI$)=(2.2\pm 0.6)\times 10^{14}$ cm$^{-2}$
which yields $\Phi _{LyC}/N($\OVI$)=(1.2\pm 0.4)\times 10^{-9}$ photons
s$^{-1}$. 

For Complex A there is a weak \OVI\ detection, 
$N($\OVI$) = (6\pm 3)\times
10^{13}$ cm$^{-2}$, and we have $I_{\alpha}=0.08\pm 0.01$ R or
$\Phi_{LyC}=1.6\times 10^{5}$ cm$^{-2}$ s$^{-1}$ giving
$\Phi_{LyC}/N($\OVI$)=(2.6\pm 1.3)\times 10^{-9}$ photons s$^{-1}$.

Both directions are consistent with $\Phi _{LyC}/N($\OVI$)\sim
(1-2)\times 10^{-9}$photons s$^{-1}$. Both the cloud evaporation model and the
TML model can attain this value for moderately low pressures, $P/k\sim 1000$
cm$^{-3}$ K.

\section{Interface Radiation as a Source of Ionizing Radiation}

While the detailed predictions for $\Phi_{LyC}$ and $N($\OVI$)$ depend on the
paricular model and its parameters, there are some general conclusions we can
draw from the existence of \OVI\ HVCs.
\begin{itemize}
\item Radiation from \OVI\ containing gas produces significant amounts
of H ionizing radiation.
\item The existence of \OVI\ in HVCs indicates either hot gas cooling
or cool gas being heated and thus we should expect warm photoionized
gas where we see \OVI .
\item Because neutral HVCs can carry the
ir ionizing radiation source with
them, the linkage between the radiation field of the Galactic disk
and the ionizing flux seen by the clouds is broken. The bad news is that
we cannot tell the distance to the HVCs based on their H$\alpha$
emission. The good news is that we have a potential explanation for the
ionization of all HVCs.
\item The flux from interfaces is unfortunately model dependent and depends on
  the pressure, temperature and abundances in the HVCs and hot gas.
\end{itemize}

\section{Conclu
sions}

\begin{itemize}
\item The HVCs that have associated H\,I and H$\alpha$ emission
appear to be in the Galactic halo and part of a multi-phase extended
hot halo.
\item Interfaces between hot halo gas and the cooler H\,I/H\,II
clouds can easily produce the required Lyman continuum fluxes needed
to provide the ionization of H; the ratio, $\Phi_{LyC}/N($\OVI$) \sim
(1-2)\times10^{-9}$ photons s$^{-1}$, can be matched both in evaporating cloud
models and in turbulent mixing layer models.
\item Observations
 of H$\alpha$ on more lines of sight with observed \OVI\ and
  observations of more ions on those lines of sight are needed to constrain
  the nature of the multi-phase HVC clouds and their relationship to hot halo
  gas and the IGM.
\end{itemize}

\bibliographystyle{kapalike}
\chapbblname{papers/slavin/slavin}
\chapbibliography{astrophys}

%
s in the interstellar and intracluster
%
%
211:135.
%
%
%
%
{\em \apj}, 407:83.
%
%
%





\part[Soft X-ray data Analysis issues]{Soft X-ray data Analysis issues}

\def\chandra{{\it Chandra}}
\def\fuse{{\it FUSE}}
\def\hubble{{\it Hubble}}
\def\rosat{{\it ROSAT}}
\def\rass{{\it ROSAT} All-Sky Survey}
\def\xmm{{\it XMM-Newton}}
\newcommand{\hi}{H{\small ~I}}
\newcommand{\hii}{H{\small ~II}}
\newcommand{\civ}{C{\small ~IV}}
\newcommand{\nv}{N{\small ~V}}
\newcommand{\ovi}{O{\small ~VI}}
\newcommand{\ovii}{O{\small ~VII}}
\newcommand{\oviii}{O{\small ~VIII}}
\newcommand{\NH}{\mbox{$N_{\rm H}$}}        
\newcommand{\oqkev}{$1\over4$~keV}
\newcommand{\tqkev}{$3\over4$~keV}
\newcommand{\sas}{{\it SAS-3}}
\newcommand{\heao}{{\it HEAO-1}}
\newcommand{\iras}{{\it IRAS}}
\newcommand{\ir}{100~$\mu$m}
\newcommand{\lsim}{\mathrel{\mathpalette\@versim<}}




\articletitle[The ISM from the Soft X-ray Background Perspective]
{THE ISM FROM THE SOFT X-RAY BACKGROUND PERSPECTIVE}

\chaptitlerunninghead{The ISM and the Soft X-ray Background}

\author{S. L. Snowden}
\affil{NASA/GSFC and USRA}
\email{snowden@riva.gsfc.nasa.gov}


\begin{abstract}
In the past few years progress in understanding the local
and Galactic ISM in terms of the diffuse X-ray background
has been as much about what hasn't been seen as it has
been about detections.  High resolution spectra of the
local SXRB have been observed, but are inconsistent with
current thermal emission models.  An excess over the
extrapolation of the high-energy (most clearly visible at 
$E>1.5$ keV) extragalactic power law down to \tqkev\
has been observed but only at the level consistent with
cosmological models, implying the absence of at least a
bright hot Galactic halo.  A very recent \fuse\ result
indicates that \ovi\ emission from the Local Hot Bubble is
insignificant, if it exists at all, a result which is
also inconsistent with current thermal emission models.  A 
(very) short review of the Galactic SXRB and ISM is presented.
\end{abstract}


\section{Introduction}

What is the soft X-ray diffuse background (SXRB), at least in relevance 
to the Galactic interstellar medium?  Phenomenologically, the  
SXRB is whatever is left over after all other distinct identified 
emission sources have been removed.  Bright point sources (or effective 
point sources) such as AGN, binary systems, black holes, and pulsars are 
first excised from the data.  Next, bright extended structures such as 
clusters of galaxies and young supernova remnants which stand out from 
the background as distinct individual objects are removed, objects such 
as the Virgo Cluster, Cygnus Loop supernova remnant (SNR), and the Vela 
SNR.  Thus far the winnowing process is simple with the only complication 
being determining the angular extent and brightness at which an object 
becomes a ``bright extended structure,'' and so is removed.  The next step 
is to account for unresolved objects which are contributing to the residual 
background, primarily stars and AGN, which is where the situation does get
complicated.

\begin{figure}[h]
\centerline{\includegraphics[width=6cm,angle=-90]{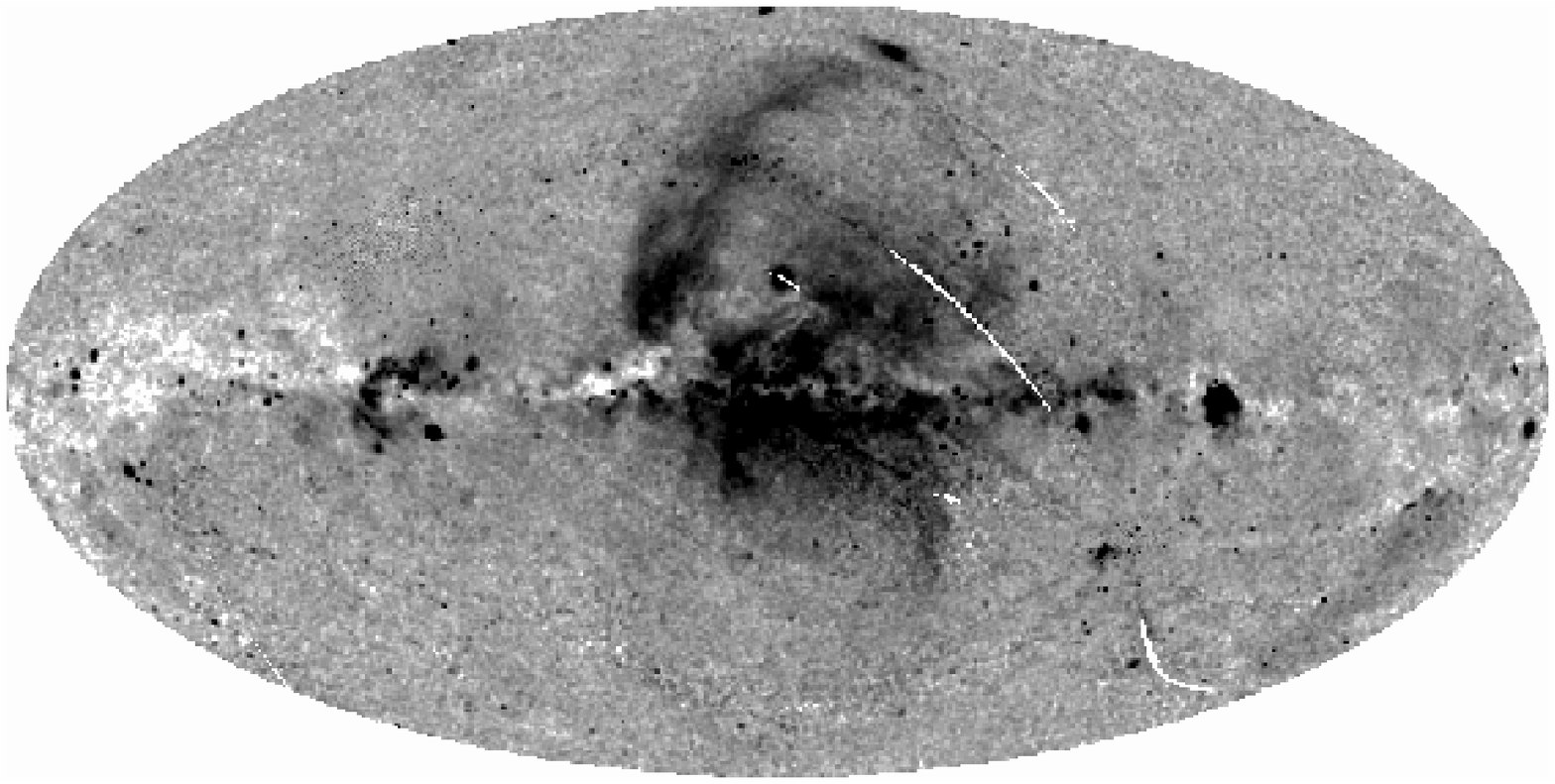}}
\caption{$ROSAT$ All-Sky Survey (RASS) image of the $1-2$ keV band 
SXRB (\cite{sea97}), displayed with an Aitoff-Hammer zero-centered 
projection in Galactic coordinates.  The data have been square-root 
scaled, longitude increases to the left, and darker shading indicates 
brighter emission.  Note the absorption minimum along the Galactic 
plane, the Loop~I and Galactic bulge emission in the direction of the 
Galactic center and rising to high latitudes. The North Polar Spur (NPS) 
is the limb-brightened edge of the Loop~I emission running diagonally 
north and then west from near the plane at $l\sim30^\circ$ to 
$l,b\sim300^\circ,70^\circ$.  Several bright but smaller objects are 
visible along the Galactic plane, e.g., the Cygnus 
Superbubble at $l\sim90^\circ$ and the Vela SNR at $l\sim270^\circ$.  The
Coma cluster is the small enhancement at the very top of the image while
the Virgo cluster is the larger enhancement at the top of the NPS.  The 
Large Magellanic Cloud is the small emission region at 
$l,b\sim270^\circ,-30^\circ$.}
\label{sn-fig:hard}
\end{figure}

The residual background component made up of cosmological objects 
(primarily AGN) 
is most clearly observed at energies above 1.5 keV where Galactic emission 
is minimal.  Except for absorption modulation by material of the Galactic
disk and Loop~I/Galactic bulge emission toward the 
Galactic center it appears isotropic 
(Figure~\ref{sn-fig:hard}).  There are several groups which have worked 
rather successfully on resolving this background by the use of deep 
surveys (e.g., \cite{hea93}; \cite{mea00}; \cite{gea01}).  This is 
convenient for those of us studying the Galactic SXRB 
as they provide highly accurate models for what must be subtracted to 
leave the truly interesting signal.  Unresolved Galactic stars also 
provide a background component but fortunately at a relatively low level 
which primarily affects the \tqkev\ band (\cite{ks01b}).

The observed Galactic SXRB covers the energy range $\sim0.1-1.5$~keV  
and originates primarily as diffuse emission from thermal processes 
with temperatures of $kT\sim0.1-1$~keV.  The lower limit of the energy band 
is set by technical reasons (typical detectors for observing 
the SXRB have very small responses approaching 0.1 keV) and by definition 
(the X-ray and EUV bands meet at about 100~eV).  The upper limit is more 
nebulous being set by the emission mechanisms, however even the 
higher temperature extended Galactic emission is still less 
than $kT\sim1$~keV.

\section{A Short Guided Tour of the SXRB}

Away from the Galactic center quadrant the SXRB in the $1-2$ keV band 
(Figure~\ref{sn-fig:hard}) is relatively isotropic illustrating the 
contaminating contribution of unresolved (and therefore unremoved) 
extragalactic sources.  Emission 
from the Galactic bulge and Loop I (and most clearly the NPS)
dominate the Galactic center region.  The Cygnus Superbubble 
at $l\sim90^\circ$ and the Vela SNR at $l\sim270^\circ$ 
are clearly seen along the Galactic plane. The effect of Galactic 
absorption is seen most clearly in the plane between $l\sim90^\circ$ 
and $l\sim180^\circ$.

Similar to the $1-2$ keV band, the \tqkev\ band 
is dominated by the unresolved extragalactic background except 
in the Galactic center quadrant and along the Galactic plane. However, 
there are a few Galactic objects which become more visible (e.g., the
Eridanus Superbubble and Monoceros/Gemini SNR, the latter also known 
as the Monogem Ring) as the energy range of the band becomes 
more compatible with the emission temperatures.  Galactic  
absorption features also become clearer as the disk becomes optically 
thick at higher latitudes (ISM absorption cross sections go roughly 
as $E^{-{8\over3}}$).

\begin{figure}[h]
\centerline{\includegraphics[width=6cm,angle=-90]{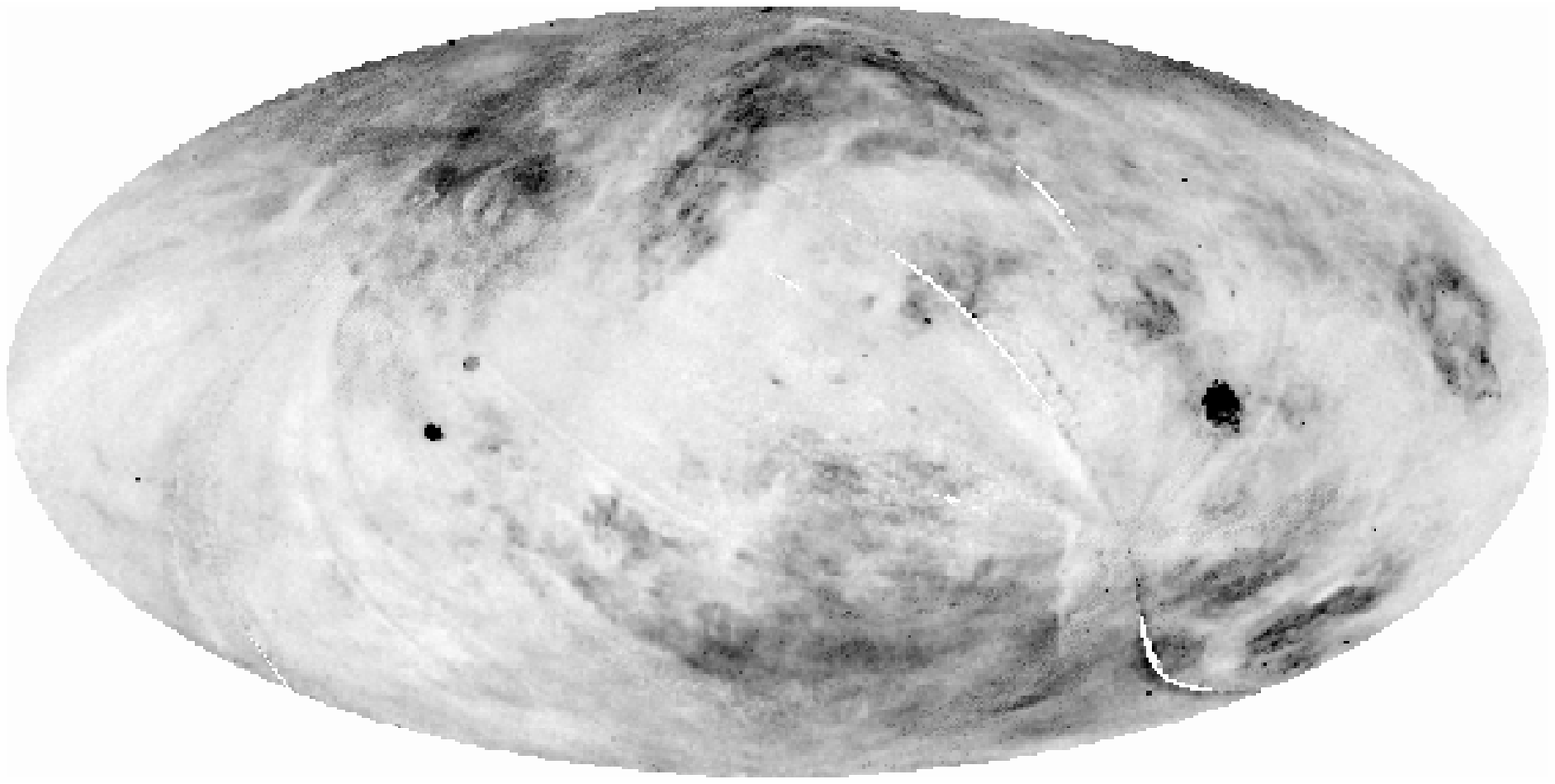}}
\caption{
Same as Figure~1 except for the \oqkev\ band 
and the data have been linearly scaled.  Galactic absorption is optically
thick to much higher latitudes ($|b|\sim30^\circ$) and obscures all but
the nearest objects in the Galactic plane.  However, a number of Galactic
features have become readily apparent, e.g., the Monoceros/Gemini SNR at 
$l,b\sim200^\circ,10^\circ$, the Eridanus (Eridion) Superbubble at 
$l,b\sim190^\circ,-30^\circ$, and the Cygnus Loop SNR at 
$l,b\sim90^\circ,-5^\circ$.  The more northerly part of the NPS is 
still visible but the southern end is obscured by absorption.
}
\label{sn-fig:soft}
\end{figure}

The structure of the \oqkev\ background (Figure~\ref{sn-fig:soft}) 
is completely different from the higher
energy bands due both to additional Galactic emission and the stronger 
effect of Galactic absorption.  This is where the subject of this paper becomes
critical for the goals of this conference.  The Galactic \oqkev\
background is much brighter than the extrapolation of the isotropic 
extragalactic background, can and does vary significantly on angular 
scales of a degree or less, and of perhaps more importance for the 
subject of this conferences can play merry havoc with observations of 
soft X-ray emission from clusters of Galaxies.

\begin{figure}[h]
\centerline{\includegraphics[width=6cm,angle=-90]{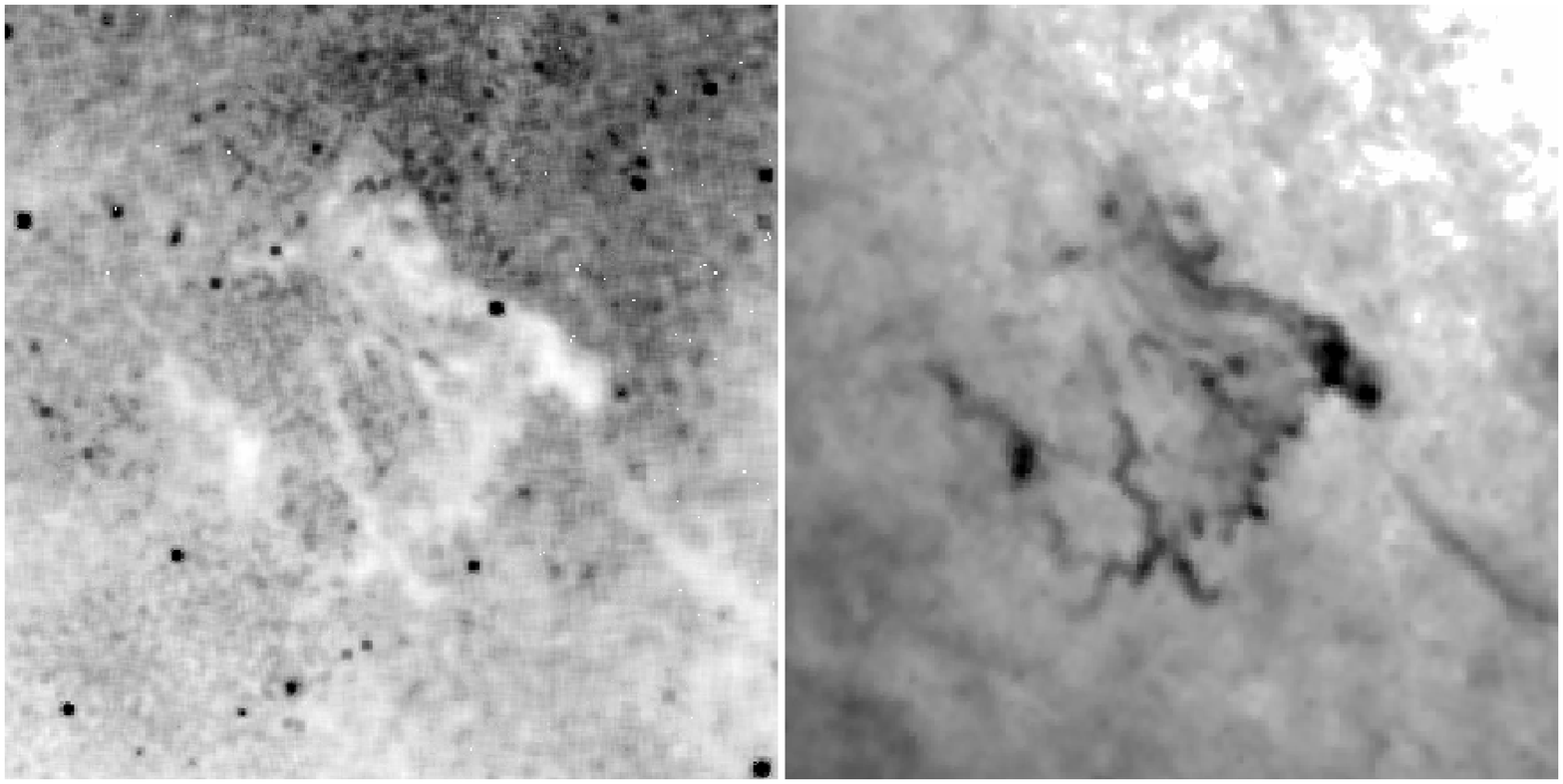}}
\caption{RASS \oqkev\ (left) and $IRAS$ 100 $\mu$m 
(right) images of the Draco Nebula. The fields are the same and are 
$12.8^\circ\times12.8^\circ$ in extent.  Galactic north is up and darker 
shading indicates higher intensity.  The bright knots in the X-ray data 
are point sources of various types. Note the detailed and fine structure 
of the X-ray data.}
\label{sn-fig:draco}
\end{figure}

The Draco Nebula provides an exceptional example of the possible structure 
of the Galactic diffuse X-ray background at \oqkev, and the coupling between
the ISM and that structure.  Figure~\ref{sn-fig:draco} shows both the \oqkev\
{\it ROSAT} All-Sky Survey (RASS) and cleaned \iras\ 100~$\mu$m (\cite{sfd98}) 
data for the Draco region where 
the detailed negative correlation between the two data sets is striking.  
The negative correlation also illustrates the primary tool for locating
the emission regions in space by use of the ``shadowing'' technique.  By
looking at the detailed negative correlation, i.e., the variation of
SXRB intensity with absorbing column density, it is possible to model 
the foreground and background (to the absorbing ISM) X-ray emission.  
With the knowledge of the distance to the absorbing material it is possible
to locate the emission in three-dimensional space.  For the case of the Draco 
Nebula, the distance and latitude places it at the upper edge of the Galactic
disk implying the existence of strong \oqkev\ emission in the Galactic halo, 
as well as significant emission in the nearest few hundred parsecs.  

\section{The Hot Phase of the Galactic ISM}

Now for the questions: What is the hot phase of the Galactic ISM? Where is
it located?  How much of it is there?  Where did it come from?  

{\it What is it?}
The hot ISM is the remnant of various energetic processes in the Galaxy
such as supernovae and stellar winds.  The X-ray emitting plasmas are 
diffusely distributed over volumes extending hundreds of
parsecs, and in the case of the Galactic bulge several kiloparsecs.  The 
plasmas have temperatures of $kT\sim0.1-1$~keV and correspondingly
low space densities ($n_e\sim0.01$~cm$^{-2}$).  Recent observations have 
confirmed that at least the \oqkev\ emission is thermal in nature and 
dominated by various lines (\cite{san01}).
The observations also showed that the emission is inconsistent with that 
from a plasma in thermal equilibrium with solar abundances.

{\it Where is the \oqkev\ emission?}
Making use of the RASS and cleaned \iras\ 100~$\mu$m 
data in a shadowing analysis it has been possible to produce a picture 
of at least the local (out to a few hundred parsecs in the plane, farther 
in the halo) \oqkev\ emission (\cite{sea98}).  From ISM 
optical absorption line studies the Sun is located in a cavity in 
the \hi\ of the Galactic disk extending from less than $\sim50$~pc from 
the Sun in the plane to well over 100~pc at high Galactic latitudes 
(e.g., \cite{sfe99}).  From the X-ray data it is clear that this cavity 
contains an extensive distribution of thermal plasma at $kT\sim0.1$~keV, 
which has been given the name the Local Hot Bubble (LHB).  A few distinct 
emission regions such as the Monoceros/Gemini SNR, Eridanus Superbubble, 
Vela SNR, and Cygnus Loop SNR mentioned above can be seen near the
plane out to several hundred parsecs, but that is about the limit as ISM
absorption cuts off observations of emission regions at greater distances.
At higher Galactic latitudes absorption optical depths for \oqkev\ X-rays 
drop significantly to a minimum of $\tau\sim0.5$ but with typical values
from $\sim1-1.5$.  The X-ray emission in the halo appears rather patchy 
with intensities ranging from near zero to many times typical values 
for the LHB.  

{\it Where is the \tqkev\ emission?}
The two largest and brightest \tqkev\ Galactic emission regions as seen 
from the Earth 
unfortunately lie in the same direction, that of the Galactic center.
The Loop~I superbubble is powered by the Sco-Cen OB associations and has 
a radius of $\sim100$~pc and is centered at a distance of $\sim150$~pc 
in the direction of $l,b\sim330^\circ,15^\circ$.  The Galactic
bulge is most clearly seen in the southern hemisphere where it is not 
confused with the Loop~I emission.  It appears to have a radial extent
of $\sim5$~kpc, a scale height of $\sim2$~kpc, a temperature of 
$kT\sim0.4-0.5$~keV, and a luminosity of $\sim2\times10^{39}$~ergs~s$^{-1}$
(\cite{sea97}).  It is not clear whether there exists a large scale-height
\tqkev\ halo such as suggested by the work of \cite{w98}.  \cite{ks01a} 
and \cite{ksm01} place restrictive limits on the amount
of halo and truly extragalactic emission in this band.  Any such extended
halo emission is going to be confused with more cosmological emission, 
at least to the extent of the Local Group of galaxies.  A recent 
high-resolution spectrum of the diffuse background places strong limits 
on the spectral distribution of the \tqkev\ background based on the 
\ovii\ and \oviii\ emission lines (\cite{mea02}).  The oxygen results
when combined with the deep survey AGN spectra show that 42\% of the 
observed flux can be attributed to unresolved AGNs and a minimum of 36\%
to thermal emission at zero redshift (which would include the contributions 
of both unresolved stars and the cooler plasma contributing to the \oqkev\ 
background).  Subtracting the high-latitude contribution of unresolved 
stars, LHB, and cooler halo leaves an upper limit of $\sim20$\% of the 
observed flux which (because of its low redshift) has either a Galactic 
halo or local group origin. 

{\it How much of it is there?} The emission at \tqkev\ from the Galactic 
bulge is well constrained, as is the hotter Galactic halo, at least by upper 
limits (see above).  At \oqkev\ the amount of emission from the LHB is fairly 
well constrained as is the nearest few hundred parsecs to kiloparsec of the 
Galactic halo.  However, the amount of $kT\sim0.1$~keV plasma in the disk of
the Milky Way as a whole is essentially unknown as it is invisible to us.
The fact that the \oqkev\ halo appears so patchy suggests that it lies at
a relatively low scale height and presumably just above the neutral gas of
the disk.  If this is the case then we know little more about the extent 
of the cooler halo as we can only sample $\sim1$\% of the Galaxy.

{\it Where did it come from?}  The bottom line is that the production of
extensive plasma with temperatures in the $kT\sim0.1-1$~keV range 
requires very energetic phenomena such as supernovae and the stellar winds 
of OB associations and star forming regions.  This is clearly the case for
many of the distinct objects mentioned above such at the Loop~I superbubble 
powered by the Sco-Cen OB associations and the Eridanus Superbubble powered 
by the Orion OB associations.  The Galactic bulge and
the hotter part of the Galactic halo (if it exists) may be powered by star
formation and perhaps infall from the IGM.  The cooler part of the Galactic 
halo can be produced either by the breakout of emission regions in the disk 
(otherwise known as Galactic fountains) 
or in-situ by halo supernovae.  The LHB appears to be an aging supernova 
remnant with no candidate for the originating object.  The physical extent
of the local cavity suggests that in was created through the effort of 
more than one supernova, and plasmas at $kT\sim0.1$~keV are relatively long 
lived with cooling times of millions of years providing a time scale where
that would not be unreasonable.  \cite{mai01} suggests that a component of 
the Sco-Cen OB associations passed through the region of the local cavity 
several million years ago and supernovae at that time could have created 
or reheated 
the plasma in an existing cavity.  The occurrence of a supernova within the 
last five million years and within 30~pc of the Sun is supported by the 
results of \cite{kea99} who analyzed $^{60}$Fe (a supernova byproduct) in 
ocean-floor sediment.

\section{Caveats and Other Considerations}

{\it The McKee and Ostriker model for the ISM:}
I raise this subject as there seemed to be some surprise at the conference
that although the model had a good run it has been effectively ruled out.  
\cite{mo77} postulated an ISM comprised of a highly clumped cooler component 
(\hi\ clouds) embedded in a hotter matrix (X-ray emitting plasma) where the 
\ovi\ observed by Copernicus could arise from the interface region between
the two.  This model solved a number of model/data inconsistencies at the 
time and proved rather popular over the years.  However, such a clumped 
distribution of cold clouds has been searched for but has not been observed 
over all reasonable angular scales in 21-cm emission and absorption.  In 
addition, the X-ray data do not show any support for the interconnecting 
matrix of hot plasma.  (See \cite{s01} for a somewhat more extended 
discussion of this subject). 

\begin{figure}[hb]
\centerline{\includegraphics[width=12cm,angle=0]{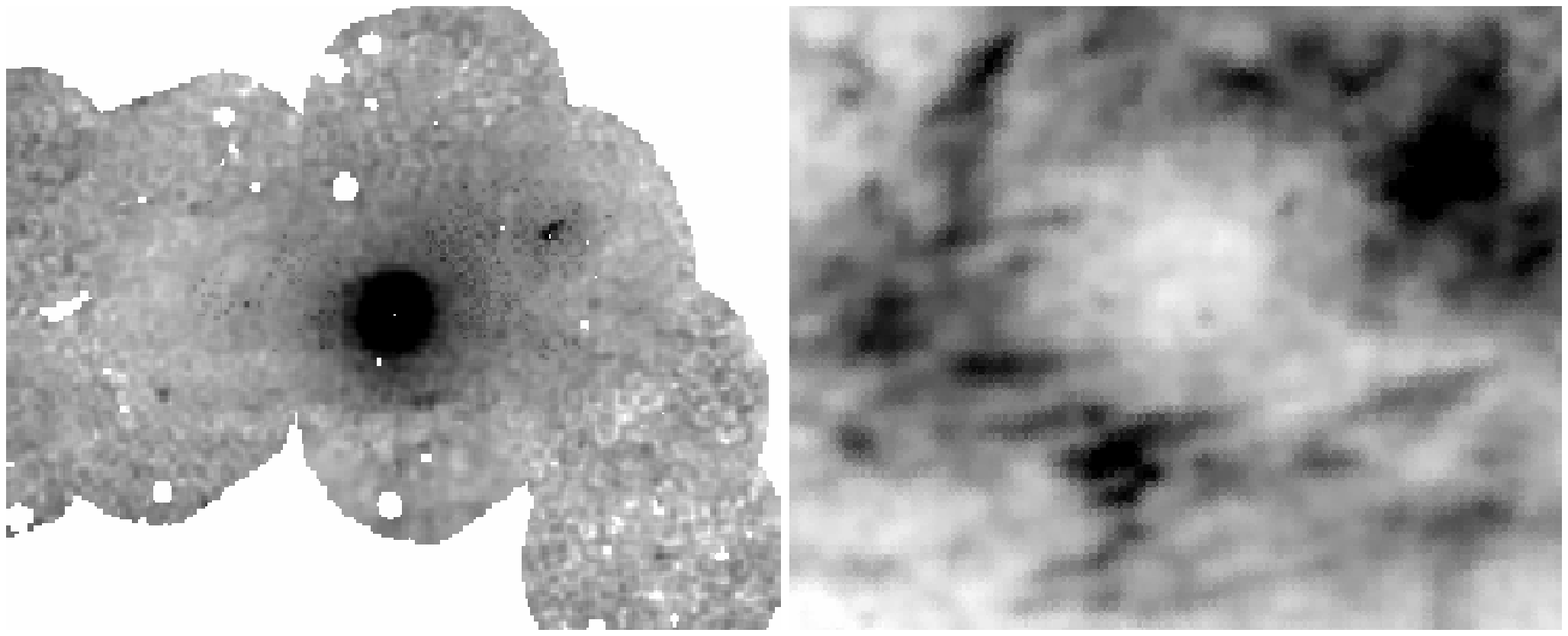}}
\caption{$ROSAT$ \oqkev\ (left) and $IRAS$ 100 $\mu$m 
(right) images of the Virgo Cluster. The fields are the same and are 
$\sim6^\circ\times5^\circ$ in extent.  Darker 
shading indicates higher intensity.  The peak $IRAS$ 100 $\mu$m 
intensities correspond to several optical depths at \oqkev.}
\label{sn-fig:virgo}
\end{figure}

{\it The pathological nature of the Virgo Cluster region:}
This digression is included as a cautionary tale about the analysis of soft 
X-ray data and the effects of ISM absorption.  Figure~\ref{sn-fig:virgo} shows
the \oqkev\ and $IRAS$ 100 $\mu$m images of the Virgo Cluster region, 
the most pathological case I know of where ISM absorption conspires
to confuse the analysis of a completely unrelated object.  The ring of
$IRAS$ cirrus surrounds the brightest region of the cluster emission with 
$\sim1-2$ optical depths of additional absorption at \oqkev.  Because the 
cirrus ring is roughly circular as well as roughly centered on the cluster it
significantly modifies the radial profile of the \oqkev\ emission.  However,
the additional absorption is relatively small at higher energies so the
radial profile of, for instance, the \tqkev\ to \oqkev\ hardness ratio 
is also affected.

{\it Where has all the \ovi\ gone?}
\cite{sh03} presents results which throw some confusion on models for the 
LHB and cold cloud/hot plasma interfaces.  Using
\fuse\ observations there appears to be little or no \ovi\ emission from 
the LHB where there should be $T\sim300,000$~K gas in interface regions
between the hotter plasma and the cooler material of the cavity walls.  
There should also be \ovi\ emission from the edge of small clouds 
located within the cavity (the Sun itself is located
in the ``Local Fluff,'' a region of partially ionized gas at a 
few thousand degrees with an extent of a few parsecs).

{\it Current spectral models don't fit the observed \oqkev\ spectra:} While there
are a few relatively good spectra of the diffuse background at \oqkev\ 
they are not well fit by current thermal equilibrium and normal abundance 
emission models, or non-equilibrium models for that matter (\cite{san01}).

{\it What is the zero level of the \oqkev\ X-ray data?} Recently the production
of soft X-rays in the solar system by charge exchange between the solar wind
and interstellar neutrals has become a hot topic.  While so far this mechanism 
has been used to explain the flaring long-term enhancement background observed 
in the RASS (\cite{crs01}), the question arises whether at quiescence it still 
produces a non-negligible flux which could significantly impact our view of 
the LHB (e.g., \cite{lal03}).





%


\bibliographystyle{kapalike}
\chapbblname{huntsville}
\chapbibliography{huntsville.bib}














\articletitle[Peering Through the Muck: the influence of the ISM on observations]
{Peering Through the Muck:  
Notes on the Influence of the Galactic 
Interstellar Medium on Extragalactic Observations}


\author{Felix J. Lockman}
\affil{National Radio Astronomy Observatory\footnote{The National Radio 
Astronomy Observatory is operated by Associated Universities, Inc., 
under a cooperative agreement with the National Science Foundation.} \\ 
Green Bank, WV USA}
\email{jlockman@nrao.edu}

\begin{abstract} 
This paper considers some effects of foreground  Galactic gas on 
 radiation received from 
 extragalactic objects, with an  emphasis on the use of the 21cm line 
to determine the total $N_{HI}$. In general, the opacity of the 21cm line 
makes it impossible to derive an accurate value of $N_{HI}$
by simply applying a formula to the observed emission, 
except in directions where there is very little interstellar matter. 
The 21cm line can be used to estimate the likelihood that there is significant 
 $H_2$ in a particular direction, but carries little or no information on 
the amount of ionized gas, which can be a major source of foreground effects. 
Considerable discussion is devoted to the importance of 
small-scale angular structure in HI, with the 
conclusion that it will rarely contribute significantly to the total  
 error compared to other factors (such as the  effects of 
ionized gas) for extragalactic sight lines  at high Galactic latitude.
 The direction of the Hubble/Chandra Deep Field North is used as an example of 
the complexities that might occur even in the absence of opacity or 
molecular gas. 

\end{abstract}


\section{Introduction}
The Interstellar Medium (ISM) regulates the evolution of the Galaxy.  
It is the source of 
material for new stars and the repository of the products of stellar 
evolution.    But it is 
a damned nuisance to astronomers seeking to peer beyond the local gas. 
 In this article I treat the ISM as if it were simply an impediment to 
knowledge, and suggest ways that one might estimate its effects. 
This topic has taken on increasing importance in recent years as 
more and more experiments are requiring correction for 
 the ``Galactic foreground'' (e.g., Hauser 2001).  Here the 
 emphasis will be on the use of the 21cm line 
to determine a total $N_{HI}$, for the 21cm line 
 is our most general tool, and $N_{HI}$ is an 
important quantity which can be used 
to estimate  $N_{He}$, E(B--V)  and $S_{100\mu}$, as well as the  likelihood 
that there is molecular hydrogen along the line of sight. 
Some of the points treated here are discussed in more detail in 
reviews by Kulkarni \& Heiles (1987),  
Dickey \& Lockman (1990; hereafter DL90), Dickey (2002), and 
 Lockman (2002).

\section{General Considerations}

Figure 1 shows the amount of neutral interstellar gas, expressed as an 
equivalent HI column density, needed to produce 
unity opacity given normal abundances.  
Below the C-band edge at 0.25 KeV the opacity results almost
entirely from photoelectric absorption by  hydrogen and  helium, which 
contribute about equally to $\tau$ (Balucinska-Church \& 
McCammon, 1992).

\begin{figure}
{\includegraphics[height=0.9\hsize,width=3in,angle=-90]{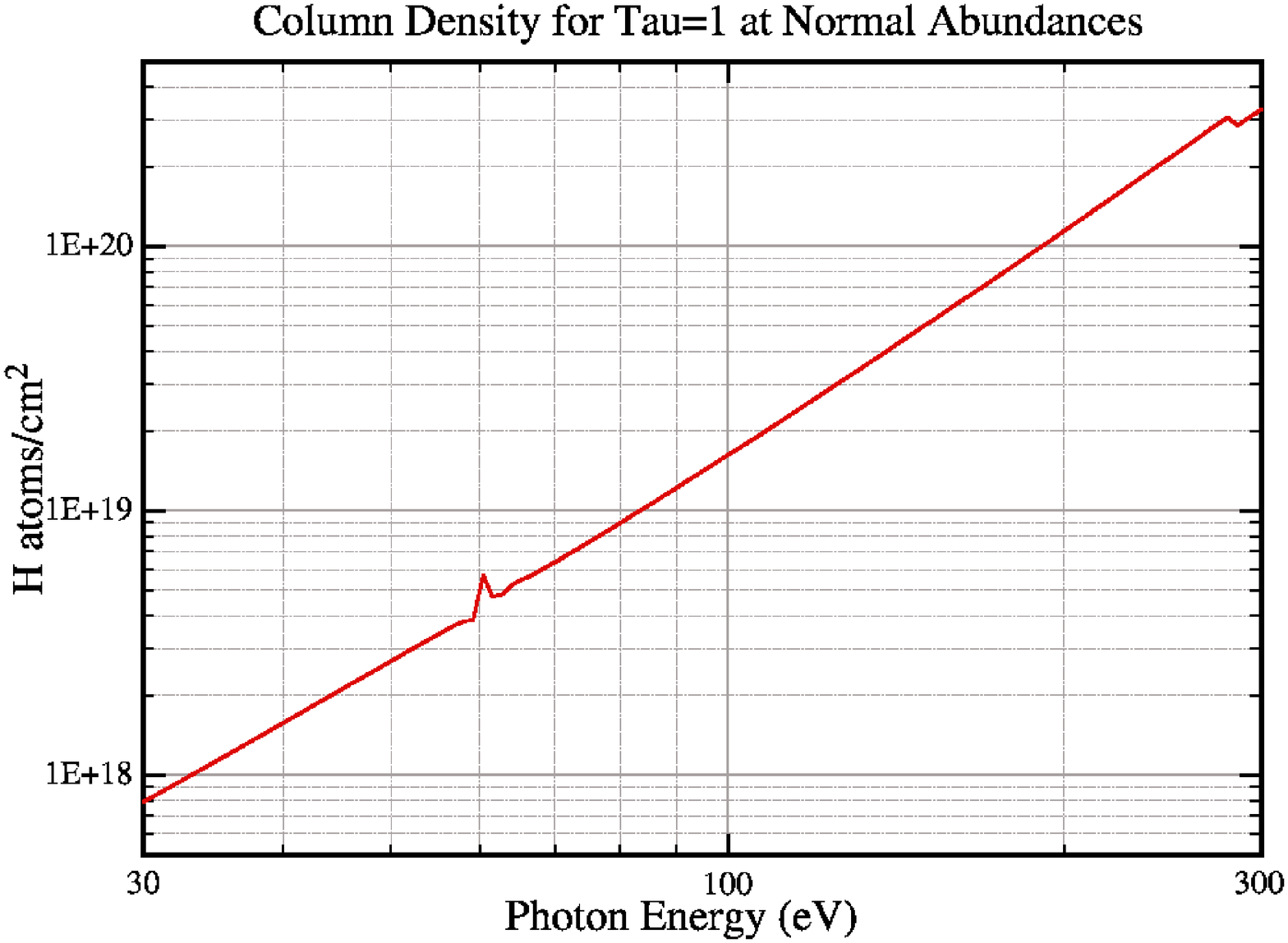}}
\caption{The equivalent $N_{HI}$ needed to produce an opacity of 
unity for the Galactic ISM at normal abundances, as a function of photon energy 
(Balucinska-Church \& McCammon 1992).
}
\end{figure}

Surveys of the sky in the 21cm line find 
 $\langle N_{HI} sin|b|\rangle  = 3 \times 10^{20}$ cm$^{-2}$ 
 where b is the Galactic latitude (DL90).   
Thus, {\it most}  sight-lines through the Milky Way  have $\tau \geq 1$ for 
$13.6 <  E < 300 $ eV, even without taking into account any contribution 
to the opacity from molecular hydrogen, $H_2$.    
Luckily, there are large areas of the sky over which $N_{HI}$ is 
a factor of several below the average, but even so,  the lowest 
$N_{HI}$ in {\it any} known direction is $4.4 \times 10^{19}$ cm$^{-2}$ 
(Lockman, Jahoda \& McCammon 1986; Jahoda, Lockman \& McCammon 1990), 
so  observations at $13.6 < E \leq  100$ 
 eV must always involve substantial corrections for the Galactic ISM. 

Except in directions where it is possible to make 
a direct measurement of $N_{HI}$ in  UV  absorption lines, 
every attempt to determine the effect of the ISM on extragalactic observations
should begin with the 21cm line.  
The Leiden-Dwingeloo (LD) 21cm survey covering $\delta > -30^{\circ}$ 
 at $35'$ angular resolution (Hartmann \& Burton 1997) 
supersedes all previous general 
surveys because of its angular and velocity resolution,
 and high quality of data.  A southern extension will be completed soon.  
Some parts of the sky of special interest 
have been mapped at higher angular resolution (e.g., Elvis et al. 1994; 
Miville-Desch\^{e}nes et al. 2002; 
Barnes \& Nulsen 2003).  The brighter emission near the 
Galactic plane is now being  measured at $1'$ resolution 
by a consortium who employ three different synthesis arrays 
(Knee 2002; McClure-Griffiths 2002; Taylor et al. 2002). 

\section{Estimating $N_{HI}$ from  21cm HI Data}

Radio telescopes measure an HI brightness temperature, $T_b$, as 
a function of velocity, but 
in general, there is no single formula that can be applied 
to derive   $N_{HI}$ from the observed  $T_b$. 
The solution to the equation of transfer for 21cm emission 
from a uniform medium is simple enough: 
\begin{equation}
T_b(v) = T_s[1-exp(-\tau(v))], 
\end{equation}
where $\tau(v) = 5.2 \times 10^{-19} N_{HI} / (T_s\ \Delta v)$ 
 for a Gaussian profile from a 
uniform cloud of linewidth  $\Delta v$ (FWHM) in km s$^{-1}$. 
$T_s$ is the excitation temperature of the transition, 
which is often, but not always, equal to the 
gas kinetic temperature (e.g., Liszt 2001).  But the real interstellar 
medium is not uniform, and the typical 21cm profile 
consists of several blended components formed in 
regions of different temperature.  If 
the line is optically thin  at all velocities 
there is no dependence of $N_{HI}$ on $T_s$ 
and $N_{HI} = 1.8 \times 10^{18} \int T_b dv$ cm$^{-2}$.  
The optically thin assumption always gives a lower limit on  $N_{HI}$.  

In directions where  part of the line has $\tau \geq 0.1$  
 the concept of a meaningful $T_s$ becomes ambiguous and 
 there is no unique solution for $N_{HI}$ from 21cm emission  data 
alone (e.g., Kalberla et al. 1985; DL90; Dickey 2002).
An HI cloud at 100 K with $\Delta v = 10$ km s$^{-1}$ has 
 $\tau = 0.1$ for $N_{HI} = 2 \times 10^{20}$ cm$^{-2}$, so 
the 21cm line in an average direction (see Fig.~2) 
should be treated as if it has components which are not optically thin.  

\subsubsection{Digression: The Two-phase ISM}

Theory tells us that under some conditions  HI can exist in 
 two stable phases at a single pressure:  a warm phase whose 
temperature is thousands of Kelvins, and a cool phase whose temperature is
$\leq100$ K (e.g., Field, Goldsmith, \& Habing 1969; 
Wolfire et al. 2003).  Observations suggest that reality is not so 
bimodal (e.g., Liszt 1983), but the 
generalization is still useful --- the ISM does contain 
cool HI with a high 21cm line opacity and warm HI with a low opacity 
(e.g., Heiles \& Troland 2003).   In the Solar neighborhood there is 
more  mass in the warm HI than the cold (Liszt, 1983; Dickey \& Brinks, 1993). 
 The cold phase fills a much smaller volume than the warm phase 
 and has a smaller scale-height as well, so at high 
Galactic latitudes many sight lines skirt the 
 clouds and intersect predominantly   ``intercloud'' medium, which 
has a low opacity because of its high temperature and turbulence.
In these directions $N_{HI}$ can be determined quite well.

\begin{figure}
{\includegraphics[height=0.6\vsize,width=3.0in,angle=-90]{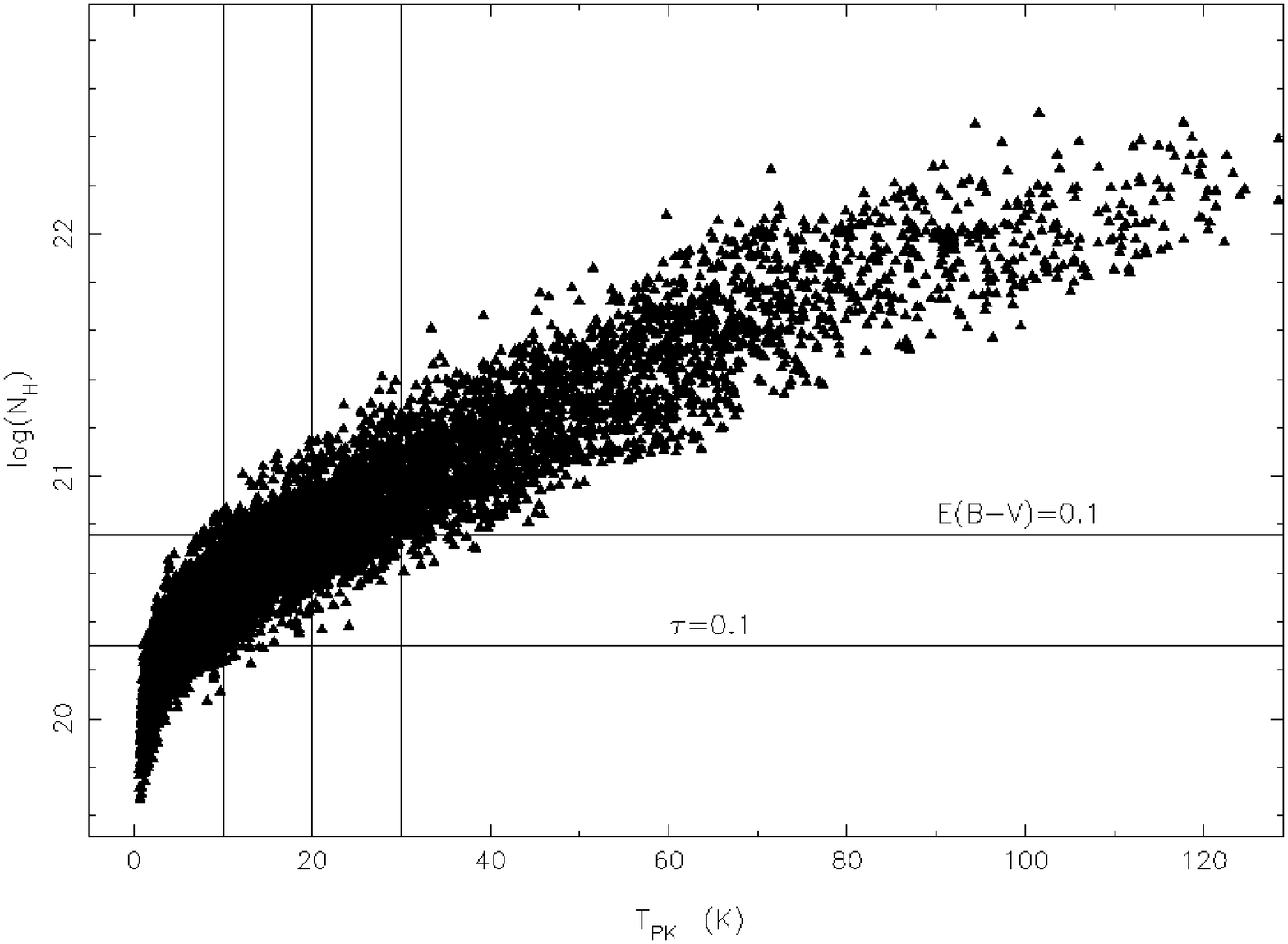}}
\caption{
The total $N_{HI}$ calculated for $T_s = 135$ K vs the peak line brightness
 temperature,  Tpk, 
for about $10^4$ 21cm spectra from the LD survey.  The sample 
includes data from $-90^{\circ} \leq b \leq 90^{\circ}$ 
 every $5^{\circ}$ in longitude between $30^{\circ}$ and 
$180^{\circ}$. Spectra above the $\tau = 0.1$ line likely contain 
components which have some opacity.  Directions with HI spectra 
above the $E(B-V)=0.1 $ line likely intersect regions of some 
molecular gas.
}
\end{figure}

\paragraph{Deriving $N_{HI}$ In Practice}
A 21cm profile should always be evaluated 
for  several excitation temperatures, 
 e.g.,  $T_s = 1000$ K and $135$ K.  If the resulting  
values of $N_{HI}$ differ significantly, 
where ``significantly'' depends on the accuracy one needs,  it 
is likely that a thorough investigation of the direction of interest 
must be made by measuring $\tau(v)$ in absorption  against nearby 
radio continuum sources ( DL90; Dickey 2002).  
A sense of how often this might be required is
given in Figure 2, where  $N_{HI}$  is 
plotted against the peak $T_b$ in the line, 
$T_{pk}$,  for a sample of $10^4$ spectra from the LD survey. 
The line  at $2 \times 10^{20}$ cm$^{-2}$ labeled $\tau = 0.1$ marks,
very approximately, the $N_{HI}$ below which 
the 21cm line is most likely optically thin.  
Conversely, Fig.~2 suggests that any  profile with $T_{pk} > 10$ K 
  should be tested for possible opacity. 
This applies to about $80\%$ of the directions in the Figure. 

\section{Angular Structure}

\begin{quote}
 The Mississippi [river] is remarkable in still another way --- its 
disposition to make prodigious jumps by cutting through 
narrow necks of land, and thus straightening and shortening itself...
In the space of 176 years the Lower Mississippi has 
shortened itself 242 miles.  That is an average of a trifle over 
one mile and a third per year.  Therefore, any calm person, who is 
not blind or idiotic, can see that in the Old Oolitic Silurian Period, 
just a million years ago next November, the Lower Mississippi River 
was upwards of 1,300,000 miles long, and 
stuck out over the Gulf of Mexico like a fishing-rod.  And by the 
same token any person can see that 742 years from 
now the Lower Mississippi will be only a mile and three quarters 
long, and Cairo [Illinois] and New Orleans will have joined their streets together,
and be plodding comfortably along under a single mayor and a 
mutual board of aldermen.  There is something fascinating about 
science.  One gets such wholesale returns of conjecture out of 
such a trifling investment of fact.

--- Mark Twain (1883), in {\it Life on the Mississippi}
\end{quote}

Use of  21cm data to determine the ISM 
toward an extragalactic object often requires extrapolation over 
orders of magnitude in solid angle:  from 
the area covered by the radio antenna beam, 
to the often infinitesimal area of the object under study. 
The highest angular resolution typically obtainable for a Galactic 
21cm HI emission spectrum (with good sensitivity) is $\sim 1'$, 
and this is very much larger than the size of an AGN.  
Hence the need for extrapolation.

This is a vexing subject which causes some quite respectable scientists 
 to loose their heads (I won't give explicit references, you 
know who you are).  A few take the situation 
as license to rearrange the ISM into whatever preposterous 
filigree of structure   simplifies their work --- 
contriving an ISM which appears smooth on scales constrained 
by the data, but which goes crazy in finer detail. They ignore 
the  wealth of data on the real small-scale structure 
of interstellar hydrogen.  The situation has been further confused by the 
reported discovery of an anomalous population of 
tiny, dense HI clouds,  whose significance, not to say reality,
 is now known to have been exaggerated.   Twain's warning 
against thoughtless extrapolation is especially appropriate to this 
topic.  

  {\it The Galactic ISM is not a free parameter!} 
Is there structure in the total Galactic $N_{HI}$  on all angular scales?
Yes!  Does this introduce errors  when 
extrapolating to small angles? Yes.  Are the errors important? 
Usually not!  From point to point across the sky HI 
clouds come and go,   and line components 
shift shape and velocity, but the dominant changes in 
{\it total}  $N_{HI}$ are usually 
on the largest linear scales, and do not cause  
 large fractional fluctuations over small angles ($\leq 0.5^{\circ}$).

Most structure in interstellar HI results from turbulence, characterized 
by a power-law spectrum with an exponent always less than $ -2$; 
this has been determined experimentally and 
is understood theoretically (e.g., Green 1993; Lazarian \& 
Pogosyan 2000; Deshpande, Dwarakanath, \& Goss 2000; 
Dickey et al. 2001 and references therein). 
Small angles in nearby gas (e.g., at high Galactic latitude) 
correspond to small linear scales where there is 
relatively little structure.  If the sight line intersects a 
distant high-velocity cloud then small angles may correspond to 
large spatial scales  and the  variations in that spectral component 
will be larger, but this is usually a problem for the 
{\it total} $N_{HI}$ only in directions  dominated 
by distant gas (see $\S 6$).  Cold HI may have more  
structure than warm HI, and molecular clouds even more, but 
as a practical matter, the extrapolation 
to small angles introduces large errors only when a significant part of the 
gas in a particular direction  is molecular or  of anomalous origin, 
e.g., comes from a high-velocity cloud.  Examination of 21cm 
spectra  around the position of interest should give 
adequate warning of possible structure which then would require 
higher resolution observations to measure.

A lack of appreciation for the effects of power-law turbulence 
at small angles  was one reason why 
 the early, high-resolution, VLBI studies were interpreted as 
 evidence for an anomalous  population of 
extremely dense  HI clouds with sizes of  tens of AU.  This in turn led 
 some to assume that there must be extreme fluctuations in the 
{\it total} $N_{HI}$ on very small angular scales.  Somewhere 
Mark Train was chuckling.  But 
with more complete observations (Faison 2002, Johnston et al. 2003) 
and a better understanding of how to interpret them (Deshpande 2000)
the anomaly has disappeared almost entirely.  The measured 
 small-scale fluctuations in $N_{HI}$ are likely to be entirely 
consistent with the known power laws
(Deshpande 2000).  Recent observations of  21cm absorption toward 
pulsars  probing linear scales of 0.005--25 AU find
no evidence of spatial structure at the 
$1 \sigma$ level of $\Delta e^{-\tau} = 0.035$ 
(Minter, Balser \& Karlteltepe 2003), and other 
careful observations toward pulsars suggest that some of the initial 
claims of small-angle fluctuations in HI absorption 
might be in error (Stanimirovic et al. 2003).  
The issue of small-scale structure in the ISM is 
interesting, and there are  anomalous 
directions, e.g.,  toward 3C 138, where clumping in cold gas appears 
to be significant (Faison 2002), though here the total $N_{HI}$ 
is $> 2 \times 10^{21}$ cm$^{-2}$ and not a typical extragalactic
sight-line.  

Recently Barnes \& Nulsen (2003)  
combined interferometric and single-dish data to measure 
21cm emission toward three high-latitude clusters of galaxies 
and found  limits on fluctuations 
in $N_{HI}$ on scales of $1'$--$10'$ of  $<3\%$ to $<9\%$ ($1\sigma$). 
DL90 had  suggested  that  on these angular scales 
$\sigma(N_{HI})/\langle N_{HI} \rangle \leq 10\%$, 
an estimate which has been controversial, but 
now appears somewhat conservative.  
I believe that, as concluded in DL90, 
 directions without significant $H_2$, 
and without significant anomalous-velocity HI, 
are unlikely to contain small-scale angular  
structure in HI that is  a major source of error 
in estimates of the effects of the Galaxy on extragalactic 
observations.

\section{Molecular Hydrogen and Helium}

\paragraph{Molecular Hydrogen}
  The 21cm data can be used to predict the likelihood 
that there is molecular 
gas along the line of sight, because there is usually 
 cool HI associated with molecular clouds (though the converse
 may not necessarily be true, see Gibson 2002).   
Direct observations of $H_2$ show   that $\geq10\%$ of the neutral ISM is in 
molecular form when  a sight line has a 
 reddening $E(B-V)\geq0.1$, equivalent to $N_{H} = 5.8 \times 10^{20}$
(Bohlin, Savage \& Drake 1978; Rachford et al. 2003).   This line is marked 
in Figure 2.   About half of the directions in 
the Fig.~2  sample lie above the 
$E(B-V)=0.1 $ line.  This suggests that for 
 $T_{pk} \geq 20$ K there may be molecular gas 
somewhere along the path, an    unfortunate circumstance, for  an 
accurate $N_{H_2}$ will then be  quite difficult to obtain in the 
absence of bright UV targets in these directions.  

\paragraph{Helium}

Interstellar $H_e^0$ and $H_e^+$ 
 both contribute to the opacity at $E > 13.6 $ eV 
but their abundance cannot be determined by simply 
 scaling  $N_{HI} $ and $N_{H_2}$, for a large fraction of the ISM is 
ionized (Reynolds et al. 1999).  Near the Sun, the mass 
  in ionized gas is about one-third the mass of HI,  
with substantial variations in different directions (Reynolds 1989).
The fractional He ionization in the medium seems low 
 (Reynolds \& Tufte 1995).  
Maps of the sky in $H_{\alpha}$ show the 
location of the brighter ionized regions, but 
the $H_{\alpha}$ intensity is proportional to $n_p n_e$, not 
to $N_{He}$.   The dispersion measure of pulsars  gives 
 $N_e$ exactly, but there are not enough pulsars to map 
out the Galactic $N_e$ to sufficient precision.  
 Kappes, Kerp and Richter (2003) 
studied the X-ray absorbing properties of a large area of the 
high-latitude sky and conclude that 20--50\% of the X-ray absorbing 
material is ionized and not traced by HI (see also Boulanger et al.  
2001 and references therein).  
Unlike molecular gas, whose effects can be neglected 
 in directions of low $N_{HI}$, the ionized component appears to 
cover the sky. Because we know so little about the detailed 
structure of the ionized component of the ISM,  {\it it probably 
contributes the most significant uncertainties in our understanding of 
Galactic foregrounds at high Galactic latitude.}

\begin{figure}[t]
{\includegraphics[height=0.6\vsize,width=3.0in,angle=-90]{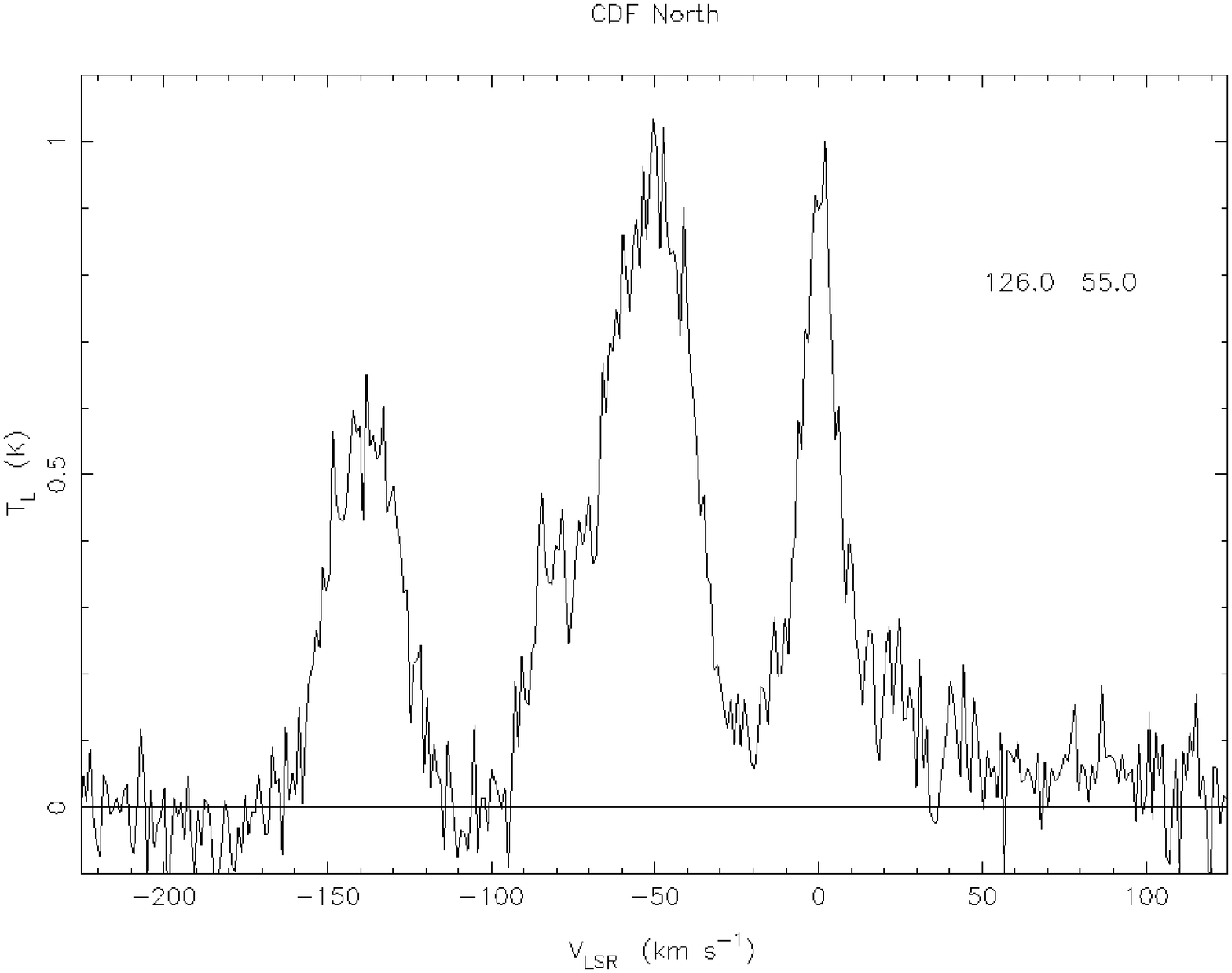}}
\caption{The 21cm spectrum toward the HDFN/CDFN at $35'$ resolution 
from the LD survey. 
}
\end{figure}

\begin{figure}[h]
{\includegraphics[height=3.5in,width=3.5in,angle=-0]{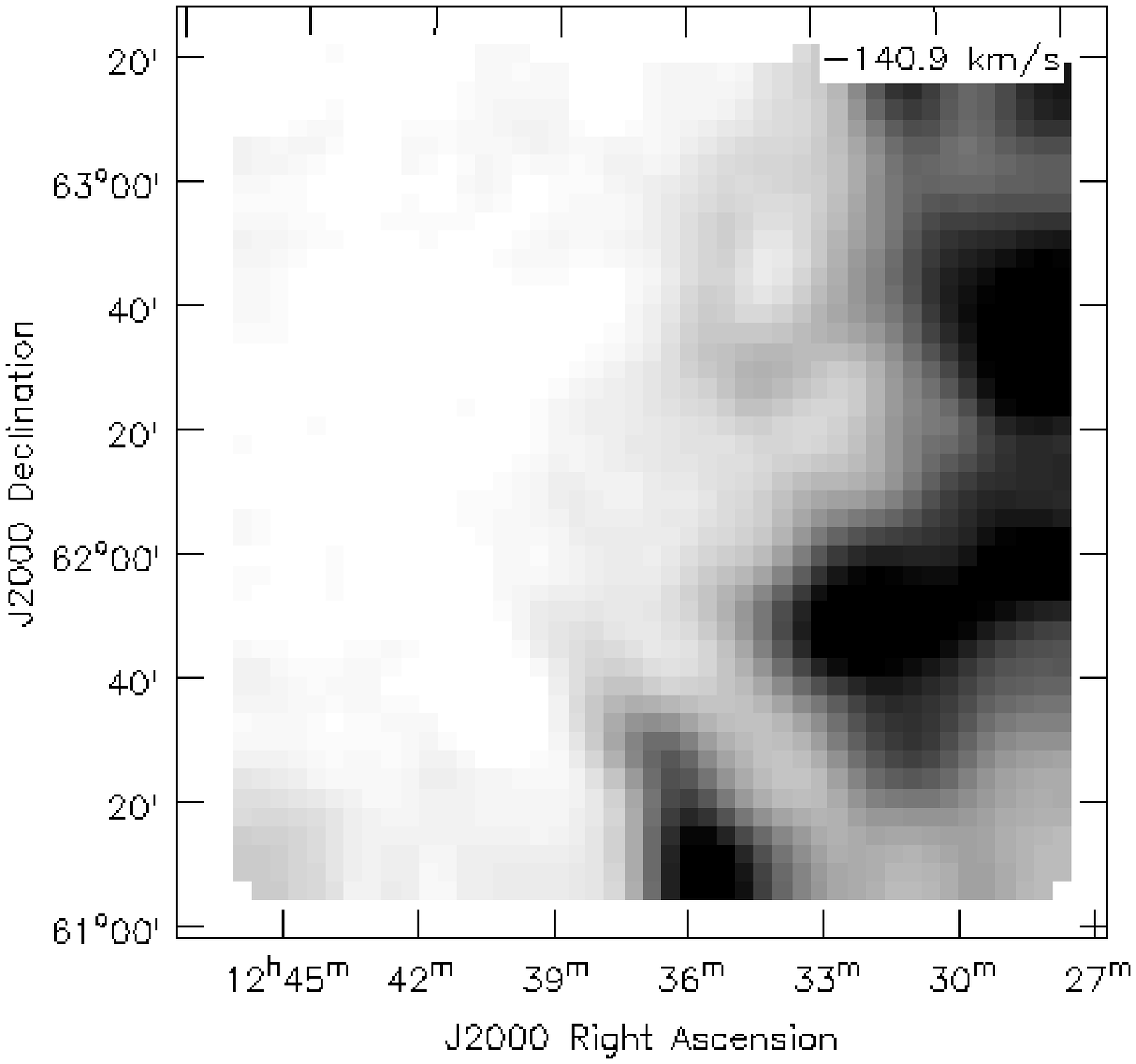}}
{\caption{The structure in the high-velocity cloud at $-140.9 $ 
km s$^{-1}$ toward 
the Chandra Deep Field North  as measured 
 at $9'$ resolution with the Green Bank Telescope (Lockman 2004, 
in preparation).  The emission line  varies from $T_{pk}>2$ K to 
$T_{pk}<0.05$ K across the field.  The CDFN is at 
 $12^h36^m50^s +62^{\circ}13'00''$ }}
\end{figure}

\section{ The Chandra Deep Field North (CDFN)}

The Chandra Deep Field North is an example of a direction 
which lacks dense gas, and thus one set of complexities, but has other 
features of interest (this direction is also the Hubble Deep Field, but 
as my comments are probably of more interest to X-ray than optical 
astronomers, I will keep the X-ray name).  
Figure 3 shows a 21cm spectrum from the LD survey at 
$35'$ resolution toward the CDFN ($\ell,b=126^{\circ}+55^{\circ}$). 
As a Galactic astronomer I find this sight-line fascinating 
because it intersects  a  high-velocity cloud,
an intermediate velocity cloud, and low-velocity ``disk'' gas, 
(from left to right in the spectrum), 
containing, respectively, 20\%, 50\% and 30\% of the total $N_{HI}$. 
This gas is almost certainly optically thin in the 21cm line. 
The low-velocity gas further shows evidence of two components, 
one warm and one cool.  
It is uncommon to find a sight line where anomalous-velocity 
gas dominates the total $N_{HI}$, but that is the case for the 
CDFN and nearby areas of the sky.  

Observations of the CDFN with the 100 meter Green Bank Telescope 
(GBT) of the NRAO at $9'$ resolution 
show a spectrum similar to Figure 3 at the low and intermediate 
velocities, but only $1/3$ as bright toward the high-velocity cloud.  
Figure 4 shows why: the high-velocity cloud 
 has a gradient of nearly two orders of magnitude 
in $N_{HI}$ near the CDFN.   It becomes the brightest component in the 
entire spectrum to the west, 
 contributing nearly $10^{20}$ cm$^{-2}$ or more than $40\%$ of
the total column density, while to the east it is almost undetectable. 
This high degree of angular structure 
is typical of high-velocity clouds, which lie far beyond the 
local disk gas (Wakker \& van Woerden 1997).  The intermediate 
velocity gas over this field has a smaller dependence on angle,
with a factor $\sim 2$ change in $N_{HI}$ and  
most structure to the south and east.

In a direction (like the CDFN) where much of the hydrogen is in 
 a high-velocity cloud, the 
 total  $N_{HI}$ is likely to be a poor predictor of many interesting 
quantities like $S_{100\mu}$ and $E(B-V)$.  
For example, high-velocity clouds have 
a lower emissivity per H atom at $100\mu$ than disk  and 
intermediate-velocity gas  because they lack of dust and/or 
have low heating (Wakker \& Boulanger 1986).  
The particular high-velocity cloud that crosses the CDFN 
has an  abundance of metals only about $10\%$ that of 
solar (Richter et al. 2001).  The intermediate velocity HI component 
also probably has different properties than disk gas at some wavelengths 
because of its different dynamical history.  
Thus, unlike most directions on the sky, where the 
 kinematics of Galactic HI is irrelevant to its
effects on extragalactic observations, the direction of the CDFN may be an 
exception, and require a specific HI component analysis.  
In contrast to the CDFN, preliminary GBT observations of 
the Chandra Deep Field South show that it has a HI 
 spectrum dominated by a single, weak, low-velocity line.  Quite boring 
compared to the CDFN.

\section{Concluding Comments}
The 21cm line can be a powerful tool for estimating the influence 
of the Galactic ISM on extragalactic observations, but it must be 
used with some thought, and it does not give the complete 
picture.  Unresolved angular structure in the 21cm data is unlikely 
to dominate the error budget in most directions.  
For $N_{HI} > 6 \times 10^{20}$ cm$^{-2}$  it is probable 
that there is some  molecular hydrogen  along the sight line and 
 the total $N_H$ will be difficult to determine.  But in my opinion, 
it is the poorly-understood ionized component of the ISM 
which introduces the most serious 
uncertainties for directions with little molecular gas,  
and it affects observations in all directions. 
We are entering an era when highly accurate 21cm data will be available 
over the entire sky, and then, (though likely even now) the limits on 
understanding the influence of the ISM on extragalactic observations 
will lie not in the uncertainties in $N_{HI}$, 
but in $N_{H_2}$ and $N_{He}$, and in the relationship between 
the dust and the gas.  Every observation of an extragalactic object is 
an opportunity to learn something about the ISM.  We should not let 
such opportunities go to waste. 

\vspace*{0.1cm}
\noindent{\bf Acknowledgements}
I thank J.M. Dickey, M. Elvis, C.E. Heiles, A.H. Minter, and R.J. Reynolds for 
comments on the manuscript.  





\part[Hard X-ray excesses and non-thermal processes]{Hard X-ray excesses and
  non-thermal processes}

\def\etal{{\it et~al.~}}
\def\bsax{{\it BeppoSAX~}}
\def\ginga{{\it Ginga~}}
\def\astroe{{\it ASTRO-E~}}
\def\constellation{{\it CONSTELLATION~}}
\def\einstein{{\it Einstein~}}
\def\exosat{{\it EXOSAT~}}
\def\tenma{{\it Tenma~}}
\def\asca{{\it ASCA~}}
\def\rosat{{\it ROSAT~}}
\def\rxte{{\it RXTE~}}
\def\euve{{\it EUVE~}}
\def\heao1{{\it HEAO-1~}}
\def\integral{{\it INTEGRAL~}}
\def\cl{clusters of galaxies~}
\def\nt{non-thermal~}
\def\syn{synchrotron~}
\def\fov{field of view~}
\def\chandra{{\it Chandra~}}
\def\xmm{{\it XMM-Newton~}}
\def\kbeta{$K_{\beta}$~}
\def\kalpha{$K_{\alpha}$~}
\def\pho{~{\rm ph~cm}^{-2}~{\rm s}^{-1}~{\rm keV}^{-1}~}
\def\erg{~{\rm erg~ cm}^{-2}\ {\rm s}^{-1}~}
\def\ergs{~{\rm erg~s}^{-1}~}
\def\h0{~{\rm H_0 = 50~km~s}^{-1}\ {\rm Mpc}^{-1}~h_{50}~}




\articletitle{Hard X-ray excesses in clusters of galaxies and
their non-thermal origin}

\chaptitlerunninghead{Hard X-ray excesses in clusters of galaxies}





\author{R.Fusco-Femiano}
\affil{Istituto di Astrofisica Spaziale e Fisica Cosmica, CNR,
Roma, Italy} \email{dario@rm.iasf.cnr.it}















\begin{abstract}
The first part of the paper reports all the results obtained by
\bsax observations concerning the search for non-thermal
components in the spectra of clusters of galaxies, while the
origin of the hard X-ray excesses detected in Coma, A2256 and, at
lower confidence level, in A754 is discussed in the second part.
\end{abstract}

 \section{Introduction}

It is well known that in addition to the hot intracluster gas
shown by X-ray measurements of the thermal bremsstrahlung emission
in the energy range 1-10 keV there is now a compelling evidence
for the existence of magnetic fields and relativistic particles in
the intracluster medium (ICM) of some clusters of galaxies. The
existence of these \nt elements is directly demonstrated by the
presence of diffuse radio emission (radio halos and relics)
detected in a growing fraction of the observed clusters.
Furthermore, FR measurements toward discrete radio sources in
clusters, coupled with the X-ray emission from the hot ICM, have
indicated the presence of $\sim \mu G$ magnetic field strengths.
The presence of \nt quantities seems to be reinforced by recent
researches that have unveiled new spectral components in the ICM
of some clusters of galaxies, namely a cluster soft excess
discovered by \euve (Lieu \etal 1966) and a hard X-ray excess
(HXR) excess detected by \bsax and \rxte.

Non-thermal HXR emission was predicted at the end of seventies in
\cl showing extended radio emission, radio halos or relics (see
Rephaeli 1979) since the same radio synchrotron electrons can
interact with the CMB photons to give inverse Compton (IC)
non-thermal X-ray radiation. Attempts to detect \nt emission from
a few \cl were performed with various experiments : balloon
experiments, \textit{HEAO-1}, the OSSE experiment onboard the
\textit{Compton-GRO} satellite and by joint analysis of \rxte \&
\asca data (Bazzano \etal 1984,90; Rephaeli, Gruber \& Rothschild
1987; Rephaeli \& Gruber 1988; Rephaeli, Ulmer \& Grubber 1994;
Delzer \& Henriksen 1998), but all these experiments reported
essentially flux upper limits. The search for \nt emission was one
of the central scientific subjects carried out by \bsax, starting
from 1997, exploiting the unique spectral capabilities of the
Phoswich Detector System (PDS) able to detect HXR radiation in the
15-200 keV energy range (Frontera \etal 1997).

\section{BeppoSAX observations}

\bsax observed seven \cl with the main objective to detect \nt
components in their X-ray spectra.

\subsection{Coma cluster}

The first cluster was Coma observed in December 1997 for an
exposure time of about 91 ksec (Fusco-Femiano \etal 1999). A
nonthermal excess with respect to the thermal emission was
observed at a confidence level of about 4.5$\sigma$ (see Fig. 1).
The $\chi^2$ value has a significant decrement when a second
component, a power law, is added. On the other hand, if we
consider a second thermal component, instead of the \nt one, the
best-fit requires a temperature of $\sim$150 keV ($>$20 keV at
90\% and $>$10 keV at 99\%). So, this unrealistic value may be
interpreted as a strong indication that the detected hard excess
is due to a \nt mechanism. In the same time a \rxte observation of
the Coma cluster (Rephaeli, Gruber \& Blanco 1999) showed evidence
for the presence of a second component in the spectrum of this
cluster. In particular the authors argued that this component is
more likely to be \nt, rather than a second thermal component at
lower temperature.

\begin{figure}
\centerline{\includegraphics[height=3.0in,angle=-90]{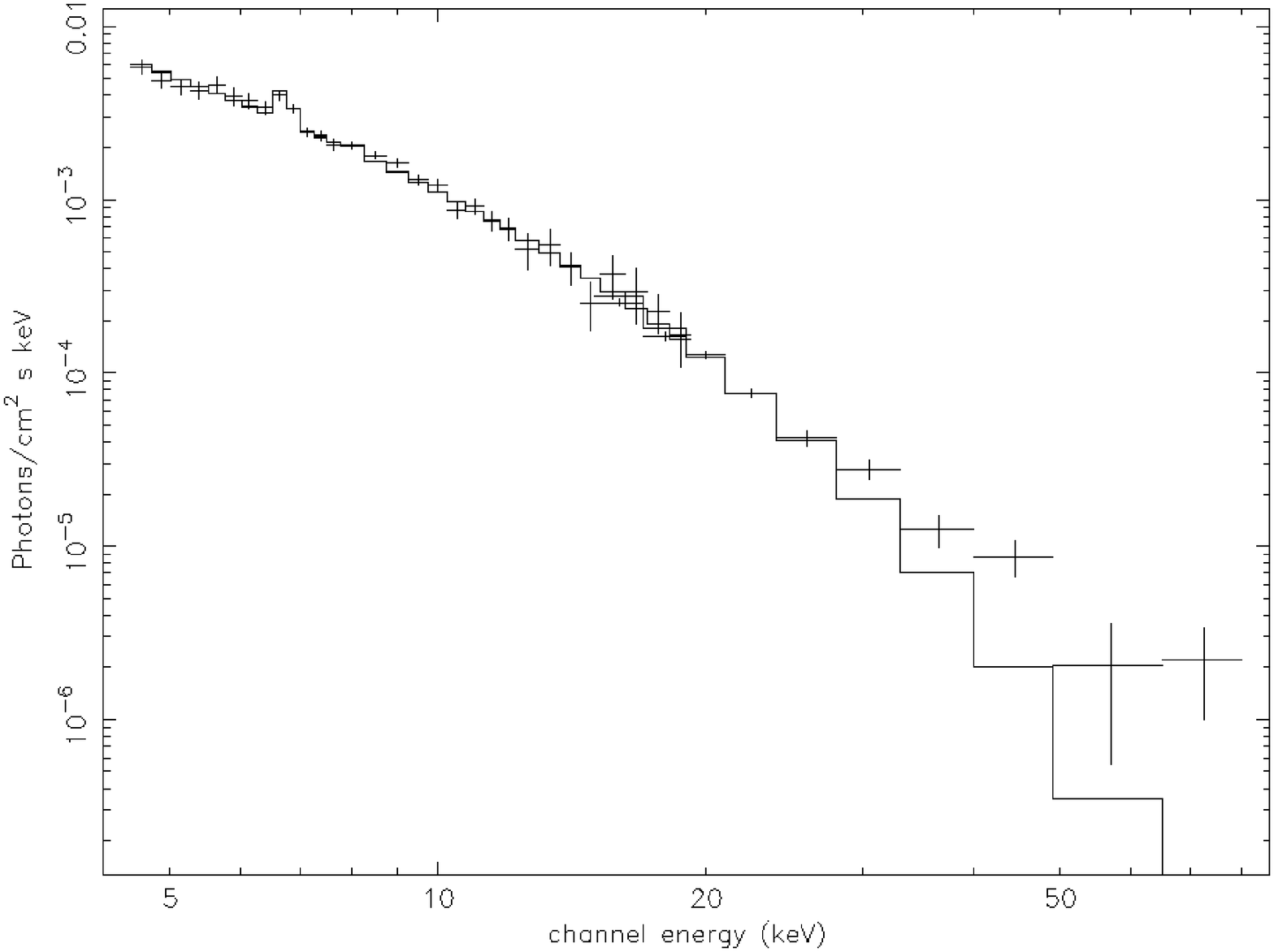}}
\caption{\protect\inx Coma cluster - HPGSPC and PDS data. The
continuous line represents the best fit with a thermal component
at the average cluster gas temperature of 8.5$^{+0.6}_{-0.5}$
keV.}
\end{figure}


\subsection{A2199}

The cluster was observed in April 1997 for $\sim$100 ksec (Kaastra
\etal 1999). The MECS data in the range 8-10 keV seem to show the
presence of a hard excess with respect to the thermal emission at
a confidence level of $\sim 3.3\sigma$. The PDS data are instead
not sufficient to prove the existence of a hard tail. We have
re-analyzed the MECS data finding that only a point is above the
thermal model at the level of $\sim 2\sigma$ (Fusco-Femiano \etal
2002). However, the cluster is planned to be observed by \xmm that
should be able to discriminate between these two different results
of the MECS data analysis, considering the low average gas
temperature of about 4.5 keV (David \etal 1993).

\subsection{A2256}

The cluster A2256 is the second cluster where \bsax detected a
clear excess (Fusco-Femiano \etal 2000) at about 4.6$\sigma$ above
the thermal emission (see Fig. 2). Also in this case the $\chi^2$
value has a significant decrement when a second component, a power
law, is added and also in this case the fit with a second thermal
component gives an unrealistic temperature of $\sim$200 keV ($>$25
kev at 90\% and $>$13 keV at 99\%) which can be interpreted as
evidence in favour of a \nt mechanism for the second component
present in the X-ray spectrum of A2256.

\begin{figure}
\centerline{\includegraphics[height=3.0in,angle=-90]{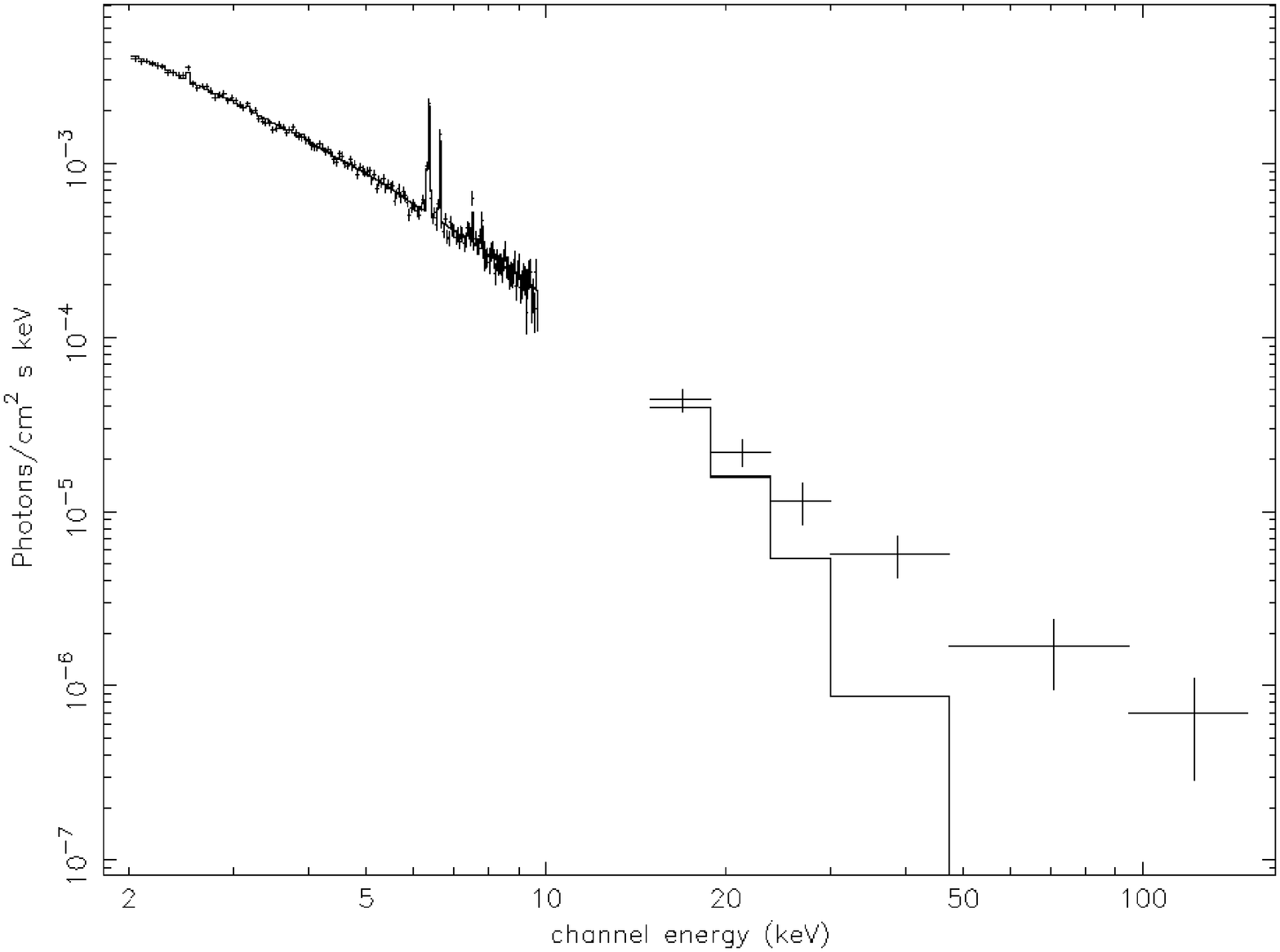}}
\caption{\protect\inx Abell 2256 - MECS and PDS data. The
continuous line represents the best fit with a thermal component
at the average cluster gas temperature of 7.47$\pm$0.35 keV.}
\end{figure}


\subsection{A1367}

A \bsax observation of Abell 1367 has not detected hard X-ray
emission in the PDS energy range above 15 keV (P.I.: Rephaeli).
A1367 is a near cluster (z=0.0215) that shows a relic at a
distance of about 22$'$ from the center and a low gas temperature
of $\sim$ 3.7 keV (David \etal 1993) that might explain lack of
thermal emission at energies above 15 keV.

\subsection{A3667}

A3667 is one of the most spectacular \cl. It contains one of the
largest radio sources in the southern sky with a total extent of
about 30$'$ which corresponds to about 2.6$h_{50}^{-1}$ Mpc. A
long observation with the PDS (effective exposure time 44+69 ksec)
reports a clear detection of hard X-ray emission up to about 35
keV at a confidence level of $\sim 10\sigma$. Instead, the fit
with a thermal component at the average gas temperature indicates
a marginal presence of a \nt component. Given the presence of such
a large radio region in the NW of the cluster, a robust detection
of a \nt component might be expected instead of the upper limit
reported by \bsax. One possible explanation may be related to the
radio spectral structure of the NW relic (Fusco-Femiano \etal
2001).

\subsection{A754}

The rich and hot cluster A754 is considered the prototype of a
merging cluster. X-ray observations report a violent merger event
in this cluster (Henry \& Briel 1995; Henriksen \& Markevitch
1996; De Grandi \& Molendi 2001), probably a very recent merger as
shown by a numerical hydro/N-body model (Roettiger, Stone, \&
Mushotzky 1998). A long \bsax observation of A754 shows a \nt
excess at energies above about 45 keV with respect to the thermal
emission at the temperature of 9.4 keV (Fusco-Femiano \etal 2003).
The excess is at a level of confidence of 3.2$\sigma$ (see Fig.
3).

\begin{figure}
\centerline{\includegraphics[height=3.0in,angle=-90]{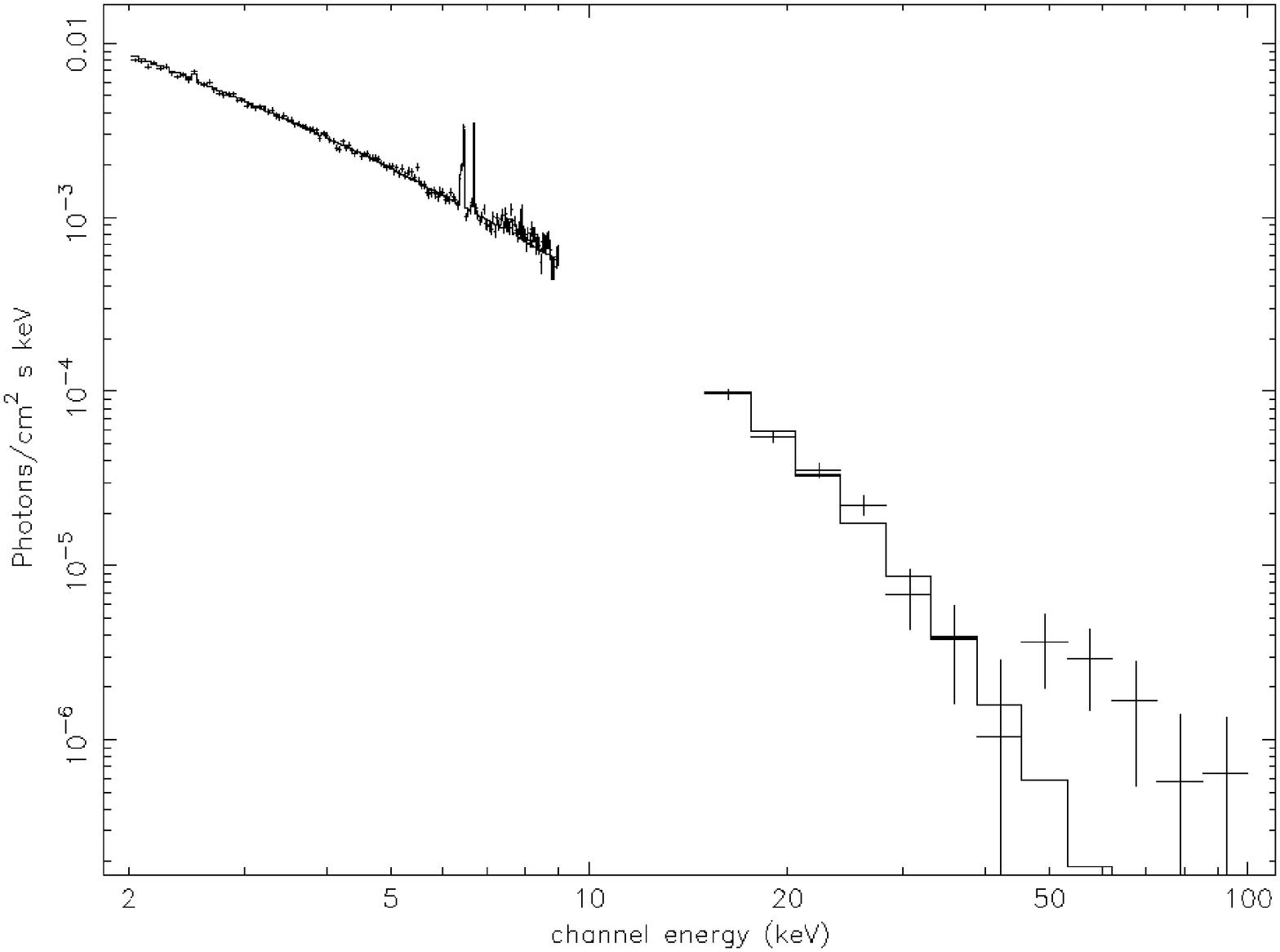}}
\caption{\protect\inx Abell 754 - MECS and PDS data. The
continuous lines represent the best fit with a thermal component
at the average cluster gas temperature of $9.42^{+0.16}_{-0.17}$
keV}
\end{figure}

\subsection{A119}

Finally, the last cluster observed by \bsax to detect \nt
component was A119. The X-ray observations have shown a rather
irregular and asymmetric X-ray brightness suggesting that the
cluster is not completely relaxed and may have undergone a recent
merger (Cirimele \etal 1997; Markevitch \etal 1998; Irwin, Bregman
\& Evrard 1999; De Grandi \& Molendi 2001). The PDS reports an
excess with respect to the thermal emission at the average gas
temperature measured by the MECS (5.66$\pm$0.16 keV) at a
confidence level of $\sim 2.8\sigma$. A119 does not show evidence
of a radio halo or relic, but the presence of a recent merger
event could accelerate particles able to emit nonthermal emission
in the PDS energy range. However, the presence of 7 QSO in the
field of view of the PDS with redshift in the range 0.14-0.58
makes very unlikely that this possible excess may be due to a
diffuse source (Fusco-Femiano \etal 2002).

\section{Possible interpretations of the observed \nt HXR
excesses in Coma, A2256 and A754}

The first possible explanation for the detected excesses is
emission by a point source in the \fov of the PDS. It has been
verified that sources like X Comae or the three quasars reported
by \xmm in Coma and the QSO in A2256 cannot be responsible for the
detected excesses (Fusco-Femiano \etal 1999; 2002), but it is not
possible to exclude that an obscured source, like Circinus (Matt
\etal 1999) a Seyfert 2 galaxy very active at high X-ray energies,
may be present in the \fov of the PDS able to simulate the
observed \nt emission. With the MECS image it is possible to
exclude the presence of this kind of source only in the central
region of about 30$'$ in radius unless the obscured source is
within 2$'$ of the central bright core. The probability to find an
obscured source in the \fov of the PDS is of the order of 10\%
(Fusco-Femiano \etal 2002). In the case of A754 the radio galaxy
26W20 (Harris \etal (1980) is present in the field of view of the
PDS that shows a X-ray bright core similar to that of a BL Lac
object. The fit with a Synchrotron Self-Compton (SSC) model to the
SED requires a flat energy index of about 0.3 to extrapolate the
flux detected by \rosat in the PDS energy range. The inclusion of
the PDS points makes difficult to fit well all the points of the
SED (Fusco-Femiano \etal 2003).

A support against the point source interpretation is given by the
spectrum of A2256 re-observed after two years from the previous
one. The two spectra are consistent (Fusco-Femiano \etal 2002) and
both observations comprise two exposures with a time interval of
$\sim$ 1 year and $\sim$ 1 month, respectively and all these
observations do not show significant flux variations. Besides, a
second recent \rxte observation of the Coma cluster (Rephaeli \&
Gruber 2002) confirms the previous one (Rephaeli, Gruber, \&
Blanco 1999) of a likely presence of a non-thermal component in
the spectrum of the cluster. These results strongly support the
idea that a diffuse non-thermal mechanism involving the ICM is
responsible for the observed excesses, also considering that Coma,
A2256 and A754 present extended radio regions.

Another interpretation is that the \nt emission is due to
relativistic electrons scattering the CMB photons and in
particular the same electrons responsible for the extended radio
regions present in these clusters. In particular, in the case of
the Coma cluster we derive a volume-averaged intracluster magnetic
field, $B_X$, of 0.15 $\mu$G, using only observables, combining
the X-ray and radio data. The value of the magnetic field derived
by this \bsax observation seems to be inconsistent with the
measurements of Faraday rotation (FR) of polarized radiation of
sources through the hot ICM that give a line-of-sight value of
$B_{FR}$ of the order of 2-6 $\mu$G (Kim \etal 1990; Feretti \etal
1995). Regarding this discrepancy it has been shown that IC models
that include the effects of more realistic electron spectra,
combined with the expected spatial profiles of the magnetic
fields, and anisotropies in the pitch angle distribution of the
electrons allow higher values of the intracluster magnetic field ,
in better agreement with the FR measurements (Goldshmidt \&
Rephaeli 1993; Brunetti \etal 2001; Petrosian 2001). One of these
models is the \textit{two-phase} model of Brunetti \etal (2001) in
which relativistic electrons injected in the Coma cluster by some
processes (starbursts, AGNs, shocks, turbulence) during a first
phase are (systematically) re-accelerated during a second phase
for a relatively long time ($\sim$ 1 Gyr) up to present time. This
model is able to explain the radio properties of the Coma cluster
: the spectral cutoff and the radial spectral steepening, and the
HXR excess present in this cluster. In this model the maximum
energy of the re-accelerated electrons with Fermi II-like
processes in the ICM is lower than $10^5$ and consequently a
cutoff in the electron spectrum might be present in the radio band
($\gamma_{Syn}\sim 10^4$). Considering that the energy of the IC
electrons is lower than that of the Syn electrons for $B<1\mu G$
($\gamma_{syn}/\gamma_{HXR}\sim 3.5\times B^{-1/2}_{\mu G}$), a
cutoff close to $\gamma_{syn}$ would reduce the synchrotron
emission without affecting the IC in the HXR band. So, for a given
ratio of IC/Syn emission the magnetic field must be higher to
obtain the same synchrotron emission. This cutoff in the spectrum
of the emitting electrons is confirmed by the spectral cutoff
observed in Coma (Deiss \etal 1997).

The other radio halo property of Coma is the spectral steepening
with the distance from the center (Giovannini \etal 1993) that can
be explained if the magnetic field strength decreases with the
distance from the center. In this case the corresponding frequency
of the cutoff in the synchrotron spectrum ($\nu_c\sim B\gamma^2$)
should decrease with the distance from the center yielding a
possible radial steepening. The fit to the overall radio \syn
properties and to the HXR excess provides a set of possible radial
profiles of the cluster magnetic field. In particular, in the
cluster core region where the RM observations are effective B is
between 0.8-2 $\mu G$ not much distant from the values inferred
from RM measurements. Besides, recently, Newman, Newman, \&
Rephaeli (2002) have pointed out that many and large uncertainties
are associated with the values of the magnetic field determined
through FR measurements (see also Govoni \etal 2001).

However, alternative interpretations to the IC model have been
proposed, some of these essentially motivated by the discrepancy
between the values of $B_X$ and $B_{FR}$. It has been proposed
that a different mechanism may be given by \nt bremsstrahlung
(NTB) from suprathermal electrons formed through the current
acceleration of the thermal gas (Ensslin, Lieu, \& Biermann 1999;
Dogiel 2000; Sarazin \& Kempner 2000; Blasi 2000; Liang, Dogiel,
\& Birkinshaw 2002).

Considering also the difficulties with the NTB model (Petrosian
2001; 2002) the presence of \nt phenomena in clusters of galaxies
(diffuse radio emission, HXR excesses and probably also EUV
emission) is usually explained in the framework of the IC model in
which the alternative between primary and secondary electron
populations represents the basis of the more recent theoretical
works developed on the argument.

\subsection{Primary electrons}

Primary electrons can be injected in the ICM by different
processes (Brunetti 2002) : one possibility is given by
acceleration by shocks, in particular merger shocks can release a
part of their energy into particle acceleration (Takizawa \& Naito
2000; Miniati \etal 2001; Fujita \& Sarazin 2001). In this case
the typical extension of $\sim$2 Mpc of radio halos requires an
unreasonably high Mach number greater than 5 considering the short
lifetime of the electrons. Furthermore, this implies a strong
field compression with the consequence that the emitted
synchrotron radiation is highly polarized, as in the case of radio
relics, but in contrast with the very low polarization found in
radio halos. Another possibility is given by re-accelerated
electrons as shown by the \textit{two-phases} model of Brunetti
\etal (2001). A number of sources of relativistic electrons in
galaxy clusters (radio galaxies, merger shocks, supernovae and
galactic winds) can efficiently inject relativistic electrons in
the cluster volume over cosmological time (e.g.: Sarazin 2002).
Relativistic electrons with $\gamma \sim$ 100-300 can survive for
some billion year and thus can be accumulated in the cluster
volume. These electrons could be responsible for the EUV emission.
To produce radio and HXR emission this population of electrons
must be re-accelerated and a possible mechanism is given by a
significant level of turbulence in the ICM produced by cluster
mergers (Brunetti \etal 2001).

\subsection{Secondary electrons}

It is well known that a secondary electron population is invoked
by the difficulty in explaining the extended radio halos by
combining their $\sim$Mpc size, and the short radiative lifetime
of the radio emitting electrons. To resolve this problem Dennison
(1980) first pointed out that a possible source of relativistic
electrons in radio halos is the continuous production of secondary
electron due to the decay of charged pions generated in cosmic ray
collisions in the ICM. This possibility has been reconsidered in
detail by various authors (Blasi \& Colafrancesco 1999; Dolag \&
Ensslin (2000); Ensslin 1999; Miniati \etal 2001)

\subsection{Primary or secondary electrons ?}

Some observational constraints seem to be able to discriminate
between primary and secondary electrons (Brunetti 2002) :

a) The comparison between radio and soft X-ray brightness of a
number of radio halos (Govoni \etal 2001; Feretti \etal 2001)
indicates that the profile of the radio emission is broader than
that of the X-ray thermal emission. This appears difficult to be
explained within secondary models which would yield narrower radio
profiles. This is due to the fact that the timescale of the p-p
collision is inversely proportional to the gas density : $\propto
n^{-1}_{th}$. As a consequence, for a given number density of the
relativistic protons, the secondary electrons are expected to be
injected in the denser regions and the radio emission would be
stronger in the cluster core. A possibility to skip this problem
is to admit an \textit{ad hoc} increasing fraction of energy
density of the relativistic protons with radius. This would imply
at least in some cases an energetics of the relativistic protons
higher than that of the thermal pool.

b) The spectral cutoff observed in Coma strongly point to the
presence of a cutoff in the spectrum of the emitting electrons and
this cutoff may be naturally accounted for if the synchrotron
emission is produced by re-accelerated electrons (Brunetti \etal
2001). Instead, this cutoff is not naturally explained by
secondary electrons unless to assume a cutoff in the energy
distribution of the primary relativistic protons. This should be
at $E_p < $ 50 GeV to have a cutoff of the spectrum from the
secondary electrons at $\sim$GHz frequencies. This seems to be in
contrast with the observations of the spectrum of Galactic cosmic
rays and with the theoretical expectations from the most accepted
acceleration mechanisms.

In conclusion the radio spectral properties cannot be satisfied in
a natural way in the case of secondary electrons. Of course, it is
important to understand if the the synchrotron spectral properties
of Coma are common among radio halos.

A confirm of the scenario described by the re-acceleration models
(e.g. the two-phases model of Brunetti \etal) seems to be given by
the radio and HXR observations of A754 (Fusco-Femiano \etal 2003).
One possible explanation of the hard excess detected by \bsax is
tied to presence of diffuse radio emission recently discovered by
Kassim \etal (2001) and confirmed by a deeper VLA observation at
1.4 GHz (Fusco-Femiano \etal 2003). The detection of the hard
excess in A754 determines, in the framework of the IC model, a
volume-averaged intracluster magnetic field of the same order
($B_X\sim 0.1\mu G$) of that determined in the Coma cluster
(Fusco-Femiano \etal 1999). This value of $B_X$ implies
relativistic electrons at energies $\gamma\sim 10^4$ to explain
the observed diffuse synchrotron emission. At these energies IC
losses determine a cut-off in the electron spectrum as confirmed
by the spectral cut-off observed in A754 obtained combining the
VLA radio observation at 1.4 GHz with the observation of Kassim
\etal (2001) at lower frequencies (a spectral index of 1.1 in the
band 74-330 MHZ and 1.5 in the higher 330-1400 MHz frequency
band). This spectral cutoff is similar to that reported in Coma.
Considering that a cutoff in the radio emitting electrons cannot
be explained in a natural way by a secondary electron population,
as discussed above, the radio and HXR results obtained for A754
seem to reinforce the scenario that primary and not secondary
electrons are responsible for \nt emission in clusters of
galaxies.

\section{Conclusions}

\bsax observed a clear evidence of \nt X-ray emission in two
clusters, Coma and A2256, both showing extended radio regions. All
the observations do not show variability in the \nt flux
supporting the idea of a diffuse \nt mechanism involving the ICM.
At a lower confidence level \nt HXR emission has been observed in
A754. The detected excess in A754, if confirmed by a a deeper
observation with imaging instruments, support the scenario that
primary re-accelerated electrons are responsible for \nt phenomena
in the ICM and reinforce the link between Mpc-scale radio emission
and very recent or current merger processes.

\bsax, as it is well known, has ceased its activity. The next
missions able to search for \nt components are : \integral ,
launched in October, \astroe, \constellation and \textit{NEXT}
with a great improvement in sensitivity. These missions will be
operative in the next years, but the energy range and the spectral
capabilities of \xmm/EPIC give the possibility to localize \nt
components in regions of low gas temperature ($\sim$ 4-5 keV). So
with \xmm we should have the possibility, by comparing the X-ray
and radio structures, to constrain the profiles of the magnetic
field and of relativistic electrons.








%


\bibliographystyle{kapalike}





\articletitle{Thermal And Non-Thermal SZ Effect In Galaxy Clusters}
\author{S.Colafrancesco}

\begin{figure}[h]
\includegraphics[width=1.17\textwidth,viewport=130 145 540 564,clip]{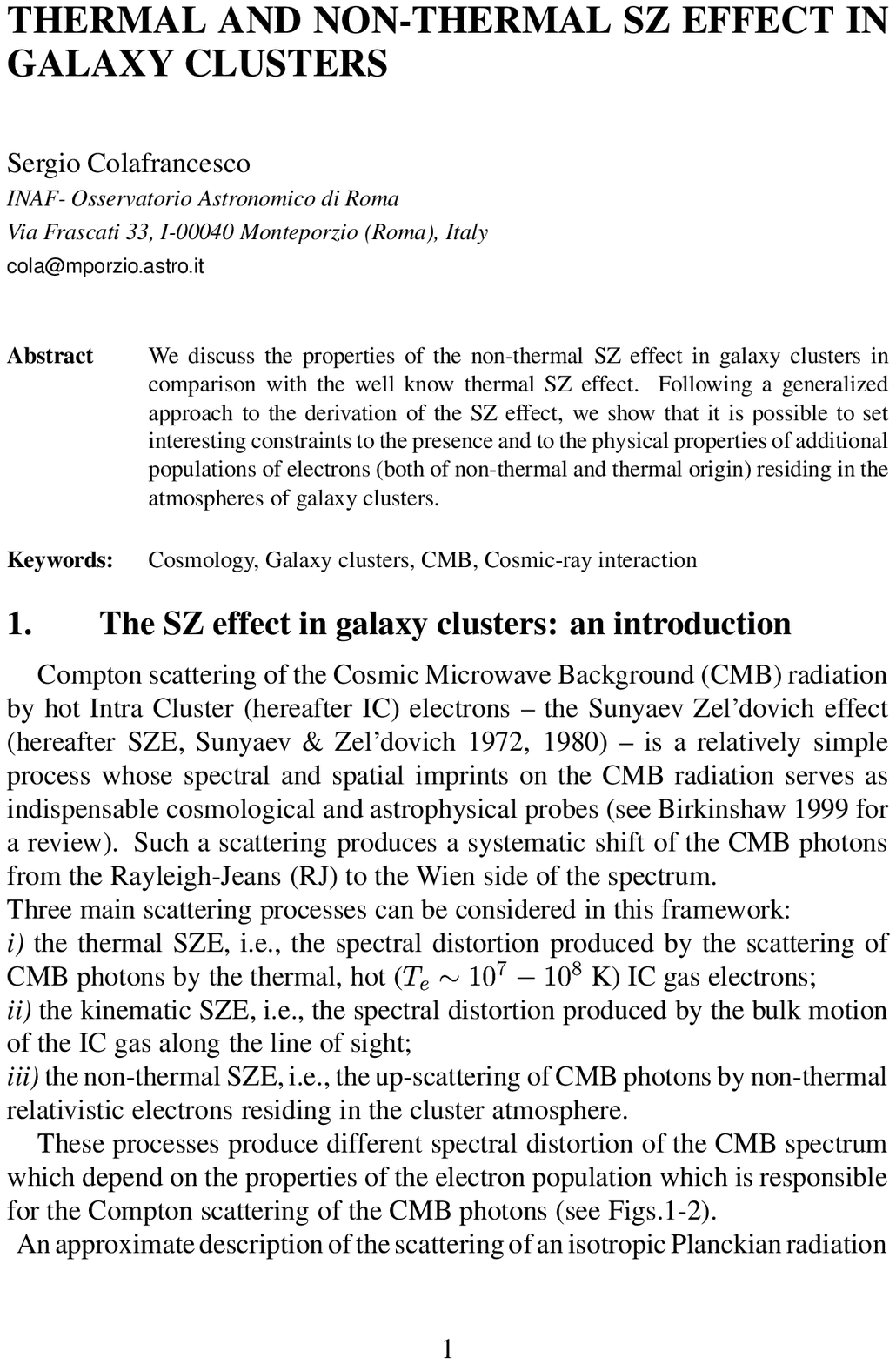}
\end{figure}
\begin{figure}[h]
\includegraphics[width=1.2\textwidth,viewport=130 120 540 690,clip]{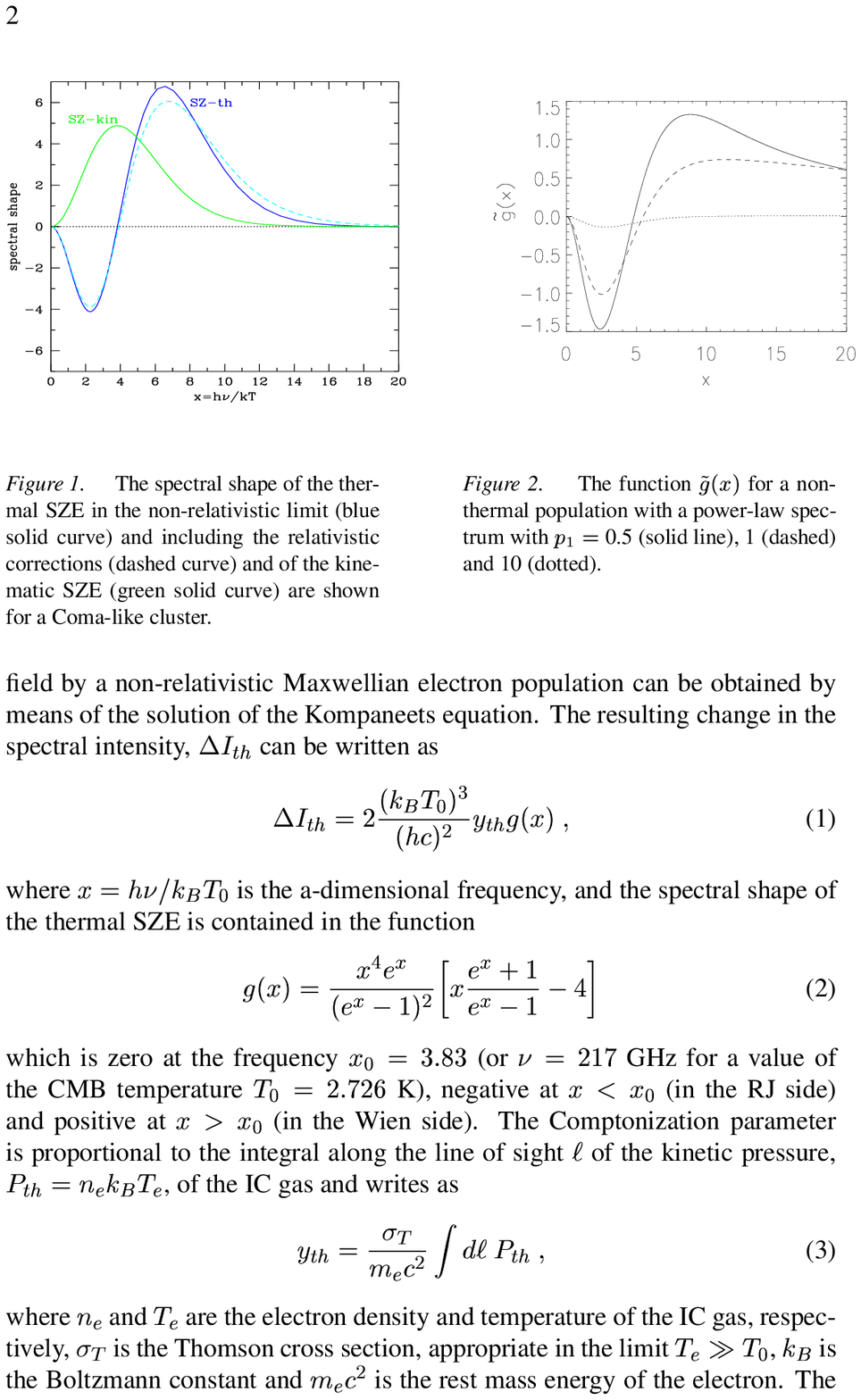}
\end{figure}
\begin{figure}[h]
\includegraphics[width=1.2\textwidth,viewport=130 120 540 690,clip]{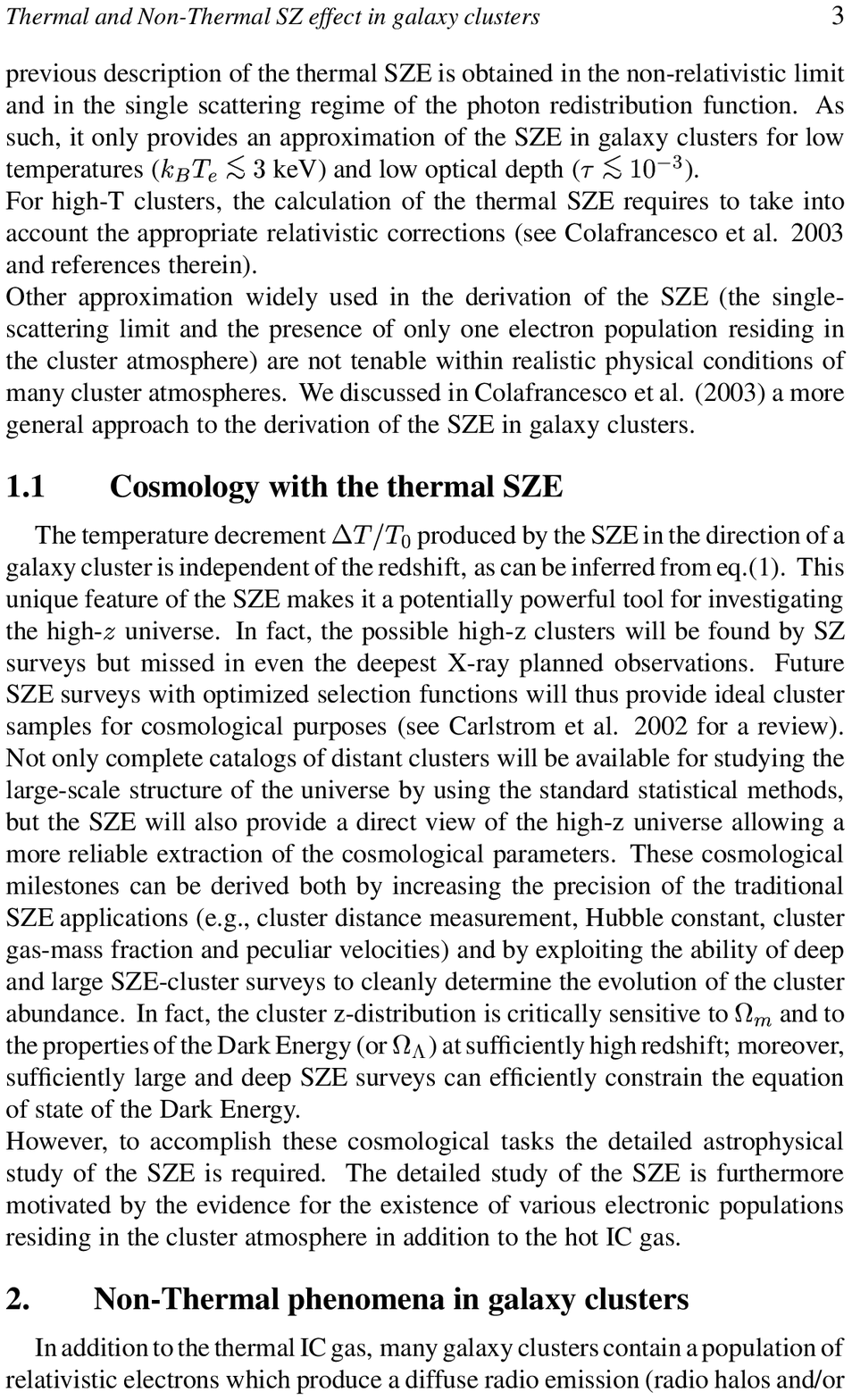}
\end{figure}
\begin{figure}[h]
\includegraphics[width=1.2\textwidth,viewport=130 120 540 690,clip]{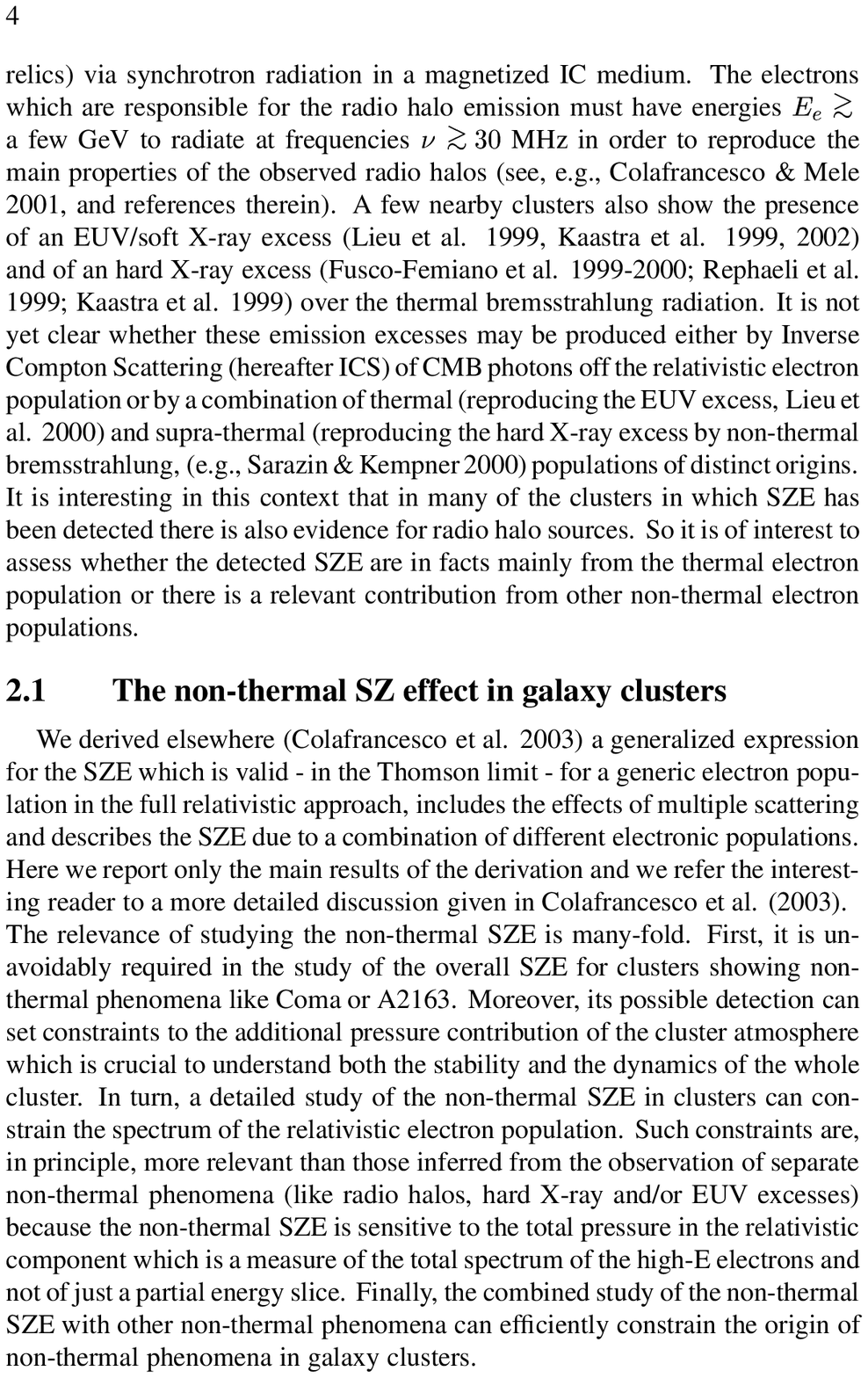}
\end{figure}
\begin{figure}[h]
\includegraphics[width=1.2\textwidth,viewport=130 120 540 690,clip]{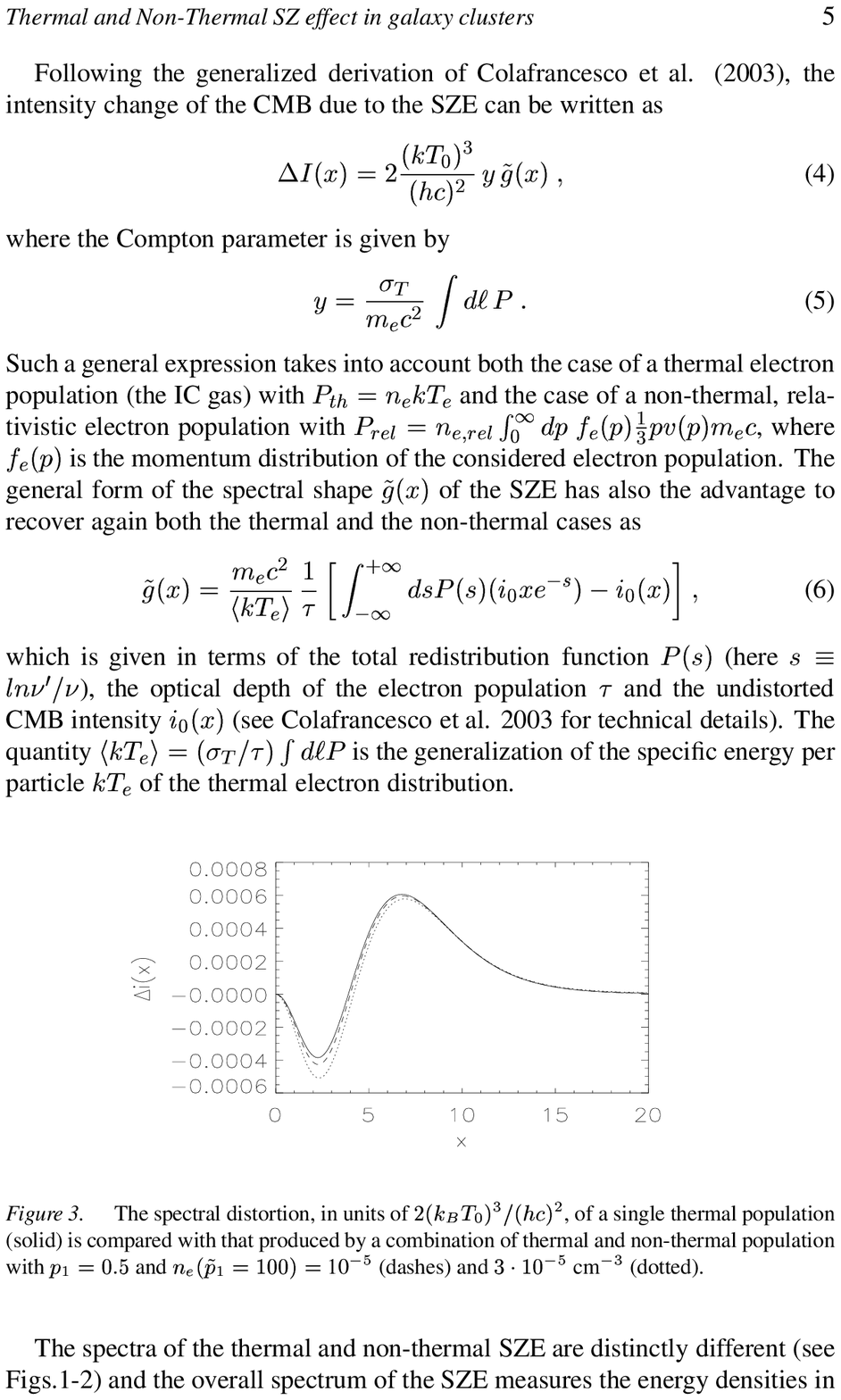}
\end{figure}
\begin{figure}[h]
\includegraphics[width=1.2\textwidth,viewport=130 120 540 690,clip]{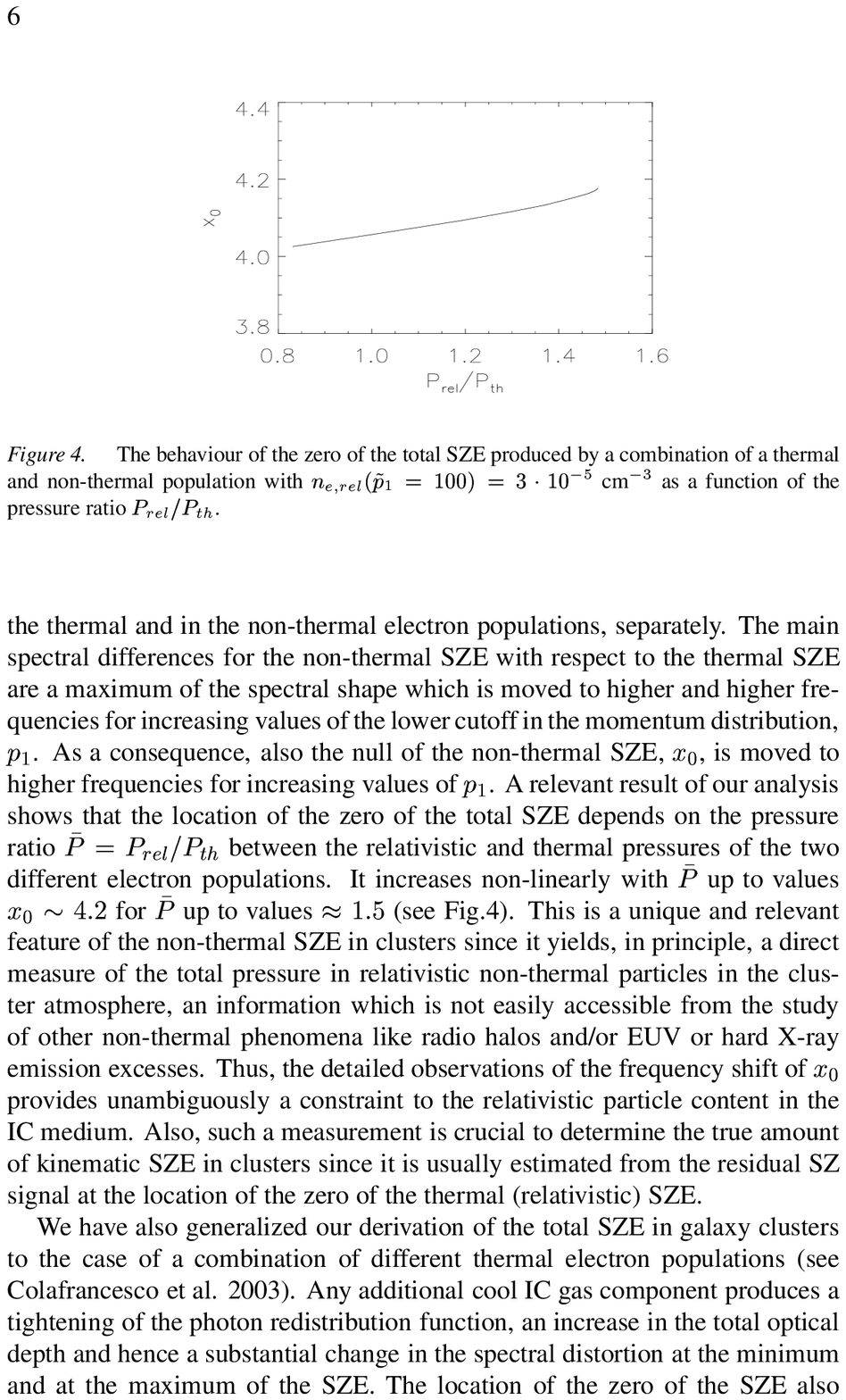}
\end{figure}
\begin{figure}[h]
\includegraphics[width=1.2\textwidth,viewport=130 120 540 690,clip]{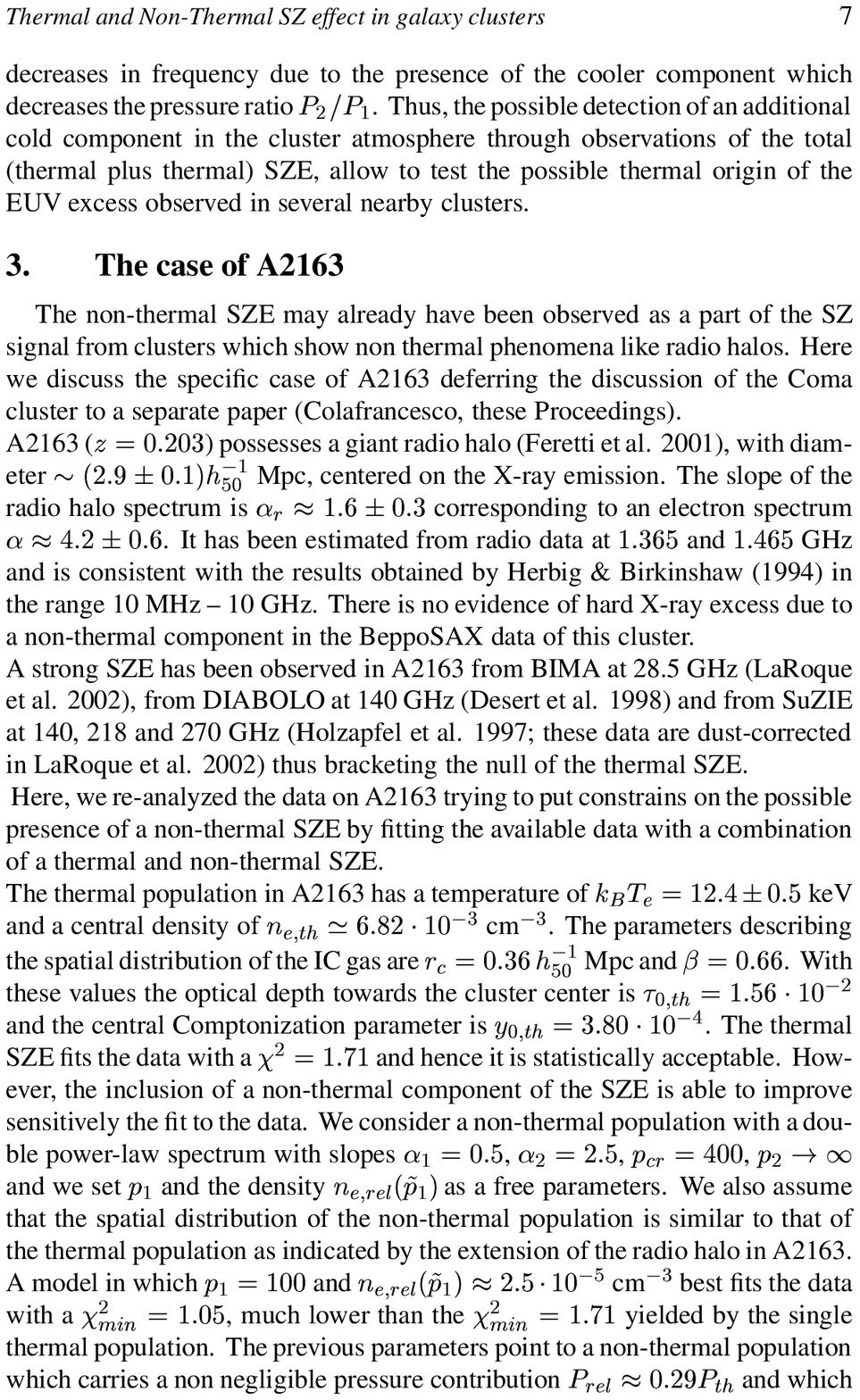}
\end{figure}
\begin{figure}[h]
\includegraphics[width=1.2\textwidth,viewport=130 120 540 690,clip]{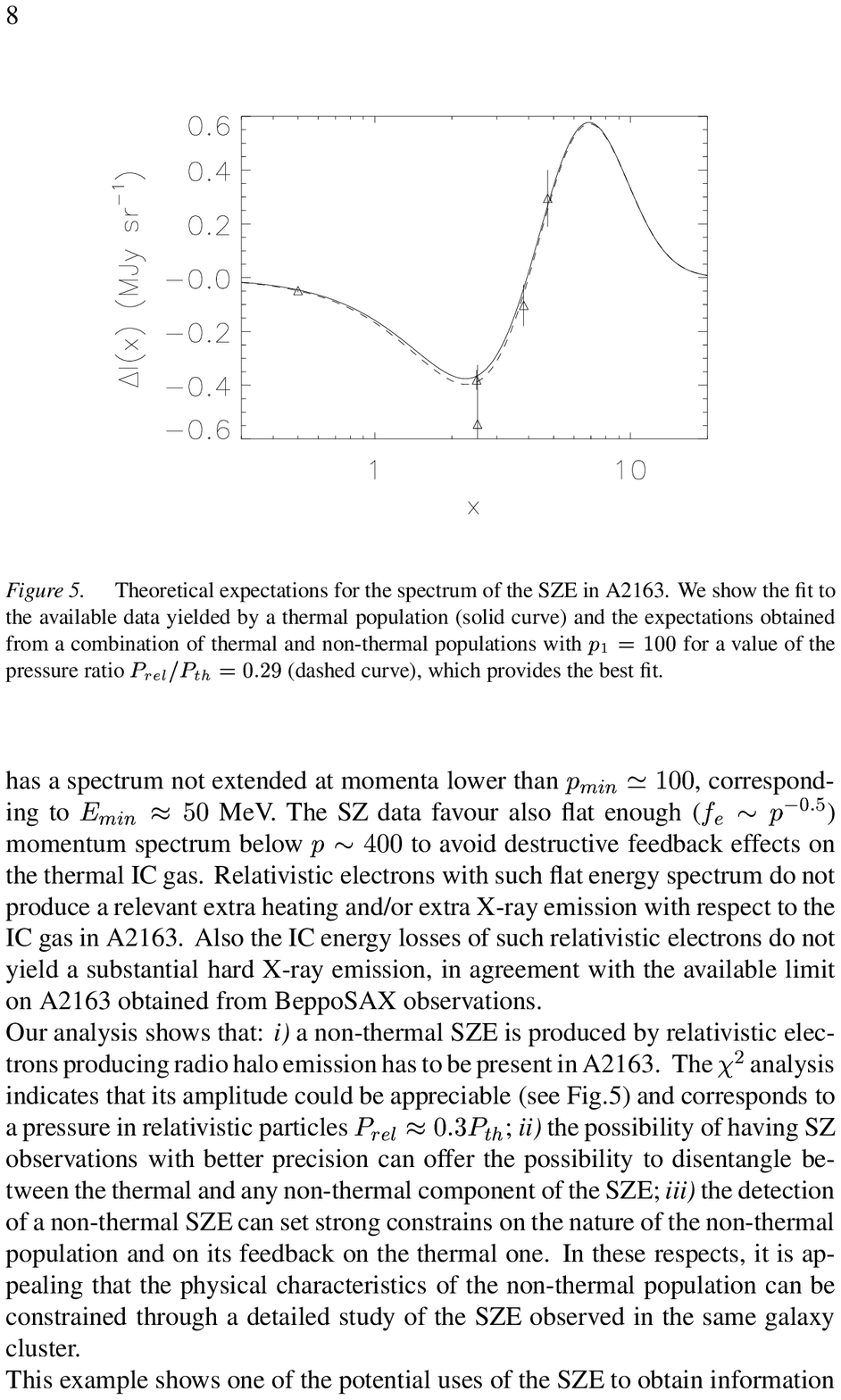}
\end{figure}
\begin{figure}[h]
\includegraphics[width=1.2\textwidth,viewport=130 120 540 690,clip]{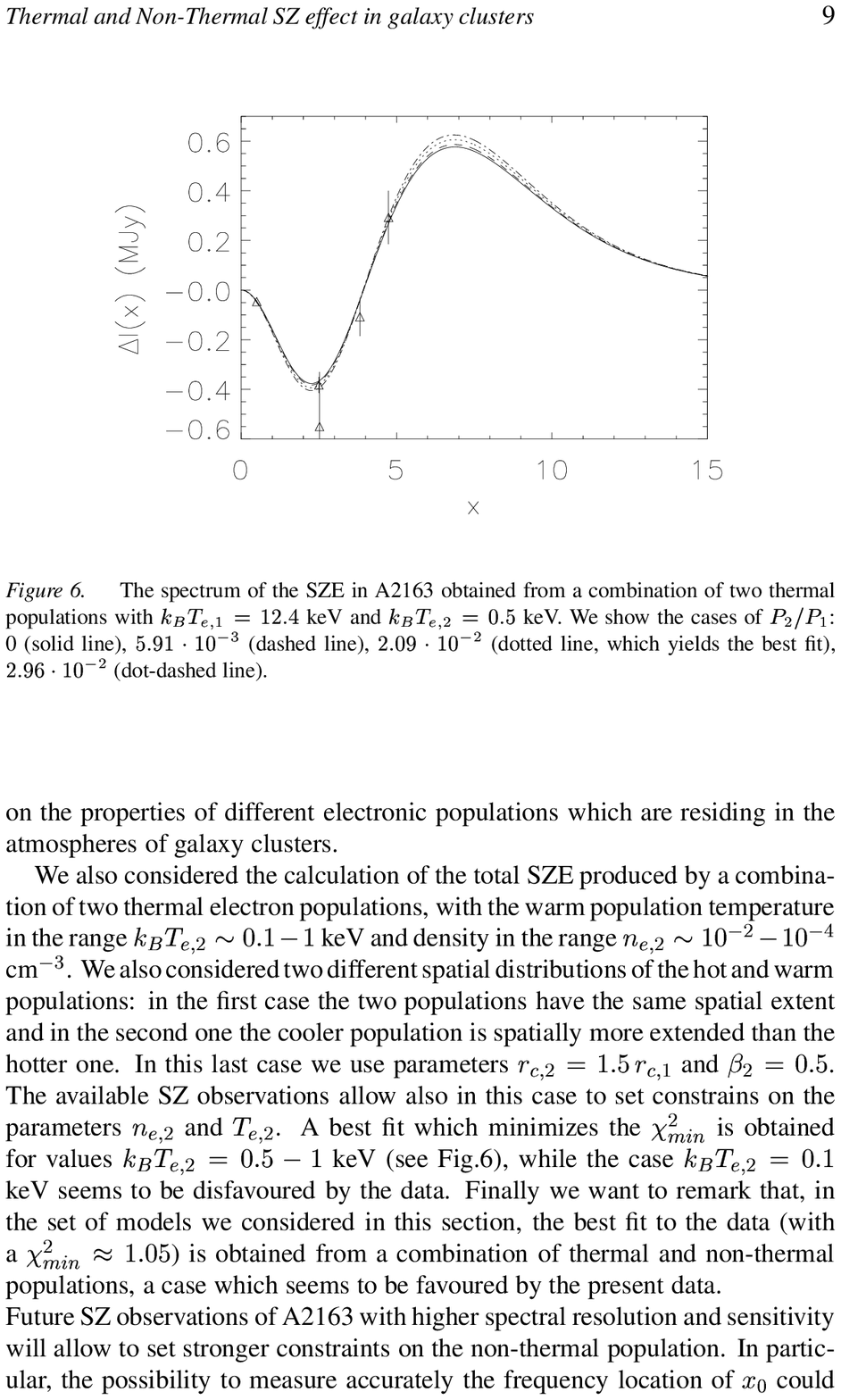}
\end{figure}
\begin{figure}[h]
\includegraphics[width=1.2\textwidth,viewport=130 120 540 690,clip]{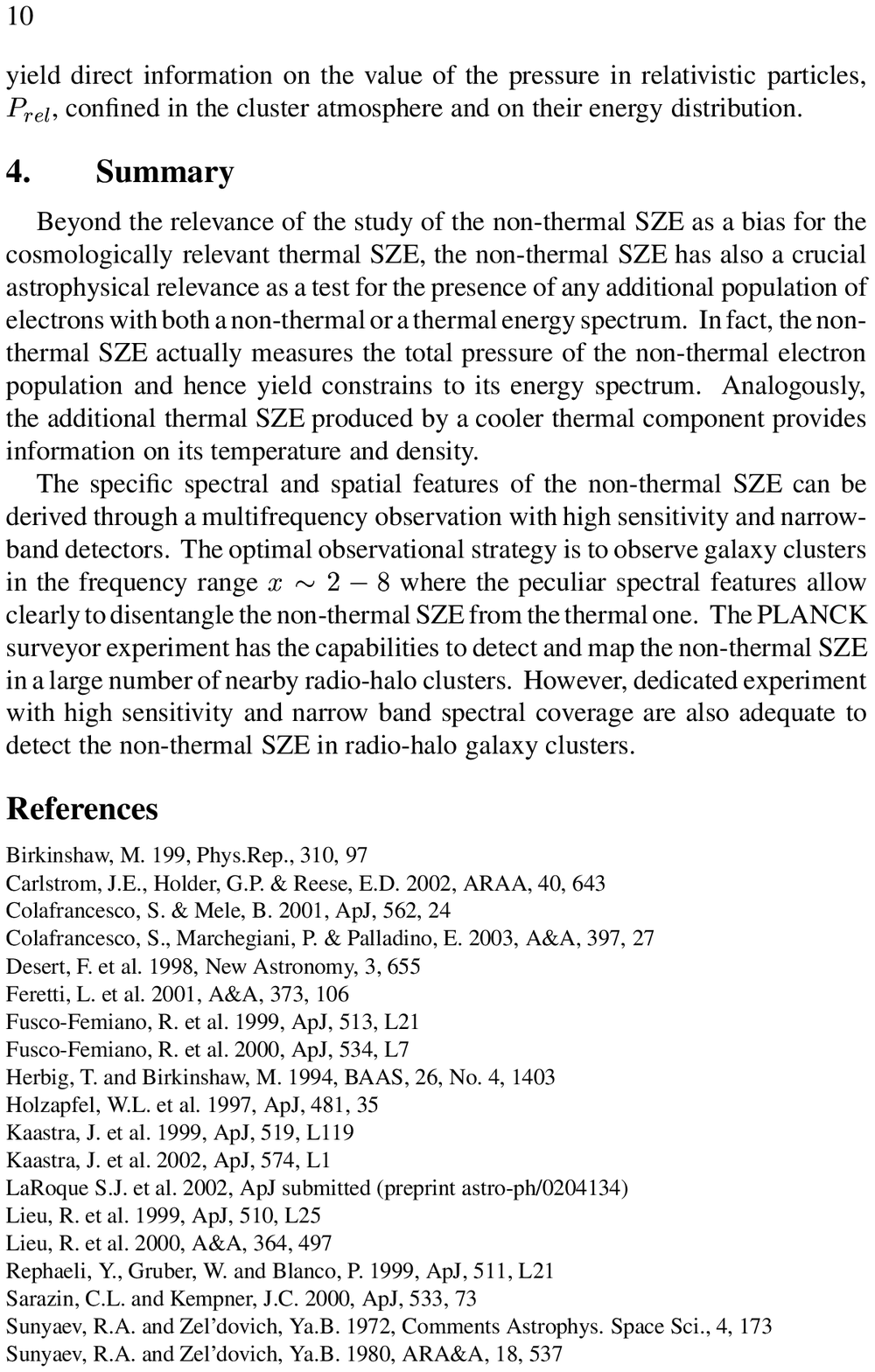}
\end{figure}

\articletitle{What The SZ Effect Can Tell Us About The Electron Population In The Coma Cluster?}
\author{S.Colafrancesco}

\begin{figure}[h]
\includegraphics[width=1.17\textwidth,viewport=130 150 540 548,clip]{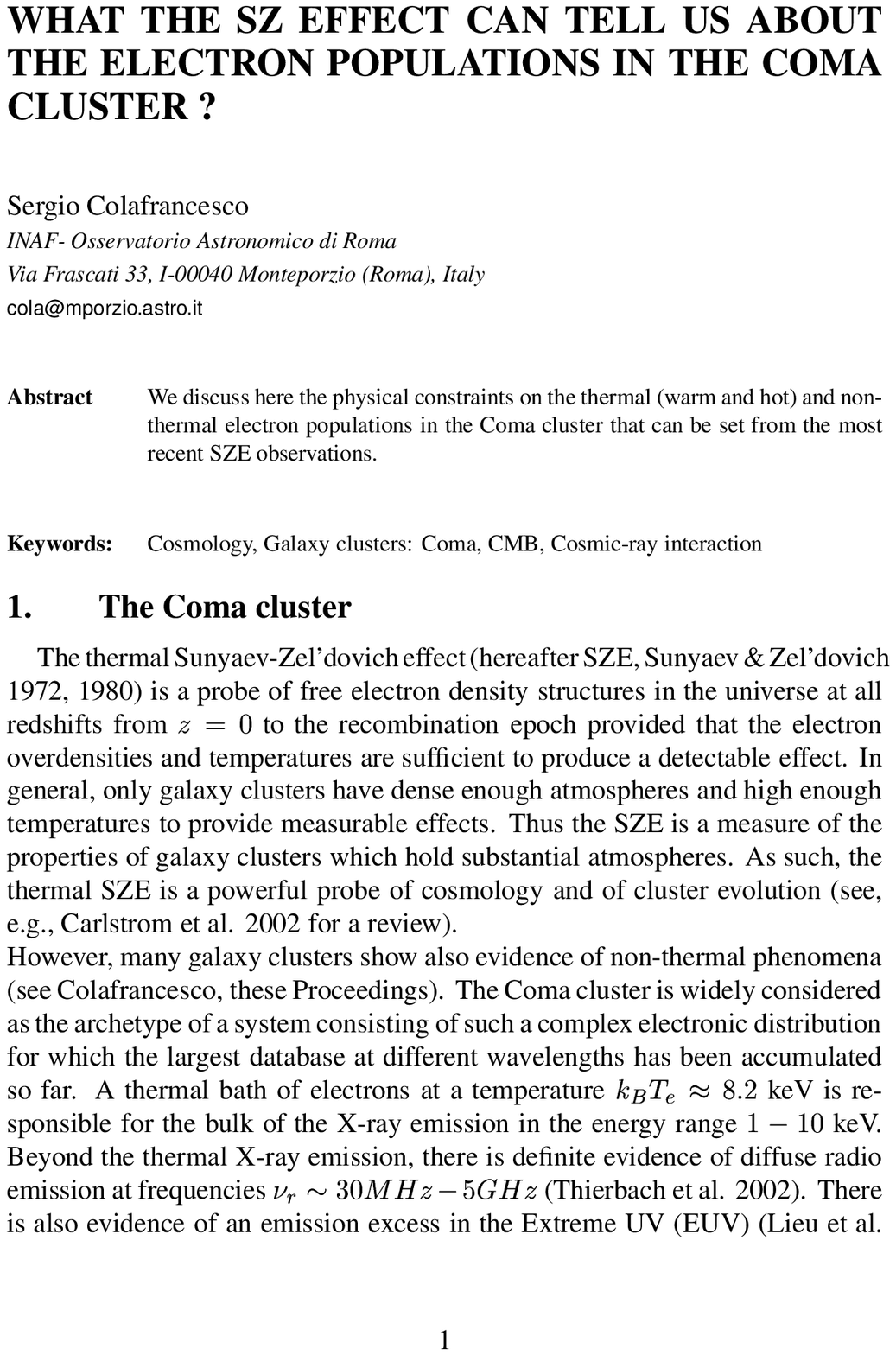}
\end{figure}
\begin{figure}[h]
\includegraphics[width=1.2\textwidth,viewport=130 120 540 690,clip]{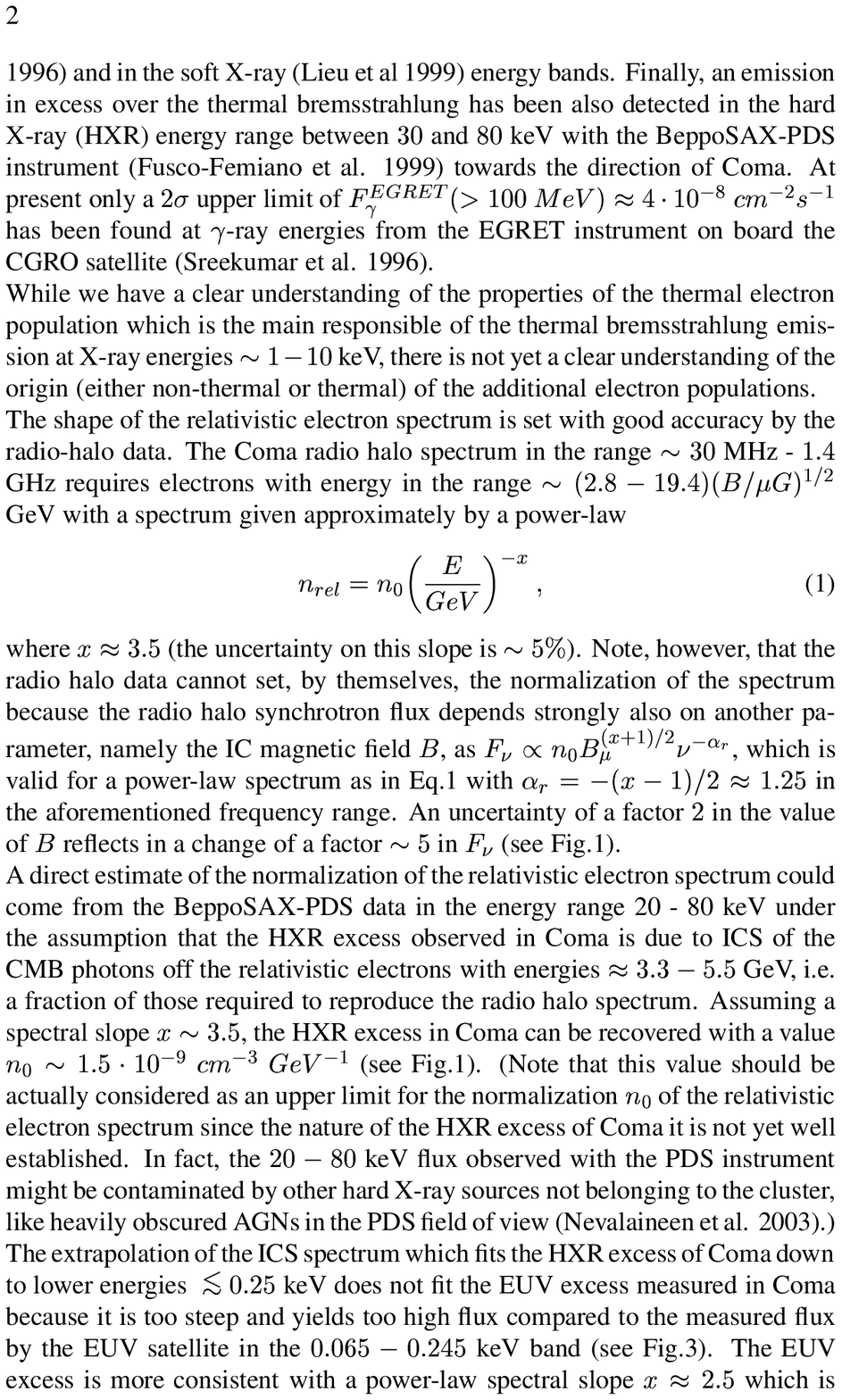}
\end{figure}
\begin{figure}[h]
\includegraphics[width=1.2\textwidth,viewport=130 120 540 690,clip]{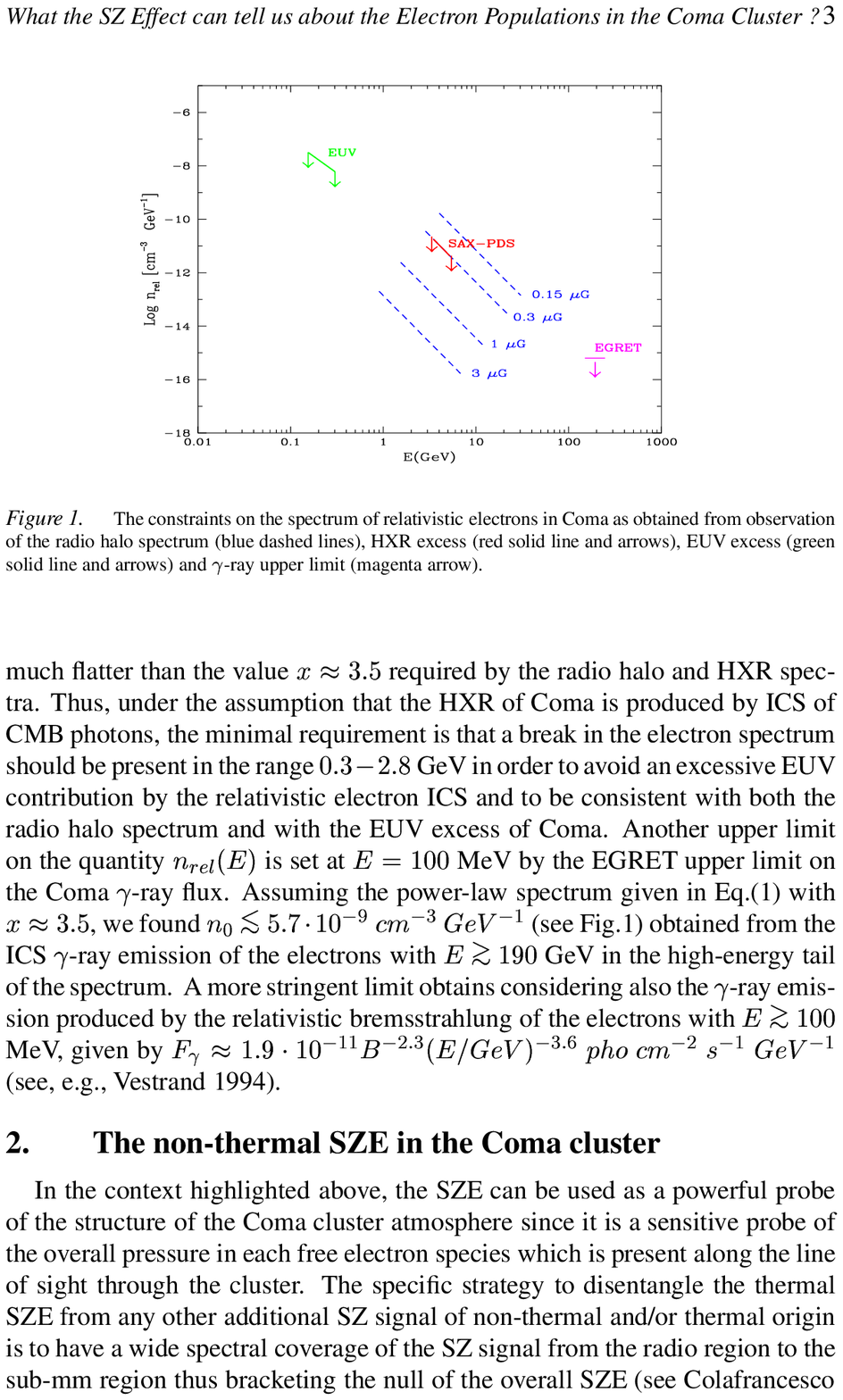}
\end{figure}
\begin{figure}[h]
\includegraphics[width=1.2\textwidth,viewport=130 120 540 690,clip]{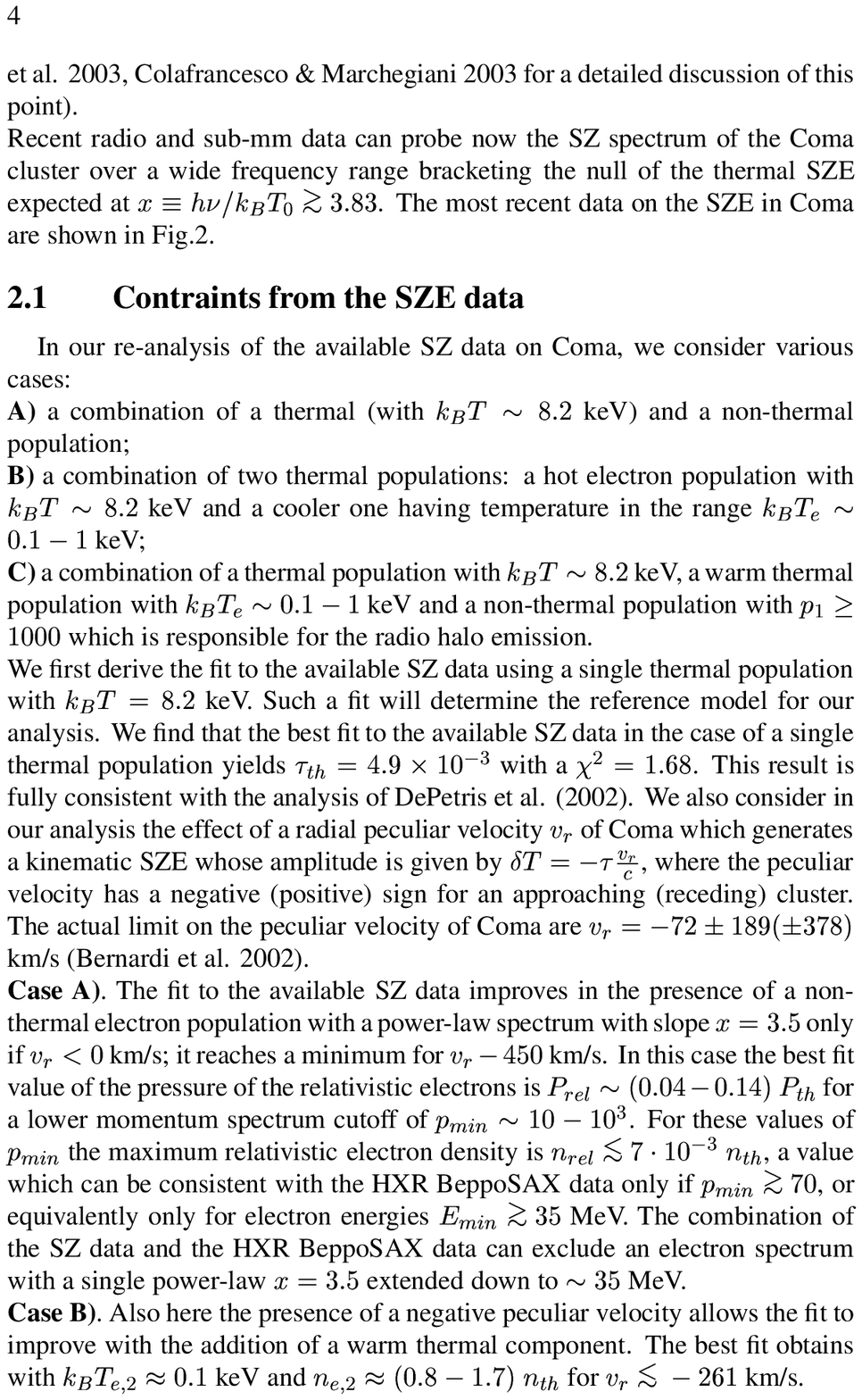}
\end{figure}
\begin{figure}[h]
\includegraphics[width=1.2\textwidth,viewport=130 120 540 690,clip]{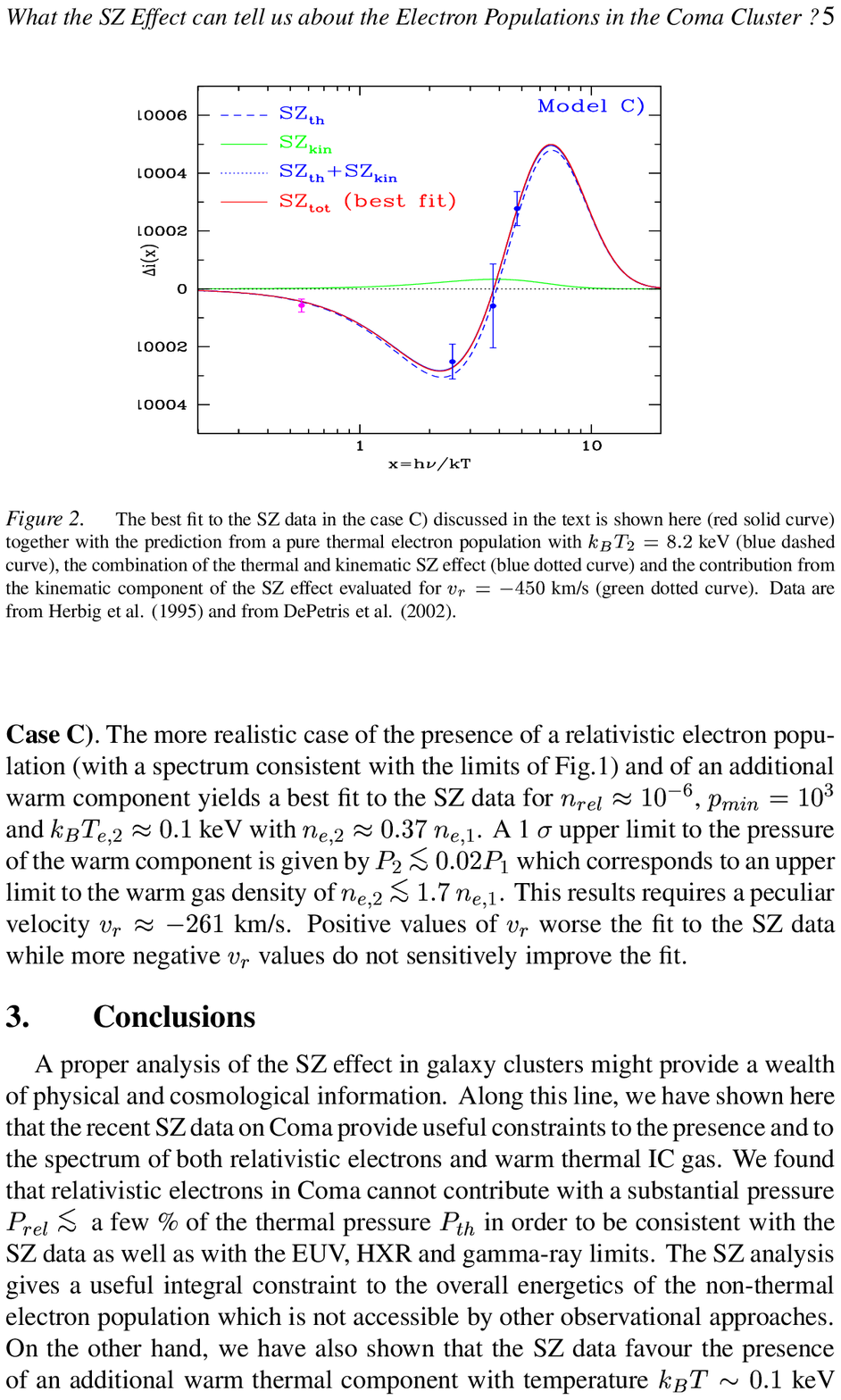}
\end{figure}
\begin{figure}[h]
\includegraphics[width=1.2\textwidth,viewport=130 120 540 690,clip]{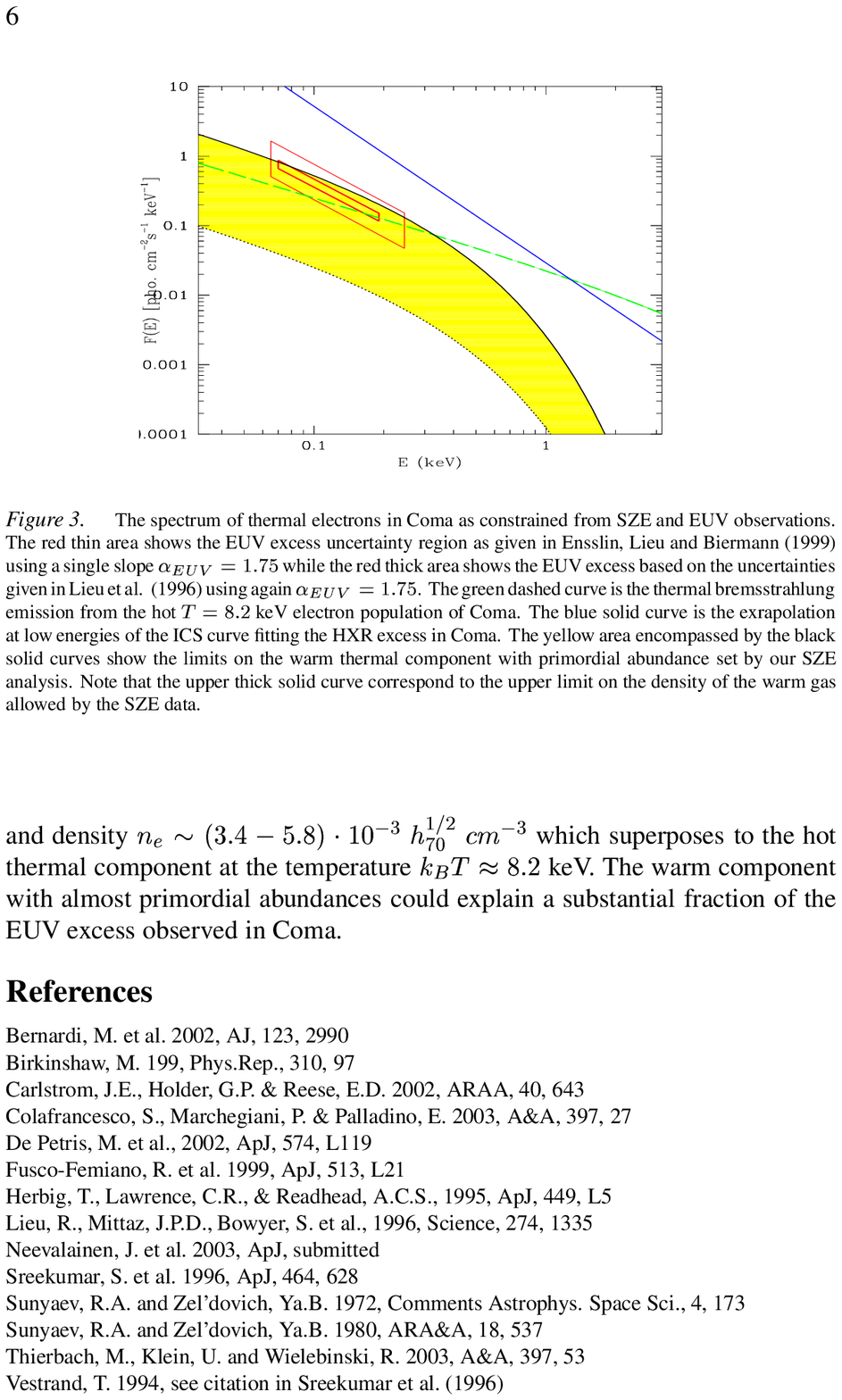}
\end{figure}

\part[Theoretical models of Clusters, the CSE and the WHIM]{
Theoretical models of Clusters, the CSE and the WHIM}


\articletitle{Observational constraints on models for the
cluster soft excess emission} 
\author{Richard Lieu \& Jonathan P.D. Mittaz} 
\affil{Department of Physics, UAH, Huntsville, AL35899}

\begin{abstract}
Although the cluster soft excess phenomenon is confirmed by XMM-Newton
observations of many clusters, the cause of this new component of radiation
remains an enigma.  The two mechanisms proposed in the late 90's, viz.  thermal
emission from a massive warm baryonic gas and inverse-Compton scattering
between cosmic rays and the microwave background, are still to date the only
viable interpretations of the soft excess.  In as much as cosmic rays cannot
exist at the vast void of a cluster's outskirts, warm gas also cannot be
present with any significant degree of abundance at the center of a cluster.  In
this sense, one could say that both models have their merits, and account for
the soft excess in different spatial regions.  There is however no clincher
evidence that points definitively to the correctness of either explanation.
Thus the door remains open for more exotic scenarios that could even consider
the detected emission as signature of some dark matter process.  In fact, the
absence of absorption lines in the spectrum of background quasars along
sightlines going through the outer radii of clusters argues against the thermal
model within the domain where it most suitably applies.  On the other hand, if
the central excesses are due to cosmic rays, the pressure of the proton
component will be large enough to choke a cooling flow, and missions like GLAST
may have the sensitivity to detect the gamma rays that ensue from proton-gas
interactions.  However the diagnosis may turn out to be, it is likely that the
new radiation represents something of cosmological importance.
\end{abstract}


\section{Introduction}\label{l-sec1}

Historically, the cluster soft excess phenomenon was discovered by the EUVE
mission from three clusters of galaxies: Virgo, Coma, and Abell 1795, when the
EUV flux was detected at a level higher than that expected from the low energy
spectral tail of the hot virialized intracluster medium (ICM) emission (Lieu et
al 1996a,b; Mittaz, Lieu, \& Lockman 1998).  In the original interpretation of
this excess, a warm intracluster gas component at $T \leq$ 10$^6$ K was
invoked, with a density radial profile at least as extended as, and a total
mass comparable to, that of the hot ICM.

Such a model was not received with equal warmth, principally because of the
high radiative cooling rate of the new gas.  In order for a warm component to
co-exist with the hot ICM, the former has to be clumped into clouds, so that
pressure equilibrium can be maintained between the two media, i.e.
\begin{equation}
P_{warm} = P_{hot}~{\rm requires}~n_w =(n_h/10^{-3}~{\rm cm}^{-3})
\left(\frac{T_h}{T_w} \right)\label{l-eq1}
\end{equation}
where $n_h \leq$ 10$^{-3}$ cm$^{-3}$ for the central and intermediate radii of
clusters.  Since the temperature ratio $T_h/T_w >$ 10, Eq. (1) implies $n_w >$
10$^{-2}$ cm$^{-3}$ we have, for these radii, a cooling time from free-free
emission of
\begin{equation}
\tau_w = 6 \times 10^8 \left(\frac{T}{10^6~{\rm K}}\right)^{\frac{1}{2}} \left(\frac{n_w}{10^{-2}~
{\rm cm}^{-3}}\right)^{-1}~~{\rm years}.\label{l-eq-cool}
\end{equation}
Hence the gas is highly unstable.  The difficulty may somewhat be alleviated by
invoking magnetic fields in a generalization of Eq. (\ref{l-eq1}):
\begin{equation}
P_w + \frac{B_w^2}{8 \pi} = P_h + \frac{B_h^2}{8 \pi},
\end{equation}
though it is still quite hard to envision field arrangements that can make a
great difference.  Another hurdle that confronts this type of scenarios is that
photo-ionization of the warm clouds by the hot ICM radiation, which occurs at
the very short timescale of:
\begin{equation}
\tau_{photo} \approx 2 \times 10^7 (F/10^4~{\rm ph}~{\rm cm}^{-2}~
{\rm s}^{-1})^{-1}~~{\rm yr}
\end{equation}
for O VII,
where $F$ is the hot ICM emitted flux.

All the above was 6-8 years ago.  Today, with the advent of XMM-Newton the
warm gas is now thought to exist at large radii, accounting for the extended
soft excess radial profile - we shall return to this point.  Historically,
however, affairs twisted and turned like a mountain road - an altogether
distinct emission mechanism was considered instead, which also turned out to
have a high probability of being relevant to the soft excess.  Several authors
independently and contemporaneously proposed the non-thermal model (Hwang 1997,
Ensslin \& Biermann 1998, Sarazin \& Lieu 1998), in which the soft X-ray and
EUV photons are produced by inverse-Compton scattering between a population of
relativistic electrons and the cosmic microwave background.  The basic tenet of
the theory is that electrons capable of doing this have a Lorentz factor
\begin{equation}
\gamma \sim {\rm 300}~(h \nu/75 eV)^{\frac{1}{2}},
\end{equation}
and their main loss mechanism electrons is indeed through this same
inverse-Compton process, which restricts the electron lifetime to:
\begin{equation}
\tau_{IC} \sim 7.7 \times 10^9 (\gamma/300)^{-1}~~{\rm years}
\end{equation}
Thus even if the electrons belong to a relic population that dates back to the
era of cluster formation when there were considerably more AGN activity and
shock acceleration, some of them survive to the present time, and can in
principle account for the observed emission.  In reality, however, the
brightness level of the soft excess necessitates a large energy density of
cosmic rays, which may continuously be produced in the centers of clusters.

\section{Thermal origin of the soft excess - the missing baryons}

From theoretical considerations (Cen \& Ostriker 1999) one expects the majority
of the baryons at the present epoch of the Universe's evolution to be `hidden'
in the form of a Warm-Hot-Intergalactic-Medium (WHIM) gas.  Crudely speaking
the argument goes as follows.  If $\lambda$ is the characteristic wavelength of
the bulk motion of the intergalactic medium at a given epoch, then when these
waves collide and break the thermal velocity of the shocked gas will typically
be $v \sim H_o \lambda$ where $H_o$ is the Hubble constant, i.e.
\begin{equation}
v \sim 100 (H_o/70) (\lambda/1.5~{\rm Mpc})~{\rm km}~{\rm s}^{-1}\label{l-eq2}
\end{equation}
where we set $\lambda$ at $\sim$ the present size (virial radius) of galaxy
clusters, i.e. $\lambda \sim$ 1-3 Mpc.  The value of $v$ as given in
Eq. (\ref{l-eq2}) corresponds to a gas temperature of 10$^{5-7}$ K.

The assumption of a WHIM origin alleviates several of the aforementioned
problems with a thermal model, notably the pressure problems and the cooling
time.  But is there observational evidence both for the WHIM and its
association with the cluster soft excess (CSE)?

\subsection{Characteristics of the soft excess as thermal emission}

If we assume that the CSE is indeed thermal and from the WHIM, what are the
constraints, if any, that we can make on the parameters for the warm baryonic
gas that resides in the vast domain of intergalactic space outside the virial
radius of clusters?  For example, it is deceptive to think that an arbitrarily
low density can be envisaged without eventually contradicting certain
observational facts.

\begin{figure}
\centerline{\includegraphics[height=3.0in,angle=-90]{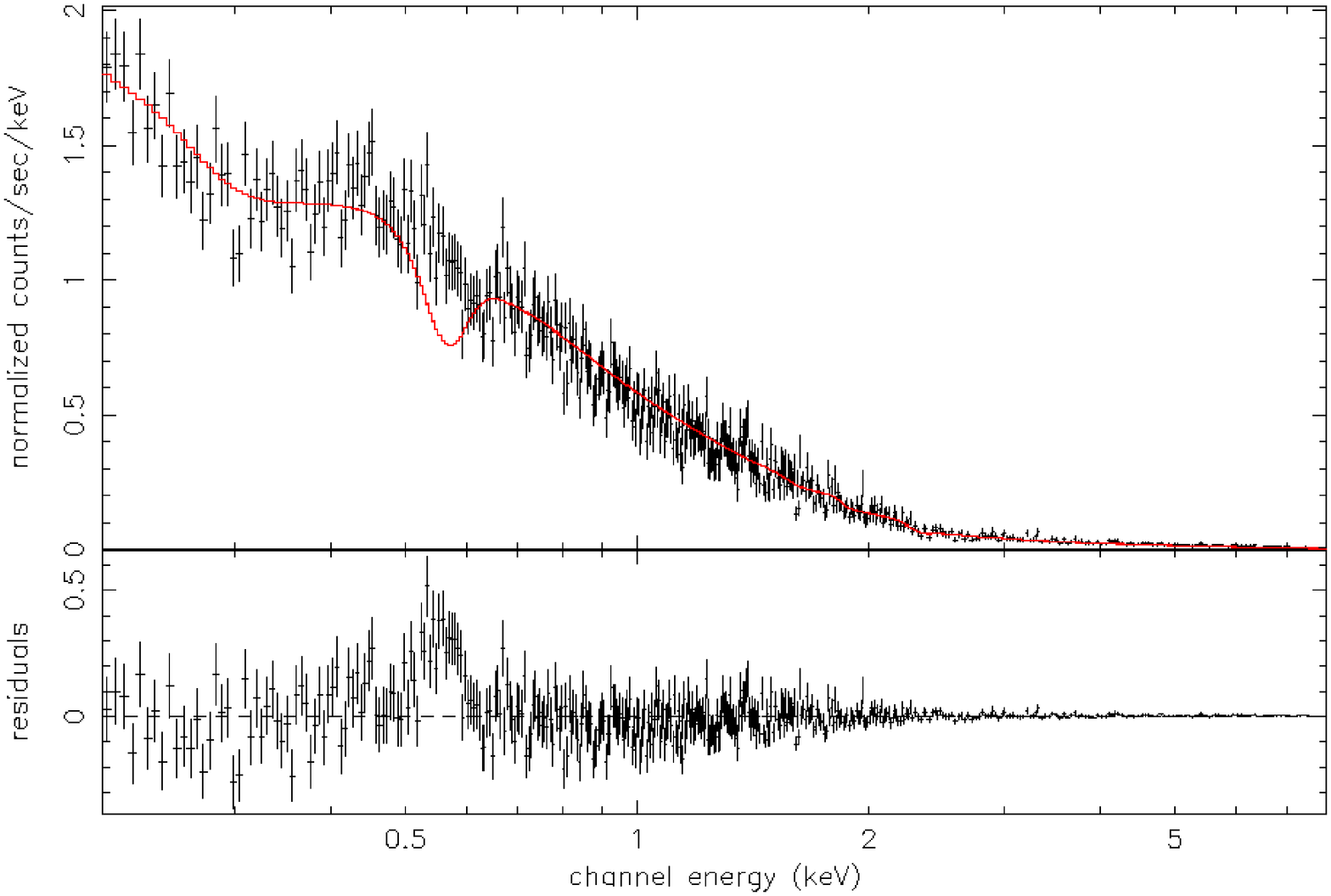}}
\caption{The EPIC PN spectrum of X-Comae together with the best fit model (a
  power-law plus Galactic absorption) and a gaussian absorption line with an
  equivalent width of 28eV.  This line is rejected at the 99\% confidence level
  by an F-test demonstrating that the strong absorption expected from a WHIM
  like model for the CSE is not seen.  }\label{l-fig1}
\end{figure}

To pursue this point further, we take as specific example the Coma cluster,
which was seen by the ROSAT/PSPC to possess a giant soft emission halo
extending to $\geq$ 1 Mpc (Bonemente et al. 2003).  There is no plausible
explanation of such a phenomenon other than the thermal model with a warm and
massive baryonic component.  The observed brightness of the soft excess may be
expressed as the projected emission integral (EI) of the warm gas column, i.e.
\begin{equation}
{\rm EI} = n^2 LA,
\end{equation}
where $n$ is the HI number density, assumed to be uniform throughout an
optically thin emission column of length $L$ and cross sectional area $A$.  For
the 0-20 arcmin central radius of Coma, $EI \sim$ 10$^{68}$ cm$^{-3}$ from the
published ROSAT data (Bonamente et al 2003), leading to:
\begin{equation}
L \approx 3(n/10^{-3} cm^{-3})^{-2}~{\rm Mpc}\label{l-length}
\end{equation}
and an equivalent HI column density $N_H = nL$ of
\begin{equation}
N_H \approx 10^{22} (n/10^{-3} cm^{-3})^{-1}~{\rm cm}^{-2}\label{l-nh}
\end{equation}
A warm column with such a value of $N_H$ already poses significant opacities to
certain EUV and soft X-ray lines that may originate from underlying emission
layers, though its continuum opacity remain small. {\it If the density $n$
assumes values smaller than $\sim 10^{-3} cm^{-3}$}, the resulting column will
be too high for consistency with observations of bright QSOs behind Coma (see
Figure~\ref{l-fig1}).

The second constraint comes from the total mass budget of the warm gas.  To
make the point, let us consider the simple picture of a warm halo as having the
uniform density $n$ and radius $L$ as above.  Then the total mass $M$ is given
by
\begin{equation}
M = \frac{4}{3} \pi L^3 n m_p f
\end{equation}
where $m_p$ is the proton mass and $f$ is the volume filling factor of the gas,
i.e. fraction of the spherical volume occupied by the warm filaments, which
might be in the form of `spaghetti' converging onto a node - the cluster.  If
there exists an inner radius $r_o$ at which the filamentary footpoints cover
the entire cluster surface, and if the filaments have constant cross sectional
area, then, by geometry, $f = (r_o/L)^2$, and
\begin{equation}
M = 10^{14} (n/10^{-3} cm^{-3})^{-1} (r_o/0.5~{\rm Mpc})^2~~M_{\odot}\label{l-mass}
\end{equation}
where in arriving at Eq. (\ref{l-mass}) use was made of the observed soft excess
emission brightness, Eq. (\ref{l-length}).  Once again, it can be seen that if
the density is too low, or the warm gas located too far out ($r_o$ large to
avoid contact with the hot ICM), the resulting mass budget will exceed even
that of the dark matter limits placed by gravitational lensing and galaxy
velocity dispersion measurements.

The final constraint comes from the cooling time of the gas.  As shown by
Eq. (\ref{l-eq-cool}) if the density becomes too high then the gas will
catastrophically cool.  For Coma we have a gas at a temperature of 0.2 keV and
an abundance of 0.1 (Bonamente et al 2003) setting a limit on the density of
the gas $n < 6,5 \times 10^{-4}$ cm$^{-3}$ for the gas to remain stable in a
Hubble time.  However, at these temperatures the cooling time is also strongly
affected by line emission.  Using the emissivities taken from Sutherland \&
Dopita (1993) we then derive a revised cooling time of
\begin{equation}
\tau_w = 1.18 \times 10^9
\left(\frac{n}{10^{-3}cm^{_3}}\right)^{-1} years
\end{equation}
This sets a tighter limit on the density of $n < 8.64\times 10^{-5}$cm$^{-3}$
and hence from Eq. (\ref{l-nh}) a limit on the column density for the CSE of
$>10^{23}$cm$^{-2}$.  Then using Eq. (8) of Nicastro et al. (1999) and the
gas parameters for Coma we can derive a limit on the equivalent width of the
OVII\ $21.6$\AA\ absorption line of $EW > 160$eV.

Such a large equivalent width should be observable.  In Figure~\ref{l-fig1} is
shown an EPIC PN spectrum of X-Comae, a quasar located 28 arcminutes from the
Coma cluster center and at a redshift of 0.091, well behind the Coma cluster.
The plot also shows the best fitting model (a power-law with Galactic
absorption) together with an absorption line at the energy of OVII\ $21.6$\AA\
with an equivalent width of 28eV.  As is apparent from the spectrum, a line
with this equivalent width is not consistent with the data.  This lack of
absorption sets a limit on the observed column density of gas to a factor of at
least 5 lower than the column needed to stabilize the gas for the duration of
a Hubble time.  If we reduce the cooling time to match up the two constraints
we then find that the warm CSE emitting component could only have been formed
$2.7$Gyr ago, or at a redshift of 0.23.  This is clearly in contradiction with
models for the formation of the WHIM (e.g. Dave et al 2001).

\subsection{Characteristics of the WHIM from UV absorption lines and its
  relationship to the CSE}\label{l-uv}

The above discussion shows that, in fact, strong constraints can be placed on
the thermal/WHIM interpretation of the CSE.  It is then reasonable to ask what
is the evidence of a WHIM at all?  There are plenty of examples of absorption
lines in the UV from the intergalactic medium in the literature, most notably
those from the Ly-$\alpha$ forest.  However, in general the lines being studied
(Ly-$\alpha$, Mg II, CIV for example) probe clouds of relatively low
temperature when compared to those of the WHIM.  With the launch of FUSE this
has changed because FUSE is able to look for absorption from e.g.  the OVI\
$1031.9,1037.6$\AA\ doublet which is sensitive to material with a temperature
typical of the WHIM ($10^5 - 10^7$K).  Current observations show that there is
indeed material in intergalactic space at a density and temperature consistent
with the WHIM, although most detections are from material within our local
group (e.g. Nicastro et al. 2003 and references).  There are also several
reported detections of OVI absorption at intermediate redshifts (Tripp \&
Savage 2000, Tripp et al. 2000).  These observations allow us to conclude that
material showing all the characteristics of the WHIM does exist.

\begin{figure}[h]
\centerline{\includegraphics[height=2in]{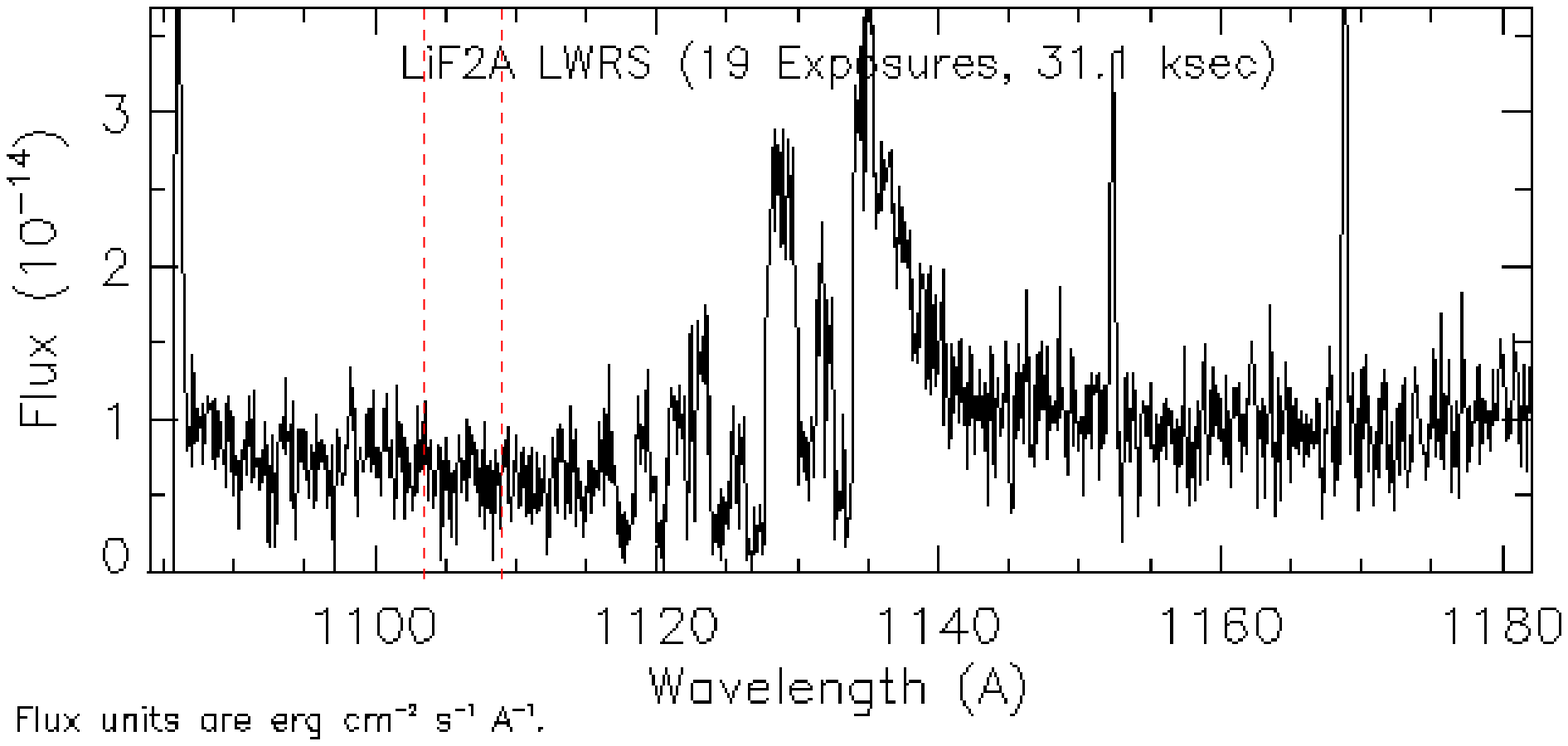}}
\caption{FUSE spectrum of IR07546+3928 which is 6.3 arcminutes away from Abell
  607.  The dashed lines show the expected position of the
  OVI$1031.9,1037.6$\AA\ absorption lines (spectrum taken from the data preview
  for the FUSE mission)}\label{l-fig-fuse}
\end{figure}

However, in terms of WHIM related to the CSE i.e. WHIM like material close to
cluster of galaxies, the situation is less than clear.  In general when
absorption is seen at the redshift of a cluster the inferred column density of
material is much lower than that required to explain the CSE.  For example, a
absorption system seen in the spectrum of 3C273 at a redshift of $z_{abs} =
0.00337$ (close to the redshift of the Virgo cluster, $z=0.0036$) but located
$> 3$Mpc from the cluster center show material with a temperature $T >
10^{5.29}$K and has a column density of $log_{10}(OVI) \sim 13.32$ (Tripp et
al. 2002).  However, the typical OVI column density inferred from the emission
properties of the CSE indicate that $log_{10}(OVI) = 16 - 18$ (e.g. Bonamente
\& van Dixon 2004).  Other searches for absorption from nearby clusters have
also met with difficulties.  In another case a possible OVI absorption line
seen at the redshift of Abell 3782 gives a column density associated with the
cluster of $log_{10} (OVI) = 13.95$, which is again much smaller (factors of
$10^2 - 10^4$) than the columns derived from the emission properties of the CSE
(Bonamente \& van Dixon 2004).  Other AGNs behind clusters show no evidence for
OVI absorption in their spectra (see for example Figure~\ref{l-fig-fuse}).  It
would thus appear that as in the the case of X-ray absorption, UV absorption
line studies in the vicinity of clusters are failing to find sufficient
material to account for the CSE.  It should be noted, however, that the number
of available observations of QSOs behind clusters is still very small, and the
temperature range accessible by these UV observations ($10^5 - 10^6$K) is lower
than that observed for the CSE (typically kT = 0.2 keV e.g. Kaastra et
al. 2003).

\section{Non-thermal processes}
 
By using the emission characteristics of the observed CSE we have shown that
stringent limits can be placed on both thermal and non-thermal models for the
CSE.  In the case of the former, cooling time and pressure balance arguments
imply that the only viable option is to associate the CSE in a WHIM like
filament external to the cluster environment.  While such an undertaking may
alleviate many problems, the allowed range of column density for the warm gas
means strong absorption lines are expected to be visible in the spectra of
background lighthouses, which have not been seen in either the UV or soft X-ray
regime.  This contradiction is a significant embarrassment for the thermal
interpretation of the CSE.
 
Turning now to non-thermal considerations, it is looking increasingly certain
that this may be the only option for the center of clusters where the extreme
brightness of the CSE necessitates an unacceptably large quantity of warm gas
if the phenomenon is to be interpreted thermally.  Also, in terms of a $\chi^2$
goodness-of-fit to the XMM spectra (which is controlled by the continuum shape
as the O VII line is too weak) there is as yet no compelling reason to prefer a
thermal model.  In fact, often for these central cluster regions the best-fit
power laws are found to secure lower $\chi^2$ values.

At the time of writing, the authors found that if non-thermal processes are
responsible for the central soft excesses of A1795 and AS1101, which are bright
enough to shine above the peaked emission of the `cooling flow' gas in the hot
ICM, the cosmic ray pressure will assume non-trivial values.  A scenario may
therefore be envisaged whereby the protons, which carry the bulk of the
pressure, are in equipartition with the hot ICM.  While these protons have the
prerequisites to choke a cooling flow, they also produce secondary electrons
(via pions) during interactions with the hot ICM gas.  It is these electrons
which undergo inverse Compton scattering with the microwave background photons,
resulting in excess EUV and soft X-ray emissions.
 
\section{Conclusion and Future prospects}
 
We conclude that while the new observations by the XMM/Newton mission have
convincingly demonstrated the existence of the CSE, the emission mechanism
remains an unanswered question.  The current picture attributes the phenomenon
to cosmic rays at the center and thermal warm gas at the outer parts, but there
are difficulties.
 
In the future, it is anticipated that the Astro-E2 and Constellation-X missions
will deliver clincher evidence, by means of their superior spectral resolution,
for or against the thermal origin of the soft excess, as the issue over the
reality of the O VII lines will be resolved after their launches.  Moreover, if
cluster cosmic rays are present in abundance, they will produce gamma rays with
the hot ICM - radiation which the GLAST mission is sensitive to.  Observations
of clusters via the Sunyaev-Zeldovich effect may also lead to a spectral
distinction between thermal and non-thermal processes, as both carry
significant SZ implications (it is sometimes argued that warm gas columns may
not contribute much to the SZ integral $\int nkT dl$ because of the lower $T$,
yet as we saw above this could be compensated by the larger column length $L$).

\def \xray {\hbox{X--ray} }
\def \hMpc      {h^{-1}{\rm\ Mpc}}
\def \kms       {\hbox{ km s$^{-1}$}}
\def\lesssim{\mathrel{\hbox{\rlap{\hbox{\lower4pt\hbox{$\sim$}}}\hbox{$<$}}}}
\def\gtrsim{\mathrel{\hbox{\rlap{\hbox{\lower4pt\hbox{$\sim$}}}\hbox{$>$}}}}
\def\apj{ApJ}
\def\aj{AJ}
\def\apjlett{ApJ}
\def\apjs{ApJS}
\def\aa{A\&A}
\def\mnras{MNRAS}
\def\araa{ARA\&A}
\def\nature{Nature}
\def\pasj{PASJ}
\def\nat{Nature}
\def\apjl{ApJ}
\def\aap{A\&A}



\articletitle{High-Resolution Simulations of Clusters of Galaxies}
\author{Daisuke Nagai\altaffilmark{1,2}, Andrey V. Kravtsov\altaffilmark{1,2}}

\altaffiltext{1}{Center for Cosmological Physics,
University of Chicago, Chicago  IL 60637}
\altaffiltext{2}{Department of Astronomy and Astrophysics,
University of Chicago, 5640 S Ellis Ave, Chicago  IL 60637}

\begin{abstract}
Recently, high-resolution {\sl Chandra} observations revealed many
complex processes operating in the intracluster medium such as cold
fronts, cooling flows with puzzling properties, and subsonic turbulent
motions.  The detailed studies of these processes can provide new
insights into the cluster gas properties and cluster evolution.  Using
very high-resolution Adaptive Mesh Refinement (AMR) gasdynamics
simulations of clusters forming in the CDM universes, we investigate
structural evolution of the intracluster gas and the role of dynamical
(e.g., mergers) and physical (e.g., cooling, star formation, stellar
feedback) processes operating during cluster evolution.  In
particular, we use the simulations to investigate the structure and
dynamical properties of cold fronts.  We also discuss the internal
turbulent motions of cluster gas induced by mergers.  Finally, we
present preliminary results from an ongoing program to carry out a
series of simulations that include cooling, star formation, stellar
feedback, and metal enrichment and discuss implications for the
cluster observables.
\end{abstract}

\begin{keywords}
cosmology: theory -- intergalactic medium -- methods: numerical -- galaxies: clusters: general -- instabilities--turbulence--X-rays: galaxies: clusters
\end{keywords}


\section{Introduction}

Cosmological $N$-body$+$gasdynamics simulations are a powerful tool for
modeling cluster formation and studying processes that determine
observable properties of intracluster medium (ICM). The simulations
start from theoretically-motivated initial conditions and follow the
dynamics of gravitationally dominant dark matter and gasdynamics of
baryons. Such {\em ab initio} simulations can capture the full
complexity of matter dynamics during the hierarchical build-up of
structures and, therefore, allow us to systematically study the
effects and the relative importance of various processes operating
during cluster evolution in a realistic cosmological setting.  

We present analysis of a suite of simulations of cluster formation 
using the newly developed Adaptive Refinement
Tree (ART) $N$-body$+$gasdynamics code \cite{kravtsov99,kravtsov02} in
the currently-favored flat $\Lambda$CDM cosmology ($\Omega_0=0.3$, $h=0.7$,
$\sigma_8=0.9$).  The code uses a combination of particle-mesh and
shock-capturing Eulerian methods for simulating the evolution of DM
and gas, respectively. High dynamic range is achieved by applying
adaptive mesh refinement to both gasdynamics and gravity calculations.
The peak spatial resolution in the cores of clusters will reach $\sim
1-5h^{-1}$~kpc and clusters will have $\gtrsim 10^6$ particles within
the virial radius.  This is more than an order of magnitude
improvement over previous studies. The simulations thus bridge
the gap between the superb resolution of current X-ray observations
and the spatial resolution of numerical simulations and, hence, enable
extensive comparisons of numerical simulations with the
high-resolution {\sl Chandra} observations.  Such comparisons will
allow us to understand underlying processes responsible for observed
properties of ICM and help facilitate the interpretation of
observations. Below, we illustrate this using three specific examples.
The details of the studies can be found in Nagai \& Kravtsov (2003)
and Nagai, Kravtsov \& Kosowsky (2003). 

\section{Cold Fronts in CDM cluster}

Recent discoveries of {\em ``cold fronts''} by {\sl Chandra}
observations have come as a surprise
\cite{markevitch00,vikhlinin01a,sun02,forman02,markevitch02}. It 
was quickly realized that the existence of such sharp features 
puts interesting constraints on the small-scale properties of the
intracluster medium (ICM), including the efficiency of energy
transport \cite{ettori00} and the magnetic field strength in the ICM
\cite{vikhlinin01b}.  However, placing meaningful constraints on the
physics of the ICM requires detailed understanding of gas dynamics in
the vicinity of cold fronts, since the power of these constraints
relies on the validity of the underlying assumptions of the dynamical
model.  It is therefore interesting to search for counterparts of the
observed {\em ``cold fronts''} in simulated clusters forming in
hierarchical models and study their structure and dynamical
properties.

\begin{figure}[t]
  \centerline{ \includegraphics[height=4.8truein]{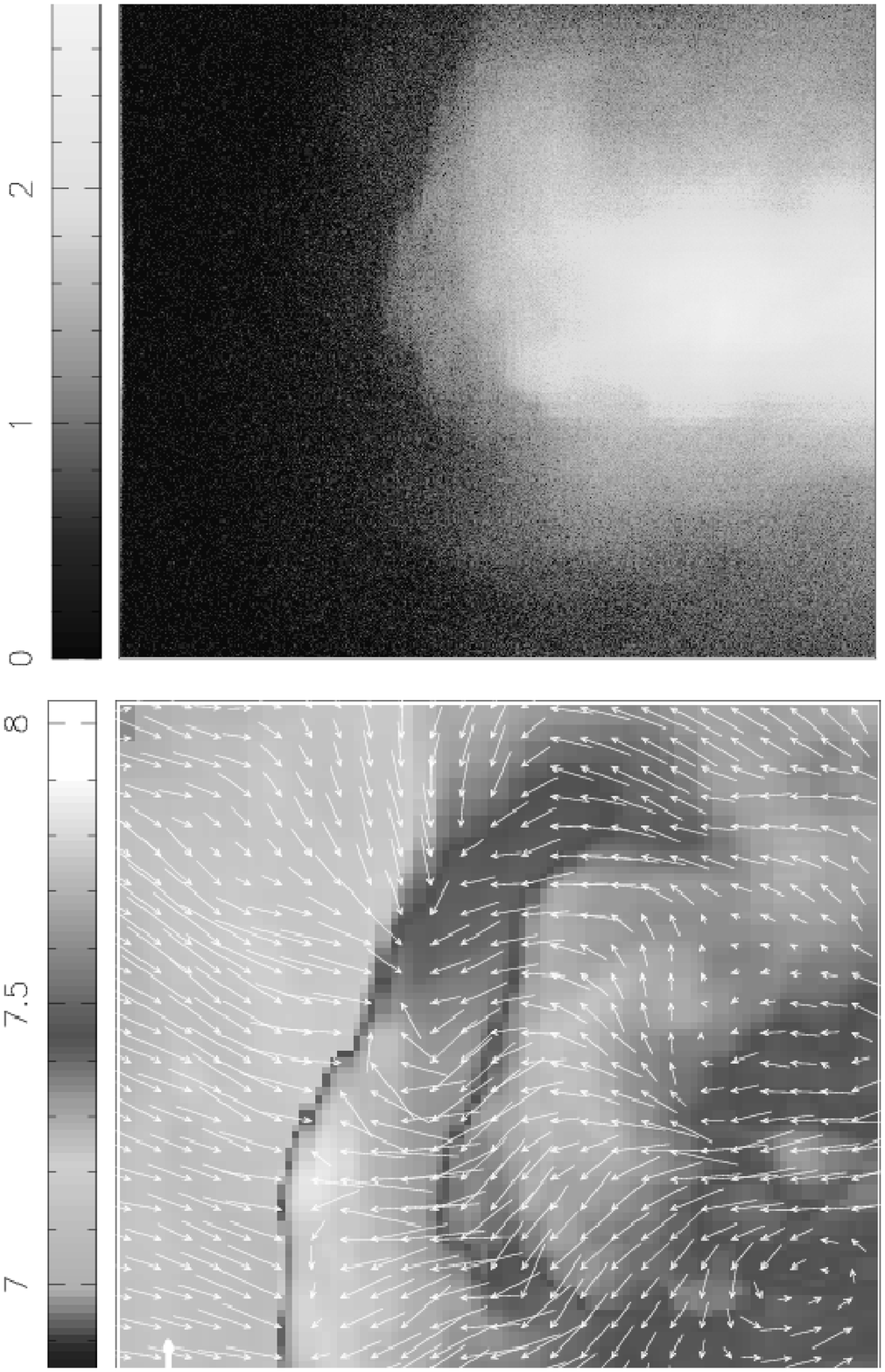}
  \includegraphics[height=4.8truein]{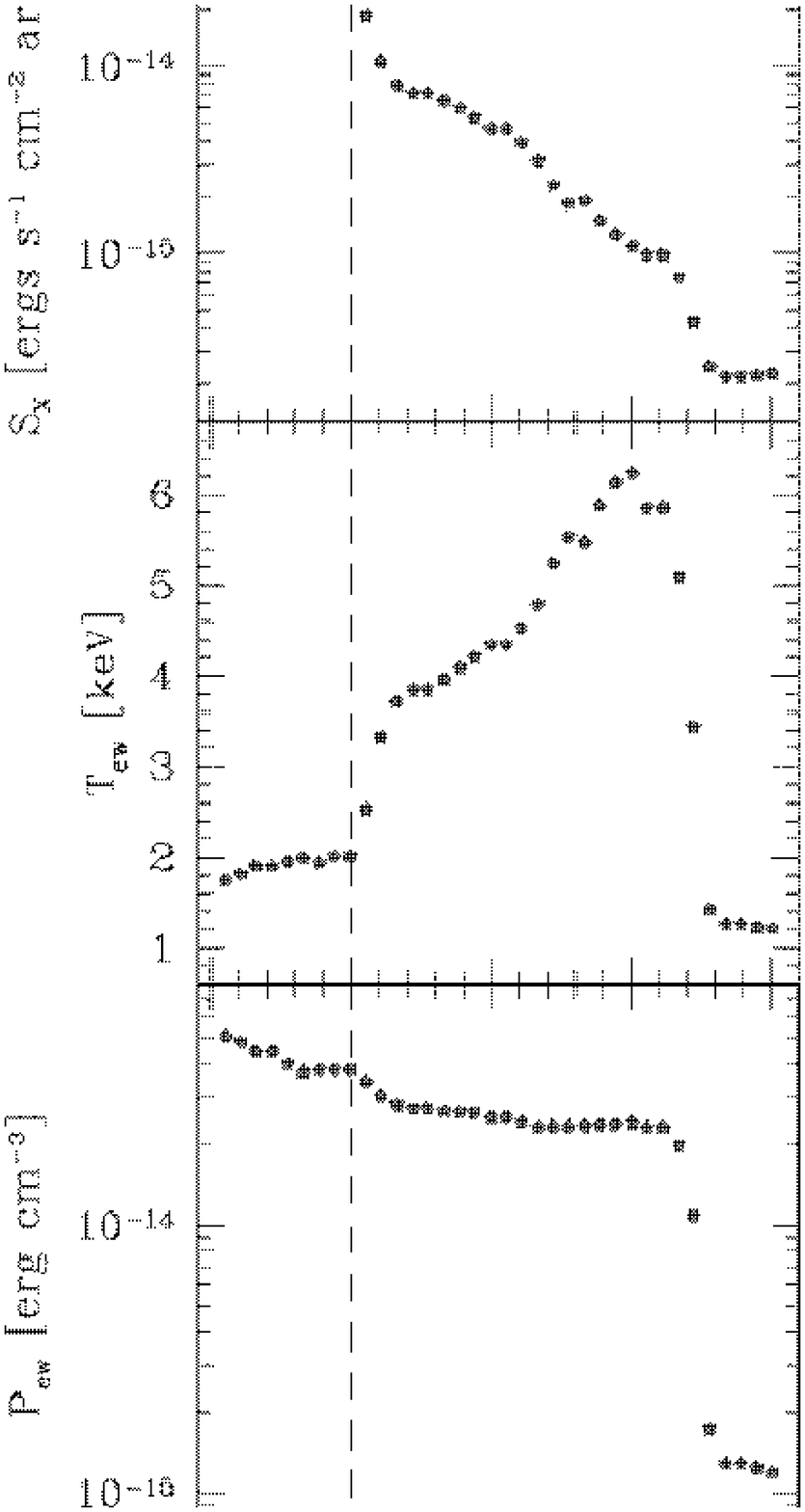} }
\caption{ 
  Cold front in the simulated $\Lambda$CDM cluster identified in a
  major merger at z=0.43. The top-left panel shows the mock
  $16^\prime\times 16^\prime$ ACIS-I {\sl Chandra} photon count image
  of the cold front. The image was constructed assuming 50~ksec
  exposure in the $0.5-4$ keV band, cluster redshift of $z=0.05$, and
  a flat background level of
  $4.3\times10^{-6}$~cnts~s$^{-1}$~pixel$^{-1}$. The bottom-left panel
  shows the velocity map overlaid on the emission-weighted temperature
  map of the same region.  The length of the thick vertical vector in
  the bottom left corner corresponds to 1000\kms.  The right panel
  shows the X-ray surface brightness, emission-weighted temperature,
  and emission-weight pressure profiles across cold fronts.  The
  distance in kpc is measured relative to the cold fronts shown by
  dashed lines.  The simulated profiles and images reproduce all the
  main features of the observed cold fronts (sizes, jumps in surface
  brightness and temperature, relatively smooth gradient of pressure
  across the front).}
\label{n-fig:cold}
\end{figure}

Sharp features similar to the
observed cold fronts are produced naturally in a cluster merger when
the merging subclump, moving slightly supersonically, is undergoing
tidal disruption.  The Figure~\ref{n-fig:cold} shows the photon count
{\sl Chandra} map (top-left) and the emission-weighted temperature map
(bottom-left) of the region around the sharp feature identified in a
simulated cluster during a major merger.  The cold front is a low
temperature region embedded in the high temperature gas just behind
the left bow shock.  The appearance and the spatial extent of the cold
front in simulation ($\sim 0.5\hMpc$) are very similar to those of two
prominent observed cold fronts in clusters A2142 and A3667
\cite{markevitch00,vikhlinin01a}.  The right panel shows that the
behavior of the profiles (opposite-sign jumps in surface brightness
and temperature profiles and relatively smooth change of pressure
across this front) reproduces those of observed cold fronts
\cite{markevitch02}.  However, the velocity map around the
simulated cold front (bottom-left) shows that the flow of gas is not
laminar and in general is not parallel to the front.  This is very
different from assumptions made by \cite{vikhlinin01b}, who assumed a
laminar flow similar to that about a blunt body and used the sharpness
of the observed front to put tight constraints on the thermal
conduction and magnetic field strength in the vicinity of the front.

\begin{figure}[t]
\centerline{ 
\includegraphics[height=2.50truein]{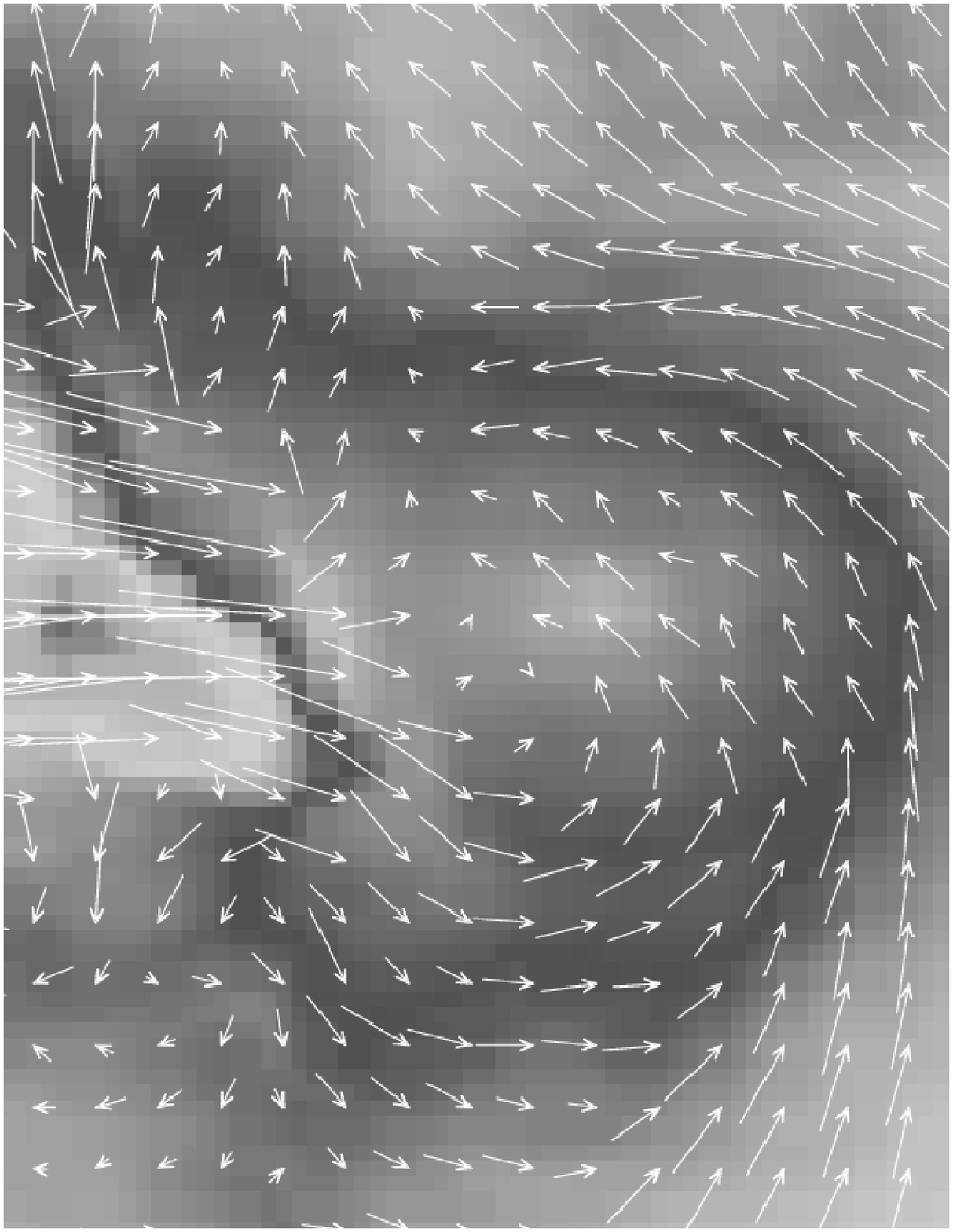}
\includegraphics[height=2.50truein]{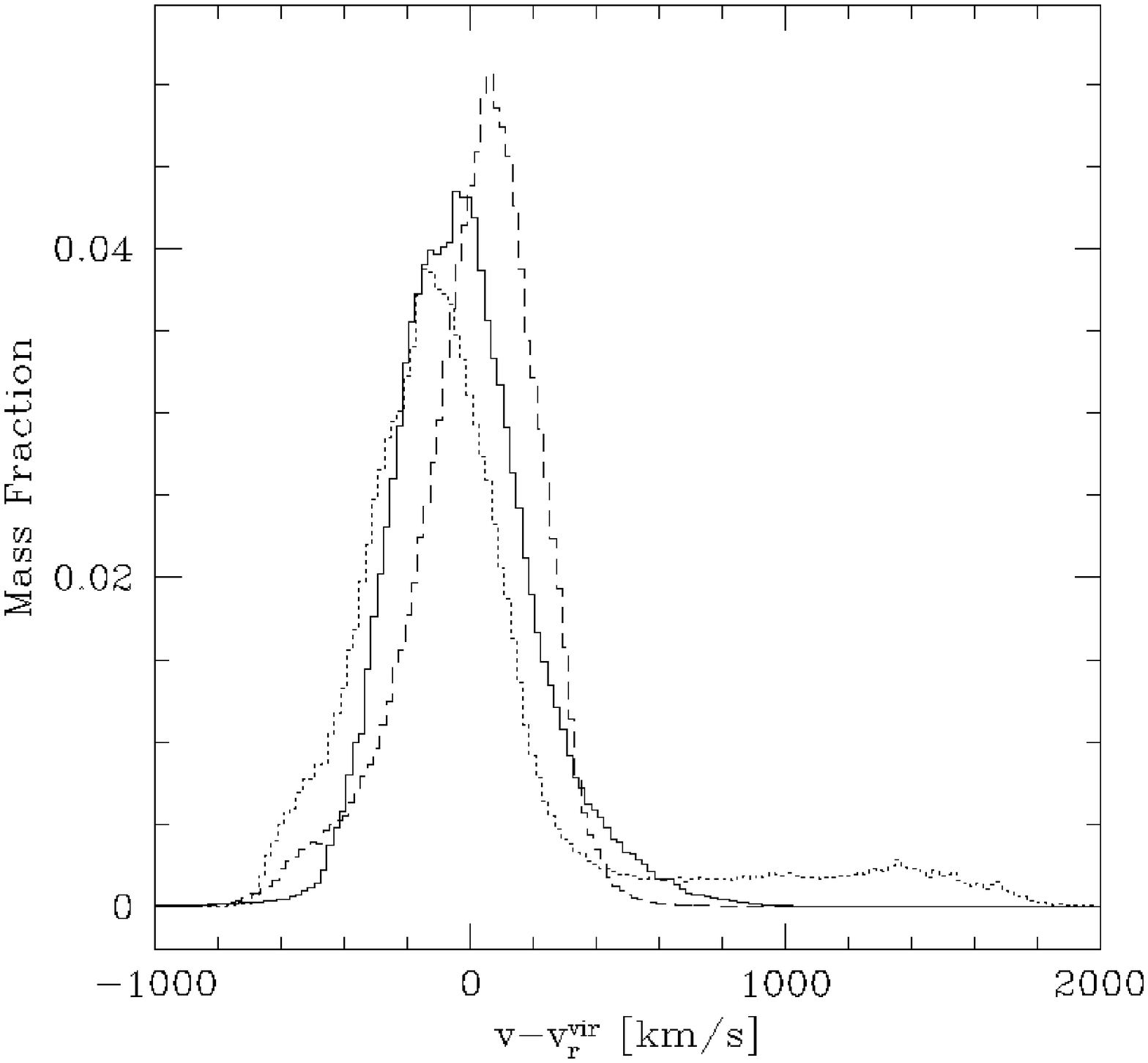}
}                                                  
\caption{ 
  Internal gas flows of the cluster in a relatively relaxed state,
  undergoing a minor merger.  Left: the velocity field is overlaid on
  the color maps of the emission-weighted temperature, color-coded on
  a log10 scale in units of Kelvin. The vertical vector in lower left
  corner correspond to 500 km/s. The size of the shown region is
  0.78$h^{-1}$~Mpc.  Right: the distribution of the gas velocity
  component along three orthogonal projections within the virial
  radius of the cluster. The gas is moving with velocities of 200-500
  km/s (typically 20-30\% of the sound speed) even in the cores of
  relaxed clusters due to frequent minor mergers. Such gas motions
  have important implications for the mixing and transport processes
  in cluster cores, mass modeling of cluster, as well as observable
  properties of ICM. From Nagai, Kravtsov \& Kosowsky (2003).  }
\label{n-fig:vel} 
\end{figure}

In the hierarchical models, ``small-scale'' cold fronts are expected
to be much more common.  Our simulations show that such cold fronts
are produced in a minor merger when a merging subclump reaches the
inner regions of the larger cluster and gas in front of the subclump
is significantly compressed by ram-pressure (shown in the left panel
of Figure~\ref{n-fig:vel}).  The compression sharpens and enhances the
amplitude of gas density and temperature gradients across the front.
The relatively cooler gas of the merging sub-clump is often trailed by
low-entropy ($\sim$1-2~keV) intergalactic gas in the direction of the
subclump's motion.  Many instances of such ``small-scale'' cold fronts
\cite{sun02,forman02,markevitch02} and diffuse \xray filament
extending beyond the outskirt of a cluster \cite{durret03} have been
seen in {\sl Chandra} and {\sl XMM-Newton} observations of nearby
clusters.

\section{Internal turbulent motions of cluster gas}

Frequent minor mergers discussed above will continuously stir the gas
and generate turbulent gas motions within the cluster.  The left panel
of the Figure~\ref{n-fig:vel} shows the internal turbulent motions of
cluster gas within the cluster in a relatively relaxed state.  In the
left panel, a small merging subclump moving with velocity $\gtrsim
1000$\kms along the filament is on the first approach to the cluster
center and has not yet suffered major tidal disruption.  When this
merging subclump reaches the cluster core it generates random,
slightly supersonic motions in which it dissipates its kinetic energy.
The velocity is quite chaotic with typical motions at a level of $\sim
200-300$\kms (typically 20-30\% of the sound speed) even in the core
of the relatively relaxed cluster due to frequent minor mergers (shown
in the right panel).  Our simulations suggest that regions of
fast-moving gas are generally present even in ``relaxed'' clusters due
to ongoing minor mergers.  Gas motions of a similar magnitude in the
cores of ``relaxed'' clusters are also implied by the {\sl Chandra}
observations \cite{markevitch02}.  This means that some processes
(e.g., motion of substructures or asymmetric inflow of gas along
filaments) in simulations continuously stir the gas even when the
cluster is in hydrostatic equilibrium globally.  Such gas motions have
very important implications for the mixing and transport processes in
cluster cores, mass modeling of cluster, as well as observable
properties of ICM.

\begin{figure}[t]
\centerline{ 
\includegraphics[height=2.30truein]{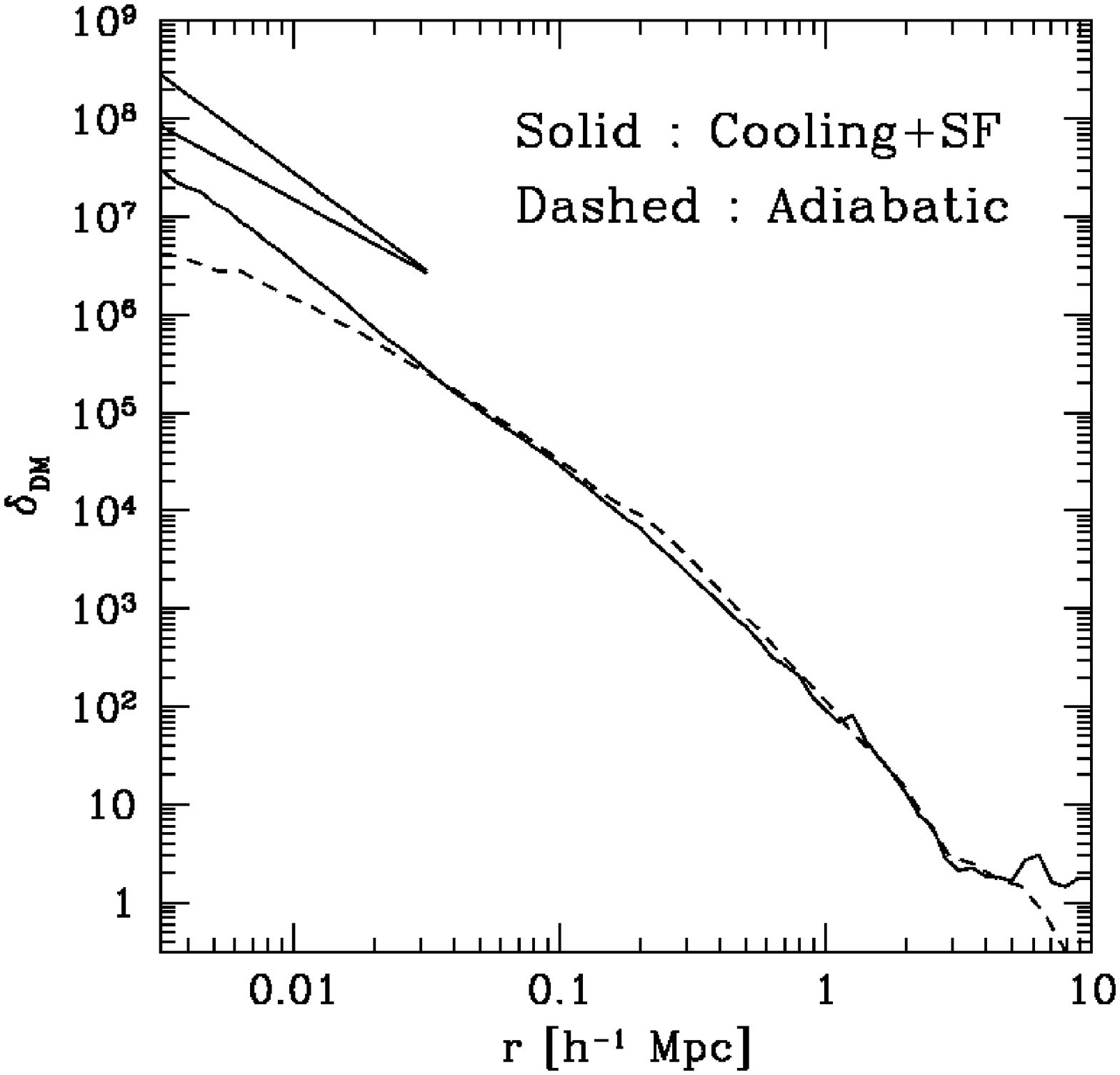}
\includegraphics[height=2.30truein]{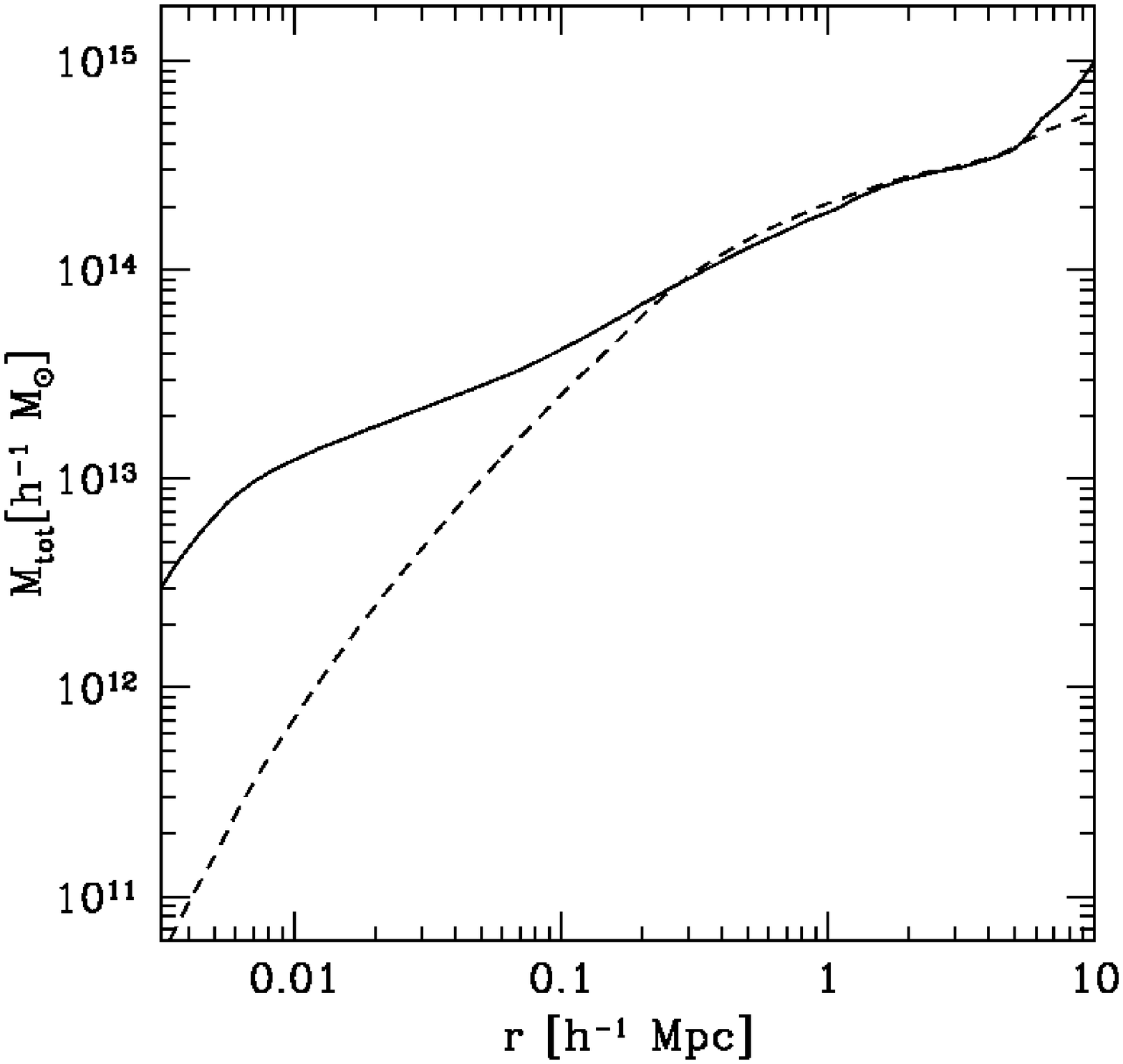}
}                         	   
\centerline{ 			   
\includegraphics[height=2.30truein]{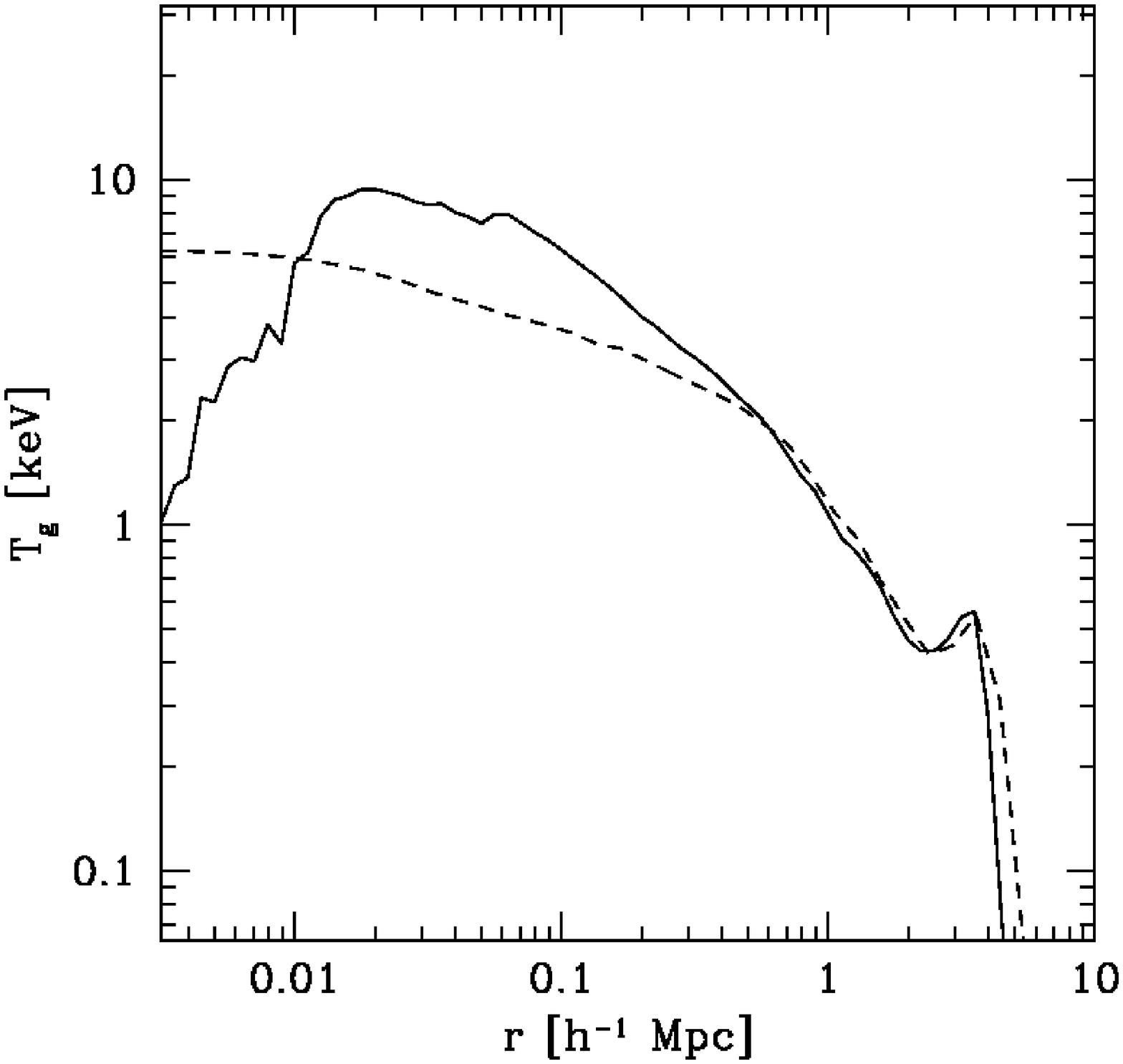}
\includegraphics[height=2.30truein]{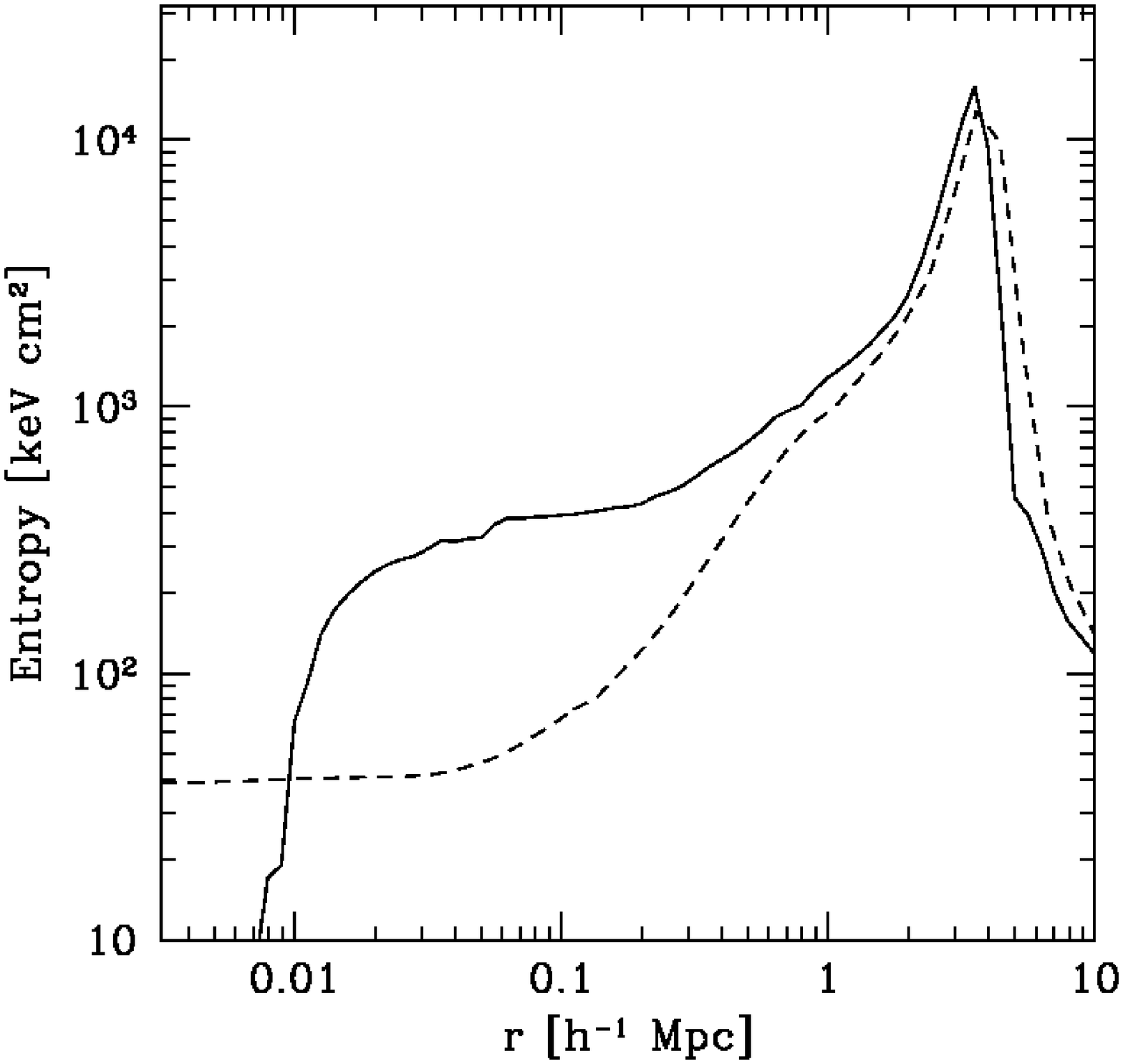}
}                         
\caption{ 
  The 3D radial profiles of over-density of dark matter, total mass,
  entropy and temperature (clockwise from top-left) of the simulated
  clusters with adiabatic gasdynamics (dashed) and the simulations
  that include radiative cooling and starformation (solid).  The
  cooled low entropy gas is replaced by the higher entropy gas, which
  causes the increase of entropy and temperature in the center within
  ~500kpc (roughly half of the virial radius).  These results support
  the suggestion of Voit and Bryan (2001) that gas cooling and star
  formation may be able to explain the observed properties of clusters.  Note
  also that the cooling of baryon results in a steepening of the dark
  matter density profile in the center of the cluster.  Two straight
  solid lines in the top-left panel indicate lines with $\sim
  r^{-1.5}$ and $\sim r^{-2.0}$.  }
\label{n-fig:pro} 
\end{figure}

\section{Effects of cooling and starformation}

The Figure~\ref{n-fig:pro} shows the three-dimensional radial
profiles of over-density of dark matter, total mass, entropy and
temperature (clockwise from top-left) of the simulated clusters with
adiabatic gasdynamics (dashed) and the simulations that include
radiative cooling and starformation (solid). The cooled
low-entropy gas is replaced by the higher entropy gas, which causes
the increase of entropy and temperature in the center within
~500kpc (roughly half of the virial radius).  Significant fraction
($\sim 80\%$) of cooled gas within the core of the cluster has been
turned into stars which formed a giant elliptical galaxy with mass of
$M\gtrsim10^{13} M_{\odot}$.  Note also that the compression of 
baryons causes steepening of the dark matter density profile from
$\sim r^{-1.5}$ to $\sim r^{-2.0}$ in the center of the cluster.  Our
simulations indicate that cooling and star formation are likely
important processes that shape the properties of ICM as well as the
dark matter profiles in the cluster core region.

\section{Summary}

We used the very high-resolution Adaptive Mesh Refinement (AMR)
gasdynamics simulations of clusters forming in the CDM universe to
investigate structural evolution of the intracluster gas and the role
of dynamical (e.g., mergers) and physical (e.g., cooling, star
formation, stellar feedback) processes operating during cluster
evolution.  

We find  that the sharp features
similar to the observed {\em cold fronts} are produced naturally in a
cluster mergers.  The simulated profiles and images reproduce all the
main features of the observed cold fronts (sizes, jumps in surface
brightness and temperature, relatively smooth gradient of pressure
across the front).  We find that the velocity fields of gas surrounding
the cold front can be very irregular which would complicate analyses
aiming to put constraints on the physical conditions of the
intracluster medium in the vicinity of the front.  

Although the most spectacular cold fronts are produced in a major
merger, ``small-scale'' cold fronts arising from minor mergers are
expected to be much more common, and many more instances of such cold
fronts are expected to be found by {\sl Chandra} observations of
nearby clusters.  The frequent minor mergers will also continuously
stir the gas and generate subsonic turbulent gas motions within the
cluster in a relatively relaxed state.  Such gas motions, if present,
have very important implications for the mixing and transport
processes in cluster cores, mass modeling of cluster, as well as
observable properties of ICM.

Finally, we presented preliminary results from an undergoing program
to carry out a series of simulations that include cooling, star
formation, stellar feedback, and metal enrichment.  Using these
simulations, we illustrated that cooling and star formation have a
significant impact on the properties of ICM (e.g., entropy and
temperature profiles) as well as dark matter distribution in the
cluster core.  Our results are consistent with the suggestion that gas
cooling and star formation together increase both entropy and
temperature in the cores of groups and clusters of galaxies (e.g.,
Voit and Bryan 2001).  The results presented here are preliminary. We
are currently undertaking convergence study and detail study of the
effects of cooling and star formation. In the near future, comparison
of the high-resolution simulations of the kind presented here to
high-resolution {\sl Chandra} and {\sl XMM-Newton} observations of
merging clusters and cores of relaxed clusters should 1) constrain the
relative importance of the physical processes and the dynamical
processes operating during different stages of cluster evolution, 2)
allow us to study metal enrichment of the ICM by supernova type Ia and
II, and mixing of metals and their radial distribution, and 3) to shed
new light on the puzzling absence of cold gas in the centers of
classic ``cooling flow'' cluster.



D.N.
  acknowledges support from NASA GSRP Fellowship NGT8-52906.  A.V.K.
  was partially supported by NSF grant No. AST-0206216 and by
  funding of the Center
  for Cosmological Physics at the University of Chicago (NSF
  PHY-0114422).










\newcommand{\vdag}{(v)^\dagger}
\newcommand{\myemail}{mittazj@email.uah.edu}
\newcommand{\ROSAT}{{\it ROSAT\ }}
\newcommand{\XMMNewton}{{\it XMM-Newton\ }}



\articletitle{WHIM emission and the cluster soft excess: a model comparison}
\author{Jonathan P.D. Mittaz\altaffilmark{1}, Richard Lieu\altaffilmark{1} \& 
Renyue Cen\altaffilmark{2}} 
\altaffiltext{1}{Department of Physics, UAH, Huntsville, AL35899}
\altaffiltext{2}{Princeton University Observatory, Princeton University, Princeton, NJ 08544}

\begin{abstract}
The confirmation of the cluster soft excess (CSE) by XMM-Newton has rekindled
interest as to its origin.  The recent detections of CSE emission at large
cluster radii together with reports of OVII line emission associated with the
CSE has led many authors to conjecture that the CSE is, in fact, a signature of
the warm-hot intergalactic medium (WHIM).  In this paper we test this scenario
by comparing the observed properties of the CSE with predictions based on
models of the WHIM.  We find that emission from the WHIM is 3 to 4 orders of
magnitude too faint to explain the CSE emission.  The only possibility is the
models if they are missing a large population of small density enhancements or
galaxy groups, but this would place have severe ramifications on the baryon
budget.
\end{abstract}

\section{Introduction}
The location of all the baryons existing at the current epoch is still somewhat
of a mystery.  Observationally, the total sum of baryons seen in stars,
galaxies and clusters of galaxies ($\Omega_b = (2.1^{+2.0}_{-1.4})
h_{70}^{-2}$\%, Fukugita, Hogan \& Peebles 1998) is only about half of the
number of baryons required by big bang nucleosynthesis models ($\Omega_b = (3.9
\pm 0.5) h_{70}^{-2}$\% Burles \& Tytler 1998) or from measurements of the
cosmic microwave background ($\Omega_b = (4.4 \pm 0.4) h_{70}^{-2}$\% Bennett
et al. 2003).  Recent cosmological hydrodynamical simulations have, however,
shown that this missing 50\% of baryons may be in the form of a warm ($10^5 -
10^7$K), tenuous medium (with overdensities between $\delta \sim 5-50$)
existing in filaments formed during the process of large scale structure
formation (e.g. Cen \& Ostriker 1999).  This medium is generally called the
'warm-hot intergalactic medium' or WHIM.

Ever since it was first proposed, the detection of the WHIM has been an
important goal in astrophysics.  To date there has been some success in finding
WHIM like material.  Far UV and soft X-ray absorption lines from the WHIM have
been reported, though the majority of detections seem to be confined to matter
in our local group (e.g. Fang, Sembach \& Canizares 2003, Nicastro et al. 2003,
Mathur et al. 2003).  At higher redshifts the situation is more controversial
with a few detections having been reported.  Possible emission from the the
WHIM has also been observed, with some weak X-ray detections (e.g. Soltan,
Freyberg \& Hasinger 2002, Zappacosta et al. 2002) which are in approximate
agreement with the predicted luminosity of the WHIM inferred from cosmological
simulations.  In general, however, these searches have not provided a strong,
unambiguous detection of the WHIM either in absorption or emission at redshifts
beyond our local group.

Recently, new observations of the cluster soft excess emission (CSE) show that
this phenomenon may also be a signature of the WHIM.  The CSE (seen as an
excess of observed flux above the the hot intracluster medium (ICM) at energies
below lkeV) has been an observational puzzle for a number of years.  First
discovered in the Virgo cluster (Lieu et al. 1996) subsequent observations have
found similar behavior in a number of different systems (e.g. Lieu et al. 1996b,
Mittaz et al. 1998, Kaastra et al. 1999, Bonamente et al. 2001a, Bonamente et al.
2001b, Bonamente et al. 2002).  Observationally the CSE shows a number of
different characteristics.  In some clusters the CSE when expressed as a
percentage is spatially constant (e.g. Coma: Lieu et al. 1998), in others there
is a marked radial dependence with the fractional soft excess being stronger in
the outer regions of the cluster (e.g. Mittaz et al. 1998). Observations have
also shown that the CSE is a relatively common phenomena - a \ROSAT study of 38
clusters showed that approximately 45\% of clusters show at least a $1\sigma$
effect (Bonamente et al. 2002) .  

On the current state of interpretation, models assuming a thermal origin of the
excess have invoked a warm gas intermixed with the hot ICM are unsatisfactory.
These have found to be unsatisfactory, however, since the cooling times are
extremely short (sometimes $\sim 10^6$ years Mittaz et al. 1998) so some kind
of heating mechanism to sustain the warm gas has to be envisaged (Fabian 1997).
On the other hand if the emission arises from an inverse-Compton process (such
as suggested by Hwang 1997, Sarazin \& Lieu 1998, Ensslin, Lieu \& Biermann
1998) then the radial dependence seen in some clusters can be explained, but
the inferred pressure of the required cosmic-rays would be perplexedly too high
(Lieu, Axford \& Ip 1999).

The picture has changed somewhat with new observations by the XMM-Newton
satellite.  These observations reveal apparent strong thermal lines of Oxygen
in the CSE spectra for the outskirts of a number of clusters (Kaastra et
al. 2003, Finoguenov et al. 2003) with the CSE continuum being fitted with a
characteristic temperature of 0.2 keV.  If established, this provides
irrefutable evidence that the CSE must be thermal in nature.  Such a finding is
also supported by \ROSAT observations of the Coma cluster, which shows a very
large, degree scale halo of soft emission (Bonamente et al. 2003).  Emission on
this scale cannot possibly be non-thermal in nature since there is no way of
confining a relativistic particle population at such distances from the cluster
center.  Therefore, given that there seems to be a large scale thermal
component at the outskirts of clusters, the CSE emitting material has to be
located predominantly beyond the cluster virial radius where we can have
sufficiently low densities to overcome any cooling time issues.  In such a
scenario the warm emitting gas is spatially consistent with the WHIM
i.e. current cosmological models should be able to predict the luminosity and
temperature of the CSE.  We have therefore undertaken a detailed comparison
between cosmological simulations of the WHIM and the observed soft excess
signal.

\section{The Model}

We have used a recent cosmological hydrodynamic simulation of the canonical
cosmological constant dominated cold dark matter model (Ostriker \& Steinhardt
1995) with the following parameters: $\Omega_m$ = 0.3, $\Omega_\Lambda$ = 0.7,
$\Omega_b$ $h^2$ = 0.017, h = 0.67, $\sigma_8 = 0.9$ and a spectral index of
the primordial mass power spectrum of n = 1.0. The simulation box has a size of
25 h$^{-1}$ Mpc comoving on a uniform mesh with 768$^3$ cells and 384$^3$ dark
matter particles giving a comoving cell size of 32.6 h$^{-1}$ kpc. This
simulation together with another similar one at lower resolution derived from
the same code and parameters have previously been used to account for a variety
of of different observational consequences of the WHIM, such as OVI absorption
line studies (e.g. Cen, Tripp, Ostriker, Jenkins, 2001) and the X-ray
background (Phillips, Ostriker \& Cen, 2001).  For a more detailed discussion
of the simulation itself see Cen, Tripp, Ostriker \& Jenkins (2001).  Here we
are primarily concerned with the emission in the vicinity of a cluster.  Thus
we focus our analysis on a cluster simulated in the high resolution mode.

\begin{figure}
\centerline{\includegraphics[height=3.0in,angle=0]{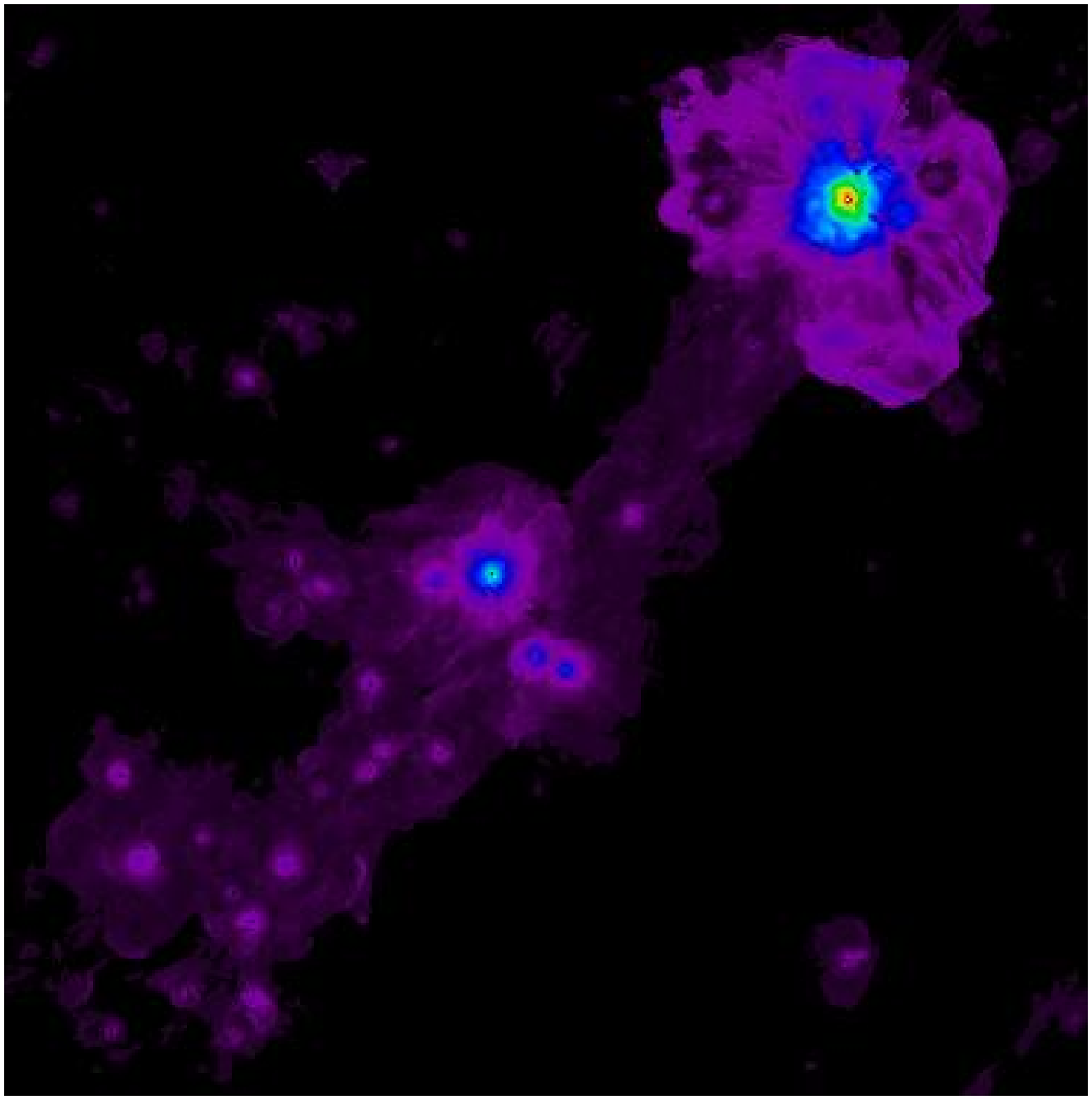}}
\caption{Emission weighted temperature map for one particular projection of the
  data volume.  The plotted temperature range is from 0.1-6.5 keV.  The
  simulated cluster can be seen in the top right of the figure}\label{m-fig1}
\end{figure}

The simulation as provided comes in the form of three data cubes containing
temperature, density and metal abundance.  Figure~\ref{m-fig1} shows an emission
weighted temperature map from one particular projection of the cube and within
the image a number of structures can be seen, including filaments, groups and
one cluster candidate at a temperature of $\sim$ 6.5keV.  For the purpose of
this paper we are going to concentrate on this cluster-like structure which is
seen in the top right corner of Figure~\ref{m-fig1}.  It has a peak mass density
of $5.7 \times 10^{13}$ M$_\odot$ Mpc$^{-2}$ placing it the regime of
virialized objects and so can be considered for our present purposes a cluster
of galaxies.

\section{Simulated X-ray spectra from the model}

\begin{figure}[t]
\begin{center}
\begin{minipage}[b]{5in}
\centering
\includegraphics[height=3in,angle=-90]{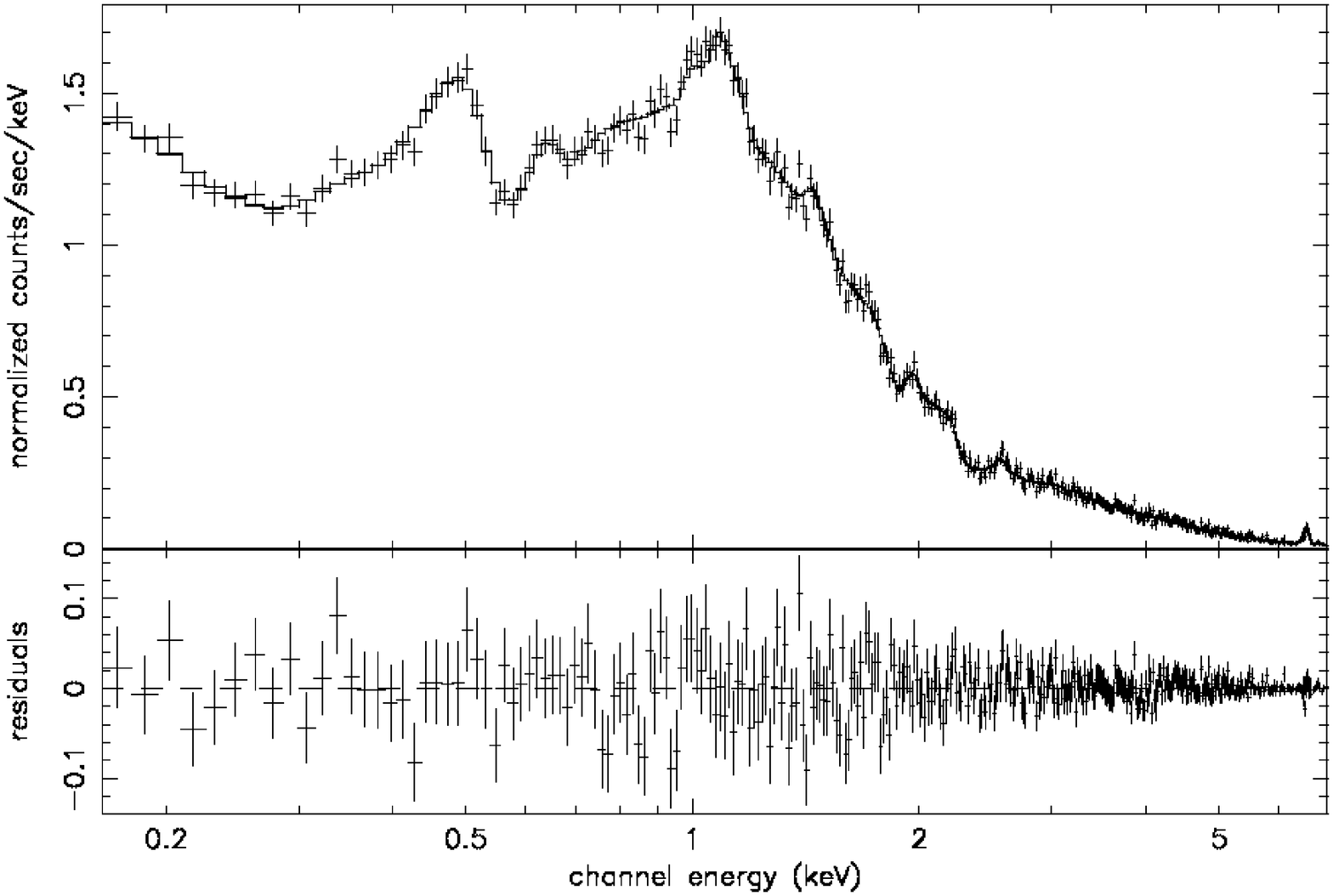}
\end{minipage}
\end{center}
\caption{The simulated spectrum from the cluster showing the spectrum
from the central 32.6 kpc region (0-2 arcminutes).  Also shown is the
best fit model (kT = 4.7 A=0.5) together with fit residuals and shows
no requirement for any cluster soft excess}\label{m-fig2a}
\end{figure}

\begin{figure}[t]
\begin{center}
\begin{minipage}[b]{5in}
\includegraphics[height=2.2in,angle=-90]{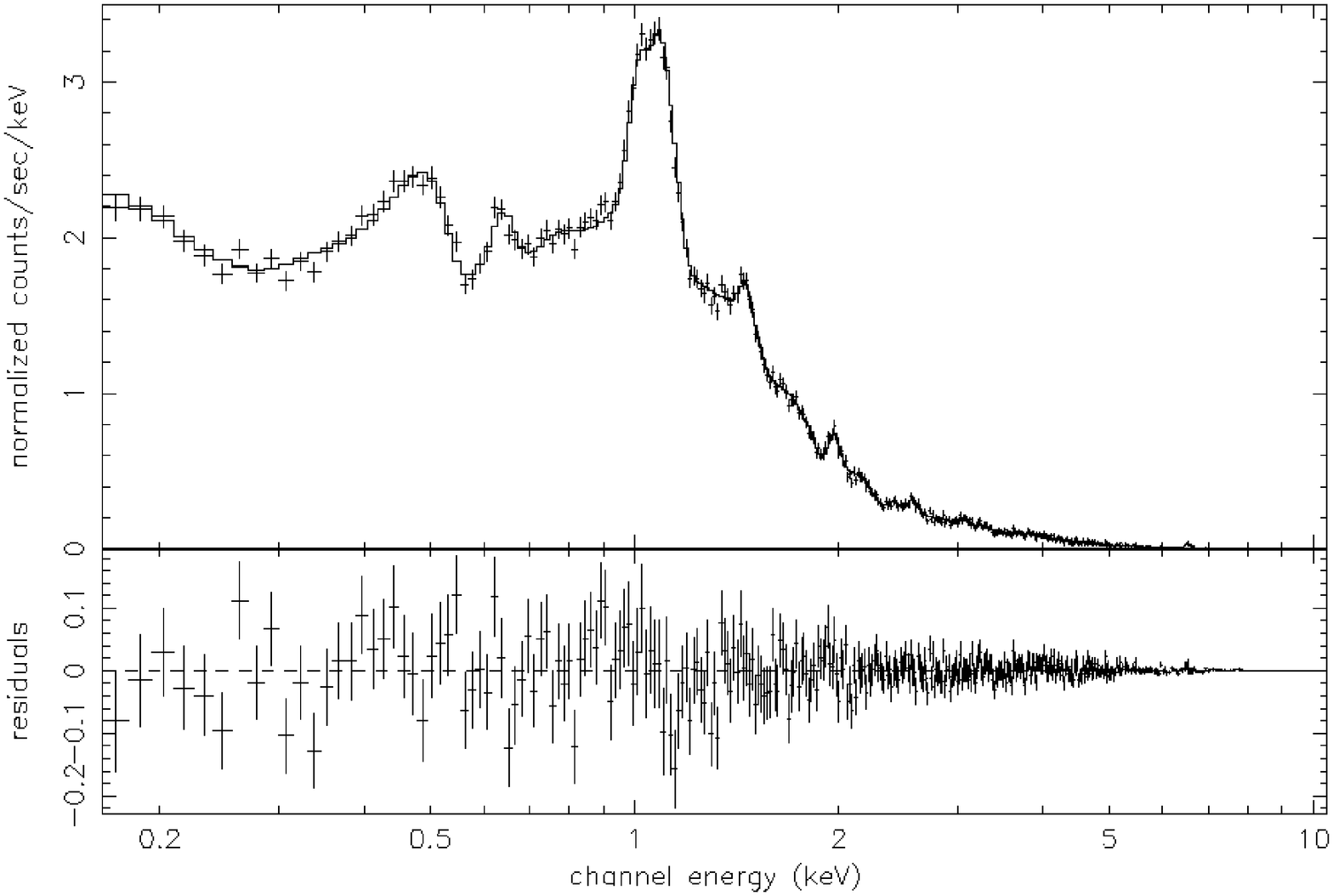}
\hspace{0.2cm}
\includegraphics[height=2.2in,angle=-90]{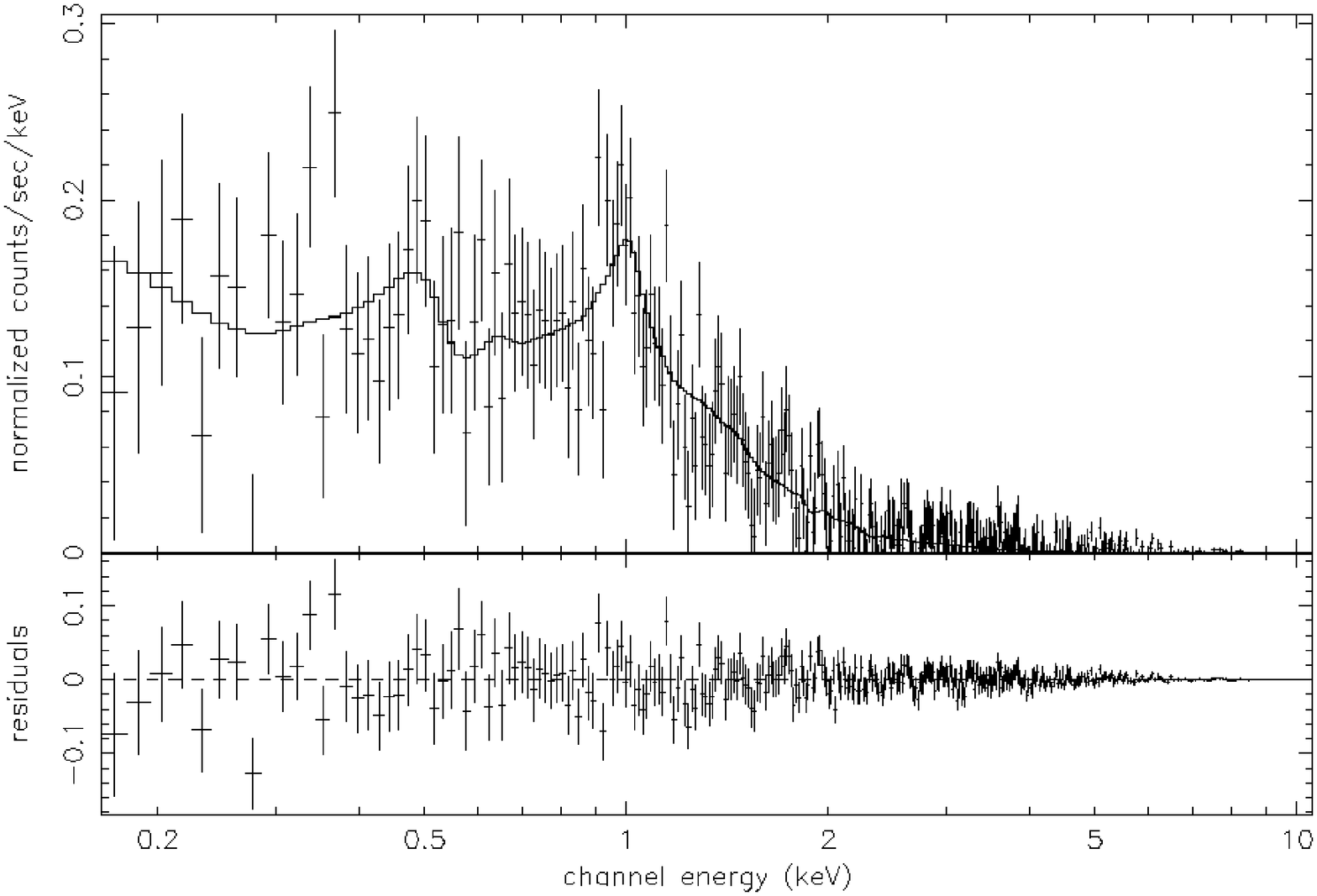}
\end{minipage}
\end{center}
\caption{Further spectra from the simulated cluster from annular
regions placed at the outskirts of the cluster.  The left panel shows
the 589-680 kpc annulus, and the right panel shows the simulated
spectrum taken from the 1.17-1.26 Mpc annulus.  Again, all spectra are
well fitted with a single temperature model (kT = 2.27 keV A=0.29
and kT=1.25 A=0.06 respectively) with no evidence for any soft excess
emission.}\label{m-fig2}
\end{figure}


The ultimate aim of this paper is to investigate the WHIM model predicted X-ray
emitting properties and compare them with current observations, particularly
those of the CSE as now observed with XMM-Newton.  To do this we have placed
the simulation at a redshift of 0.02, similar to that of the Coma cluster whose
CSE has been extensively studied (Bonamente et al. 2003, Finoguenov et al. 2003).
We further assumed that the gas is in thermal and ionization equilibrium.
Since the simulation provides the temperature, density and metal abundance at
each point, for each cell we can derive an optically thin X-ray spectrum.  To
do this we have used the mekal code for optically thin plasmas (Kaastra et
al. 1992, Leidhal et al. 1995) to generate the average bremsstrahlung and
emission-line spectra.  In order to avoid possible extreme abundance values
that can sometimes be found in the densest regions of the simulation we do not
allow the metal abundance to exceed a value of 0.5 solar.  Then after adding
the appropriate galactic absorption (N$_H = 9 \times 10^{19}$ cm$^{-2}$) and
folding the resultant spectrum through the XMM-Newton MOS1 response we can
generate an XMM-Newton counts spectrum from each cell.  To obtain the overall
spectrum we simply sum these spectra along our given line of sight.  Finally,
in order to get a reasonable estimate of the noise, an astrophysical background
spectrum was added (taken from Lumb et al. 2002) together with Poisson
fluctuations appropriate for the given exposure time (in our case 50 ksec).

\begin{figure}[t]
\centering
\includegraphics[height=2.2in,angle=-90]{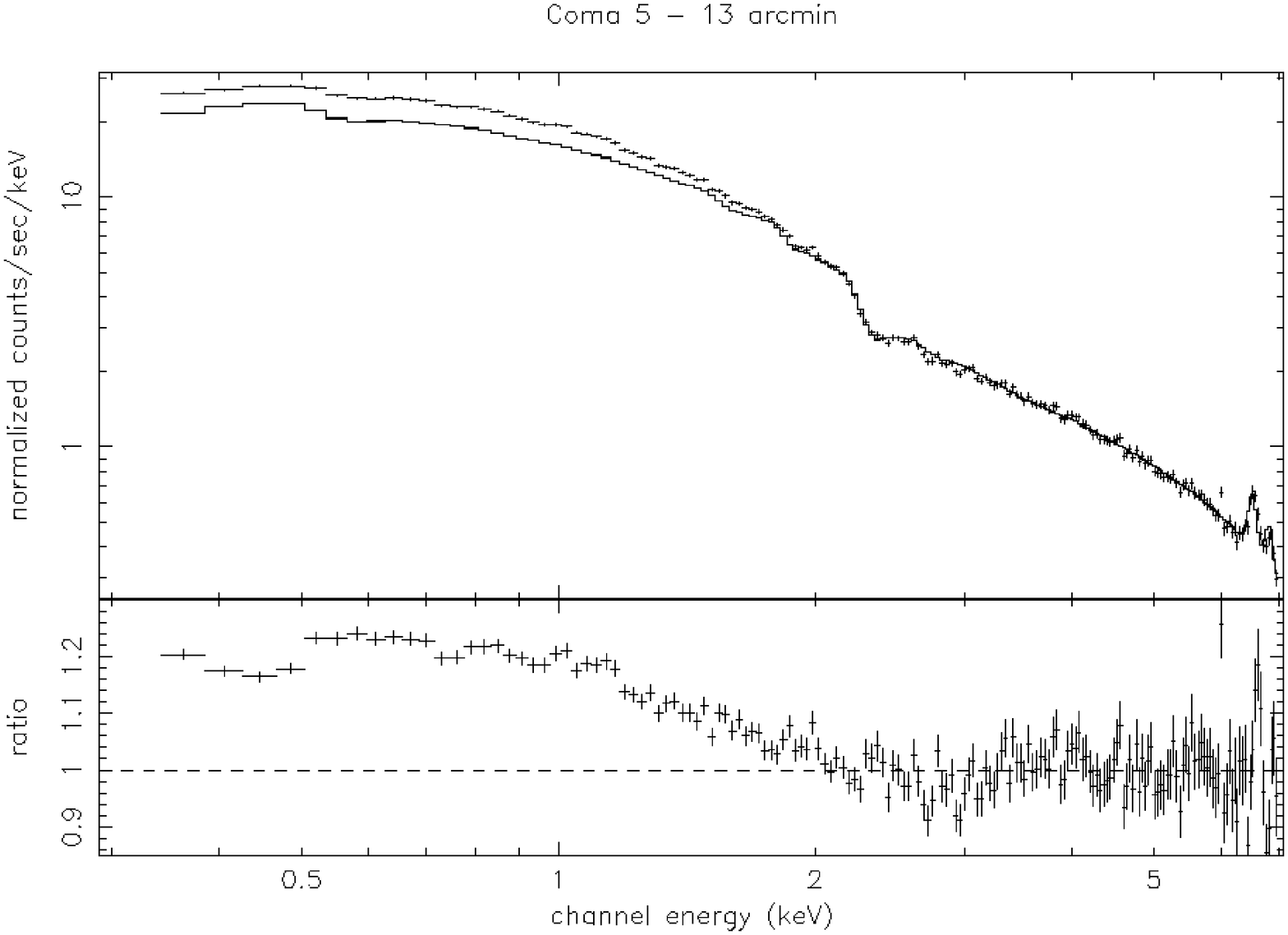}
\hspace{0.2cm}
\includegraphics[height=2.2in,angle=-90]{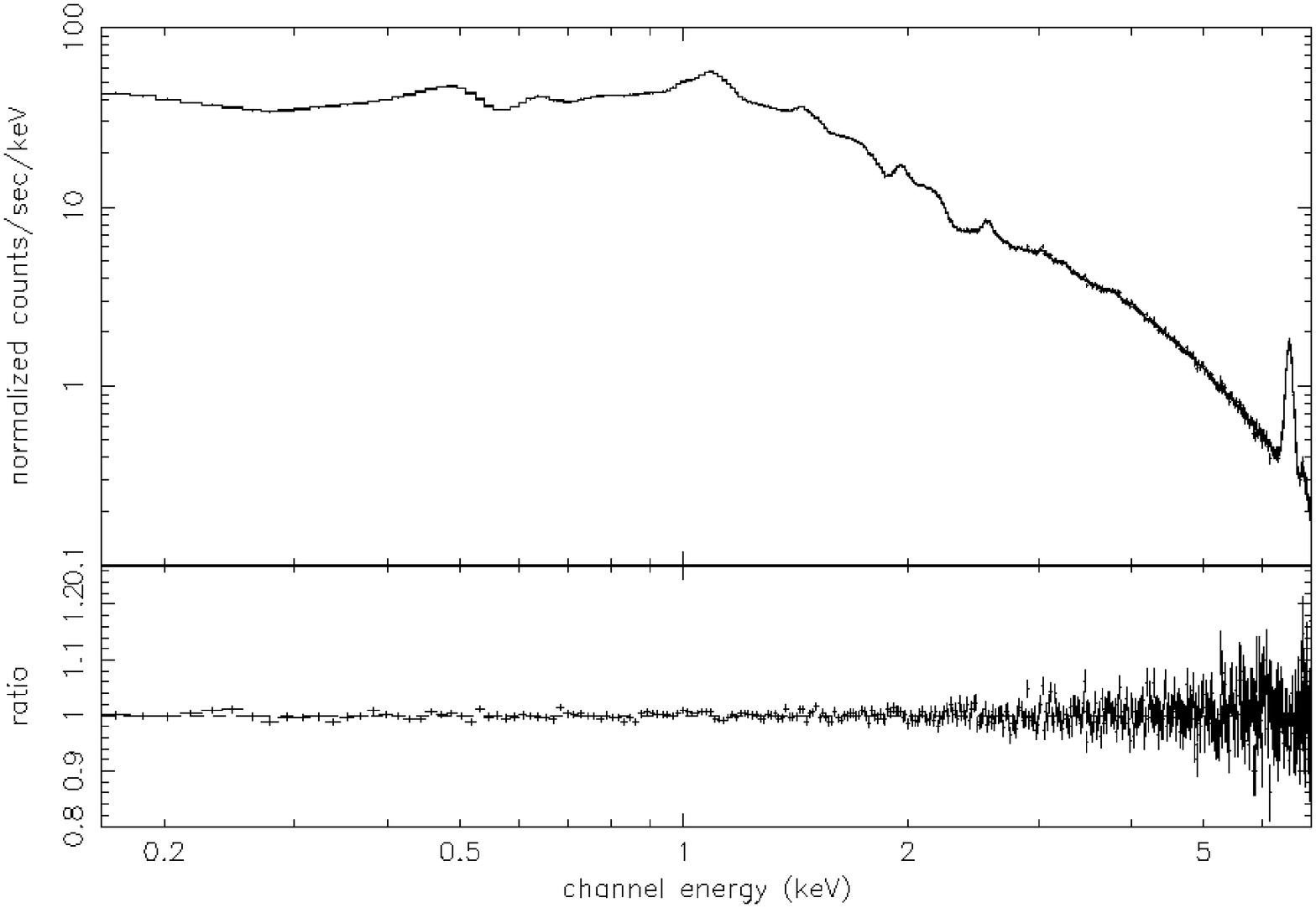}
\caption{The left panel shows the XMM-Newton spectrum of Coma extracted in the
  5-13 arcminute region (from Nevalainen et al. 2003), and the right panel
  shows a simulated spectrum from the model extracted from a similarly sized
  region.  While the observed data shows a clear soft excess, the spectrum from
  the simulation show no need for any extra components above the hot
  ICM emission.}\label{m-fig5}
\end{figure}

Figure~\ref{m-fig2a} shows an example of a simulation of the central 40 kpc
(i.e. one cell which corresponds to 0-2 arcminutes) region of the cluster.
Also shown is the best fit performance of the single temperature mekal emission
model to the data with parameters of 4.7 keV and abundance of 0.5.  As can be
seen from the residual plot, there are no strong deviations of the data from
the fit. This is consistent with general properties of clusters inferred from
XMM-Newton and Chandra data (e.g. Molendi \& Pizzolato 2001) where
observationally there is no requirement for a multi-temperature fit, even
within the so called cooling radius of the cluster where there is presumed to
be a wide range of temperatures along the line of sight.  As pointed out by
Molendi \& Pizzolato (2001) and Ettori (2002) this is to be expected for a
single phase medium which further supports our model choice.

For a more in depth understanding, the reason why we see such a good fit to the
data even though there are multiple temperatures and densities along the line
of sight, is due to the weighting effect of density.  Since emission goes as
$n_e^2$, the spectrum is heavily biased towards the densest material along the
line of sight.  Therefore the spectrum is dominated by the emission from a
narrow range of densities and temperatures.  Other components, such as may give
rise to a CSE, will not be visible unless they exists in substantial quantities
relating to the principle component.

\subsection{Looking for the cluster soft excess}

The sufficiency of a single temperature fit to the X-ray spectrum is seen at
almost all radii ranging from the center to the outermost regions. This is
illustrated in Figure~\ref{m-fig2} which shows examples from two outer annuli
situated at 580-680 kpc and 1.17-1.26 Mpc from the center where the one
temperature fits (with kT = 2.27 keV \& kT = 1.25 keV respectively) are
perfectly adequate.  In the context of a WHIM explanation of the CSE this is
interesting, because if the cluster soft excess is due to overlying filaments
from the WHIM it is exactly in these outer regions (where the cluster emission
is weak) that we would expect to see an effect.  The fact the we see no
evidence for a soft excess implies that the emission from the filaments in this
simulation must be much weaker than that of the observed CSE.

\begin{figure}
\centering
\includegraphics[height=3in,angle=-90]{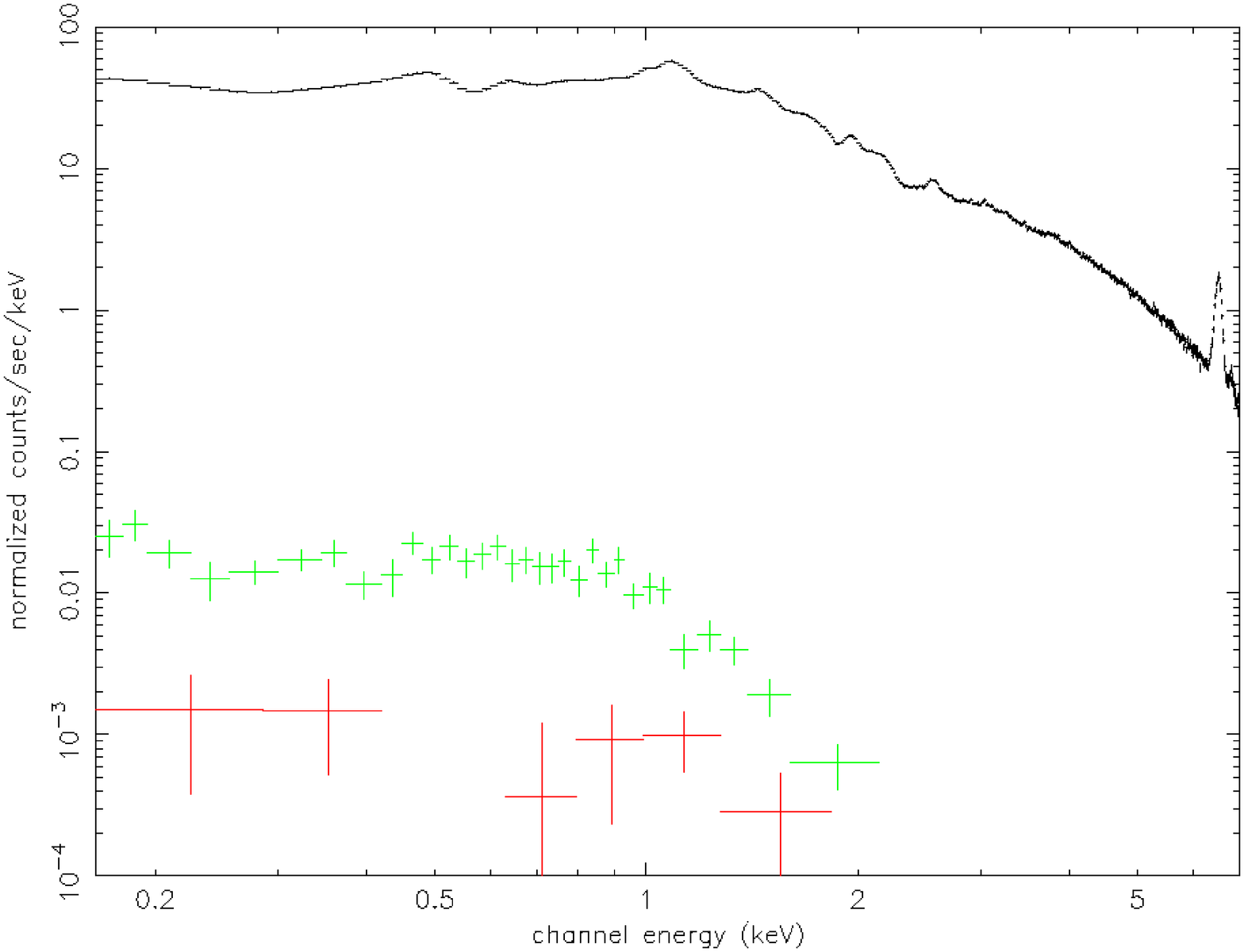}
\caption{The total simulated spectrum from figure~4 (top curve)
  together with spectra derived from cells with temperatures below 1keV.  The
  lowest spectrum is that derived from the same line of sight used for the
  total spectrum, while the middle spectrum is taken from a line of sight
  chosen to maximize the soft excess emission.}\label{m-fig6}
\end{figure}

The difference between observations and theory is even more clearly seen in
figure~\ref{m-fig5} which shows two spectra: the spectrum on the left is taken
from Nevalainen et al. (2002) and is of the 5-13 arcminute region of the Coma
cluster, the spectrum on the right is taken from the simulation covering a
similar annular region.  The first point to note is the relatively good
agreement in the flux levels - the 0.2 - 2keV luminosity from the simulation is
$\sim 4.5 \times 10^{44}$ ergs/s while the real observations shows a luminosity
of $\sim 8.8 \times 10^{44}$ ergs/s, giving us confidence that the cross
normalizations are realistic.  However, there are also significant differences.
Below 2 keV the XMM-Newton observation of Coma shows a clear soft excess at a
25\% level above the hot ICM emission, whereas the simulated spectrum show no
soft excess emission.  We therefore conclude that the model does not include
material at the right temperature and density to account for the thermal origin
of the CSE.

Such a conclusion is further underpinned when the emissivity of material at a
temperature consistent with the CSE is studied.  Figure~\ref{m-fig6} shows the
same simulated spectrum as the one in figure~\ref{m-fig5} but with the addition
of spectra derived only from those cells with temperatures below 1\ keV
i.e. just those cells from which we would expect the cluster soft excess
emission to arise.  The two lower spectra in figure~\ref{m-fig6} show the
expected emission from all components below 1keV as taken from two different
lines of sight.  The lower spectrum is derived from the same sight line used to
create the total spectrum (i.e. the top spectrum).  The brighter of the two
spectra is taken from a sight line that has the most CSE effect, because along
it the emission from cells with a temperature less the 1\ keV is maximized i.e
a line of sight that maximizes the CSE.  Note in both cases there {\it is}
emission in the crucial $< 1$\ keV regime only it is much fainter (by a factors
of $10^4$ or $10^{3}$ respectively) than the hot ICM emission.  It then becomes
obvious why no cluster soft excess was seen in the simulated spectra: the
emission from cells capable of accounting for the soft excess is very small
compared with the soft flux from the hot ICM emission lying behind it.  We can
further quantify this effect by comparing our model prediction with the soft
excess emission seen in a large sample of clusters.  Bonamente et al. (2002)
studied 38 clusters and listed the strength of any soft excess detected with
\ROSAT.  The simulated CSE luminosities fall short of the values for the
majority of these cases.



\subsection{A possible detection of a soft excess and the importance of small groups}\label{m-group}

\begin{figure}[t]
\centering
\includegraphics[height=2.2in,angle=-90]{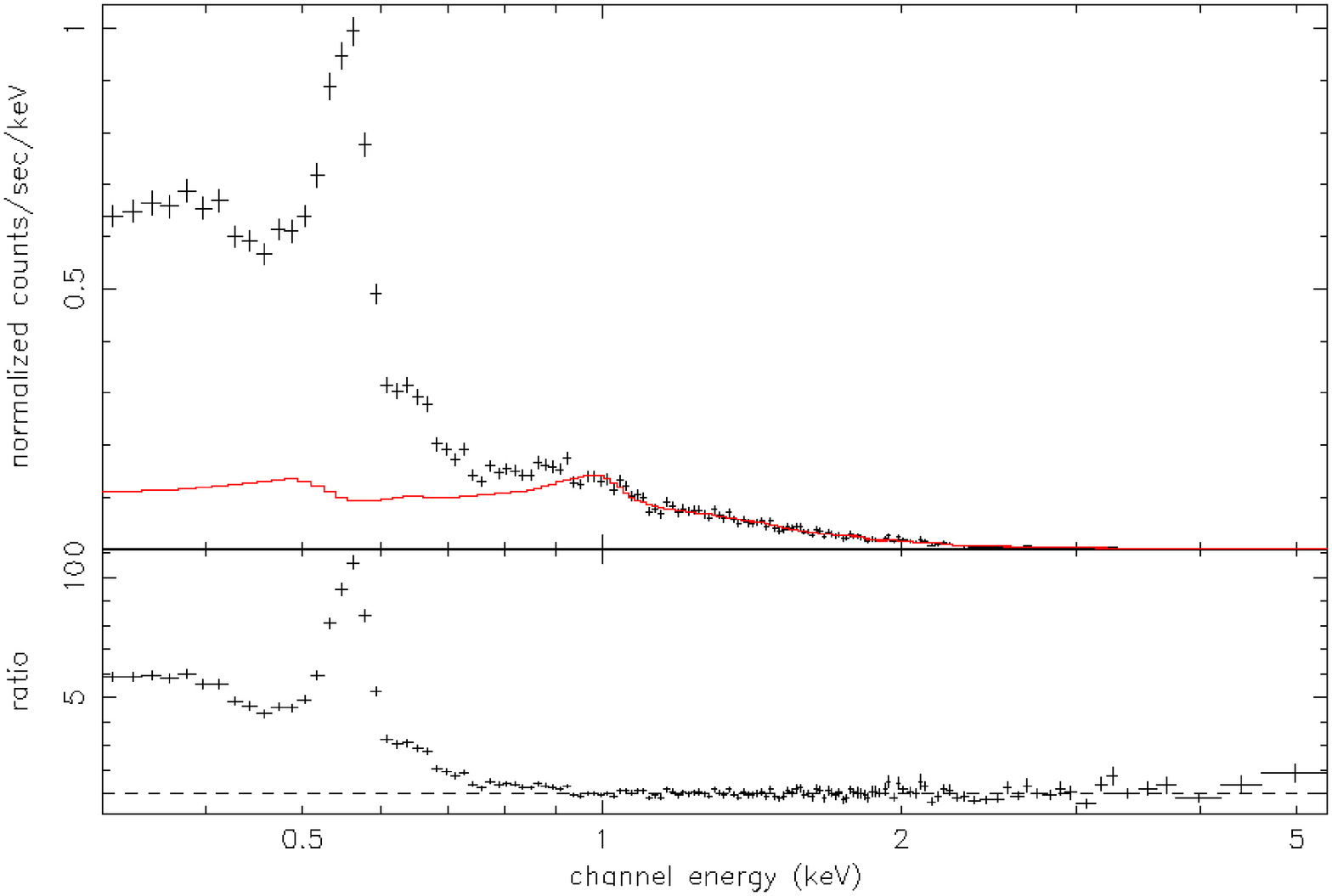}
\hspace{0.2cm}
\includegraphics[height=2.2in,angle=-90]{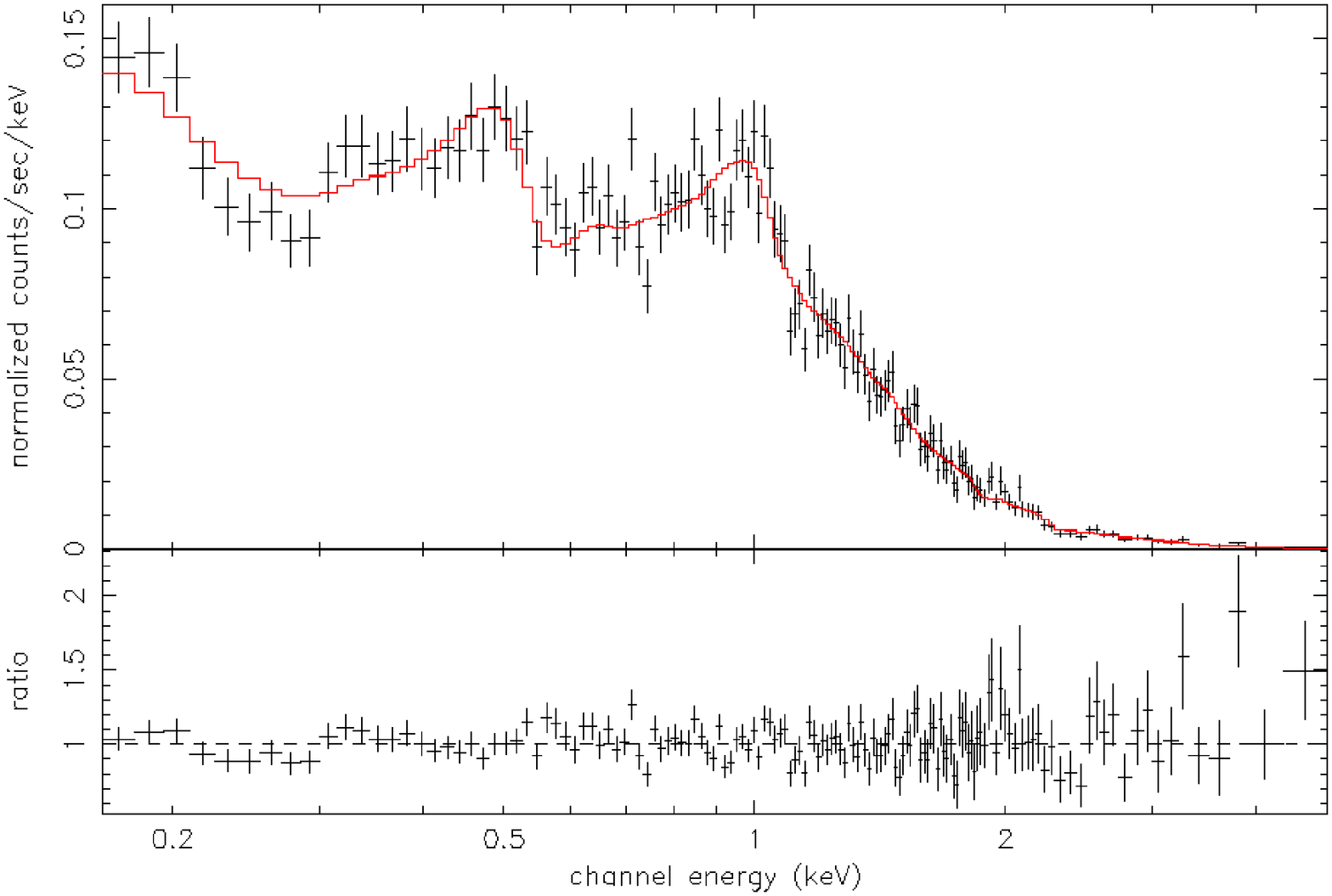}
\caption{The left panel shows the simulated XMM-Newton spectrum for the annular
  region 978-1467 kpc showing a very strong soft excess signal.  The right
  panel shows the same region if a small region containing a strong density
  enhancement is excluded.  The fit now is acceptable over the entire
  XMM-Newton energy range ($chi^2_{\nu} = 1.18$) with a temperature of 1.07
  keV}\label{m-fig9}
\end{figure}

\begin{figure}[t]
\centering
\includegraphics[height=2.2in]{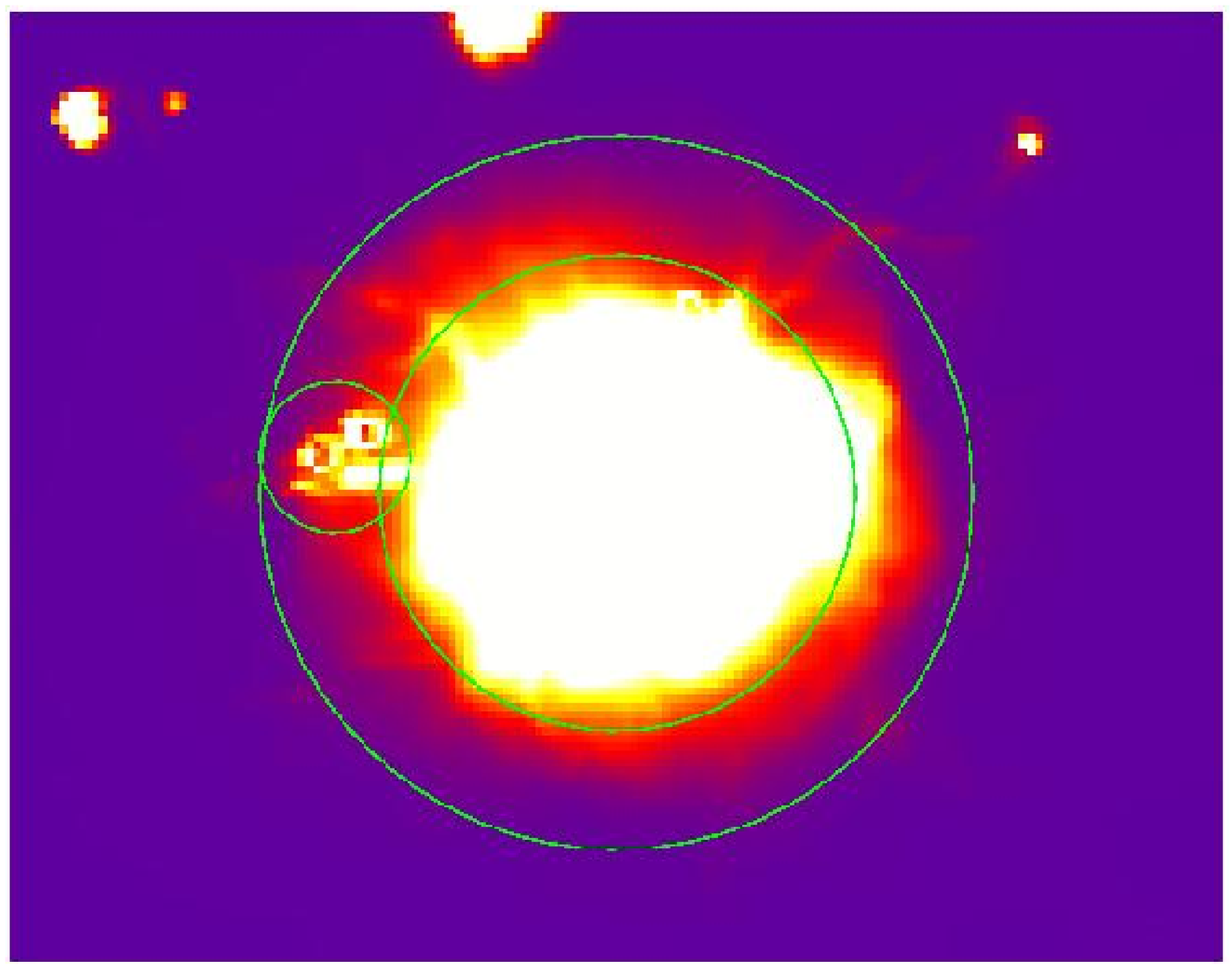}
\caption{An image of the simulated cluster showing the 978-1467 kpc
  extraction annulus used to extract the spectrum from
  figure~6 together with the excluded region that contains
  all of the soft excess emission.  The maximum count rate shown in
  the image is 0.01 cts/sec/pixel.}\label{m-fig10}
\end{figure}

Under certain circumstances it {\it is} possible to emulate a soft excess.  For
example from an extraction from the 30 - 45 cell (978 - 1467 kpc) annular
region one notices a strong excess at the lowest energies.  The left panel of
Figure~\ref{m-fig9} shows the simulated spectrum, and in this case a single
temperature fit is unacceptable with a $\chi^2_{\nu}$ of 42.5.  Following the
analysis techniques of Nevalainen {\it et al.}  (2002) and Bonamente {\it et
al.} (2003) of applying a one temperature model to the spectrum above $\sim 1$
keV we find a reasonable fit with a single temperature of $1.08 \pm 0.05$ keV
($\chi^2_{\nu} = 1.28$).  Below 1 keV there is then a very strong soft excess
including a very strong OVII line.  Interestingly the temperature of this
excess is 0.2 keV, exactly the same temperature as has been reported for a
number of clusters (eg. Kaastra {\it et al.} 2003).  However, it turns out that
this excess arises from a very small number of cells within the extraction
region.  In Figure~\ref{m-fig10} an image of the outskirts of the simulated
cluster together with the annulus used to extract the original spectrum.  Also
shown is a small circle on the left hand side containing the region of bright
pixels i.e. high densities, that actually gives rise to the soft excess.  If
this small region is removed, the resultant spectrum (the right hand panel of
Figure~\ref{m-fig9}) yields no evidence of a CSE.

We therefore conclude that there is very little evidence to link emission from
the WHIM with the cluster soft excess.  Our investigation has revealed the
possibility, however, that small density enhancements (which may be associated
with small galaxy groups) can give a signal that mimics the cluster soft
excess.  However, in the real observations the cluster soft excess seems
relatively smooth and does not seem to be confined to a few small locations
(see for example the \ROSAT results on the Coma cluster Bonamente {\it et al.}
2003).  One possibility is that the models to date do not have sufficient
spatial resolution to emulate a large population of small density perturbations
which could give rise to the observed soft excess signal.  Additional physics
related to a more complex environment, such as would exist in a supercluster,
may also be required to achieve this - Kaastra et al. (2003) discussed the
possibility, which clearly awaits further work.  An overall fundamental problem
exists, though, and it pertains to the acute lack of mass in gas at any
temperature range to account for the soft excess.  To get the required factor
of $10^3$ increase in brightness from the WHIM would imply an increase in the
density by a factor of 30.  Since the majority of the WHIM has overdensities
between 10-30 we then require overdensities greater than 300.  However, only
$<20$\% of the WHIM bave overdensities greater than 200 (e.g. Dav\'e et
al. 2001) and it is clear that the models simply do not contain material at the
densities and temperatures to give rise to the CSE signal as observed.

\section{Other models}

All of the above discussion has been based on the analysis of one particular
model.  It is reasonable to ask if other models show the same effect.  This is
actually quite a difficult question to answer since there are only a few models
with all the required information freely available.  A larger volume, lower
resolution model that that studied in this paper is obtainable from Renyue
Cen's web page (http://www.astro.princeton.edu/~cen/\newline
PROJECTS/p1/p1.html) and we also utilized this version of the model.  In
general we find the same result - it is very difficult to reproduce the
observed soft excess signal.  With a larger volume there is, of course, a
slightly higher probability of having galaxy groups located along any sight
line which can give rise to excess soft flux in a similar fashion to the case
studied in section~\ref{m-group}.  However, even by taking this effect into
account we still arrive at a similar conclusions to those obtained form the
higher resolution model - the WHIM itself cannot reproduce the observed CSE.
We have also studied one of the clusters available in the Laboratory of
Computational Astrophysics (LCA) simulated cluster archive
(http://sca.ncsa.uiuc.edu) with no major change in the conclusions.

\section{Discussion}

From the above discussion it would appear that within our current understanding
one cannot assign sufficient luminosity and low temperature material ($< 1$\
keV) to account for an observable CSE using a WHIM filament.  What are the
possible reasons for this discrepancy?  One possible problem with the high
resolution simulation is that may not cover enough filaments/structures to
emulate the CSE in the majority of cases.  However, the discrepancy between
model and observation is unlikely to be due simply to the scale of the
simulation since the box side of 25 $h^{-1}$ Mpc equals approximately that of
the proposed filament needed to explain the Coma soft excess (Finoguenov et
al. 2003).  Therefore the scale of any filament in this simulation is not
orders of magnitude discrepant from that required by observation.  Further, a
comparison of the temperatures and densities in cells in the vicinity of the
cluster shows a distinct lack of cells with an overdensity $>200$ (the value
required by observations see e.g.  Nevailenen et al 2002) and temperatures
commensurate with that of the cluster soft excess ($T<0.6$keV).  It is not
surprising then that the predicted soft emission is so weak - material at the
required temperature and density is not present.  Other simulations also seem
to have the same absence of cells with the correct temperature/density ratio.
In fact, the only time that a soft excess exists is during a strong density
enhancement along the line of sight, which may be interpreted as a small galaxy
group.

There are other possibilities that may point to this particular simulation as
not representative of the clusters showing a CSE.  Kaastra {\it et al.}
(2003,2004) have proposed a correlation between the cluster soft excess and a
supercluster environment.  They arrive at this conclusion based on two
arguments.  \ROSAT all-sky survey maps on degree scales around soft excess
clusters seem to show a large scale excess of soft emission which is claimed to
be related to a supercluster.  The authors also claimed that in the model of
Fang et al. (2002), regions where there are numerous structures (i.e. a
potential simulated supercluster) can also be associated with a OVII column
density similar to that inferred from observations ($\sim 4 - 9 \times 10^{16}$
cm$^{-2}$) thereby implying a causal link between superclusters and the
presence of a CSE.

\begin{figure}[t]
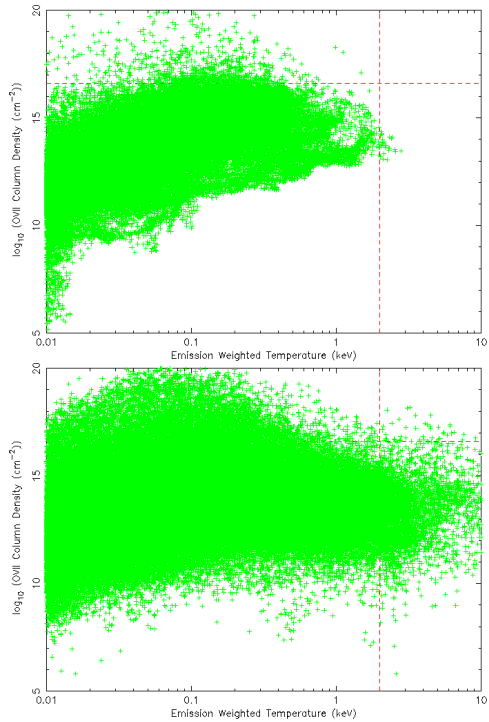

\centering
\includegraphics[height=2.5in,angle=-90]{papers/mittaz/model_scatter2.cps}
\hspace{0.2cm}
\includegraphics[height=2.5in,angle=-90]{papers/mittaz/model_scatter.cps}
\caption{The left panel and right panels show the distribution of OVII column
  density as a function of the projected emission weighted temperature for both
  the small and large simulations respectively, In both plots the horizontal
  dotted line shows the observed lower limit to the OVII column density derived
  from XMM-Newton observations of the CSE.  The vertical line shows the 2 keV
  point above which we have assumed that that particular location will
  correspond to a cluster.  A CSE observation will correspond to point to the
  right and above the two lines but in both models there are far fewer cases
  (0\% and 1.26\% for the small and large model respectively) than would
  satisfy the observations}\label{m-ovii}
\end{figure}

The first possibility, that of large scale soft emission, has already been
demonstrated as being inconsistent with the models since the density of the
WHIM and hence the emissivity is too low.  We investigated the second argument
using both the small and large scale simulations of Cen (scale 25 and 100 Mpc
respectively) and have calculated the OVII column densities projected along the
line of sight.  In both cases we then compared the emission weighted
temperature with the OVII column density to see where and how often the OVII is
consistent with the strength of the line emission reported by Kaastra et al.
The two models are shown in figure~\ref{m-ovii} and indicate that at locations
where the projected emission weighted temperature is greater than 2 keV (which
corresponds to locations where a cluster would exist) very few cells have the
required OVII column.  Indeed, for the smaller of the two models (the model
with the higher resolution) there were no areas with an OVII column $> 4 \times
10^{16}$ cm$^{-2}$ (the density required by Kaastra et al. 2003) and a
temperature above 2 keV.  For the larger, low resolution model 45 cells out of
a total of 3568 consistent (or 1.26\%) with a cluster had an OVII column large
enough.  While this may indicate that there is some relationship between a CSE
the density of structures in the model, 1.26\% is a very small fraction
compared to the size and scale of the observed soft excess seen in real
observations.  From the point of view of absorption line studies, these plots
also show that in the vicinity of clusters we would expect an OVI absorbing
column of order $10^{14}$cm$^{-2}$ rather than the value predicted by Kaastra
et al. ($4 \times 10^{16}$cm$^{-2}$).

The only remaining option is that the XMM-Newton observations observe regions
where there are a large number of small overdense, group like structures.
However, in the current simulations a large number of such regions do not
exist.  We are then left with the following options.  There could still be
problems with the reality of the OVII detection by XMM-Newton and the excess is
non-thermal.  This, however, cannot alleviate the difficulty posed by the
\ROSAT Coma results of Bonamente et al. (2003).  Therefore, there still may be
something missing in the simulations, such as insufficient resolution.  What
seems clear is that a thermal interpretation of the CSE calls for more
material, by a factor of at least 30, than is currently included in the models.
If this is true, then the implication to the final overall budget for baryons
is daunting






\part[New instrumentation and the future]{New instrumentation and the future}





\articletitle{Observing the Warm-Hot Intergalactic Medium with XEUS}

\author{Jelle S. Kaastra\altaffilmark{1}
and Frits B.S. Paerels\altaffilmark{2}}

\altaffiltext{1}{SRON National Institute for Space Research, Utrecht, The Netherlands}
\altaffiltext{2}{Columbia University, New York, USA}

\begin{abstract}
We discuss the potential for XEUS for observing the Warm-Hot Intergalactic
Medium. After a short description of the mission we demonstrate the power of
XEUS for detecting the WHIM both in emission and absorption.
\end{abstract}


\section{Introduction}

Soft excess X-ray emission in clusters of galaxies and related phenomena is the
topic of this conference.  In several contributions the role of non-thermal
emission in clusters has been discussed.  Here we focus upon the emission and
absorption properties of the Warm-Hot Intergalactic Medium (WHIM) as it can be
observed with XEUS.

XEUS is a project currently under study by ESA and Japan.  A detailed account
of several proposed aspects of the mission, both instrumental and scientific,
can be found in the proceedings of a XEUS workshop (Hasinger et al.  2003, see
http://wave.xray.mpe.mpg.de/conferences/xeus-workshop for an electronic
version).  A good overview of the potential for XEUS for studying the WHIM is
given by Barcons (2003) and Paerels et al.  (2003).  In the present paper we
present spectral simulations in order to demonstrate how XEUS can contribute to
this field of research.

\section{The XEUS mission}

XEUS is a mission concept that is currently being studied by ESA and Japan.  It
can be viewed as a successor of XMM-Newton but with an order of magnitude
better performance in several aspects.  At the heart of the mission is a large
X-ray telescope with a focal length of 50 m.  Due to this length, the
instruments are distributed over two satellites, one containing the mirror
module and the other the detector module.  Its limiting sensitivity is 200
times deeper than XMM-Newton.  Following the launch of the initial spacecraft
configuration, an observation period of 4--6 years is planned.  After this
period, the satellites will have a rendez-vous with the International Space
Station for refurbishment and addition of extra mirror area.  Also the detector
spacecraft can be replaced at that time.  The satellite is then ready to
observe for another period in this enhanced configuration.

\begin{figure}[!thb]
\resizebox{\hsize}{!}{\includegraphics[angle=-90]{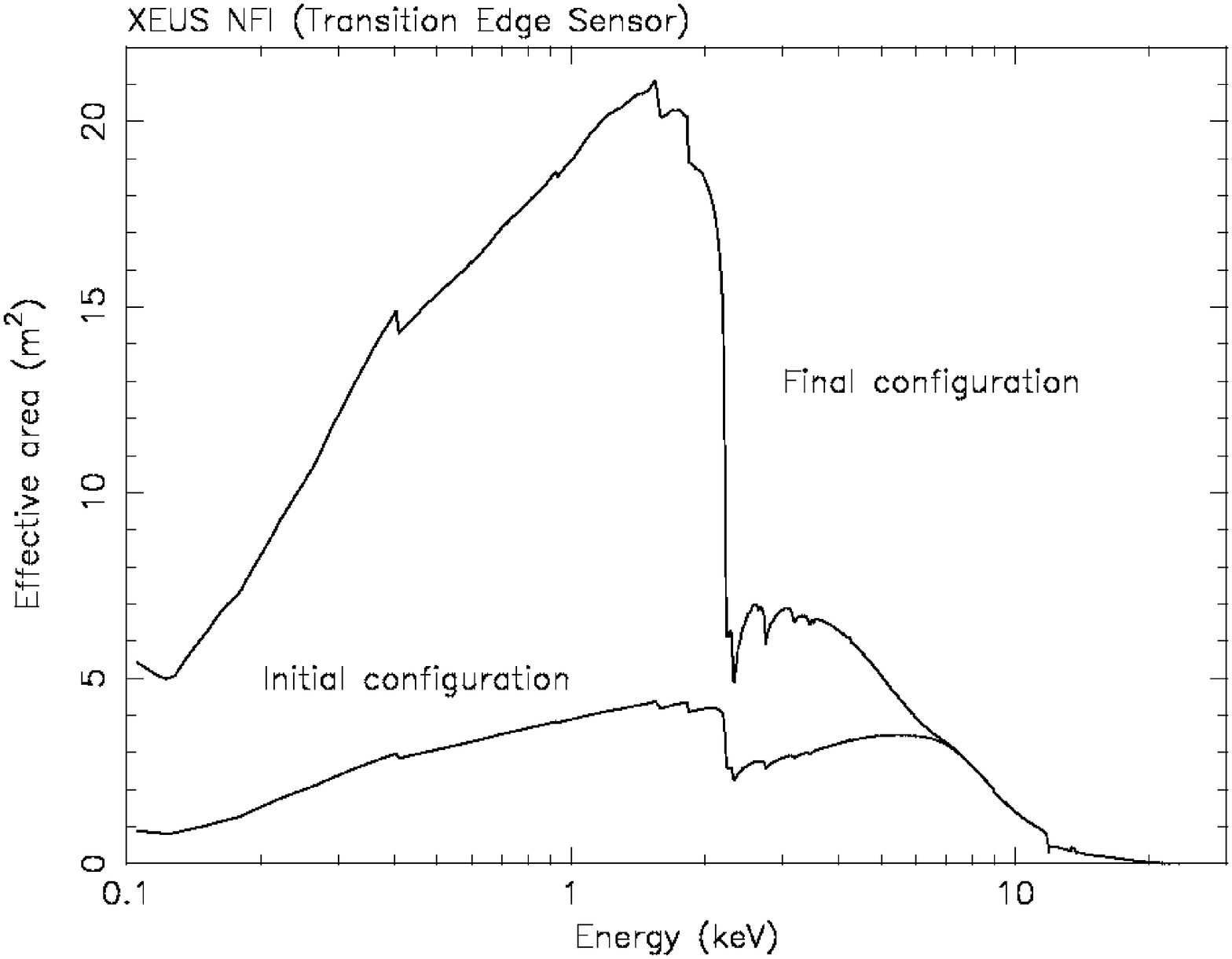}}
\caption{Effective area of XEUS (including mirror effective
area, detector quantum efficiency and filter transmission, both for the
initial (lower curve) and final (upper curve) configuration. The calculation
is made for a Transition Edge Sensor.}
\label{k2-fig:area}
\end{figure}

The collecting area of the XEUS mirrors is very large, at 1~keV the area is
6~m$^2$ for the initial and 30~m$^2$ for the final configuration.  The total
effective area is shown in Fig.~1 for both configurations.  Even in the initial
configuration, the effective area is 20 times larger than the combined
effective area of the three XMM-Newton EPIC camera's.

The imaging resolution is 2 arcsec Half Energy Width, with a field of view of 1
arcmin for the narrow field instruments (NFI) and 5 arcmin for the wide field
instruments (WFI).  For the narrow field instruments two types of detectors are
being developed: Superconducting Tunnel Junctions (STJs) and Transition Edge
Sensors (TESs).  For the wide field instruments large area, small pixel size
CCDs are envisaged.

The spectral resolution of these instruments varies between 1--5~eV for the NFI
(between 1--8~keV) to 50--100~eV for the WFI over the same energy range.
Currently studies are made in enhancing the sensitivity at higher energies even
further.  The nominal energy range is 0.05--30~keV.

In this paper we make simulations of emission and absorption spectra with XEUS.
In all simulations, we have taken the final configuration with a TES detector,
adopting a spectral resolution of 2~eV below 1~keV.  The integration time for
all simulations is taken to be 40\,000~s.

\section{The X-ray background}

\begin{figure}[!thb]
\resizebox{\hsize}{!}{\includegraphics[angle=-90]{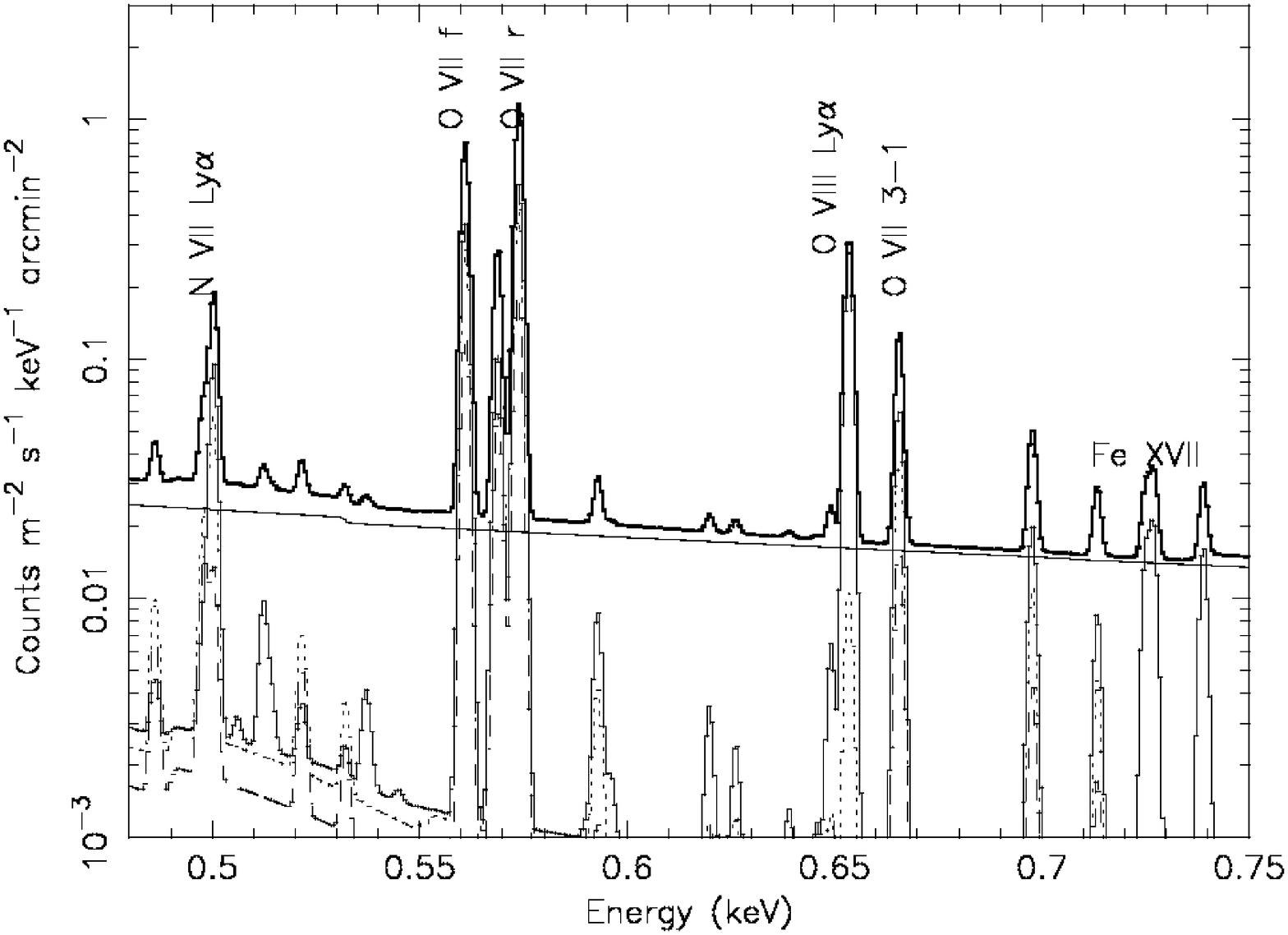}}
\caption{Portion of the predicted background spectrum for XEUS using the
background model of Kuntz \& Snowden (2000). Upper thick curve: total
background; solid thin curve: extragalactic power law component; dotted line:
Local Hot Bubble; dashed line: soft distant component; thin solid line with
spectral lines: hard distant component.}
\label{k2-fig:back}
\end{figure}

In all spectra, but in particular in emission spectra of extended sources (due
to a larger extraction radius), the Galactic foreground emission and
extragalactic background emission must be taken into account in the effective
"background" for the spectrum of the source that is studied.  In Fig.~2 we show
a portion of this simulated background spectrum, which represents a typical
high Galactic latitude pointing.  The model is based upon Kuntz \& Snowden's
(2000) decomposition of the soft X-ray background as estimated from Rosat PSPC
data.  These authors decompose the background into four components: the Local
Hot Bubble with a temperature of 0.11~keV; a soft distant component with a
temperature of 0.09~keV; a hard distant component with a temperature of
0.18~keV; and the extragalactic power law with photon index 1.46.  The last
three components are affected by Galactic absorption.  From Fig.~2 we see that
while in the continuum most of the background around the important oxygen lines
is caused by the extragalactic power law component, line emission from the
other three components can be a factor of 10 stronger than the continuum at
specific energies.  In particular the hard distant component contributes a lot
of flux in the background oxygen lines.  At energies below 0.35~keV, most of
the background flux is caused by the soft distant component and the Local Hot
Bubble, the latter one dominating all other components below 0.2~keV.  At these
energies (below 0.2~keV) the spectral resolution of XEUS becomes poorer
($E/\Delta E < 100$) as compared to for example the LETGS of Chandra, but the
effective area is orders of magnitude larger and detailed spectral analysis of
the X-ray background (and any X-ray source) is still possible using the
relative strengths of several line blends in a global fitting approach.

\begin{figure}[!thb]
\resizebox{0.9\hsize}{!}{\includegraphics[angle=-90]{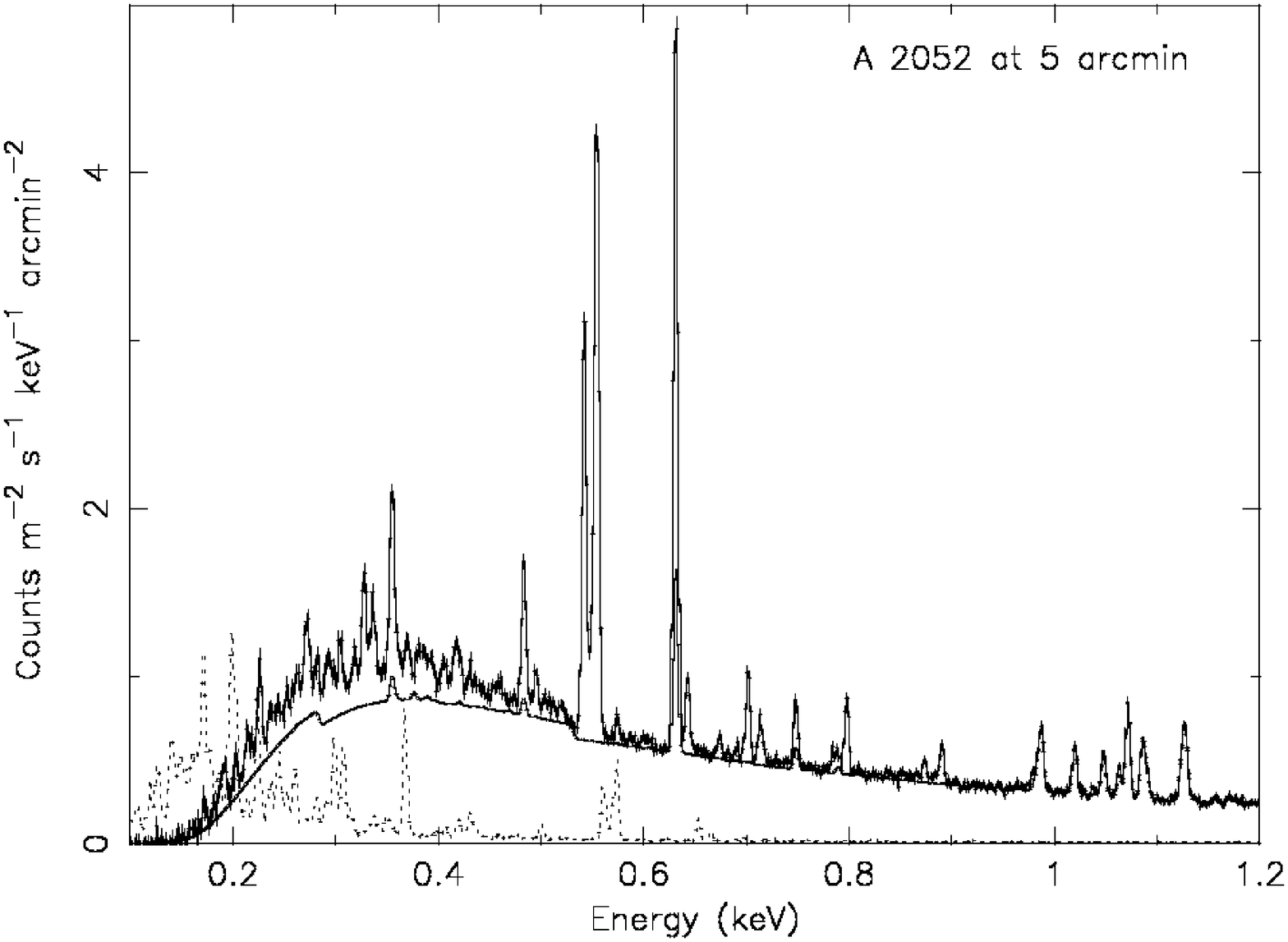}}

\caption{Simulated XEUS spectrum of the cluster of galaxies A~2052 (redshift
  0.036) at 5 arcmin off-axis.  The dashed line shows the subtracted background
  emission while the lower solid curve gives the contribution from the hot
  intracluster medium. The upper solid curve is the total (hot plus warm)
  background subtracted spectrum.}

\label{k2-fig:a2052}
\end{figure}

\section{Emission spectra from the WHIM}

With XMM-Newton the presence of thermal soft X-ray emission from the WHIM has
been deduced from the presence of redshifted O~VII line emission from a plasma
with a temperature of 0.2~keV (Kaastra et al.  2003a; see also these
proceedings: Kaastra et al.  2003b).  Although the EPIC camera's of XMM-Newton
have a much better spectral resolution than the PSPC detector used earlier in
studying soft excess emission from clusters of galaxies, the resolution (60~eV)
was still insufficient to resolve the O~VII triplet into the resonance,
intercombination and forbidden lines.  In fact, the spectral resolution
$E/\Delta E \sim 10$ at the oxygen lines only allowed a clear discrimination
between a Galactic and extragalactic origin of the emission line by measuring
its redshift, which was possible thanks to the sensitivity of XMM-Newton.

With XEUS and its high spectral resolution (2~eV) it is much easier to detect
line emission and moreover, we can resolve the O~VII triplet.  This allows us
to investigate the physical state of the warm gas in more detail, for example
we can investigate the relative contribution from collisional ionization and
photoionization by looking to the line ratio's of the triplet.

In Fig.~3 we show a simulated spectrum of the cluster A~2052 at 5 arcmin from
the center of the cluster.  The parameters for this simulation were taken from
the best fit of Kaastra et al.  (2003a).  It is evident that it is very easy to
detect the WHIM in or near such bright and nearby ($z=0.036$) clusters, even
taking into account that the simulation was done for a square box of $1\times
1$ arcmin$^2$.  Of course, the small field of view of the NFI makes it
impossible to make detailed maps of such large clusters (radii of the order of
10--20 arcmin), but the enormous amount of information that is present in
detailed investigations of selected pencil beams through the cluster compensate
for this loss of field of view.  Interestingly, with the XEUS sensitivity we do
not only detect the O~VII triplet from the WHIM, but are also sensitive to
O~VIII, Fe~XVII as well as C~VI and N~VII line emission.  In this way we can
investigate the ionization structure as well as the chemical enrichment of the
WHIM.

\begin{figure}[!thb]
\resizebox{0.9\hsize}{!}{\includegraphics[angle=-90]{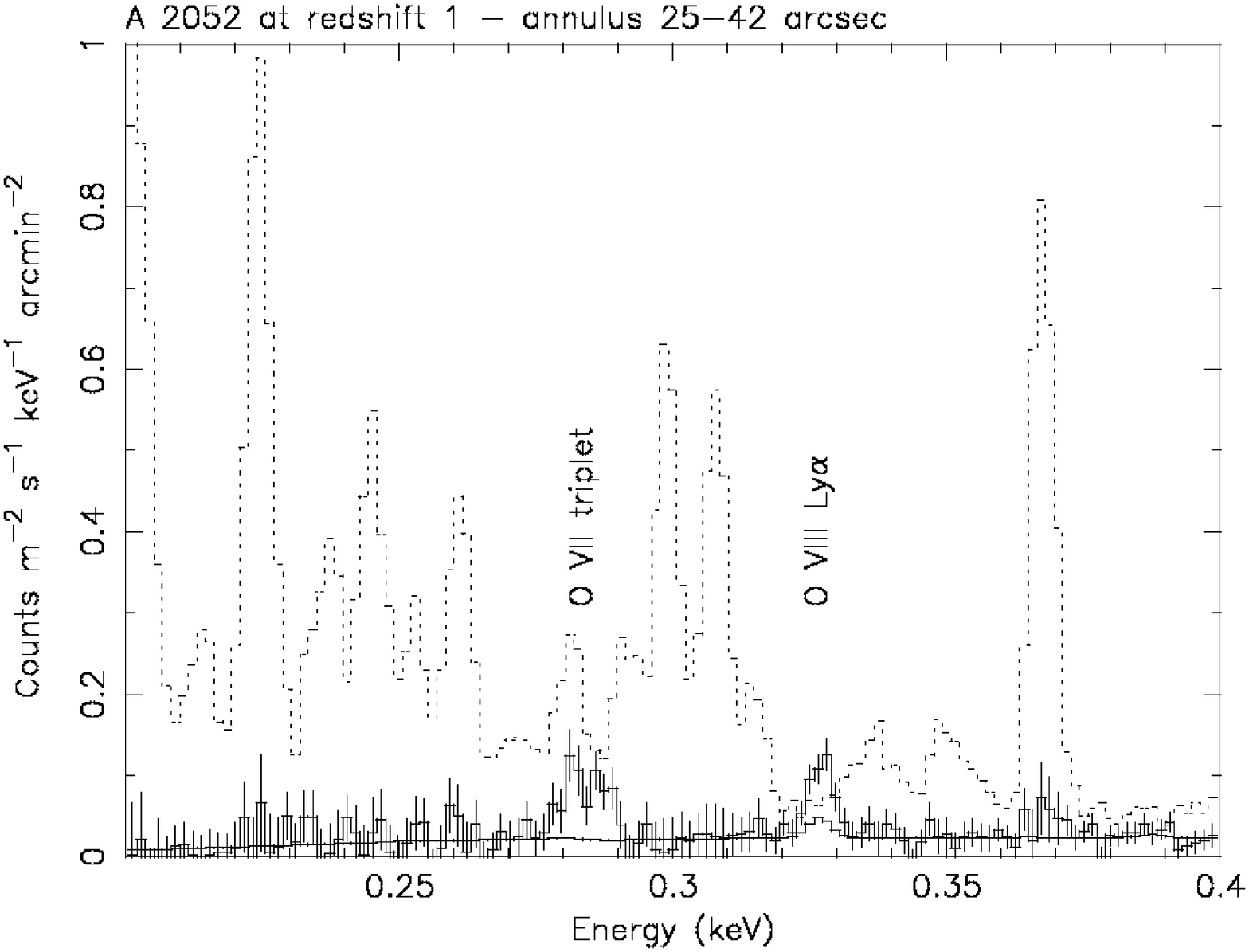}}
\caption{Simulated XEUS spectrum of the cluster of galaxies A~2052,
but now put at a redshift of 1. The dashed line shows the subtracted background
emission while the lower solid curve gives the contribution from the
hot intracluster medium.}
\label{k2-fig:az1}
\end{figure}

How far out can we do these kinds of studies?  In order to simulate this, we
have put A~2052 at a redshift of 1.  The spectrum is now extracted over the
entire annulus between 25--42 arcsec, which is smaller than the $1\times 1$
arcmin$^2$ box of Fig.~3 but covers a relatively larger fraction of the cluster
flux, due to the higher redshift.  Note that the oxygen lines are now
redshifted to $\sim$0.3~keV where the relative spectral resolution of XEUS is
somewhat poorer; and due to the $1/r^2$ dependence of the flux, the cluster
spectrum is now weaker than the subtracted background.  Due to these effects,
it is harder to disentangle the spectral components in the cluster spectrum.
Detailed spectral fitting shows that we can detect the WHIM in emission with
good confidence up to $z\sim 0.7$, and $z=1$ is the real limit.

\section{Observing the WHIM in absorption}

\begin{figure}[!thb]
\resizebox{0.95\hsize}{!}{\includegraphics[angle=-90]{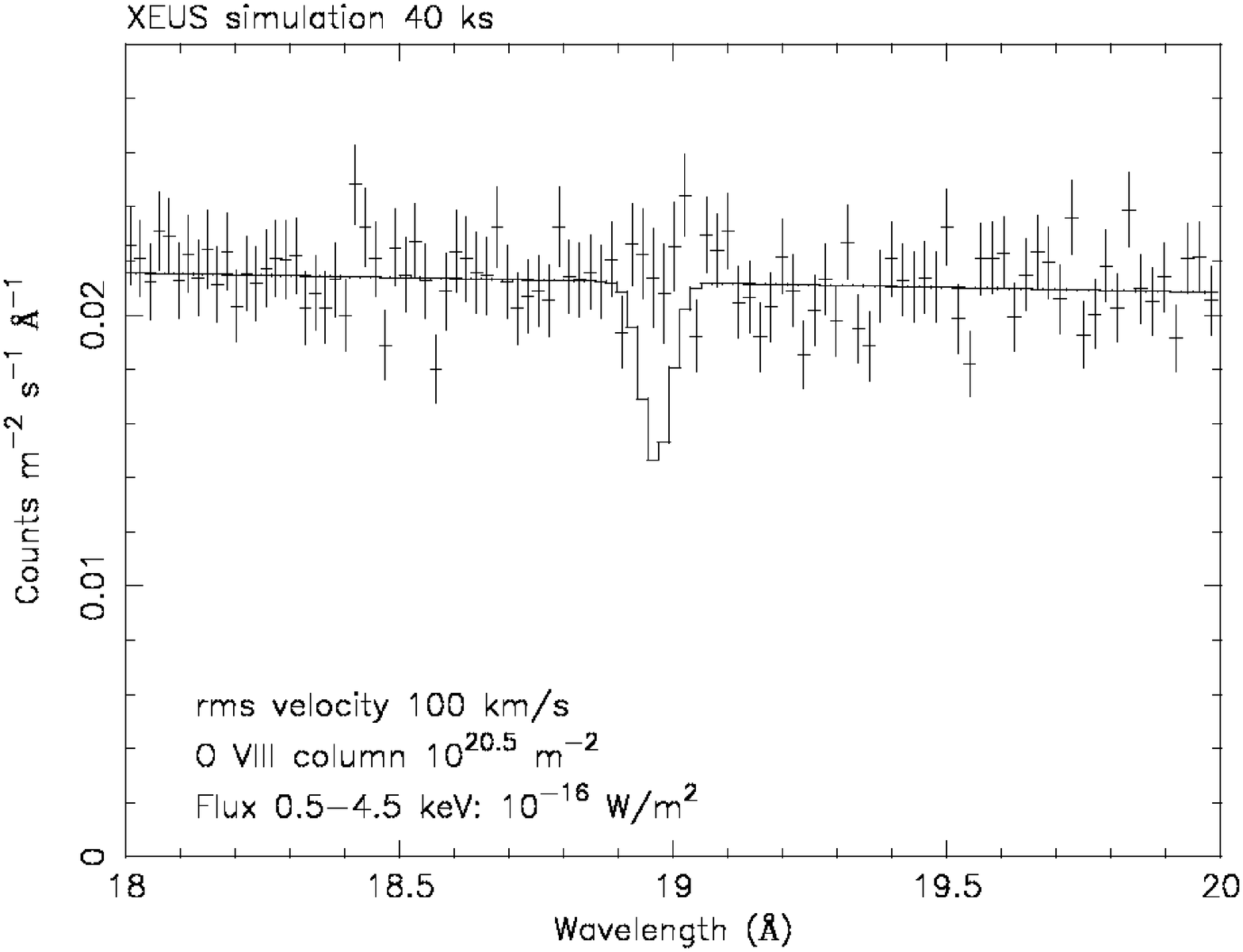}}
\caption{Simulated X-ray spectrum for a pure power law  with a
0.5--4.5~keV flux of $10^{-16}$~W\,m$^{-2}$ and Galactic absorption.  The solid
line indicates the model spectrum, but now absorbed by a slab of pure O~VIII
with a column density of $10^{20.5}$~m$^{-2}$.}
\label{k2-fig:o16}
\end{figure}

In the previous section we focussed upon observing the WHIM in emission.  Here
we study the potential for XEUS for observing absorption lines.  Simulations of
the WHIM (for example Fang et al.  2002) show that at each random line of sight
one expects to observe tens of absorbers with $z<3$ and an O~VIII column
density larger than $10^{20}$~m$^{-2}$.  Also a few absorbers with even higher
column density ($>10^{20.5}$~m$^{-2}$) are expected to occur along such lines
of sight.  So there are sufficient absorbers present.  The question is if we
have a sufficient number of bright background sources.  Taking a flux limit of
$10^{-16}$~W\,m$^{-2}$ in the 0.5--4.5~keV band, we expect about 10 sources per
square degree with this brightness.  Hence this is a very common flux level and
at almost any region of the sky that is of interest, XEUS will have at least
one such source relatively nearby to be used as a background lamp.  These
sources have on average a redshift of 0.5--1.0, and according to the numerical
simulations mentioned above they should therefore have several absorbers in the
line of sight.  We made a simulation of such a source with a photon index of 2
and absorbed by a typical Galactic column density ($2\times 10^{24}$~m$^{-2}$).
The result is shown in Fig.~5.  It is evident that XEUS can easily detect
O~VIII columns of $10^{20.5}$~m$^{-2}$ in these relatively weak sources in
reasonable integration times.  For smaller column densities
($10^{20}$~m$^{-2}$) larger exposure times than the current 40~ks are needed.

\begin{figure}[!thb]
\resizebox{0.95\hsize}{!}{\includegraphics[angle=-90]{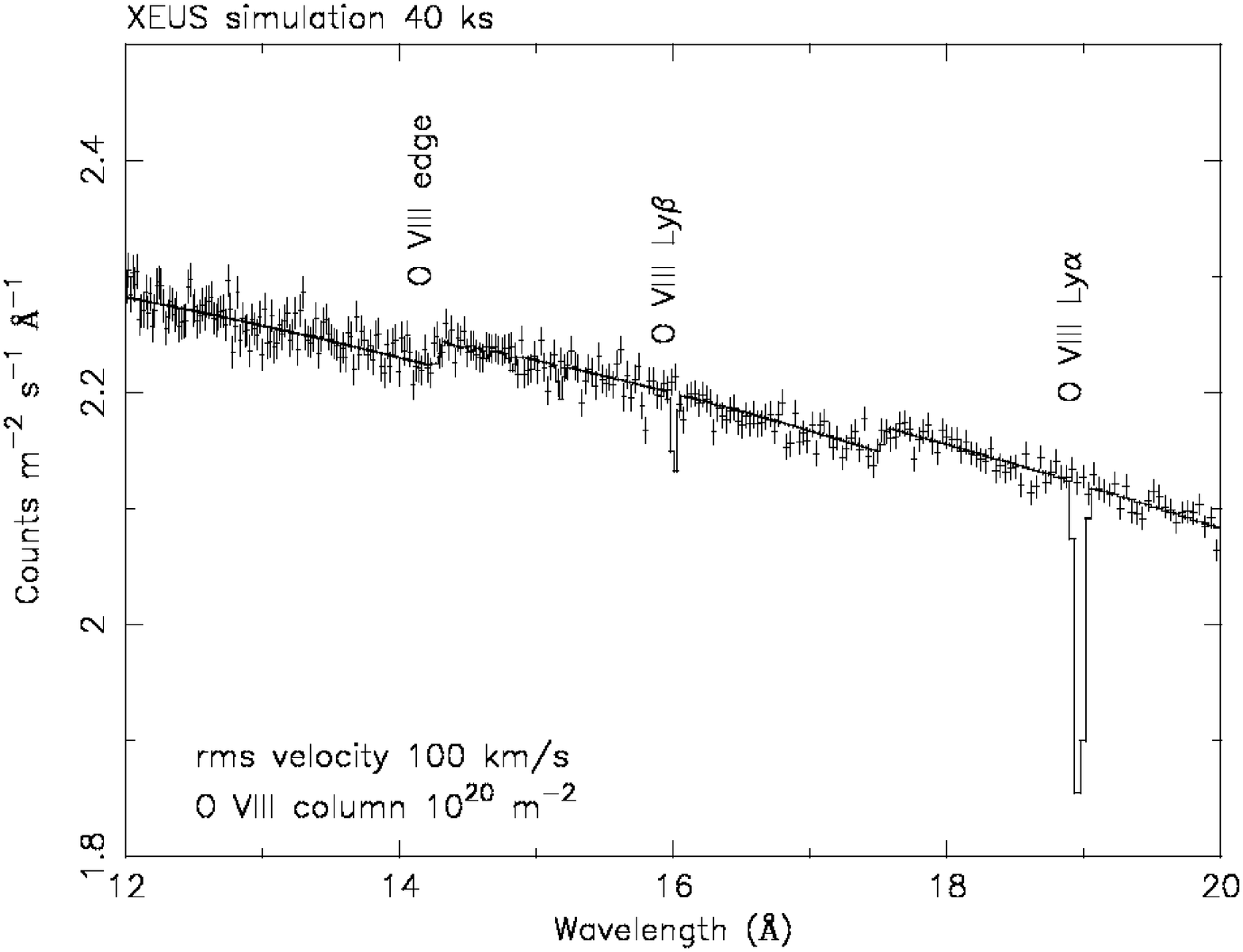}}
\caption{Simulated X-ray spectrum for a pure power law with a 0.5--4.5~keV flux
of $10^{-14}$~W\,m$^{-2}$ and Galactic absorption.  The solid line indicates
the model spectrum, but now absorbed by an additional slab of pure O~VIII with
a column density of $10^{20}$~m$^{-2}$.}
\label{k2-fig:o14}
\end{figure}

In brighter sources XEUS is even more sensitive.  Fig.~6 shows a similar
simulation as Fig.~5, but now for a source that is 100 times brighter.  There
are tens of these objects on the sky to study.  In these objects, O~VIII column
densities of $10^{20}$~m$^{-2}$ are easily detected and in fact the detection
limit is at around $10^{19}$~m$^{-2}$ in 40~ks.  Hence, for these lines of
sight it is possible to obtain detailed spectral information on the physical
parameters of all relevant absorption systems.  This is evident for example
from the fact that we also see higher order lines such as Ly$\beta$.

\section{Conclusions}

XEUS is the most sensitive instrument proposed thus far that can detect the
WHIM, both in emission and absorption.  Due to the fact that the emission
scales with the square of the density, of course the densest regions are most
easy detected.  Emission studies are mostly hampered by the narrow field of
view (1 arcmin).  XEUS has a very high sensitivity for absorption lines, it can
detect columns of $10^{20}$~m$^{-2}$ of O~VIII in relatively weak sources.
With XEUS it will be possible to obtain deep snapshots of the WHIM, both in
emission and absorption.


\vspace{0.1cm}
\noindent{\bf Acknowledgements}
SRON is supported financially by NWO, the Netherlands
Organization for Scientific Research.




\end{document}